\documentclass[12pt,preprint]{aastex}
\usepackage{graphicx}
\usepackage{textcomp}
\usepackage{amssymb}
\usepackage{amsmath}
\usepackage{subfigure}
\usepackage{natbib}
\usepackage{multirow}\usepackage[section] {placeins}
\begin{document}
\pagenumbering{arabic}

\title{The Abundance of Atmospheric CO$_2$ in Ocean Exoplanets: A Novel CO$_2$ Deposition Mechanism}
\author{A. Levi}
\affil{Harvard-Smithsonian Center for Astrophysics, 60 Garden Street, Cambridge, MA 02138, USA}
\email{amitlevi.planetphys@gmail.com}
\author{D. Sasselov}
\affil{Harvard-Smithsonian Center for Astrophysics, 60 Garden Street, Cambridge, MA 02138, USA}
\and
\author{M. Podolak}
\affil{Dept. of Geosciences, Tel Aviv University, Tel Aviv, Israel 69978}

\maketitle

\section*{ABSTRACT}

We consider super-Earth sized planets which have a water mass fraction that is large enough to form an external mantle composed of high pressure water ice polymorphs and that lack a substantial H/He atmosphere.  We consider such planets in their habitable zone so that their outermost condensed mantle is a global deep liquid ocean.  For these ocean planets we investigate potential internal reservoirs of CO$_2$; the amount of CO$_2$ dissolved in the ocean for the various saturation conditions encountered, and the ocean-atmosphere exchange flux of CO$_2$. We find that in
steady state the abundance of CO$_2$ in the atmosphere has two possible states. When the wind-driven circulation is the dominant CO$_2$ exchange mechanism, an atmosphere of tens of bars of CO$_2$ results, where the exact value depends on the subtropical ocean surface temperature and the deep ocean temperature. When sea-ice formation, acting on these planets as a CO$_2$ deposition mechanism, is the dominant exchange mechanism, an atmosphere of a few bars of CO$_2$ is established.  The exact value depends on the subpolar surface temperature. Our results suggest the possibility of a negative feedback mechanism, unique to water planets, where a reduction in the subpolar temperature drives more CO$_2$ into the atmosphere to increase the greenhouse effect.

\section{INTRODUCTION}

Recent observations of exoplanets have shown that Super-Earths are common \citep{Batalha2014,Dressing2015}, and water is expected to be a major bulk constituent for many of them. Water and CO$_2$ have been found to be common both in protoplanetary disks \citep{Pontoppidan2014} and in comets in our own solar system \citep{Bockelee2004} so it is natural to assume that they will be important components in water planets as well. Geochemically, CO$_2$ in water planets has largely been treated in the framework of silicate weathering, representing direct analogies to the Earth \citep[e.g.][]{Abbot2012,Alibert2013,Wordsworth2013}. However, for a planet to be analogous to the Earth, its water mass fraction must be kept very small. Therefore the majority of water planets are probably not Earth-like, and the geochemistry of CO$_2$ needs further study. 

We consider water planets with masses similar to the Earth and lacking a substantial hydrogen atmosphere. For such bodies, if the mass fraction of water is greater than $\sim 1\%$ the pressure at the bottom of the water layer will be high enough so that high-pressure ice polymorphs will form \citep{Levi2014}. As a result there would be no direct contact between the liquid ocean and the silicate interior \citep[as in Type $1$ planets discussed by][for modelling the Kepler-$62$e,f exoplanets]{Kaltenegger2013}, and the ocean would have very low total alkalinity. This would limit the formation of bicarbonate and carbonate ions \citep{CarbonCycle}, making dissolved CO$_2$ the dominant carbon bearing molecule in the water planet's ocean.

In this paper we consider the case of a secondary atmosphere outgassing, in particular the outgassing of CO$_2$. 
In section $2$ we discuss the solubility of freely dissolved CO$_2$ in water, in the entire pressure-temperature domain expected in water planet oceans. The solubility is derived both outside the SI CO$_2$ clathrate hydrate thermodynamic stability field and when in equilibrium with this phase. 
In section $3$ we model the thermodynamic stability field for the SI clathrate hydrate of CO$_2$, for the entire parameter space relevant to water planet oceans, and compare it with the most up to date data.
In section $4$ we explore the different potential reservoirs for CO$_2$ at the ocean bottom in water planets. In section $5$ we calculate the power required to maintain an oceanic overturning circulation, and estimate the feasibility of vertical ocean mixing in water planets. In section $6$ we investigate the ocean-atmosphere flux of CO$_2$, and derive steady state values for the partial atmospheric pressure of CO$_2$. The effect of the wind-driven circulation is the subject of subsection $6.1$ and the effect of sea-ice forming at the poles is quantified in subsection $6.2$. The results are discussed in section $7$ and a summary is given in section $8$.

\section{HIGH PRESSURE CO$_2$ SOLUBILITY}

A warm water planet represents a planetary case where the outermost layer is mostly liquid water, i.e. an ocean. For a $2$M$_E$ super-Earth whose water mass fraction exceeds a few percent this ocean may have a bottom made of high pressure water ice polymorphs \citep[see table 1 in][for water-rock boundary pressures]{Levi2014}. Because the ocean will be separated from the silicate interior its alkalinity will be low. Therefore, even for a low oceanic carbon abundance the freely dissolved CO$_2$ would represent the dominant dissolved inorganic carbon species. For lower planetary water mass fractions the ocean may be shallower having a rocky bottom. In this case the ocean may have a higher alkalinity which may turn a larger fraction of the oceanic carbon abundance to carbonate and bicarbonate. For example, in Earth's ocean the latter are the dominant dissolved inorganic carbon species. In this work we concentrate on the first water planet composition case. Therefore, investigating carbon dioxide deposition in our studied planets' deep oceans requires an estimation of the solubility of CO$_2$ in water at both low and high pressures (from ocean surface pressures to approximately $1$\,GPa). As a preliminary to solving the solubility problem in the presence of CO$_2$ SI clathrate hydrate, we shall first solve for the solubility outside of the stability field of this phase.

\subsection{CO$_2$ Solubility Outside Its Clathrate Hydrate Stability Field}\label{subsec:SolubilityOutside}

The study of the H$_2$O-CO$_2$ mixture is very important for understanding Earth's geochemistry. Therefore, there has been much effort in deriving equations of state for the mixture \citep[e.g.][]{Duan2006b}. However, the parameter space occupied by Earth's crust and upper mantle spans temperatures much higher than those expected for water planet's oceans in the habitable zone, leaving our parameter space of interest largely unexplored.   
Up until very recently the highest pressure CO$_2$ solubility experiments, at temperatures more relevant to our case of study, were those of \cite{Todheide1963}, reaching pressures up to $0.35$\,GPa and a minimal isotherm of $323.15$\,K. \cite{Diamond2003} listed the experimental solubility data known up to that time in the temperature range of $271.6$\,K to $373$\,K and evaluated the level of confidence that ought be given to any one of the data sets. \cite{Diamond2003} further showed that Henry's law can accurately describe these experimental data sets in the pressure range up to $100$\,MPa. 

More recently \cite{Bollengier2013} researched water rich systems at CO$_2$ saturation conditions in the temperature range of $250$\,K to $330$\,K and pressures up to the melting pressure of water ice VI. They found that the melting temperature of water ice VI is depressed by a few degrees when in saturation with CO$_2$. Converting melt depression data into solubility requires saying something about deviations of the mixture from ideality. For the H$_2$O-CO$_2$ system an ideal solution is a good first approximation as long as the solubility does not exceed about $2$\,mol\% \citep{Diamond2003}. \cite{Bollengier2013} assumed an ideal solution and the model of \cite{Choukroun2007} and reported that their melt depression data suggests the solubility of CO$_2$ along the melt curve of ice VI is a mole fraction of only a few percent ($\approx 4$\%). This find is consistent with an earlier experiment by \cite{Qin2010} that found an upper bound of $5$\% for the solubility of CO$_2$ at $293$\,K and $1.26$\,GPa. A small depression of the melt curve of water ice VI, when in saturation with CO$_2$, is contradictory to the findings of \cite{Manakov2009} who argued for much higher melt depressions ($30-60$\,K). We refer the reader to \cite{Bollengier2013} for a consideration of this discrepancy. 

If $T^{VI}_m(P)$ is the melt curve of water ice VI for a pure water system and $\Delta T$ is the melt depression due to saturation with CO$_2$, the solubility of carbon dioxide in mole fraction, $X_{CO_2}$, may be arrived at using the relation:
\begin{equation}\label{VImeltdepression}
\Delta T\left[-S_{VI}(T^{VI}_m(P))+S_{liq}(T^{VI}_m(P))\right]=k(T^{VI}_m(P)+\Delta T)\ln\left\lbrace\gamma_{H_2O}(1-X_{CO_2})\right\rbrace
\end{equation}
where $k$ is Boltzmann's constant and $\gamma_{H_2O}$ is the activity coefficient for water in the liquid phase. Below we will discuss the activity coefficient used in this work. 
 
In the brackets on the LHS we have the entropy difference between pure water ice VI and liquid water along the melt curve. The solubility of CO$_2$ depends exponentially on this entropy of fusion. There are various values reported for the latter in the literature. Calorimetric measurements of the entropy or enthalpy of fusion of ice VI along its melt curve are scarce. One may derive the entropy of fusion from the gradient of the Clausius-Clapeyron equation in case the volume difference at the phase transition is known. Since the latter involves the difference of two numbers that are similar, the volumes of the two phases must be known to high precision. Recent results show that the enthalpy of fusion of D$_2$O differs substantially from that of H$_2$O which may explain part of the scatter in the literature \citep{Fortes2012b}.

\begin{figure}[ht]
\centering
\includegraphics[trim=0.15cm 4cm 0.2cm 3cm , scale=0.60, clip]{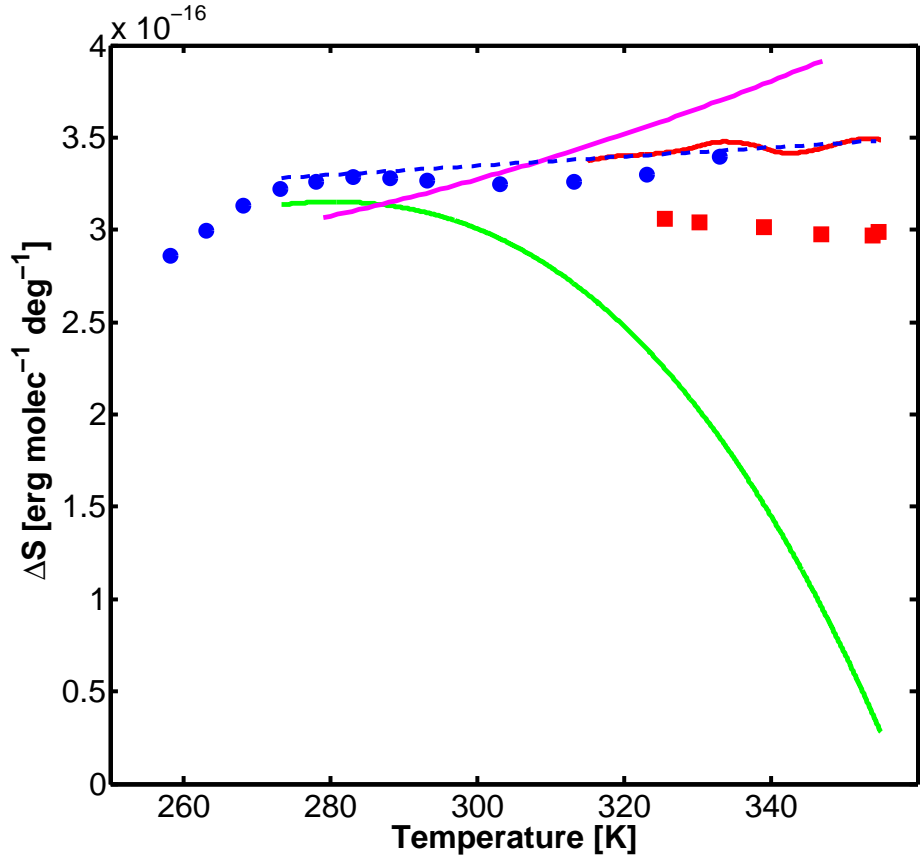}
\caption{\footnotesize{Entropy of fusion of ice VI along its melt curve. Blue circles are data points from \cite{Bridgman1912}. Red squares are data points from \cite{Bridgman1937}. Solid magenta curve is the model of \cite{Dunaeva2010}. Solid green curve is our model I and solid red curve is our model II, refer to text for model explanation. Dashed blue curve is our linear fit.}}
\label{fig:DelEntropy}
\end{figure}

In fig.\ref{fig:DelEntropy} we plot the entropy of fusion of ice VI along its melt curve. The data points from \cite{Bridgman1912} and \cite{Bridgman1937} give an entropy of fusion which is relatively constant along the melt curve of ice VI. The model of \cite{Dunaeva2010} yields the highest values for the entropy of fusion at high temperature. We further plot two models of our own. We use the melt curve equation suggested by the IAPWS for ice VI to derive its gradient. We further adopt the IAPWS equation of state for liquid water \citep{wagner02} in order to derive its volume. In model I (see solid green curve) the volume for ice VI is taken from \cite{Choukroun2007}. In model II (see solid red curve) the volume for ice VI is from the equation of state given by \cite{Bezacier2014}. Clearly model I deviates substantially from all other results. Model II coincides with the data reported in \cite{Bridgman1912} and is derived with the most up to date equation of state for ice VI. We therefore use the data from Bridgman and our model II to create a linear fit: 
\begin{equation}
S_{liq}(T^{VI}_m(P))-S_{VI}(T^{VI}_m(P)) = 2.4775\times 10^{-19}T+2.604\times 10^{-16}\quad\left[\frac{erg}{molec\times deg}\right]
\end{equation} 
where $T$ is the temperature in K.          

\begin{figure}[ht]
\centering
\includegraphics[trim=0.15cm 4cm 0.2cm 5cm , scale=0.60, clip]{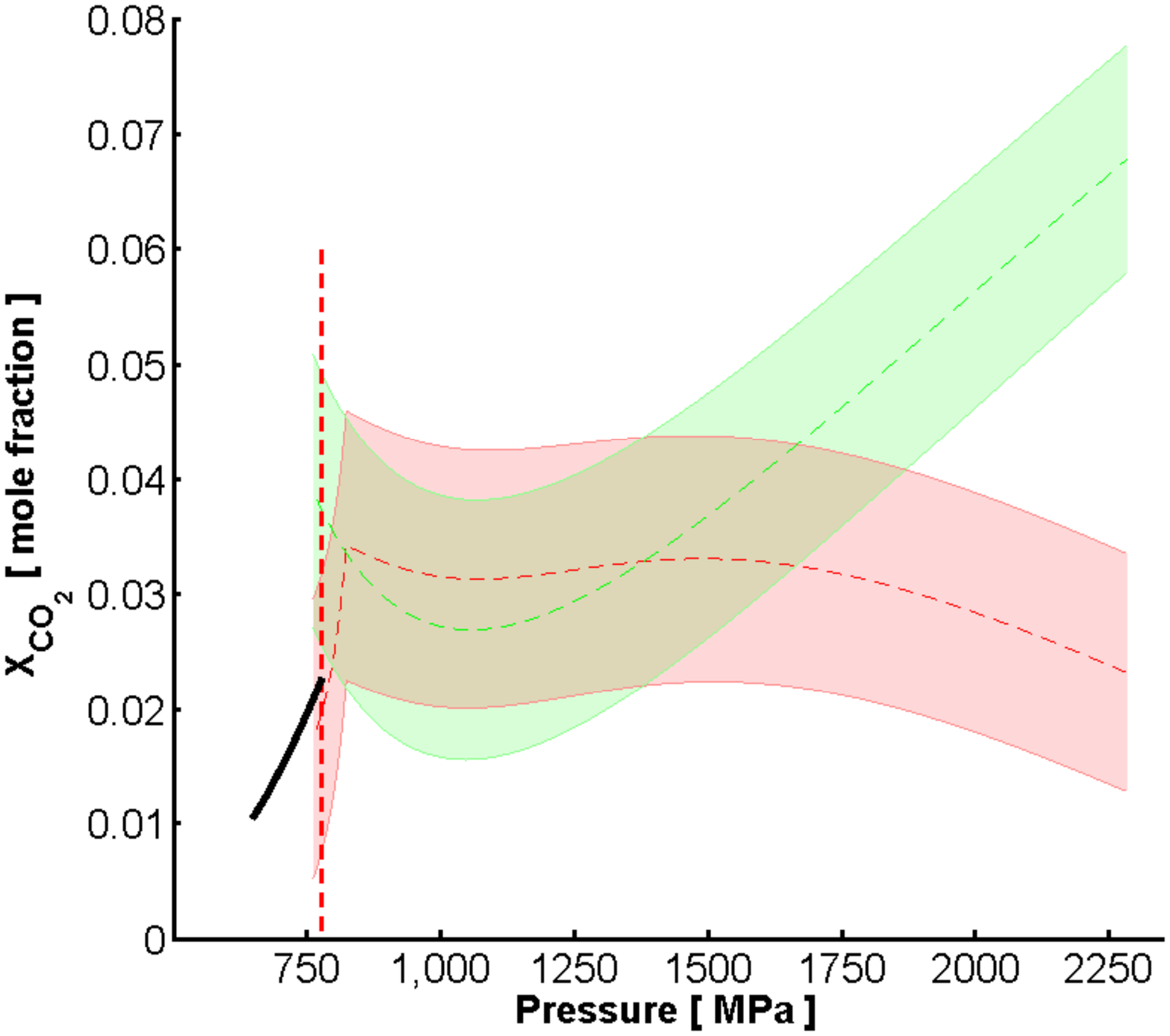}
\caption{\footnotesize{Solubility (mole fraction) of carbon dioxide in liquid water along the depressed melt curve of water ice VI. Green shaded area is the solubility assuming the binary H$_2$O-CO$_2$ is an ideal solution. The red shaded area accounts for the non-ideal behaviour of the solution. To the left of the vertical dashed red line the conditions along the ice VI melt curve coincide with the thermodynamic stability field of SI clathrates of CO$_2$. Therefore, to the left of this vertical separator these clathrates govern the solubility of CO$_2$ in the water. To the right of the vertical dashed red line the melt curve of ice VI is outside of the thermodynamic stability field of CO$_2$ SI clathrates. Thick solid black curve is our prediction for the solubility of CO$_2$ in water, as governed by the presence of SI CO$_2$ clathrates (see subsection $2.2$), for the P-T conditions along the ice VI melt curve.}}
\label{fig:SolubilityMeltDep}
\end{figure}

Using the melt depression data for ice VI from \cite{Bollengier2013} we derive the solubility of CO$_2$ in conditions along the depressed melt curve with the aid of eq.($\ref{VImeltdepression}$).
In fig.\ref{fig:SolubilityMeltDep} we present the resulting solubility for two cases. In the first case (green shaded area) the mixture H$_2$O-CO$_2$
is assumed ideal. In the second case (red shaded area) we model the non-ideal behaviour with activity coefficients from \cite{Abrams1975}. The shaded area is a result of the error in the measurement of the temperature in the experiment of \cite{Bollengier2013}. The vertical dashed red line separates the melt curve of ice VI between the segment that is inside and the segment that is outside of the CO$_2$ SI clathrate hydrate stability field. In this subsection we are solely interested in the part of the figure to the right of this vertical line.   
We note here that the much higher melt depression suggested by \cite{Manakov2009} would have resulted in a solubility along the same melt curve in the range of $20-35$\%, in case an ideal solution is assumed, which is incorrect for such a high solubility.

An interesting feature present in fig.\ref{fig:SolubilityMeltDep} is the existence of a minimum in the solubility along the melt curve of ice VI, at about $1100$\,MPa. Neither the pressure nor the temperature are constants along the melt curve of water ice VI, however, it is interesting to note here a high pressure phenomenon found for isobaric solubilities. It is common knowledge that the solubility decreases with increasing temperature. This behaviour though is pressure dependent. It is experimentally known, that outside of the clathrate stability field, at high pressures the isobaric solubility versus temperature has a minimum \citep[e.g.][]{Wiebe1940}. This phenomenon may be partly responsible for the minimum in solubility found by \cite{Bollengier2013} along the melt curve.        
       
Making extrapolations beyond the experimental data using Henry's law for the solubility is risky due to: it having several free parameters, the exponential term (i.e. Poynting correction) and mostly due to the ill constrained behaviour of the volume of infinite dilution at extreme conditions.
However, the recent experimental data of \cite{Bollengier2013} and abundant low pressure experimental data \citep[e.g.][]{Diamond2003} confine the solubility of CO$_2$ in liquid water at both the high and low pressure ends of interest for water planet oceans in the appropriate temperature range.   
In this work we can therefore use Henry's model for the solubility with relative confidence since it is used only to perform interpolations over the experimental data.

\begin{figure}[ht]
\centering
\includegraphics[trim=0.15cm 4cm 0.2cm 3cm , scale=0.60, clip]{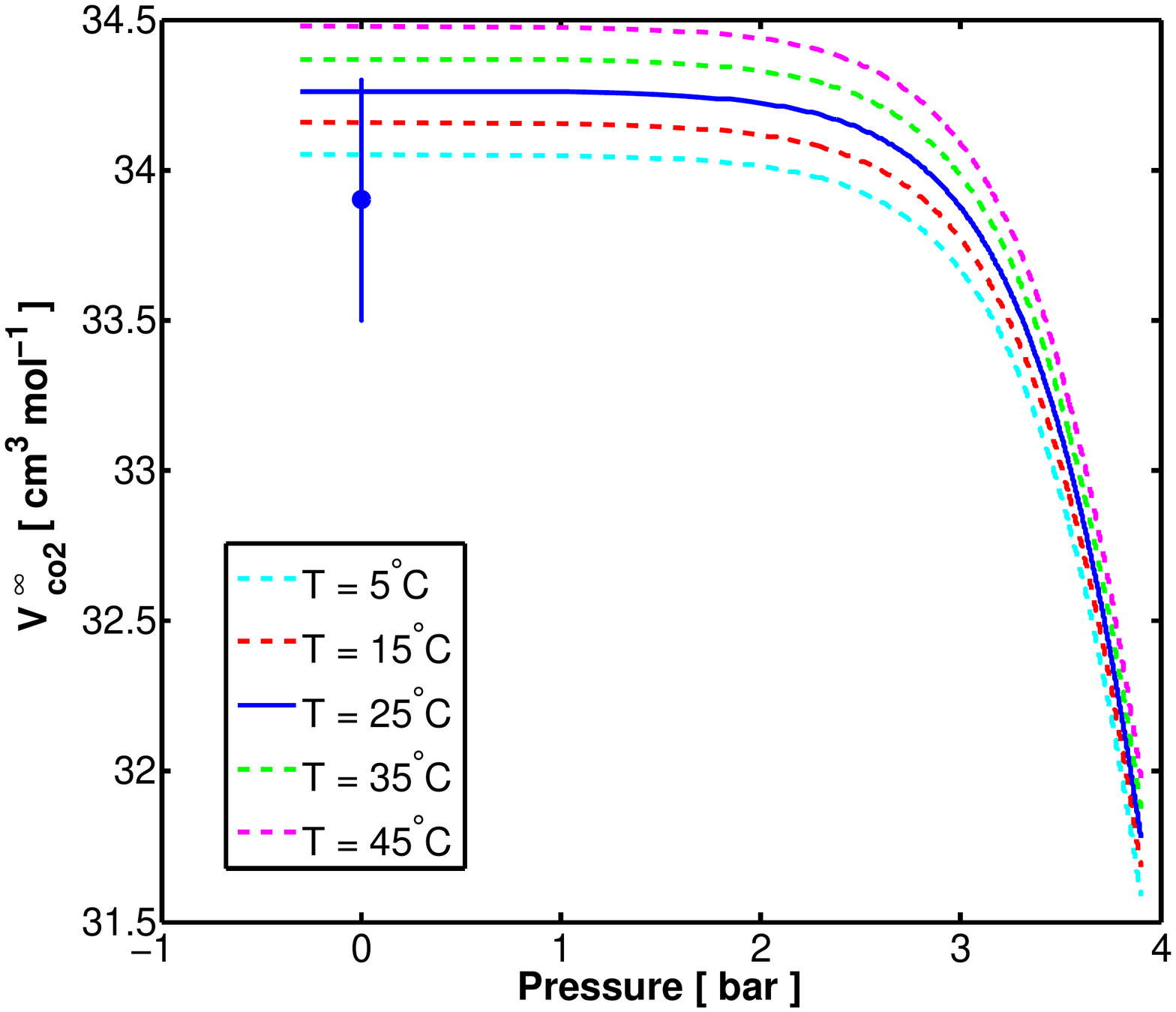}
\caption{\footnotesize{The volume of infinite dilution of CO$_2$ in water used in this work as a function of pressure for five isotherms. Data point is from \cite{Moore1982} and is for $25^\circ$C and $1$\,bar, plotted with its associated error.}}
\label{fig:VolInfDilution}
\end{figure}

The classic thermodynamic approach, using Henry's law, gives for the mole fraction of CO$_2$ ($X_{CO_2}$) in solution with water the following form \citep{Carroll1992}:
\begin{equation}\label{Henry}
X_{CO_2}=\frac{\hat{f}_{co_2}}{H\gamma_{co_2}}\exp\left\lbrace-\int^{P}_{P_{w,vap}}\frac{v^{\infty}_{co_2}dP}{kT}\right\rbrace
\end{equation}
where $\hat{f}_{co_2}$ is the fugacity of fluid carbon-dioxide in mixture, derived using the Soave-Redlich-Kwong equation of state \citep{soave}, or of solid CO$_2$ above its melt curve (see appendix \ref{subsec:AppendixA}). 
$H$ is Henry's constant derived from a fit to experimental data in the limit of a very dilute solution of CO$_2$.
We fitted all the tabulated data from \cite{Dhima1999} to the following polynomial:
\begin{equation}
H(T)=  -6.589x^5+46.67x^4-73.37x^3-145x^2+320.4x+417.4
\end{equation}
where
\begin{equation}
x\equiv\frac{T-354[K]}{80.63}
\end{equation}
This polynomial can explain $96.6$\% of the total variation in the data about the average and is good between $278$\,K and $643$\,K.  
$P_{w,vap}$ is the vapour pressure of water, here taken from the NIST Chemistry WebBook \citep[see][]{Liu1970,Bridgeman1964,Gubkov1964}. 
$P$ is the total pressure, $T$ is the temperature and $k$ is Boltzmann's constant.
The volume of infinite dilution, $v^{\infty}_{co_2}$, is theoretically not well constrained. Based on the similarity between hydration shells and clathrate cages \citep{Glew1962} we model the volume of infinite dilution with an equation of state for the SI clathrate hydrate of CO$_2$ (see eq.\ref{CellVolume}) using a volume of $34$\,cm$^3$\,mol$^{-1}$ at $273.15$\,K and $1$\,bar. This value was found by \cite{Carroll1992} to be appropriate for temperatures below $100^\circ$C. \cite{Moore1982} also reported a value of $33.9\pm 0.4$\,cm$^3$\,mol$^{-1}$. See fig.\ref{fig:VolInfDilution} for the variation of the volume of infinite dilution with pressure and temperature, as modelled in this work.

The activity coefficient for carbon dioxide in water is $\gamma_{co_2}$.
In a thorough investigation one should not approximate ideality for the solution, i.e. $\gamma_{co_2}=1$, as is assumed in the Krichevsky-Kasarnovsky equation. In order to account for the non-ideality of the CO$_2$-H$_2$O system we adopt the universal quasi-chemical (UNIQUAC) formalism for the activity coefficients \citep{Abrams1975}.
Care is taken to insure that $\gamma_{co_2}$ approaches unity in the limit $X_{CO_2}\rightarrow 0$, i.e. infinite dilution. Both the
UNIQUAC and Functional-group activity coefficients (UNIFAC) of \cite{Fredenslund1975} describe the activity coefficient dependency on temperature and composition, although its dependency on pressure is still not known. 
The activity coefficient we adopt introduces a free parameter. Activity coefficients require, among other things, an estimation for the solute-solvent energy of interaction. In the theory of \cite{Abrams1975} the interaction between molecules of types A and B in a binary mixture is modelled by taking the geometric mean of the pure components' enthalpy of sublimation:
\begin{equation}
U_{AB}=\hat{\alpha}\sqrt{U_{AA}U_{BB}}
\end{equation}
where $\hat{\alpha}$ is an adjustable free parameter. The geometric mean is often used to estimate the interaction energy between unlike molecules from data derived for homogeneous systems \citep{hirschcurbird}. To the zeroth approximation $\hat{\alpha}=1$ \citep{Abrams1975}. Empirical potential energy functions take no account of the electronic structure of matter. Therefore, they are non-transferable. Their free parameters need to be adjusted so as to fit experimental data, and as is often the case the value of these free parameters has to be changed for different P-T-x regimes. It is therefore reasonable to expect that $\hat{\alpha}$ depends on the pressure.
Up to a pressure of $825$\,MPa we fit $\hat{\alpha}$ using a combined fit to both high pressure solubility data and the dissociation curve of the CO$_2$ SI clathrate hydrate, thus maintaining consistency. In this pressure regime we find it has the following form:
\begin{equation}
\hat{\alpha}=-5.961\times 10^{-10}P^3+5.027\times 10^{-7}P^2-0.0001757P+1.071
\end{equation}
where P is pressure in MPa.
For yet higher pressures we use $\hat{\alpha}$ to somewhat improve the fit between Henry's solubility model and the experimentally inferred solubility along the melt curve of ice VI. We find that above $825$\,MPa it has the following form:
\begin{equation}
\hat{\alpha}=\frac{T^{VI}_m(P)}{288.18[K]}\left(1.271-0.00021337P[MPa]\right)
\end{equation}
We find that for the entire pressure range of interest to water planet oceans $\hat{\alpha}$ falls between $1$ and $1.1$.

\begin{figure}[ht]
  \begin{minipage}{\textwidth}
  \centering
    \includegraphics[width=.4\textwidth]{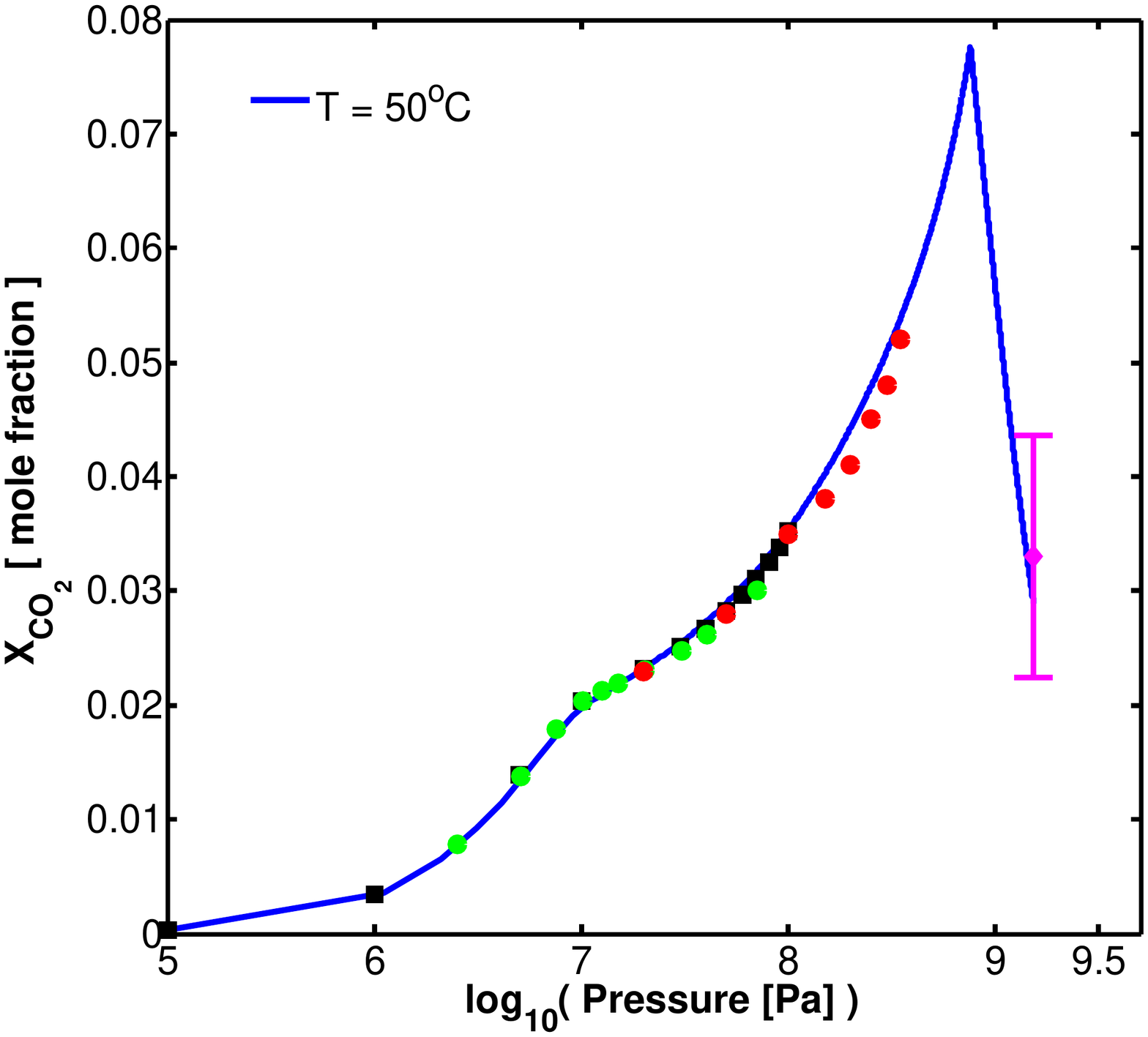}\quad
    \includegraphics[width=.4\textwidth]{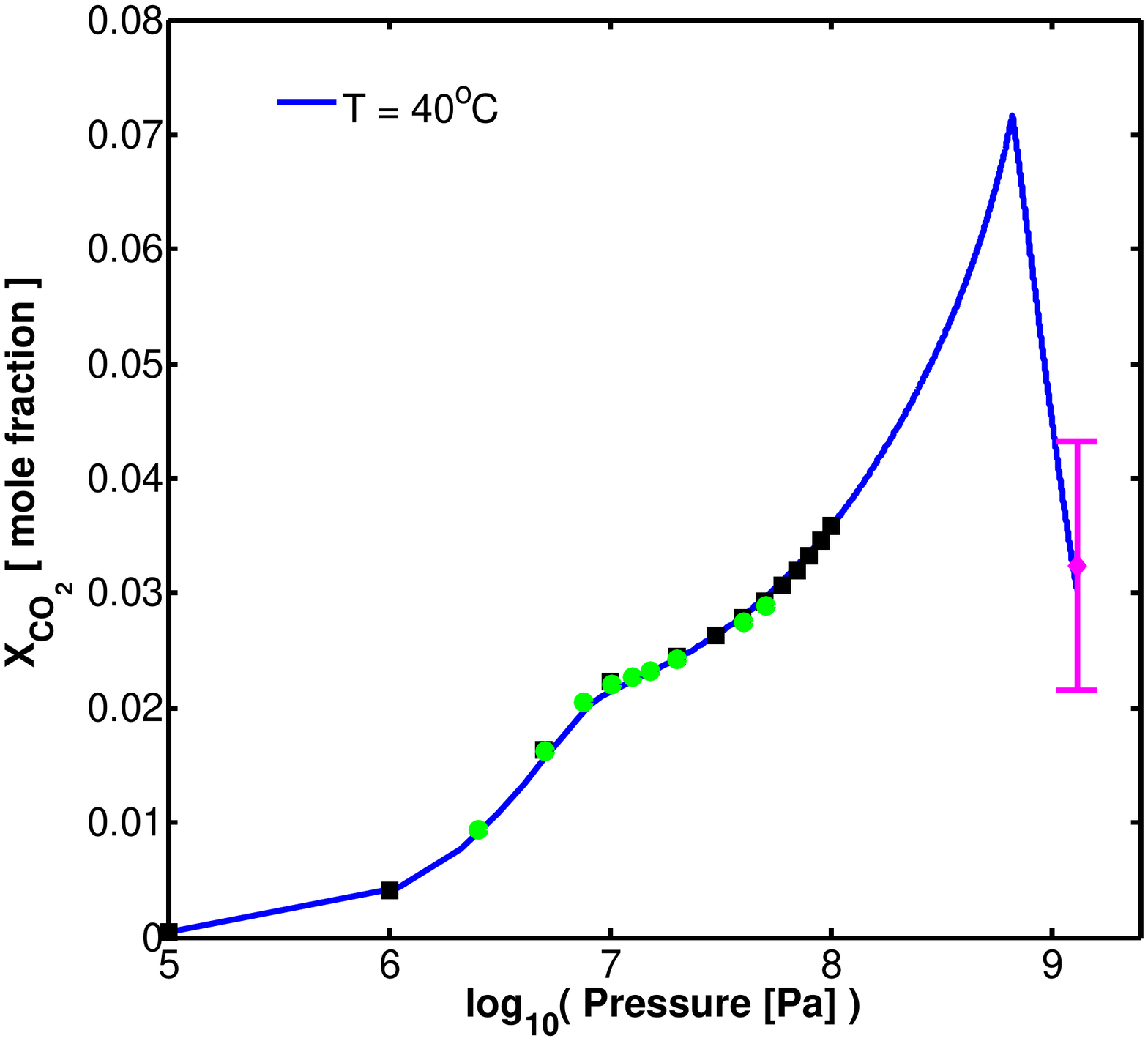}\\
    \includegraphics[width=.4\textwidth]{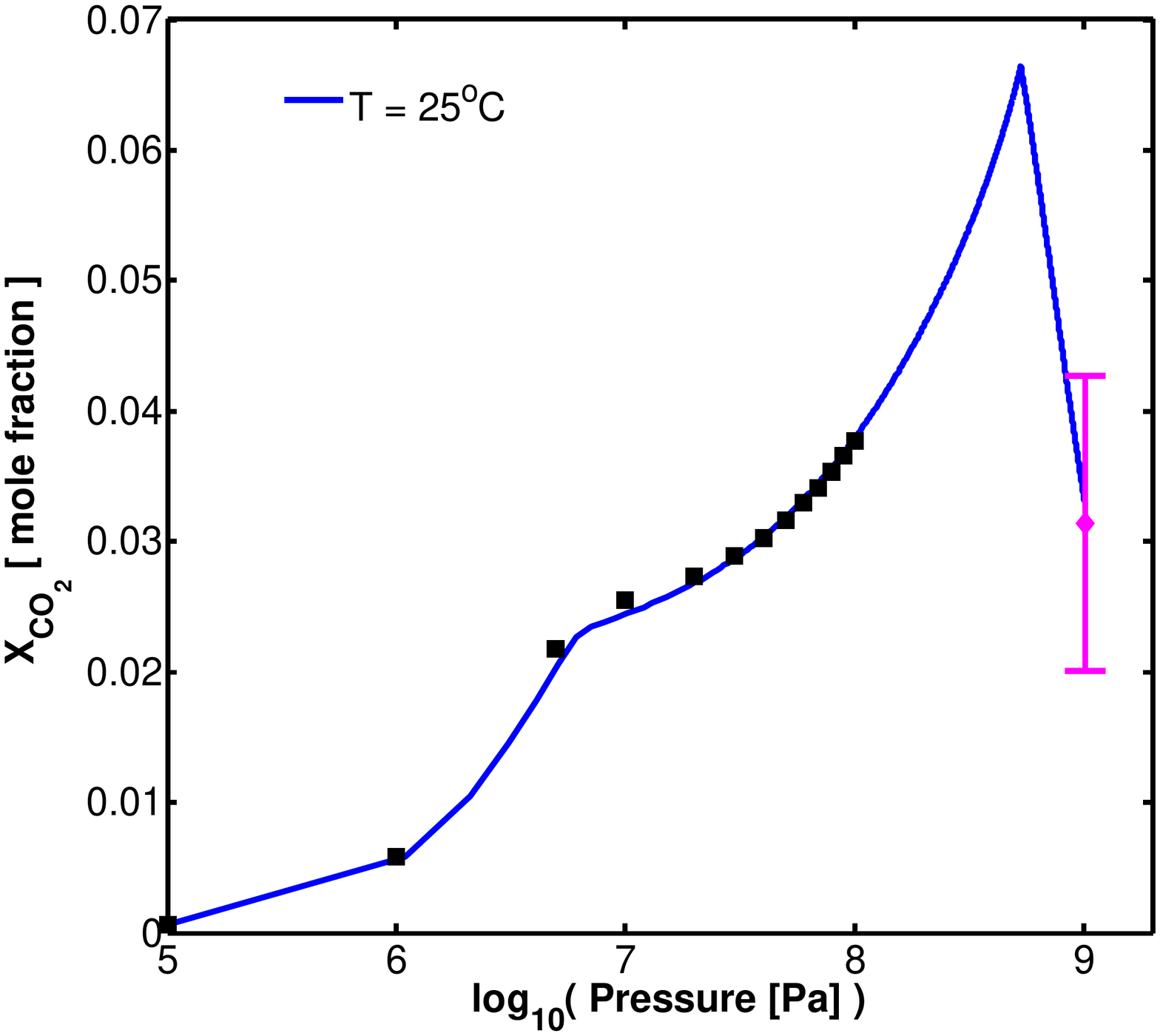}\quad
    \includegraphics[width=.4\textwidth]{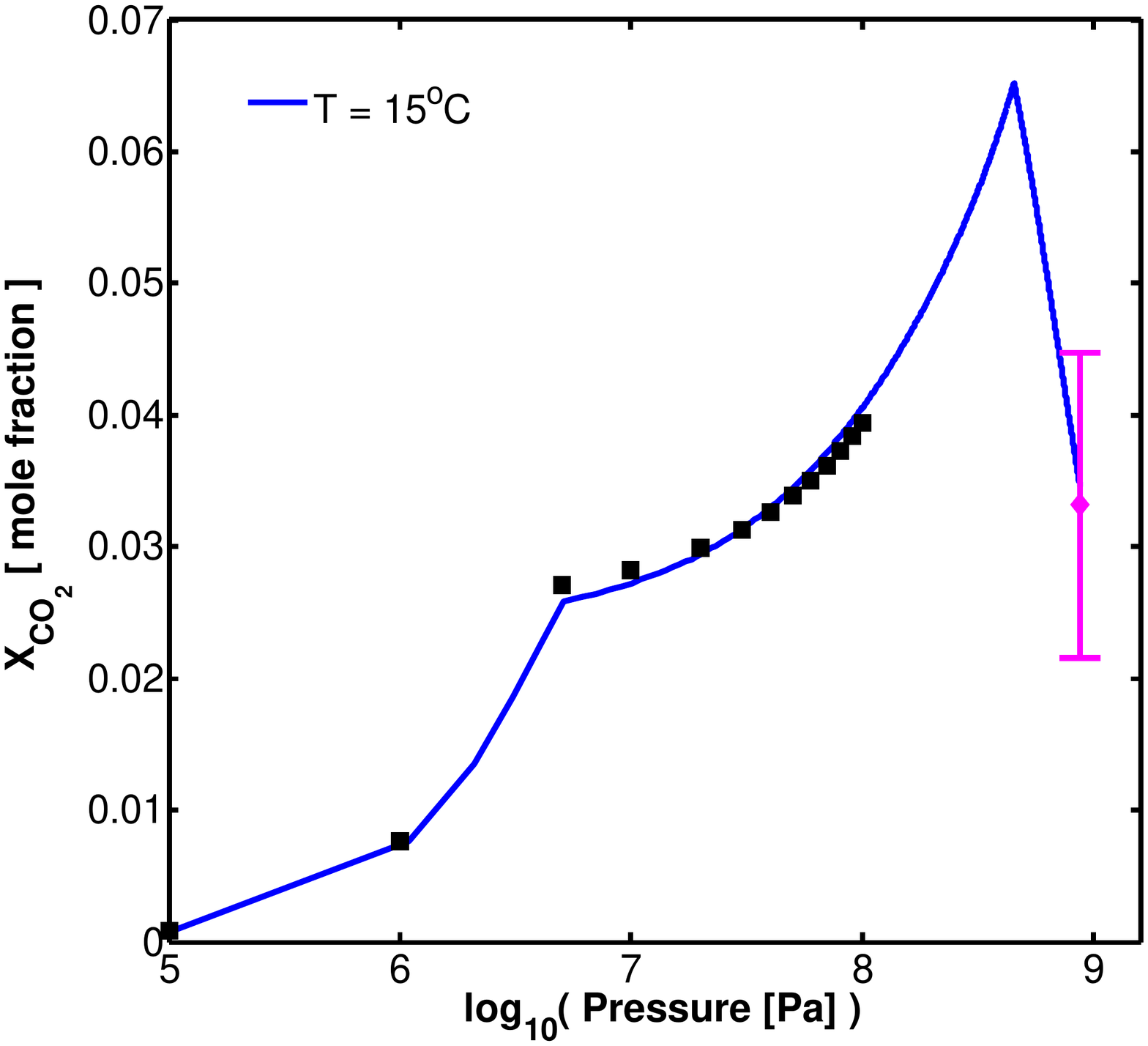}
    \caption{\footnotesize{Carbon-dioxide solubility (mole fraction) in solution with liquid water, outside the clathrate stability field, for the isotherms $50^\circ$C (upper left panel), $40^\circ$C (upper right panel), $25^\circ$C (lower left panel) and $15^\circ$C (lower right panel). Solid (blue) curve is Henry's model (see eq.$\ref{Henry}$) used in this work to make interpolations. Solid (black) squares are data points generated from the software of \cite{Diamond2003}. Circles (green) are experimental data from \cite{Wiebe1939} and \cite{Wiebe1940}. Circles (red) are experimental data from \cite{Todheide1963}. Rhombus (magenta) is the inferred solubility from the melt depression data of \cite{Bollengier2013} with its associated error. Theoretical interpolations are truncated at high pressure due to the transition to high pressure water ice poly-morphs (i.e. bottom of the ocean).}}
    \label{fig:SolubilityH50and40deg}
  \end{minipage}\\[1em]
\end{figure}

In fig.\ref{fig:SolubilityH50and40deg} we show theoretical solubility interpolations for four isotherms: $50^\circ$C, $40^\circ$C, $25^\circ$C and $15^\circ$C as a function of pressure truncated at the transition to the depressed melt curve for water ice VI. From the experimental data for the $50^\circ$C isotherm it is clear that the isothermal solubility has a maximum (the solubility measurements of \cite{Todheide1963} at lower pressures exceed the solubility estimates inferred at higher pressures from the \cite{Bollengier2013} dataset).
A maximum in solubility may be understood in the context of Eq.(\ref{Henry}) if at low pressure the influence of the fugacity is dominant increasing the solubility with pressure, while the exponential factor (i.e. the Poynting correction) gains dominance at higher pressures therefore decreasing the solubility.  
For example, the solubility of diatomic nitrogen in liquid water exhibits a maximum at around $0.27$\,GPa \citep[see][and references therein]{Prausnitz} and such is the case for several aromatic hydrocarbons, as was shown by \cite{Sawamura2007}, reaching maximum solubilities around a pressure of $0.1$\,GPa. We find for the case of the H$_2$O-CO$_2$ system that the sharp maximum is a result of the sharp change in the CO$_2$ fugacity gradient with pressure during the phase transition from fluid to solid CO$_2$. 

Our interpolations fit the experimental data for the $50^\circ$C, $40^\circ$C, $25^\circ$C and $15^\circ$C isotherms with absolute average deviations of: $3.3$\%, $2.2$\%, $2.5$\% and $3.4$\% respectively. For the $50^\circ$C isotherm the maximum solubility is $7.8$\,mol\% reached at a pressure of $0.75$\,GPa. For the $40^\circ$C isotherm the maximum solubility is $7.2$\,mol\% reached at a pressure of $0.66$\,GPa. For the $25^\circ$C isotherm the maximum solubility is $6.6$\,mol\% reached at a pressure of $0.53$\,GPa. For the $15^\circ$C isotherm the maximum solubility is $6.5$\,mol\% reached at a pressure of $0.45$\,GPa.

When pressures are high, packing efficiency becomes a consideration. As long as carbon dioxide is a fluid the volume occupied by a CO$_2$ molecule in its fluid rich phase is $\approx 65$\AA$^3$ (from liquid CO$_2$ bulk density). The volume added to the solution due to the expansion associated with the formation of hydration shells can be estimated from the partial molar volume at infinite dilution. Values from the literature suggest an added volume that could be as high as $63$\AA$^3$ per added CO$_2$ molecule \citep{Anderson2002}. 
However, the bulk density of solid CO$_2$ is much higher than that of the fluid of CO$_2$. In other words, the volume occupied by a CO$_2$ molecule in its solid ($\approx 45$\AA$^3$) is much smaller than the volume added due to the formation of a hydration shell. This means that beyond the CO$_2$ melt condition the packing of the molecules under pressure would drive a rapid reduction in the solubility.   

It is of interest to qualitatively examine the reasons the solubility of carbon dioxide behaves as described in fig.\ref{fig:SolubilityH50and40deg}, from a molecular point of view. 
In the process of dissolution of hydrophobic molecules they become encapsulated in cavities made by water molecules, i.e. hydration shells. Let us first consider a CO$_2$ molecule which is part of a CO$_2$ rich environment, for example a CO$_2$ gas or a condensed particle. This CO$_2$ molecule may reach a boundary surface with a water rich environment on the other side. With some probability an opening may form in the water hydrogen bonds forming this boundary surface, through which the CO$_2$ molecule can thermally jump, then when the hydrophobic solute molecule is in the water bulk a water hydration shell should form around it to finalize the solvation. Each such step in the process happens with some probability that is governed by an activation energy.

Activation energies for the first steps of the process of dissolution may be estimated using simulations of SI CO$_2$ clathrate. This is based on the clathrate cage-like geometry of hydration shells encapsulating dissolved hydrophobic molecules \citep{Glew1962}.
Using molecular dynamics and Monte Carlo simulations \cite{Demurov2002} found that an opening in the water rings forming clathrate cages must exist if CO$_2$ diffusion between cages via thermal hopping is to be enabled. The activation energy for forming such an opening in the hydrogen bond network of water molecules was found to be $1.31$\,eV.

\cite{Demurov2002} further suggested four possible thermal jumping routes for the carbon dioxide molecule in the SI clathrate hydrate. Two of which go through a pentagonal water ring, once from the small cage ($5^{12}$, made of twelve pentagons) to the large cage ($5^{12}6^2$, made of twelve pentagons and two hexagons), and the other way around. Two other routes represent jumps between two adjacent large cages, once via a pentagonal water ring and once via a hexagonal water ring. Each of the four different routes has a unique activation energy. \cite{Sato2000} have shown that CO$_{2}$ is more soluble than CO in water, in contradiction to the rule of thumb that "like dissolves like", due to CO$_2$ ability to form two weak hydrogen bonds with water. Therefore, during the process of dissolution the CO$_2$ molecule passes from a non-hydrogen bonded state to a weak hydrogen bonded state. \cite{Khan2003} has shown that in the SI small cage a CO$_2$ molecule forms no hydrogen bonds with its enclathrating water molecules while forming two weak hydrogen bonds with its water surroundings in the large cage. The thermal jumping of CO$_2$ from the small to large cage in the SI clathrate hydrate, with an activation energy of $0.1$\,eV, thus bears the strongest resemblance to CO$_2$ transitioning from a CO$_2$-rich phase into a water liquid cavity.     
       
A crucial part is played by the activation energy associated with the CO$_2$ encapsulation in a liquid water hydration shell. It is the sum of the work required to create the water hydration shell and the solute-solvent potential of interaction.
For a solid sphere molecular model the work necessary for creating a cavity in liquid water depends linearly on the pressure \citep{Graziano1998}. 
We note that this linear dependence on the pressure is greatly enhanced since the activation energy sets the thermal probability of the process in an exponential manner via the Boltzmann factor.
Therefore, at low pressures the work required to create a hydration shell is relatively low, making the probability of cavity creation high, and the solubility simply increases with the increasing fugacity. For high pressure on the other hand this work is large and may cause the solubility to decrease. The CO$_2$ potential of interaction with its liquid water surrounding is $-0.31$\,eV \citep{Sato2000}.

We now turn to consider the solubility of carbon dioxide in water in the presence of carbon dioxide clathrates.

\subsection{CO$_2$ Solubility Inside Its Clathrate Hydrate Stability Field}\label{subsec:SolubilityInside}

We now wish to address the solubility of CO$_2$ in liquid water while in equilibrium with CO$_2$ clathrate grains. In particular we are interested in the solubility value at the bottom of the wind driven circulation, where pressures are on the order of $100$\,bar, and at the bottom of the ocean, where the pressure is approximately $10$\,kbar. Due to the importance of this issue we will deal with it at length. 

From a thermodynamic perspective calculating solubilities is straightforward. One has to equate the chemical potentials of the different solution constituents between the various phases they occupy and which are in diffusional contact. This leads to some formulation that relates the solubility to the exponential of the constituents' partial volumes. This line of reasoning was adequately executed by various authors in order to derive the solubility of CH$_4$ and CO$_2$ while in equilibrium with their respective clathrate hydrate phase \citep[e.g.][]{Handa1990,Bergeron2010,Tsimpanogiannis2014}. Implementation of this approach however relies on the availability of an equation of state that can accurately describe the mixture. When clathrate hydrates are a part of the modelled system there is an additional complication where the equation of state of the empty clathrate hydrate (a metastable phase) is also required. Therefore, models for the solubility in equilibrium with clathrate hydrate have thus far been implemented for pressures up to a few hundred bars. 

A considerable effort was made to formulate the equation of state of the H$_2$O-CO$_2$ binary mixture. Such an equation of state for the temperature range of $273-533$\,K and up to $2$\,kbar was formulated by \cite{Duan2003} and \cite{Duan2006c}. \cite{Mao2010} developed an equation of state for the system CO$_2$-CH$_4$-C$_2$H$_6$-N$_2$-H$_2$O-NaCl for the temperature range from $273-1273$\,K and for pressures up to $5$\,kbar. This is somewhat lower than the pressures prevailing at the bottom of our studied oceans. In addition, the equation of state parameters were obtained using a regression to experimental data. However, the data used to perform this fit does not extend above $350$\,bar for our temperature range of interest, $270$-$300$\,K. These elaborate equations of state are very important. However, they have a complex and non-intuitive form and they require a large number of parameters obtained by fitting to experimental data. Therefore, it is hard to assess their performance outside of the parameter space where experimental data exists.
Below we try to overcome this problem by adopting a semi-microscopic approach to estimate the solubility.  
   
If the concentration of CO$_2$ molecules in the clathrate grain and the surrounding liquid water differs the system will tend to balance this difference. This tendency however will be restrained by the different potential well depths the CO$_2$ molecule occupies in the liquid water and in the clathrate hydrate structure. The transfer of a carbon dioxide molecule between the clathrate grain and the surrounding liquid water requires a strong enough thermal agitation to overcome the potential barrier characterizing this process. Thermal agitation or coupling to a heat bath may also be described as the action of Brownian forces. In other words, the migration of a CO$_2$ molecule between the clathrate and the liquid water may be described using Kramers theory of a Brownian particle escaping a potential well \citep{Kramers1940}. 

The flux of particles under a concentration gradient and the action of an external force is, in one dimension \citep{Kramers1940}:
\begin{equation}
j=-\frac{D}{kT}\frac{dU}{dx}n-D\frac{\partial n}{\partial x}
\end{equation}  
Here $D$ is the molecular diffusion coefficient, $k$ is Boltzmann's constant, $T$ is the temperature, $U$ is the external potential and $n$ is the number density of the Brownian particles. When a dissolved carbon dioxide molecule joins the clathrate grain it experiences an energetic change as well as the water molecules that compose its hydration shell in the liquid. Those must reorganize in order to form a clathrate cage from the liquid cavity. The Brownian particle we are considering is therefore a combination of the CO$_2$ molecule and its surrounding water molecules. We shall return to this point later.

Assuming an equilibrium between the clathrate grains and the surrounding liquid the net flux of CO$_2$ between the two environments ought vanish, $j=0$. Further integrating the last equation between the two states of entrapment for the CO$_2$ we obtain:  
\begin{equation}\label{densityratio}
\frac{n^{co_2}_{liq}}{n^{co_2}_{clath}}=e^{-\frac{1}{kT}\left(U_{liq}-U_{clath}\right)}
\end{equation}
where $n^{co_2}_{liq}$ and $n^{co_2}_{clath}$ are the number densities of CO$_2$ dissolved in the liquid water and clathrate respectively.  
$U_{liq}$ and $U_{clath}$ represent the potential wells trapping the Brownian particle in the liquid water and clathrate respectively.

The solubility of carbon dioxide in water in terms of abundance, in both the liquid water and clathrate may be written as: 
\begin{equation}\label{abundancedef}
X^{co_2}_{liq}=\frac{n^{co_2}_{liq}}{n^{co_2}_{liq}+n^{H_20}_{liq}}\qquad ; \qquad X^{co_2}_{clath}=\frac{n^{co_2}_{clath}}{n^{co_2}_{clath}+n^{H_20}_{clath}}
\end{equation}
where $n^{H_20}_{liq}$ and $n^{H_20}_{clath}$ are the number densities of water molecules in the liquid and clathrate structure respectively. With the definitions of eqs.($\ref{abundancedef}$), eq.($\ref{densityratio}$) may be written as:
\begin{equation}
\frac{n^{co_2}_{liq}}{n^{co_2}_{clath}} = \frac{n^{H_20}_{liq}}{n^{H_20}_{clath}}\frac{X^{co_2}_{liq}}{X^{co_2}_{clath}}\frac{1-X^{co_2}_{clath}}{1-X^{co_2}_{liq}} = e^{-\frac{1}{kT}\left(U_{liq}-U_{clath}\right)}
\end{equation}
After some algebraic steps the last relation yields for the solubility, in abundance, of carbon dioxide in liquid water in equilibrium with carbon dioxide clathrate grains the following form:
\begin{equation}\label{solubilityeqclathrate}
X^{co_2}_{liq}=\frac{1}{\left(\frac{n^{H_20}_{liq}}{n^{H_20}_{clath}}\right)\left(\frac{1-X^{co_2}_{clath}}{X^{co_2}_{clath}}\right)e^{\frac{U_{liq}-U_{clath}}{kT}}+1}
\end{equation}
We now turn to estimate the different variables in the last equation.

The ratio of the number densities of water molecules in the liquid and clathrate phases should be estimated using the equations of state for water at the two phases. For liquid water we use the equation of state of \cite{wagner02}. 
We estimate the number density of water molecules in the clathrate hydrate of CO$_2$ by dividing the number of water molecules in a SI clathrate unit cell, $46$, by the unit cell volume:
\begin{equation}\label{densityratioeos}
n^{H_20}_{clath}=\frac{46}{V_{cell}(T,P)}
\end{equation}
Here the CO$_2$ SI clathrate hydrate unit cell volume, as a function of temperature and pressure, is modelled as:
\begin{equation}\label{CellVolume}
V_{cell}(T,P)= V_{cell}(T_0,P_0)\left(\frac{B+\tilde{B}P}{B+\tilde{B}P_0}\right)^{-1/\tilde{B}}exp \left(\int_{T_0}^T\chi(T,P_0) dT\right)
\end{equation} 
Experimentally clathrate mechanical properties are difficult to derive. This is mainly due to the need to stay in the clathrate hydrate stability field during the experiment in addition to the difficulty in forming and then experimenting on a pure clathrate hydrate sample \citep{Fulong2012}. Therefore, molecular dynamic simulations are an important tool for calculating pure sample characteristics.
For the volume thermal expansivity we adopt the formulation from \cite{Hansen2016} obtained from diffraction experiments. 
From the molecular dynamics work of \cite{Fulong2011} we derive for the CO$_2$ SI clathrate hydrate a zero pressure bulk modulus of $B=8.5$\,GPa and for its pressure derivative a value of $\tilde{B}=5.7$. This is in agreement with the general value suggested for the bulk modulus of clathrate hydrates of $9\pm 2$\,GPa \citep{Manakov2011}. The volume thermal expansivity from \cite{Fulong2011} is only about $17$\% larger than what is reported in \cite{Hansen2016}. This may provide an estimation for the level of confidence in the calculation of \cite{Fulong2011}.
The reference temperature, $T_0$, and pressure, $P_0$, should be taken to be $271.15$\,K and $0.12$\,MPa respectively. For these reference values an edge for the cubic SI unit cell of $11.98$\AA\, is adopted \citep{Fulong2011}. 

\begin{figure}[ht]
\centering
\includegraphics[scale=0.6]{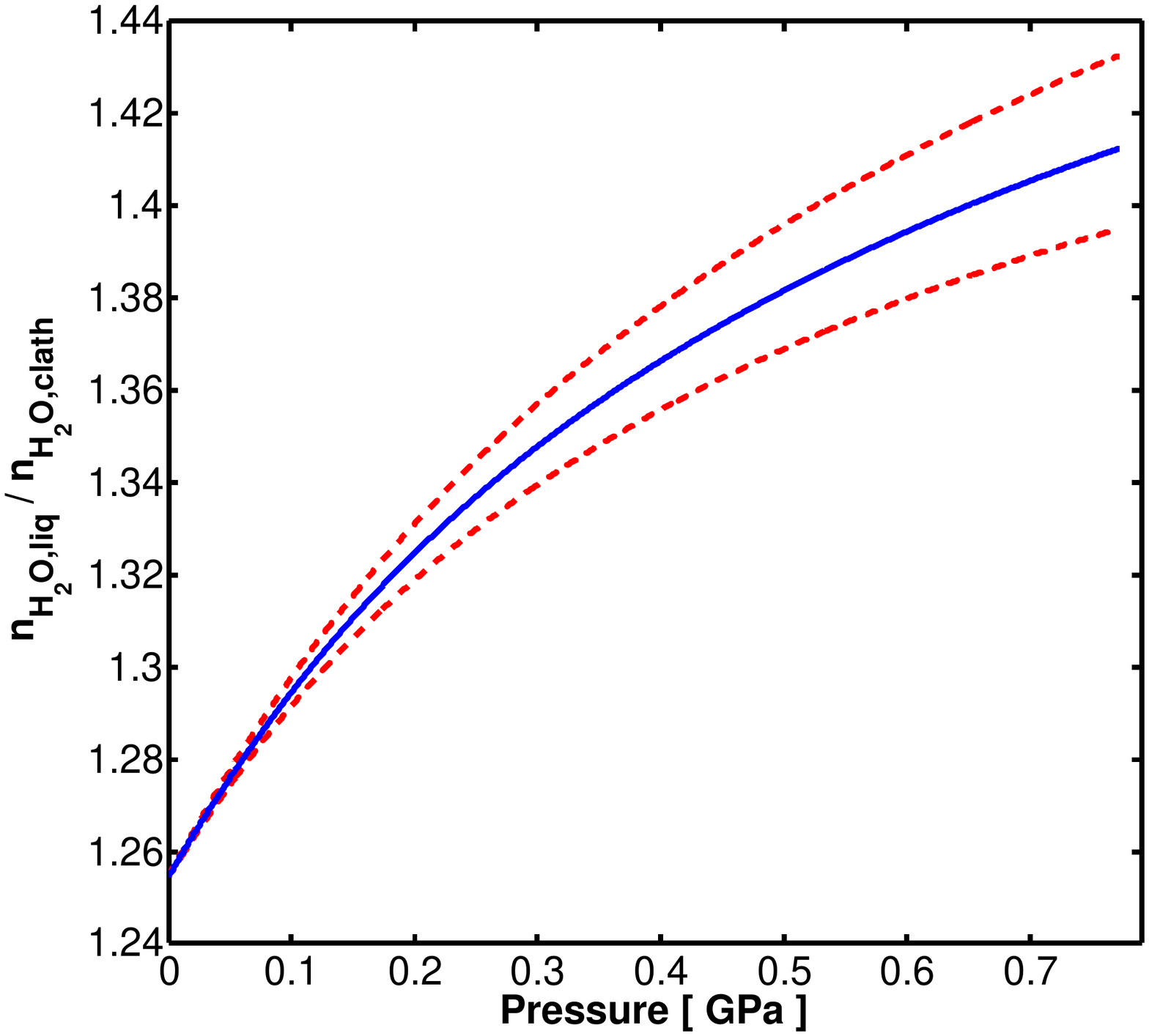}
\caption{\footnotesize{The number density ratio of water molecules in the liquid phase over the carbon dioxide clathrate phase versus pressure, for an isotherm of $280$\,K. The curve ends at the transition to water ice VI (i.e. bottom of the ocean) for a system in saturation with carbon dioxide. The solid (blue) curve is for a clathrate bulk modulus of $8.5$\,GPa. The upper and lower boundaries (dashed red curves) are for clathrate bulk moduli of $11$\,GPa and $7$\,GPa respectively. }}
\label{fig:DensityRatio}
\end{figure}

In fig.\ref{fig:DensityRatio} we plot the water number density ratio (liquid over clathrate) as a function of pressure for a $280$\,K isotherm.  \cite{Hansen2016} report an uncertainty in the lattice parameter measurement of about $0.013$\AA. This gives an uncertainty in the volume thermal expansivity of about $3\times 10^{-7}$\,K$^{-1}$. We find this error to be too small to have an effect on the number density ratio. We also vary the bulk modulus between $7$\,GPa  and $11$\,GPa. This produces an uncertainty that increases with pressure, reaching a maximum of approximately $3$\% at the bottom of the ocean. This results in a $2-3$\% uncertainty in the derived solubility at the ocean's bottom. The low error is due to the low pressure at the bottom of the ocean relative to the probable bulk modulus of the clathrate.   

The abundance of carbon dioxide in the clathrate is estimated by:
\begin{equation}
X^{co_2}_{clath}=\frac{2y^{co_2}_{small}+6y^{co_2}_{large}}{2y^{co_2}_{small}+6y^{co_2}_{large}+46}
\end{equation}
where we consider that each SI unit crystal is composed of $46$ water molecules and eight cages. Two small cages that are singly occupied by carbon dioxide at a probability of $y^{co_2}_{small}$ and six large cages also singly occupied at a probability of $y^{co_2}_{large}$. The probability of CO$_2$ entrapment in a cage of type $i$ obeys \citep{waalplat}: 
\begin{equation}\label{OccupancyProbability}
y^{co_2}_i=\frac{\hat{f}_{co_2}C^{co_2}_i}{1+\hat{f}_{co_2}C^{co_2}_i}
\end{equation}

The fugacity $\hat{f}_{co_2}$ is again derived using the Soave-Redlich-Kwong (SRK) equation of state \citep{soave}. It was shown by \cite{yoon02} and \cite{Yoon2004} that this equation of state in conjunction with the van der Waals and Platteeuw model for clathrates can accurately predict the dissociation curve of various clathrates, including SI CO$_2$ clathrates. These authors were primarily concerned with low to medium pressures, up to approximately $1000$\,bar. At higher pressures $\hat{f}_{co_2}C^{co_2}_i>>1$ and hence the form of Eq.($\ref{OccupancyProbability}$) would tend to minimize the effect of errors in the fugacity. Since for our clathrate thermodynamic stability regime we are below the critical point for CO$_2$ one has to model the fugacity of CO$_2$ both as vapour and liquid. This is accomplished within the SRK equation of state by taking the smaller (larger) compressibility root of the cubic polynomial to represent liquid (gaseous) carbon dioxide. We set the transition between liquid and gaseous carbon dioxide in the presence of water by adopting \cite{Wendland1999} experimentally derived Clausius-Clapeyron equation for CO$_2$ vapour pressure in mixture with water.
In case CO$_2$ is in its phase I solid state the SRK equation of state is no longer applicable and we turn to use a more appropriate form for the solid CO$_2$ fugacity given in appendix \ref{subsec:AppendixA}.    
  
$C^{co_2}_i$ is the Langmuir constant for CO$_2$ in the clathrate cage of type $i$, defined for clathrates in \cite{waalplat}.
For the small cage we adopt the guest-host potential of interaction formalism given in \cite{mckoysinan}. This formalism accounts for non-covalent forces only modelled by a Kihara potential. Potential parameters are obtained from the second virial coefficients tabulated in \cite{hirschcurbird} using empirical combining rules \citep[see][pages 222-223]{hirschcurbird}. Applying density functional theory \cite{Khan2003} found that the CO$_2$ molecule also forms two weak hydrogen bonds with the water lattice when it is entrapped in the large cage. Each such weak hydrogen bond contributes an additional $2$\,kcal\,mol$^{-1}$ to the depth of the potential well occupied by the CO$_2$ molecule in the large cage. Intermolecular potential energies are often taken as the sum of two terms: a non-covalent and an electrostatic contribution \citep[e.g][]{Abascal2012,Horn2004,Manesh2009}. Therefore, for the large cage we again employ the model of \cite{mckoysinan} but add to it the contribution of the weak hydrogen bonding (i.e the electrostatic contribution).   

\begin{figure}[ht]
\centering
\includegraphics[scale=0.6]{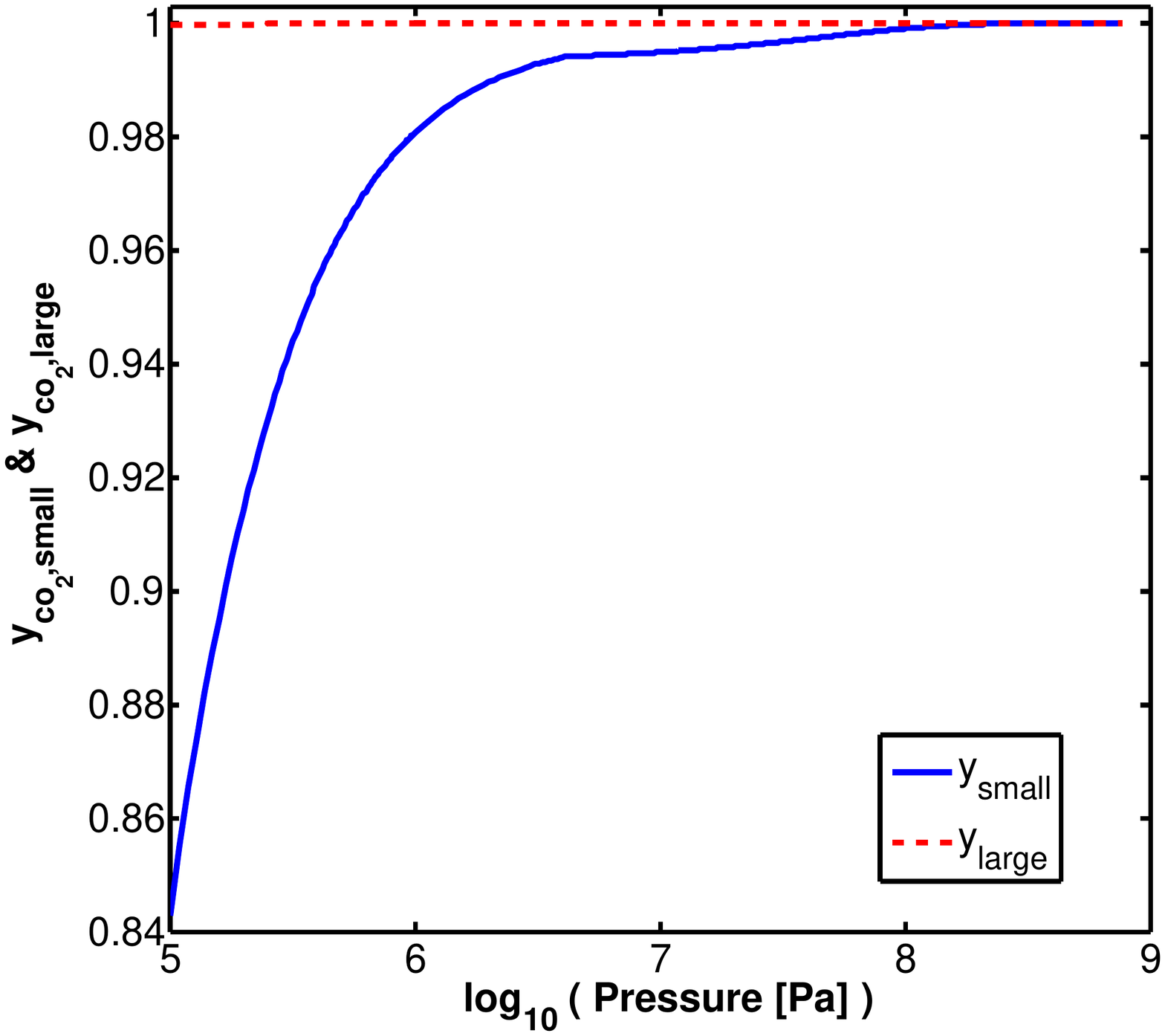}
\caption{\footnotesize{Dashed (red) curve is the probability a carbon dioxide molecule will occupy the large cage of a SI clathrate hydrate versus pressure. Solid (blue) curve is the probability a carbon dioxide molecule will occupy the small cage of a SI clathrate hydrate versus pressure. Calculation is for an isotherm of $280$\,K. The curves end at the transition to water ice VI (i.e. bottom of the ocean) for a system in saturation with carbon dioxide.}}
\label{fig:OccupPorb}
\end{figure}

In fig.\ref{fig:OccupPorb} we give the occupancy probabilities for a CO$_2$ molecule in the small ($y^{co_2}_{small}$) and large ($y^{co_2}_{large}$) cages of its SI clathrate hydrate versus pressure and for a $280$\,K isotherm. Our derived probabilities are in accordance with experimental data. One should keep in mind, however, that when forming clathrates in the laboratory supersaturation (i.e. disequilibrium) is required in order to initiate the clathrate formation process. The measure of disequilibrium may influence the final measured clathrate composition, thus producing some scatter in the published clathrate compositions among the different experiments \citep[see discussion in][]{Cicone2003}. Therefore, according to eq.($\ref{solubilityeqclathrate}$) this should also introduce some scatter between the different solubility experiments if their disequilibrium conditions were not similar.

\begin{figure}[ht]
\centering
\includegraphics[scale=0.6]{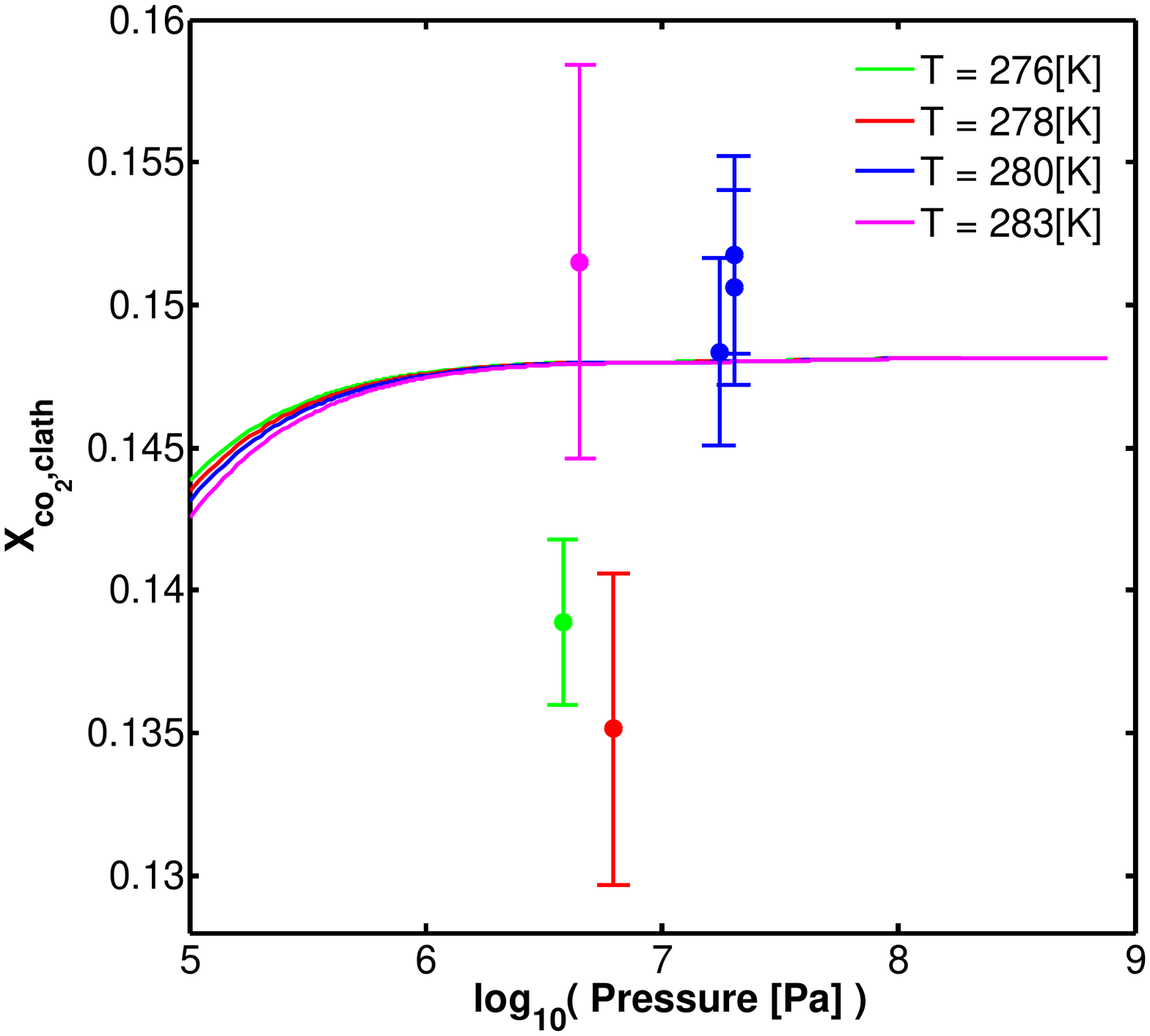}
\caption{\footnotesize{CO$_2$ abundance in its SI clathrate versus pressure, calculated for four isotherms. Green data point with its associated error is from \cite{Udachin2001}. Red data point with is associated error is from \cite{Henning2000}. Magenta data point with its associated error is from \cite{Anderson2003} and the blue data points with their associated errors are from \cite{Cicone2003}.   For high pressure the system asymptotically approaches full occupancy where all clathrate cages are occupied by CO$_2$ molecules. In this case the abundance is $8/(8+46)$, representing eight full cages for every unit cell composed of $46$ water molecules.}}
\label{fig:CO2AbundClath}
\end{figure}

In fig.\ref{fig:CO2AbundClath} we show the variation with pressure of the abundance of carbon dioxide in its SI clathrate hydrate, $X^{co_2}_{clath}$, for four isotherms. In the pressure regime of a few tens of bars the data from the literature is quite scattered. There is a general consent that the large cage is fully occupied, though for the degree of occupancy of the small cage the reported data varies widely. Diffraction experiments, though on deuterated rather then hydrogenated clathrates, find for the small cage a degree of occupancy in the range of $60\%-80\%$ \citep{Udachin2001,Henning2000} and as high as $90$\% \citep{Ikeda1999}. Analysing the dissociation curve \cite{Anderson2003} found both the large and small cages to be fully occupied at $44$\,bar. For a pressure of $180$\,bar \cite{Cicone2003} reported their sample of CO$_2$ SI clathrate was fully occupied. In this work we are mostly interested in the solubility in equilibrium with clathrates for pressures above $100$\,bar. Therefore, the uncertainty manifested by the scatter in the data for lower pressures should not result in substantial errors in the geophysical model developed below.         

We now turn to estimate, $U_{liq}-U_{clath}$, the difference in the potential wells trapping our Brownian particle in the liquid and clathrate water phases. As mentioned above our Brownian particle is a combination of the CO$_2$ molecule and its surrounding water molecules. The difference in the potential wells is therefore a superposition of both: the difference in the CO$_2$ interaction with the surrounding water structure between the liquid and clathrate phases ($\Delta{E}_{co_2-h_2o}$) and the difference in the structural energy ($\Delta{E}_{shell}$) of the water-made shell influenced by the CO$_2$ transition between the two phases:
\begin{equation}\label{BrownPotentialDif}
U_{liq}-U_{clath}=\Delta{E}_{co_2-h_2o}+\Delta{E}_{shell}
\end{equation}
where the indices $liq$ and $clath$ stand for liquid and clathrate water phases respectively. 

When an enclathrated CO$_2$ molecule transfers to the liquid water its solid hydration shell (i.e. clathrate cage) restructures to form the hydration shell in the liquid. The change in the hydration shell energy is the work invested in this restructuring. The total work, $\Delta{W}_{shell}$, is the sum of the work required to change the orientational configuration of the water molecules composing the shell, $\Delta\acute{W}_{shell}$, and work done due to a possible volume change in a mechanical contact with a pressure bath:
\begin{equation}
\Delta{E}_{shell}=\Delta{W}_{shell} = \Delta\acute{W}_{shell}-P\Delta{V}
\end{equation} 
It was already suggested by \cite{Glew1962} that the hydration shell in liquid water resembles clathrate cages. \cite{Bowron1998} used fine structure x-ray absorption spectroscopy to probe the structure of a liquid water hydration shell and a clathrate hydrate cage for the case of a Kr solute. It was found that the first peak in the radial distribution function occurs at the same distance both in the liquid and solid hydration shells. Based on this result we will assume there is no volume change when the hydration shell transitions between the two phases, thus $\Delta{V}=0$. However, a change in the water molecules' orientation is clearly seen, therefore:
\begin{equation}
\Delta{E}_{shell} = \Delta\acute{W}_{shell}
\end{equation}
Thermodynamics tells us that the element of total work on a system in contact with a heat bath is equal to the difference in Helmholtz's free energy:
\begin{equation}
dW=dU_{int}-TdS=d(U_{int}-TS)\equiv dF
\end{equation}
where $U_{int}$ is the internal energy.  
Let's consider a system in mechanical equilibrium with a pressure bath. The element of work on the system, not including expansion or contraction against the external pressure obeys:
\begin{equation}
d\acute{W}=dW-(-PdV)=dF+PdV=d(F+PV)\equiv dG
\end{equation} 
where $G$ is Gibbs free energy, which for a system in mechanical equilibrium with a pressure bath obeys:
\begin{equation}
dG=-SdT
\end{equation}
this means that:
\begin{equation}
\Delta{E}_{shell} = \Delta\acute{W}_{shell}=G_{liq}-G_{clath}
\end{equation}             

To estimate the Gibbs free energy difference of a water molecule between liquid and solid we draw an analogy to the theory of homogeneous nucleation and ice surface phenomena as described by \cite{Fletcher1962,Fletcher1968}. In the picture he describes, liquid water is made up of "flickering" molecular clusters, each made up of tens of molecules. In a thermodynamic regime where liquid water is stable the clusters with minimum free energy are not the ones whose structure resembles that of ice. Ice-like clusters are therefore rare in this thermodynamic regime. On the melting curve the free energy of a water molecule in an ice-like and non ice-like clusters becomes equal, except that the surface molecules on an ice-like cluster keep its total free energy high. When entering a state of supercooling the free energy of a water molecule is lower in an ice-like cluster than in a non ice-like cluster. Though, only when fluctuations create an ice-like cluster which is big enough so that the overall effect of the surface molecules is sufficiently diminished can such a cluster become stable and initiate a rapid phase transformation. We argue that in the transformation of a CO$_2$ molecule between clathrate and liquid water several water molecules will also have to transform between a non ice-like arrangement in the liquid state and an ice-like cluster in the clathrate solid. For a constant pressure, remembering the free energy difference should vanish for melting conditions, we may therefore write:
\begin{equation}
\tilde{G}_{liq}-\tilde{G}_{clath}=-\int_{T_{melt}}^T\left(\tilde{S}_{liq}-\tilde{S}_{solid}\right)dT\approx -\tilde{S}_{f}\left(T-T_{melt}\right)
\end{equation}  
where the \textit{tilde} means the variable is per water molecule. The entropy difference between the two phases is estimated as the entropy of fusion, $\tilde{S}_{f}$. For ice Ih the entropy of fusion is equal to $3.65\times 10^{-16}$\,erg\,K$^{-1}$\,molec$^{-1}$ \citep[see][for $0^\circ$C]{CRC}. In fig.\ref{fig:DelEntropy} we have plotted the entropy of fusion for ice VI. Clearly, to a good approximation the entropy of fusion is the same even though ice Ih at $1$\,bar and the melt curve of ice VI span four orders of magnitude in pressure. Also from fig.\ref{fig:DelEntropy} it seems the entropy of fusion is insensitive to the temperature. We will therefore take the entropy of fusion to be a constant equal to the value above.   
In accordance with our analogy to the theory of homogeneous nucleation the clathrate grain is basically a super-heated ice and the melting temperature is the ice to clathrate+liquid transition. We estimate the latter by the first quadruple point temperature taken to be $272.12$\,K \citep{yoon02}.   

The difference in the potential wells, see Eq.($\ref{BrownPotentialDif}$), may thus be written as:     

\begin{equation}
U_{liq}-U_{clath} = \Delta{E}_{co_2-h_2o}-\eta\tilde{S}_f\left(T-T_{melt}\right)
\end{equation}
where $\eta$ is the number of water molecules included in a single Brownian particle, containing a single carbon dioxide molecule.
Considering that in a fully occupied SI clathrate hydrate there are $5.75$ water molecules per every carbon dioxide molecule we expect $\eta\geq 5.75$. Full occupancy though is not always achieved in the laboratory and values as high as $7$ water molecules per every carbon dioxide molecule have been reported \citep[see][and references therein]{Cicone2003}.

The interaction of the CO$_2$ molecule with its surrounding water structure depends on the volume of the hydration shell, for both liquid and solid. Therefore, the difference in this interaction between the two phases may be written as:  
\begin{equation}
\Delta{E}_{co_2-h_2o}(P,T) = \Delta{E}_{co_2-h_2o}(P_0,T_0)+\left(\frac{\partial\Delta{E}_{co_2-h_2o}}{\partial P}\right)_{T_0}(P-P_0) + \left(\frac{\partial\Delta{E}_{co_2-h_2o}}{\partial T}\right)_{P}(T-T_0)
\end{equation}
Using the definitions for the bulk modulus, $B$, and for the volume thermal expansivity, $\chi$, one may obtain after a few algebraic steps:
\begin{equation}
\Delta{E}_{co_2-h_2o}(P,T) = \Delta{E}_{co_2-h_2o}(P_0,T_0)-\left(\frac{V}{B}\frac{\partial\Delta{E}_{co_2-h_2o}}{\partial V}\right)_{T_0}(P-P_0)+ \left(V\chi\frac{\partial\Delta{E}_{co_2-h_2o}}{\partial V}\right)_P(T-T_0) 
\end{equation}
The volume thermal expansivity times a temperature difference of $10$\,K gives a dimensionless number of the order of $10^{-3}$. A pressure difference  spanning our water planet ocean (about $1$\,GPa) divided by a bulk modulus appropriate for clathrates gives a dimensionless number of the order of $0.1$. Thus, corrections to the CO$_2$ interaction with its hydration shell (in either phase) due to thermal expansion are negligible in comparison to high pressure compressional effects. We may therefore write:
 \begin{equation}\label{pressurecorrection}
\Delta{E}_{co_2-h_2o}(P,T) = \Delta{E}_{co_2-h_2o}(P_0,T_0)-\left[\left(\frac{V}{B}\frac{\partial{E}^{liq}_{co_2-h_2o}}{\partial V}\right)_{liq} - \left(\frac{V}{B}\frac{\partial{E}^{clath}_{co_2-h_2o}}{\partial V}\right)_{clath}\right]_{T_0}(P-P_0)
\end{equation}    
 
We adopt for $E^{liq}_{co_2-h_2o}$ a value of $-4.97\times 10^{-13}$\,erg.  This value was derived by \cite{Sato2000} using the polarizable continuum model describing the interaction of the CO$_2$ molecule with its liquid water surroundings, the latter described as a dielectric continuum. This value for $E^{liq}_{co_2-h_2o}$ includes the electrostatic interaction and the dispersion and repulsion free energies. 
For the potential of interaction of the CO$_2$ molecule with its water surroundings in the clathrate, $E^{clath}_{co_2-h_2o}$, we average the values for the small and large cages weighted by their relative abundance in the unit cell. We find for the small cage a potential of interaction of $-4.71\times 10^{-13}$\,erg and for the large cage a value of $-6.17\times 10^{-13}$\,erg which gives for $E^{clath}_{co_2-h_2o}$ a value of $-5.80\times 10^{-13}$\,erg. The energies of interaction differ by about $15$\% between the two phases. The difference in the interaction energy is therefore $\Delta{E}_{co_2-h_2o}(P_0,T_0)=8.30\times 10^{-14}$\,erg. This is of the same order of magnitude as the contribution from the Gibbs free energy difference derived above.

We now wish to estimate the correction to $E^{clath}_{co_2-h_2o}$ due to the high pressure in the deep ocean. The clathrate equation of state (eq.\ref{CellVolume}) gives a relative volume decrease of $7.3$\% over the depth of the ocean. This means the relative clathrate cage radii decrease by about $2.4$\%. Solving for the solute-solvent interaction using the model of \cite{mckoysinan} once for the low pressure cage radii and once for the high pressure values gives an energy difference of $4.78\times 10^{-14}$\,erg. Therefore, the pressure correction is:
\begin{equation}\label{ClathShellPressCorrection}
\left(\frac{V}{B}\frac{\partial{E}^{clath}_{co_2-h_2o}}{\partial V}\right)_{clath}(P-P_0)\approx \frac{P[GPa]}{1.29\times 10^{13}}\quad[erg]
\end{equation} 
At the bottom of the ocean this is approximately $6\times 10^{-14}$\,erg, an order of magnitude less than $E^{clath}_{co_2-h_2o}$. However, as is clear from Eq.$\ref{pressurecorrection}$ we are interested in the \textit{difference} of the pressure corrections between the two phases. Unfortunately, we do not have an equation of state for the hydration shell in the liquid for our pressure range of interest. Although, if the findings of \cite{Bowron1998} may indeed be extended for the case of CO$_2$ there is reason to believe the two corrections ought be very similar, resulting in their difference being a small number. If this assumption is valid we end up with the approximate form:
\begin{equation}
U_{liq}-U_{clath} = \Delta{E}_{co_2-h_2o}(P_0,T_0)-\eta\tilde{S}_f\left(T-T_{melt}\right)
\end{equation} 
As we discuss below we compare the predictions of our model with high pressure solubility inferred from experiments. We find that our model agrees with experiments, within the experimental uncertainty, as long as the pressure correction to $E^{liq}_{co_2-h_2o}$ is not more than $70$\% lower than the correction in Eq.($\ref{ClathShellPressCorrection}$) or $20$\% higher than said pressure correction.      

It is interesting to note that since the fugacity times the Langmuir constant is much larger than unity \citep{Levi2013} the probability for CO$_2$ entrapment in the clathrate is not a strong function of pressure. For pressures lower than the bulk moduli of water in liquid and clathrate phases the water number density ratio may also be estimated as independent of pressure. \textit{Therefore, the solubility of carbon dioxide in liquid water in equilibrium with its clathrate (eq.$\ref{solubilityeqclathrate}$) should be a weak function of pressure}. This is corroborated by several experimental works \citep[e.g.,][]{Servio2001,Yang2000,Aya1997,Zatsepina2001,Kim2008}. 

We tested our theory against several experimental data sets, which are usually reported in isobaric form. We searched for the value of $\eta$ that gives a minimum absolute average deviation (AAD) when compared to the data sets chosen. The $61$\,bar and $104$\,bar data sets of \cite{Yang2000} were well fitted with $\eta$ of $6.23$ and $6.35$ with AAD of $2.01\%$ and $0.52\%$, respectively. These are well within the $\eta$ criterion mentioned above. The data sets for $20$\,bar, $42$\,bar, $50$\,bar and $60$\,bar of \cite{Servio2001} were fitted with $\eta$ of $3.17$, $4.11$, $5.32$ and $5.48$ with AAD of $1.07\%$, $2.55\%$, $3.11\%$ and $1.76\%$ respectively. These values for $\eta$ are below the theoretical minimum of $5.75$. One possible explanation is that our chosen parameters may be in error. If, however, it is due to an issue in the experiment then it is pressure related,  since when a larger isobar is tested the value for the fitted $\eta$ increases, approaching the minimum of $5.75$.
For the $300$\,bar data set reported in \cite{Aya1997} we find a value of $\eta=6.11$ fits the experimental data with AAD of $2.15\%$. Again complying with the theoretical requirement. Finally, the four salt-free isobars of $101$\,bar, $121$\,bar, $151$\,bar and $201$\,bar, reported in \cite{Kim2008}, were best fitted with $\eta$ values of: $5.90$, $5.60$, $5.62$ and $5.93$ respectively, the AADs' for these four fits are: $1.27\%$, $1.19\%$, $1.84\%$ and $2.33\%$ respectively.   
We argue that our theory may provide an indirect method for approximating the clathrate composition from solubility data. 

\begin{figure}[ht]
  \begin{minipage}{\textwidth}
  \centering
    \includegraphics[width=.4\textwidth]{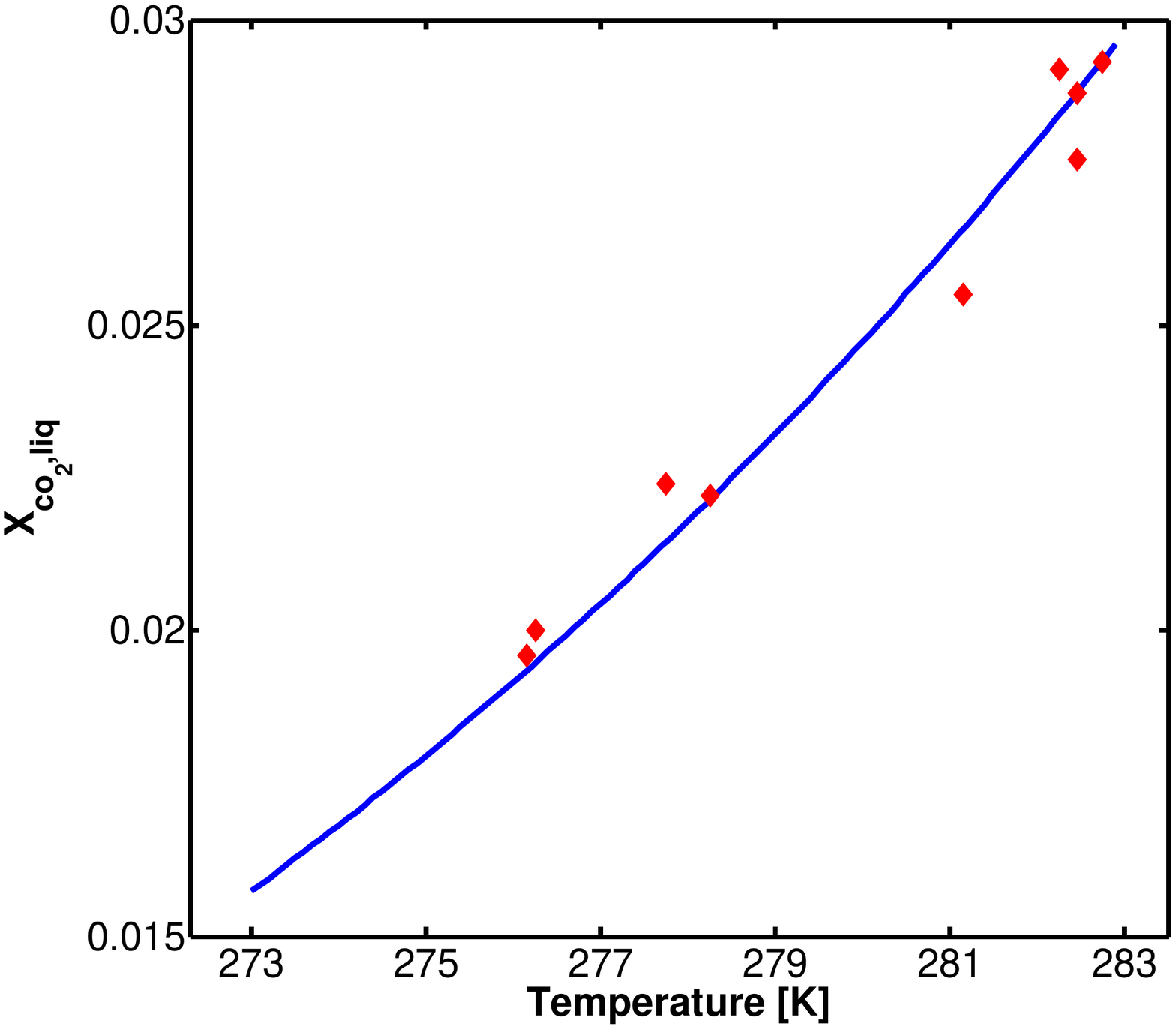}\quad
    \includegraphics[width=.4\textwidth]{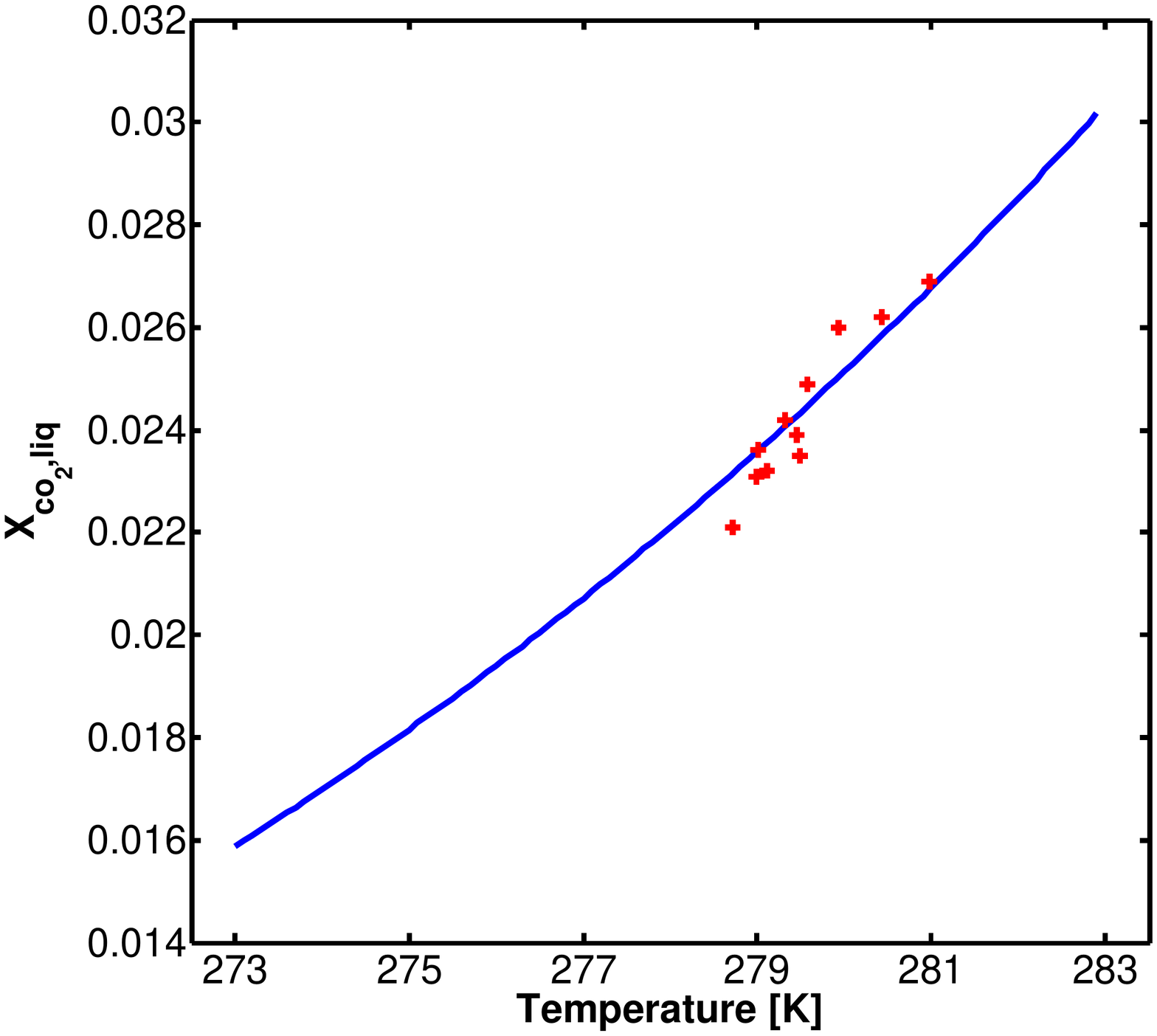}\\
    \includegraphics[width=.4\textwidth]{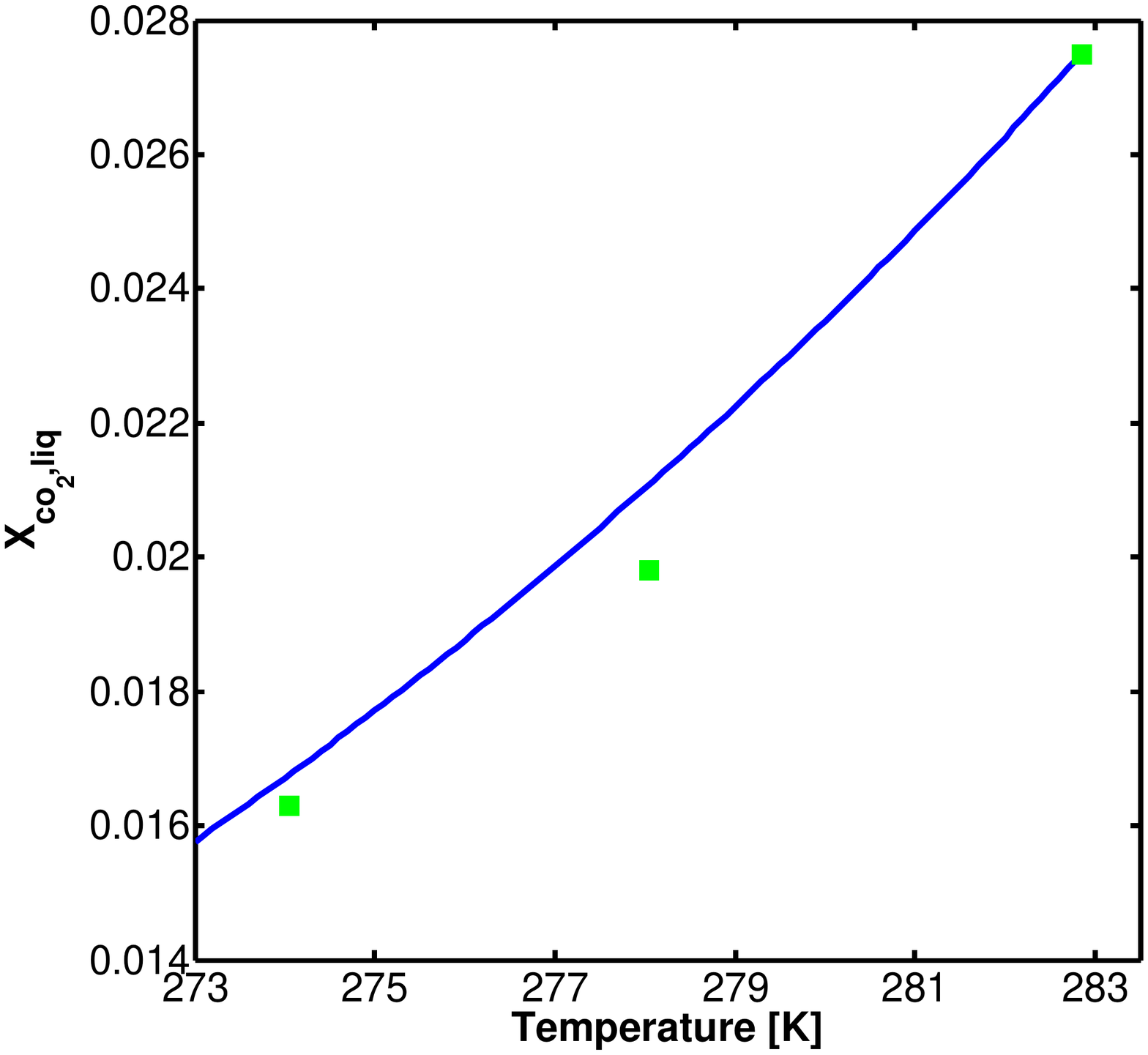}\quad
    \includegraphics[width=.4\textwidth]{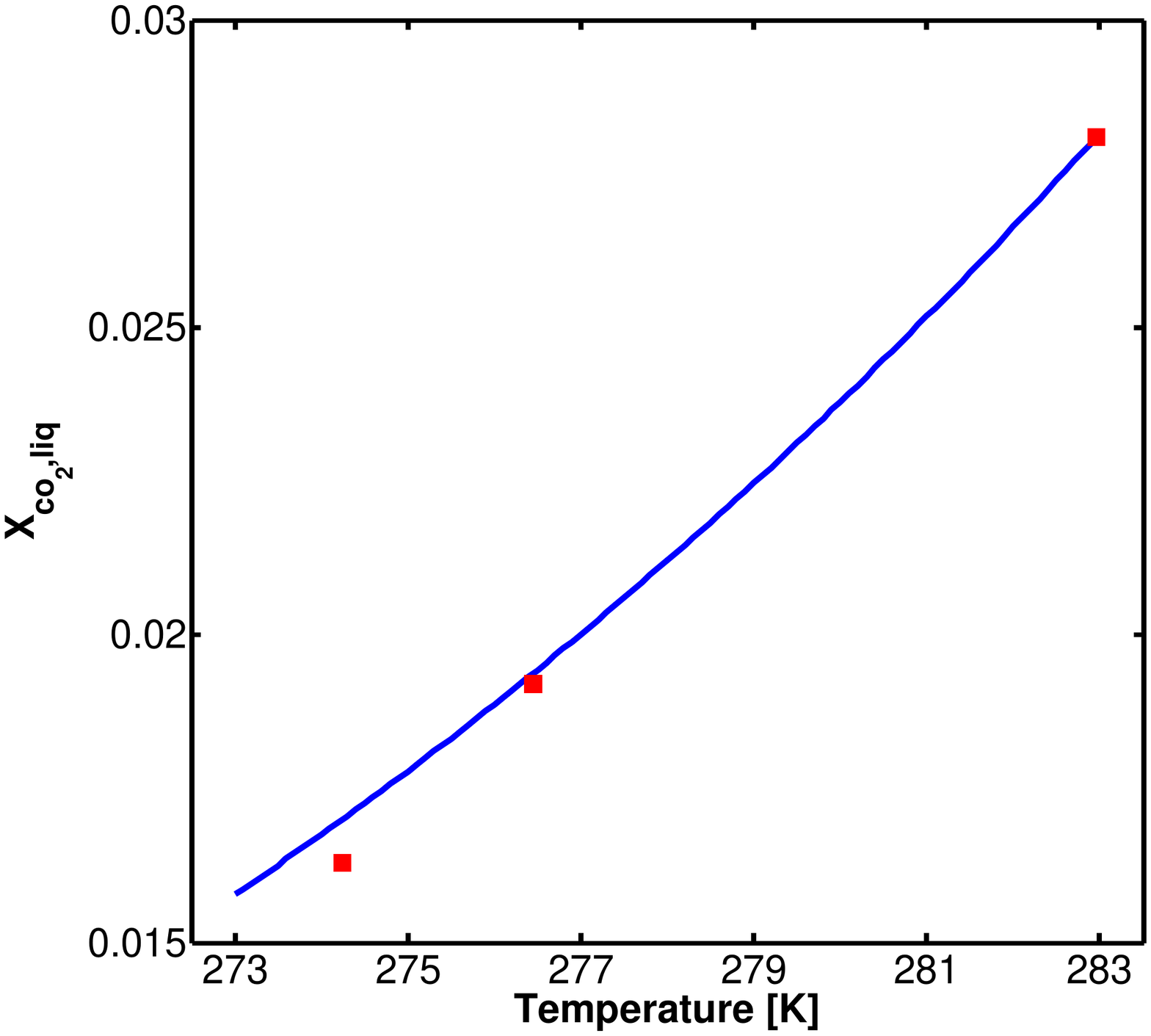}
    \caption{\footnotesize{Theoretically predicted isobaric carbon dioxide solubility (mole fraction) in equilibrium with its clathrate hydrate versus temperature (solid blue curves), compared with experimental data sets. $300$\,bar isobaric data set from \cite{Aya1997} (upper left panel), $61$\,bar isobaric data set from \cite{Yang2000} (upper right panel), $50$\,bar isobaric data set from \cite{Servio2001} (lower left panel) and $60$\,bar isobaric data set from \cite{Servio2001} (lower right panel).}}
    \label{fig:SolubilityEquiClath}
  \end{minipage}\\[1em]
\end{figure}

In figs.\ref{fig:SolubilityEquiClath} we plot our theoretically predicted isobaric carbon dioxide solubility in equilibrium with its clathrate hydrate versus temperature. In each panel the theoretical solubility shown is the one that gave the lowest absolute average deviation, by adjusting $\eta$ as explained above, in comparison to the specific experimental data set also shown in the same panel.   

The above comparison with experimental values for the solubility indicates our theory is capable of modelling the solubility in equilibrium with clathrates up to pressures of a few hundred bars. This is, as we will discuss below, sufficient for modelling the bottom of the wind driven circulation. 

In fig.\ref{fig:SolubilityMeltDep} we have estimated the solubility of CO$_2$ along the melt curve of ice VI from the experiments of \cite{Bollengier2013}. As is shown in fig.\ref{fig:SolubilityMeltDep}, in their experiments \cite{Bollengier2013} reach and may even cross into the CO$_2$ SI clathrate hydrate thermodynamic stability field (left of the vertical red dashed line). Within this narrow domain our model predicts a solubility which is within the experimental error. 

As shown in fig.\ref{fig:SolubilityMeltDep}, there is a clear trend in the solubility in the domain of the clathrate stability field. 
Clearly there is a particular trend in the solubility characterising the immediate domain around the vertical red dashed line. This trend is obtained for the case of the non-ideal solution and using Henry's law for the solubility. In equilibrium, on the phase boundary of the clathrate stability field (on the vertical red dashed line) Henry's law for the solubility should hold. It is reasonable that kinetic inhibition ought widen the phase transition boundary. Therefore, in the immediate region around the vertical red dashed line in fig.\ref{fig:SolubilityMeltDep} Henry's law should hold true and properly represent the solubility in equilibrium with the appropriate clathrate hydrate. The trend in the solubility in this immediate region is thus probably real and a consequence of the behaviour of the solubility when in equilibrium with clathrates.        
Although the model presented in this subsection for the solubility in equilibrium with clathrates is derived independently of Henry's law, we see in fig.\ref{fig:SolubilityMeltDep} that it predicts the same trend in the solubility along the ice VI depressed melt curve.
We therefore conclude that our model for the solubility in equilibrium with clathrates also agrees with experiments at pressures prevailing at the bottom of our water planet oceans.     

To summarize, in this subsection we have attempted to model the solubility of CO$_2$ in water while in equilibrium with its clathrate grains. The model of eq.($\ref{solubilityeqclathrate}$) predicts two general behaviours: the first is that the solubility decreases with decreasing temperature and the second is that the solubility along an isotherm is relatively constant (does not increase much) with increasing pressure when in equilibrium with clathrates. Our model predicts these behaviours should hold true in the range from the low pressure end of the clathrate dissociation curve and up to the pressures prevailing at the bottom of a water planet ocean. We show that these predicted behaviours are verified experimentally at the lower pressure end of the desired regime, as shown in figs.\ref{fig:SolubilityEquiClath}. They are also verified using data inferred from experiments at the high pressures prevailing at the ocean bottom. Therefore, our model ought be considered interpolative rather than extrapolative. 

\begin{figure}[ht]
\centering
\includegraphics[scale=0.6]{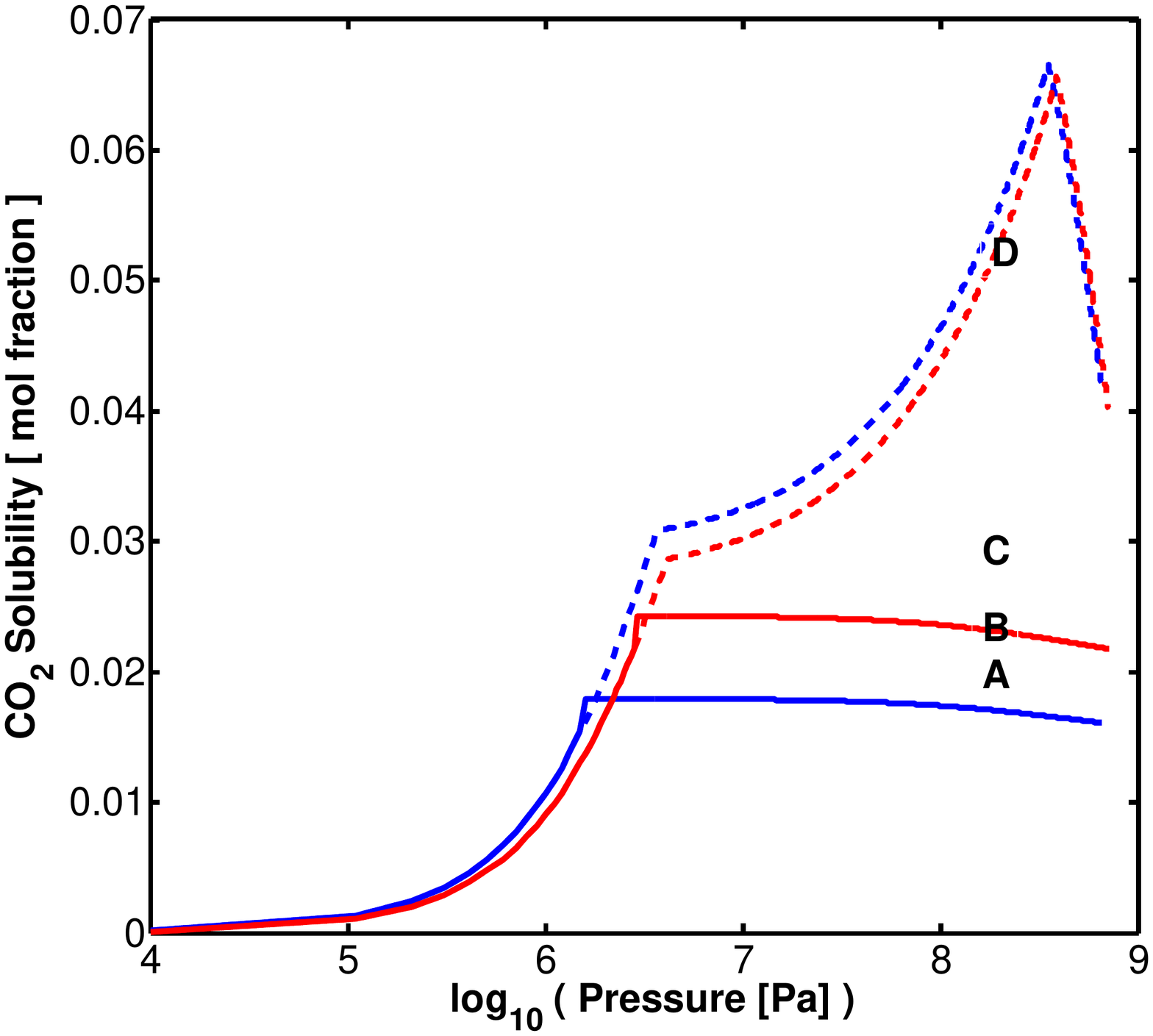}
\caption{\footnotesize{ CO$_2$ solubility [mole fraction] profiles versus pressure for two isotherms: $275$\,K (solid and dashed blue curves) and $280$\,K (solid and dashed red curves). The jump in solubility in the solid curves is at the clathrate dissociation pressure for each of the two temperatures. The jump represents the transition into the clathrate thermodynamic stability field where the solubility is fairly constant with pressure. For each isotherm the dashed curve is the continuation of the solubility assuming no clathrates are present (see eq.$\ref{Henry}$) into the clathrate thermodynamic stability field. Refer to text for explanation of points A,B,C, and D. The pressure range examined spans the conditions prevailing in our studied oceans.}}
\label{fig:IsothermSolubilityProfile}
\end{figure}

For purposes of clarity, and to be used later in the paper, we plot in fig.$\ref{fig:IsothermSolubilityProfile}$ CO$_2$ solubility profiles versus pressure for two isotherms. The solid and dashed blue curves are for the $275$\,K isotherm, and the solid and dashed red curves are for the $280$\,K isotherm. The two dashed curves represent the solubility when in equilibrium with liquid/solid CO$_2$ and are derived by solving eq.($\ref{Henry}$) for the entire oceanic pressure range, for the two isotherms. They represent a continuation of Henry's law for the solubility into the clathrate hydrate stability field. The solid curves span the low pressure solubility outside of the CO$_2$ SI clathrate thermodynamic stability field (solved by eq.$\ref{Henry}$) and the solubility in the presence of CO$_2$ SI clathrates when entering their thermodynamic stability field (solved by eq.$\ref{solubilityeqclathrate}$).
The jump seen in the solubility in each of the solid curves is at the clathrate dissociation pressure for the given isotherm. 
As discussed above, the CO$_2$ solubility in the presence of clathrates remains fairly constant with pressure. It is also clear from the figure that \textit{while outside of the clathrate stability field (low pressure end of the solid curve) the solubility increases with decreasing temperature, in the presence of clathrate grains the solubility increases with increasing temperature}. In addition, within the deep ocean, clathrates tend to keep the level of solubility of CO$_2$ much lower than what is predicted by assuming equilibrium with either liquid or solid of CO$_2$ (point D gives a solubility which is higher than at point B by a factor of about $2.2$).  

We also wish to note the required formation conditions within the ocean of CO$_2$ SI clathrate hydrate grains. The points $A$ through $D$, in fig.$\ref{fig:IsothermSolubilityProfile}$, all sit along an isobar and therefore represent some depth level in our approximated $280$\,K isothermal ocean. Although the temperature and pressure conditions, shared by all these four points, fall in the thermodynamic stability field of CO$_2$ SI clathrate this does not mean a clathrate grain placed under such conditions would necessarily be stable. For example, a CO$_2$ clathrate grain placed under conditions represented by point $A$ would see a CO$_2$ subsaturated (with respect to clathrates) liquid water environment and would diffuse its CO$_2$ to the surrounding water and dissociate. In other words, if the CO$_2$ abundance in liquid water is below the saturation value for equilibrium with clathrates (solid red curve for the $280$\,K isotherm) then clathrate grains will not form from the CO$_2$ dissolved in the ocean. If conditions in the ocean were perturbed to equal that of point $A$ for example no clathrates would form at the reference depth.
For CO$_2$ abundances above point $B$ the depth level examined becomes over-saturated with respect to clathrates and these begin to form as grains directly from the CO$_2$ dissolved in the ocean.
In case the ocean was perturbed so that conditions equalled those represented by point $C$ clathrate grains would readily form. Those would sink to the bottom of the ocean taking along local CO$_2$ and decreasing the abundance of CO$_2$ at the examined depth back towards point $B$.

\begin{figure}[ht]
\centering
\includegraphics[scale=0.6]{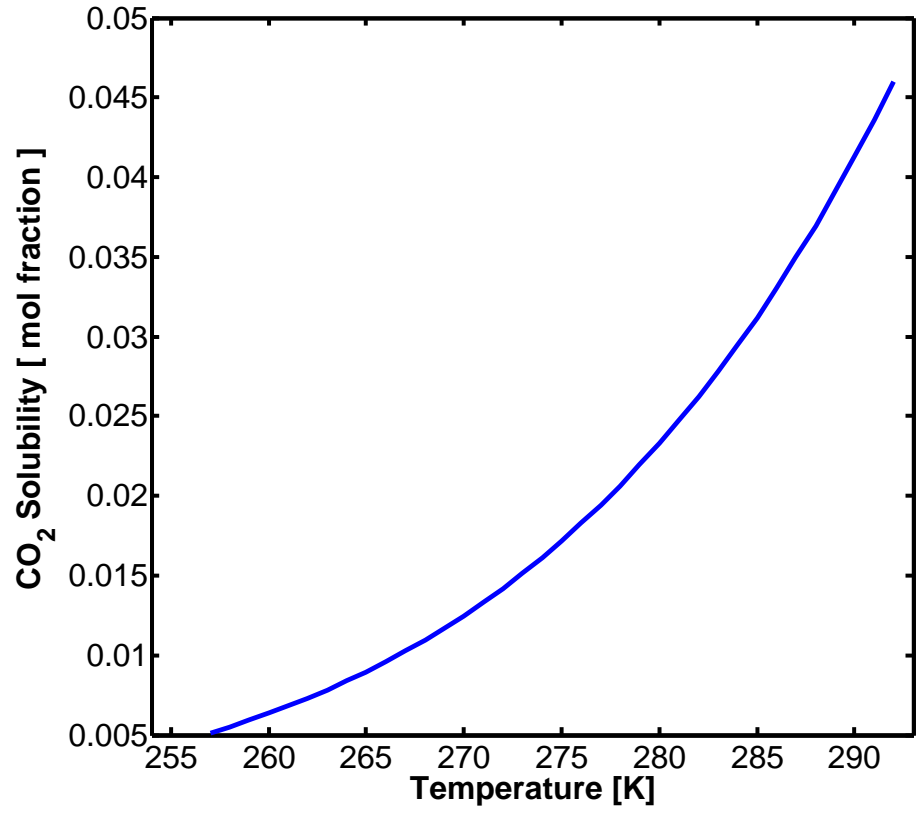}
\caption{\footnotesize{Solubility of CO$_2$ [mole fraction] in equilibrium with its clathrate grains versus temperature. The curve is for an isobar of $0.16$\,GPa and the temperature range is that spanning the clathrate thermodynamic stability field for this pressure level.}}
\label{fig:SolResTemp}
\end{figure}

Finally, in fig.$\ref{fig:SolResTemp}$ we plot the solubility of CO$_2$, in abundance, as a function of temperature in equilibrium with CO$_2$ SI clathrate grains. The plot is for an oceanic depth matching an isobar of $0.16$\,GPa. The temperature range spans the minimum and maximum temperatures for which a CO$_2$ SI clathrate is thermodynamically stable for the given isobar. According to this figure if a water planet's ocean was warmer at an earlier stage of its life then that ocean was able to dissolve more CO$_2$ before clathrate grain formation ensues. In addition, as the ocean cools and CO$_2$ solubility with respect to clathrates decreases any excess in the dissolved CO$_2$ with respect to the lower solubility would form clathrate grains and sink to the bottom.

With the solubility of freely dissolved CO$_2$ in water analysed we turn to build the phase diagram of the SI CO$_2$ clathrate hydrate, spanning conditions appropriate for water planets.

\section{THE CO$_2$ SI CLATHRATE PHASE DIAGRAM}\label{sec:PhaseDiagram}

We adopt the theory of \cite{waalplat}, based on the theory of solid solutions, in order to derive the phase diagram for the SI CO$_2$ clathrate hydrate.
On the boundaries of the thermodynamic stability field of a clathrate hydrate with either ice Ih or liquid water the chemical potential of the clathrate equals that of the other water phase with which it is in contact.
\begin{equation}
\mu_{H_2O}^{clathrate}=\mu_{H_2O}^{\alpha}
\end{equation}
where the $\alpha$ phase represents either ice Ih or liquid water.
For the chemical potential of water ice Ih we adopt the formalism of \cite{feistel06}. The chemical potential of liquid water in solution with CO$_2$ may be written as a superposition of two terms: one for the pure liquid water and a correction for it being in a non-ideal solution \citep{Denbigh}:
\begin{equation}
\mu_{H_2O}^{liq}=\mu_{H_2O}^{liq,pure}+kT\ln{\left(\gamma_{H_2O} X_{H_2O}\right)}
\end{equation}
The chemical potential of pure liquid water is accurately given in \cite{wagner02} and is adopted here. In the solubility correction term: $k$ is Boltzmann's constant, $T$ is the temperature, $\gamma_{H_2O}$ is the activity coefficient for water in solution derived using the UNIQUAC method of \cite{Abrams1975}, the latter method was discussed in subsection $2.1$. The abundance of water in solution, $X_{H_2O}$, is the difference from unity of the abundance of CO$_2$ in solution. The latter calculated using eq.($\ref{solubilityeqclathrate}$), since the water solution is in equilibrium with the clathrate phase on its dissociation curve. 

According to the theory of \cite{waalplat} the chemical potential of a clathrate may be represented as a sum of two terms, the chemical potential of the empty clathrate hydrate (i.e. the $\beta$ phase) and the contribution of the stabilizing guest molecules. We may therefore write for a pure clathrate, where carbon dioxide is the sole guest species, the following:       
\begin{equation}
\mu_{H_2O}^{clathrate}=\mu_{H_2O}^{\beta}+kT\sum_{i=1}^2\nu_i\ln{\left(1-y_i^{co_2}\right)}
\end{equation}
where $\nu_i$ is the ratio between the number of $i$ type cages to water molecules per cubic unit crystal. The probability a CO$_2$ molecule occupies a type $i$ cage, $y_i^{co_2}$, was given explicitly in eq.($\ref{OccupancyProbability}$). The summation is carried out over the two types of cages formed in the SI clathrate crystal.  

The dependency of the empty clathrate hydrate chemical potential on both the pressure and temperature was first given by \cite{Holder1980} in terms of a difference between the empty clathrate hydrate and the other water phase in contact. This method of difference does not account for the extensive experimental knowledge accumulated for liquid water and ice Ih as compared to that accumulated for clathrates. We therefore write for the $\beta$ phase alone:
\begin{equation}
\frac{\mu_{H_2O}^{\beta}(T,P)}{kT}=\frac{\mu_{H_2O}^{\beta}(T_0,P_0)}{kT_0}-\int_{T_0}^T\frac{H^{\beta}(T',P_0)}{kT'^2}dT'+\frac{1}{kT}\int_{P_0}^PV^{\beta}(T,P')dP'
\end{equation}
where:
\begin{equation}\label{enthalpy}
H^{\beta}(T',P_0)=H^{\beta}(T_0,P_0)+\int_{T_0}^{T'}C^{\beta}_p(\tau,P_0)d\tau
\end{equation}
Here $H^\beta$ is the enthalpy of the empty hydrate, $V^\beta$ is its volume and $C^{\beta}_p$ its isobaric heat capacity. $T_0=273.15$\,K and $P_0=0.135$\,MPa are our reference temperature and pressure respectively. 
The $\beta$ phase is not stable thus one cannot characterise it experimentally. In addition each kind of guest molecule distorts the water clathrate hydrate lattice surrounding it somewhat differently \citep{klaudasand}. Therefore, both the empty clathrate chemical potential and its enthalpy at the reference temperature and pressure are taken to be free parameters. A fit to the experimental dissociation pressure data sets of the CO$_2$ clathrate yields values of $\mu_{H_2O}^{\beta}(T_0,P_0)=5.937\times 10^{-14}$\,erg\,molec$^{-1}$ and $H^{\beta}(T_0,P_0)=-5.186\times 10^{-14}$\,erg\,molec$^{-1}$. Estimating the enthalpy of the empty clathrate hydrate as prescribed in eq.($\ref{enthalpy}$) we also need to estimate its isobaric heat capacity. A good approximation for the latter is the isobaric heat capacity of ice Ih \citep{avlonitis1994}. 
Finally, the CO$_2$ SI clathrate equation of state for the crystal cell volume (see eq.$\ref{CellVolume}$) is used for estimating $V^\beta$. 

Regarding our adopted values for the reference empty clathrate hydrate parameters, there are references in the literature for the empty clathrate hydrate chemical potential \citep[see discussion in][]{Dharmawardhana1980}. Our suggested value is larger than given in \cite{Dharmawardhana1980} by a factor of $2.75$. This is a result of the approach we adopt to modelling clathrates, where instead of optimizing the guest-host potential of interaction parameters we explicitly account for the weak-hydrogen bonding between CO$_2$ and the water host lattice. As a consequence the empty clathrate hydrate reference chemical potential and enthalpy need to be optimized. This is based on the idea that every guest specie has its own reference empty clathrate hydrate lattice. The value in \cite{Dharmawardhana1980}, which is often adopted, is from measurements for the clathrate hydrate of cyclopropane. This bigger guest molecule only occupies the large cage of the SI clathrate and thus does not distort the SI small cage, contrary to the case when CO$_2$ is the guest molecule. Thus for the case of CO$_2$ its empty reference clathrate hydrate should be even less stable than the reference lattice for the case of cyclopropane. This is manifested in our adopted larger value for $\mu_{H_2O}^{\beta}(T_0,P_0)$.     

\begin{figure}[ht]
\centering
\includegraphics[trim=0.2cm 4cm 0.2cm 4.5cm , scale=0.6, clip]{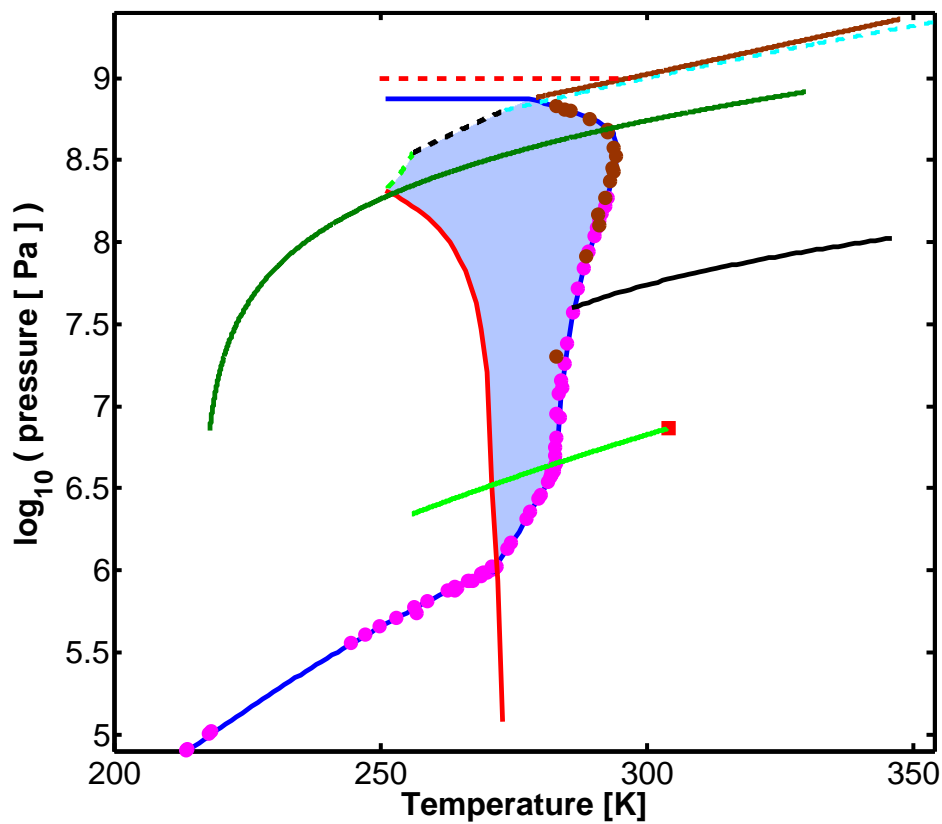}
\caption{\footnotesize{CO$_2$ SI clathrate hydrate phase diagram. Solid red curve is the melting curve of water ice Ih including the melt depression due to carbon-dioxide. Dashed light green curve is the melting curve for water ice III. Dashed black curve is the melting curve for water ice V. Dashed cyan curve is the melting curve for water ice VI. Solid brown curve is the depressed melt curve for water ice VI. Dashed red curve is the probable upper limit on the stability of CO$_2$ filled-ice. Solid blue curve is the boundary of the thermodynamic stability regime of the CO$_2$ SI clathrate hydrate, i.e. the dissociation curve. The clathrate hydrate survives to the left of this curve. The solid dark green curve is a segment of the pure phase I solid CO$_2$ melting curve. Solid light green curve is the vapour curve for CO$_2$ in the presence of water and the red square is its critical point. The solid black curve is where the bulk mass density of water rich liquid and fluid carbon-dioxide equals. The magenta circles are CO$_2$ SI clathrate dissociation data points from several groups: \cite{Takenouchi1965}, \cite{Ng1985}, \cite{Wendland1999}, \cite{Mohammadi2005}, \cite{Mohammadi2009}, \cite{Yasude2008}, references $21$, $27$ and $58$ in \cite{Fray2010}, \cite{Adamson1971}, \cite{Miller1970} and \cite{Bollengier2013} (data points from the latter are coloured brown). The shaded area emphasizes the region where the clathrate can coexist with a liquid solution of water and carbon-dioxide.}}
\label{fig:PhaseDiagram}
\end{figure}

In fig.$\ref{fig:PhaseDiagram}$ we plot the CO$_2$ SI clathrate hydrate phase diagram. The solid red curve is the melting curve of water ice Ih including the melting point depression due to the effect of CO$_2$ on the liquid water chemical potential. In calculating this melting point depression care was taken in choosing the appropriate solubility model when crossing into the CO$_2$ SI clathrate hydrate stability field.  
The dashed light green curve is the melting curve for water ice III, the dashed black curve is the melting curve for water ice V and the dashed cyan curve is the melting curve for water ice VI (taken from the IAPWS, \textit{Revised Release on the Pressure along the Melting and Sublimation Curves of Ordinary Water Substance}, September 2011), all in the pure water system. The solid brown curve is the depressed melt curve of water ice VI when in contact with an aqueous solution saturated in CO$_2$ from the experiments of \cite{Bollengier2013}. 
The solid blue curve is the boundary of the thermodynamic stability regime of the CO$_2$ SI clathrate hydrate. The clathrate hydrate is thermodynamically stable to the left of this curve. At temperatures below the ice Ih melting temperature the solid blue curve represents the three phase of: H-ice Ih-CO$^{vap}_2$. For temperatures higher than the ice Ih melting temperature the solid blue curve represents three different three phase curves of: H-Lw-CO$^{vap}_2$, H-Lw-CO$^{liq}_2$ and H-Lw-CO$^{solid}_2$ in succession of increasing pressure. Here H stands for clathrate hydrate, Lw stands for liquid water solution with CO$_2$ and CO$^j_2$ is the $j$ phase of CO$_2$. 
At pressures above the melting curve of water ice VI, the blue line denotes the three phase curve H-water ice-CO$^{solid}_2$.
The solid dark green curve is a segment of the pure phase I solid CO$_2$ melting curve \citep{Span1996}. The solid light green curve is the vapour pressure curve for CO$_2$ in the presence of water and the red square is its critical point \citep[see][]{Wendland1999,Diamond2003}. The high pressure arm of the dissociation curve of the CO$_2$ SI clathrate hydrate (solid blue) and the dashed red curve confine the probable stability field of a newly discovered phase called CO$_2$ filled ice \citep{Bollengier2013,Tulk2014,Hirai2010}, whose structure was only recently analysed \citep[see][]{Tulk2014}. The shaded area emphasizes the region where SI CO$_2$ clathrate hydrates can coexist with a solution of liquid water and dissolved CO$_2$.

Still in fig.$\ref{fig:PhaseDiagram}$, the solid black curve is where the bulk mass density of our water rich liquid equals that of fluid carbon-dioxide. In obtaining the latter curve the mass density of fluid carbon-dioxide was modelled using the formulation in \cite{Span1996}. We have assumed the solubility of water in the fluid of carbon-dioxide to be negligible \citep[see discussion in][]{Teng1997}. The water rich liquid was modelled using the equation of state for pure water \citep{wagner02}, and corrected for the effect of the solubility of CO$_2$ on the density using \cite{Teng1997}. We assume our water liquid is saturated with CO$_2$. The results of \cite{Teng1997} are in agreement with solution densities from \cite{Hebach2004} and \cite{Garcia2001}. We see from fig.\ref{fig:IsothermSolubilityProfile} that clathrates tend to keep the solubility of CO$_2$ lower than the value when in equilibrium with the pure fluid of CO$_2$. Therefore, the pure fluid of CO$_2$ is not stable within the stability field of CO$_2$ SI clathrate hydrates. Consequently the black curve is derived only outside of the clathrate stability field, and the solubility of CO$_2$ there is modelled using Henry's law (see Eq.$\ref{Henry}$).

We see from fig.$\ref{fig:PhaseDiagram}$ that phases which are potential rich reservoirs of CO$_2$ are in direct contact with the bottom of the ocean. It is therefore likely that deep mantle CO$_2$ enters the ocean. How this CO$_2$ can be stored in the deep ocean, and its transport to the atmosphere are the subjects of the following sections.

\section{DEEP RESERVOIRS FOR CO$_2$ IN WATER PLANET OCEANS}\label{sec:Reservoirs}

\begin{figure}[ht]
\centering
\includegraphics[trim=0.2cm 4cm 0.2cm 4.5cm , scale=0.6, clip]{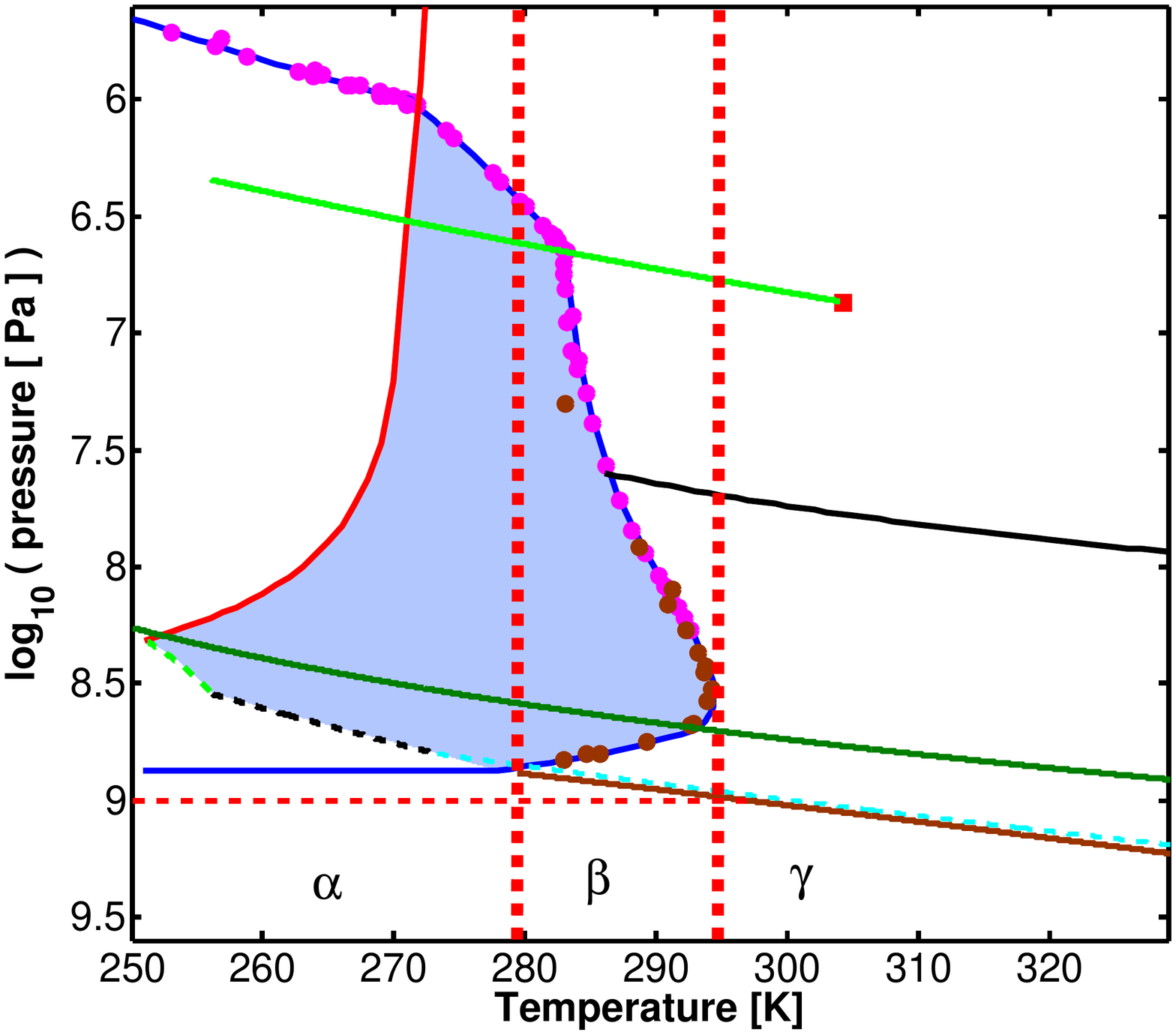}
\caption{\footnotesize{In this figure we divide (using vertical thick dashed red lines) the phase diagram of CO$_2$ SI clathrate hydrate into three domains, each having a different stratification for the sinks of carbon dioxide at the bottom of the ocean. In the $\alpha$ domain the carbon dioxide sink is solely in the form of a clathrate hydrate layer. In the $\beta$ domain a mid-ocean SI CO$_2$ clathrate hydrate layer forms, while beneath it a strata composed of phase I solid CO$_2$ may accumulate. In the $\gamma$ domain the first layer is composed of phase I solid CO$_2$ followed by liquid CO$_2$, ending when the latter becomes less dense than liquid water. The solid red curve is the melting curve of water ice Ih including the melt depression due to carbon-dioxide. Dashed: light green, black and cyan curves are the melting curves for water ice III, V and VI, respectively. Solid brown curve is ice VI depressed melt curve. Solid blue curve is the boundary of the thermodynamic stability regime of the CO$_2$ SI clathrate hydrate, i.e. the dissociation curve. The hydrate survives to the left of this curve. The solid dark green curve is a segment of the pure phase I solid CO$_2$ melting curve. Solid light green curve is the vapour curve for CO$_2$ in the presence of water and the red square is its critical point. The solid black curve is where the bulk mass density of water rich liquid and fluid carbon-dioxide equals. Horizontal thin red dashed curve is a possible high pressure boundary for CO$_2$ filled-ice stability. The shaded area is as in fig.\ref{fig:PhaseDiagram}. Solid circles are data points (see fig.\ref{fig:PhaseDiagram}).}}
\label{fig:alphabetagamma}
\end{figure}

Considering a secondary atmospheric outgassing, most of the CO$_2$ outgassing occurs at a later stage in the planet's history. In this case as solid state convection is initiated, along with outward transport of CO$_2$, the ocean may be initially subsaturated with respect to CO$_2$. Making its way into the ocean the sinks available to CO$_2$ depend primarily on the thermal profile in the deep to mid ocean. 
In fig.\ref{fig:alphabetagamma} we show that there are three possible stratification cases (denominated as: $\alpha$, $\beta$ and $\gamma$) and in this section we shall deal with each one of them separately. For each of the three cases we present a quantitative example solution for the reservoirs' capacity to store carbon using an isothermal profile in the ocean. The real thermal profile will not be isothermal. For example, an ocean thermal profile with a surface temperature beginning in the $\beta$ domain may largely fall in the type $\alpha$ domain in case the temperature decreases with depth in the ocean. Though the isothermal profile is a good first approximation for our analysis of the deposition budget of carbon at the bottom of the ocean, the more exact thermal profile in the ocean ought be derived when considering a particular water planet by using its particular energy balances. 

\subsection{The $\alpha$ Domain}

In this case the bottom of the ocean is composed of either water ice V or VI if no CO$_2$ is present. In the presence of CO$_2$ it is composed of a SI CO$_2$ clathrate hydrate layer. The clathrate hydrate layer may overlie a CO$_2$ filled-ice layer in this case, though more experimental data is needed to verify this. However, it is the clathrate layer that will be in direct contact with the overlying ocean, controlling the chemical and physical interaction between mantle and ocean.
 
The flux of CO$_2$ from the ice mantle and into the ocean is dependent on its ability to incorporate into and be transported with the mantle water ice convection cell. The flux also depends on the geological behaviour of the ice boundary layer composing the ocean's bottom surface. For example, it is important to know whether this ice boundary layer is rigid and internal CO$_2$ has to diffuse through it to reach the ocean or is it breakable directly exposing the ocean to internal CO$_2$. 

\begin{figure}[ht]
\centering
\includegraphics[trim=0.2cm 11cm 0.2cm 4.5cm , scale=0.60, clip]{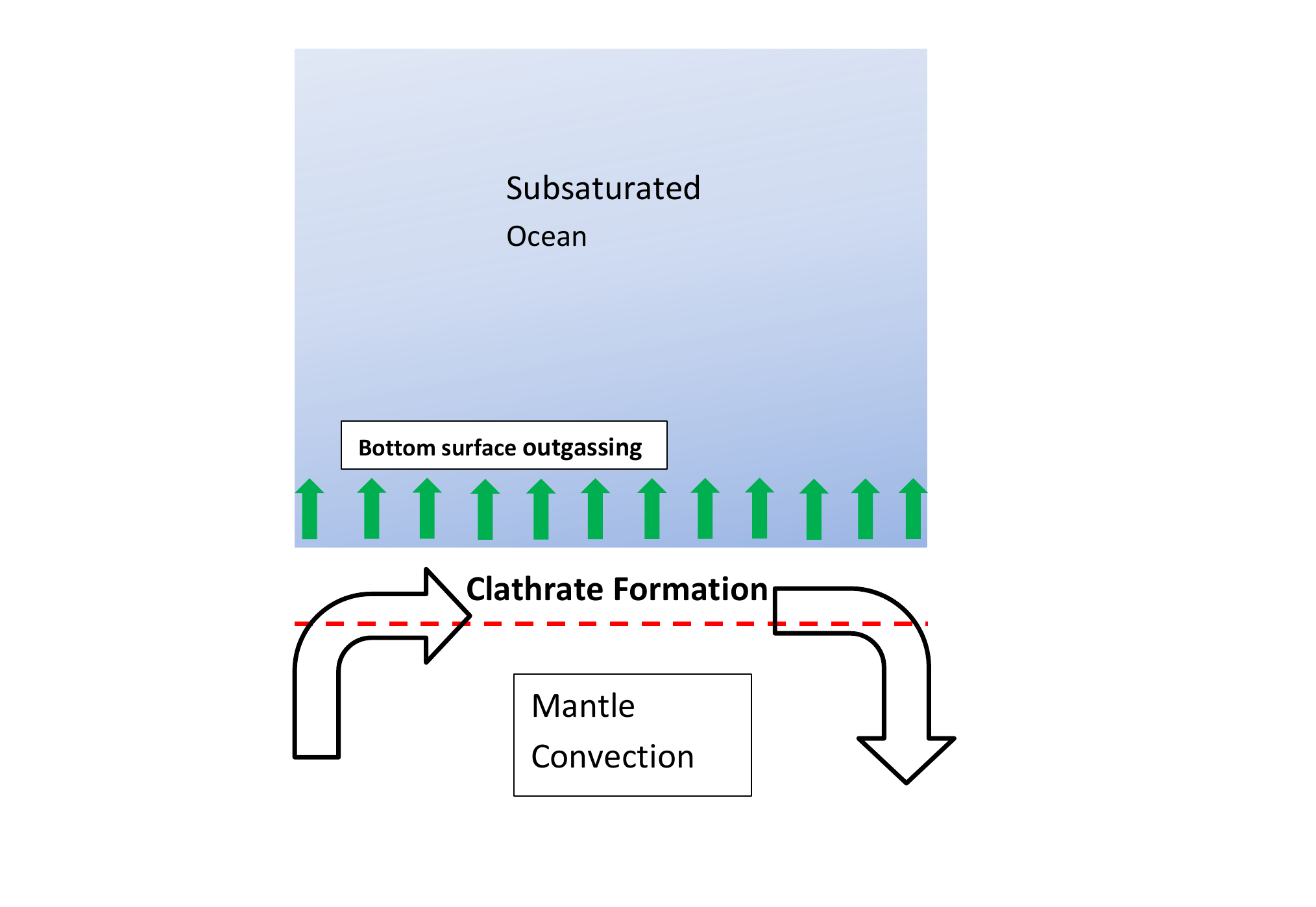}
\caption{\footnotesize{CO$_2$ embedded in the ice mantle is transported outward along with the ice convection. Upon entering the clathrate thermodynamic stability field it transforms into CO$_2$ SI clathrate hydrate. This clathrate layer then becomes the ocean's bottom surface, consequently making physical contact with the overlying ocean. In case the ocean is subsaturated with respect to CO$_2$ then the CO$_2$ from the clathrates diffuses into the ocean. We refer to this mechanism as the "gentle" outgassing mechanism.}}
\label{fig:OceanOutgassing}
\end{figure}

In the case of the $\alpha$ domain, deep ice mantle CO$_2$ transported outward should transform together with the water ice surrounding it into CO$_2$ SI clathrate hydrate upon entering the latter thermodynamic stability field (see fig.\ref{fig:OceanOutgassing}). In case the ocean is initially subsaturated with respect to CO$_2$ this clathrate layer at the top of the convection cell (also composing the ocean's bottom icy surface) would spontaneously revert back to ice V (or VI) and release its CO$_2$ into the ocean. This mechanism for releasing CO$_2$ into the ocean may be regarded as "gentle", meaning it requires no violent geological mechanisms that would break up the ice forming the ocean's bottom in order to directly inject CO$_2$ into the ocean. 
This mechanism can only strive to saturate the ocean with CO$_2$. Once the ocean approaches saturation (CO$_2$ concentration approaches the solubility value in equilibrium with clathrates) the CO$_2$ SI clathrate ice layer composing the ocean bottom begins to stabilize and the "gentle" mechanism shuts off. Consequently, additional mantle CO$_2$ transported outward by the convection cell would experience no forcing to enter the ocean and would simply continue to cycle internally in the mantle along with the high pressure water ice convection cell.  

Experiments show that the dissolution of clathrate hydrate in seawater is diffusion limited. In an interesting experiment \cite{Rehder2004} placed blocks of CO$_2$ SI clathrate hydrates on the bottom of the ocean, at a depth of $1028$\,m. With the aid of underwater cameras they measured the dissolution rates of the clathrate hydrate blocks. The clathrates dissolved due to their placement in an environment which is subsaturated in CO$_2$ with respect to clathrates. This field experiment clearly shows that the dissolution rate depends on the ability of CO$_2$ to diffuse away from the surface of the clathrate hydrate block and into the bulk ocean. 
For an ocean that energetically cannot maintain a general circulation and that does not establish convection cells, the extent of the diffusive boundary layer right above its bottom is of the order of magnitude of the ocean's depth. Under such circumstances, the "gentle" mechanism would require a time scale of
$L^2_{ocean}/D_{eddy}$
to bring the ocean close to saturation.  
Here $D_{eddy}$ is the vertical eddy diffusion coefficient for the deep ocean and $L_{ocean}$ is the ocean's depth. We shall return to elaborate on this point in the following sections.

One though has to bear in mind, that this time scale requires that the underlying mantle convection cell be able to transport CO$_2$ with enough efficiency to constantly maintain a clathrate hydrate layer at the bottom of the ocean (top of the convection cell). A full investigation of the ability of the convection cell to transport CO$_2$ outward is in order, but this will depend on the particular characteristics of a given planet.

\begin{figure}[ht]
\centering
\includegraphics[trim=0.2cm 4cm 0.15cm 5cm , scale=0.60, clip]{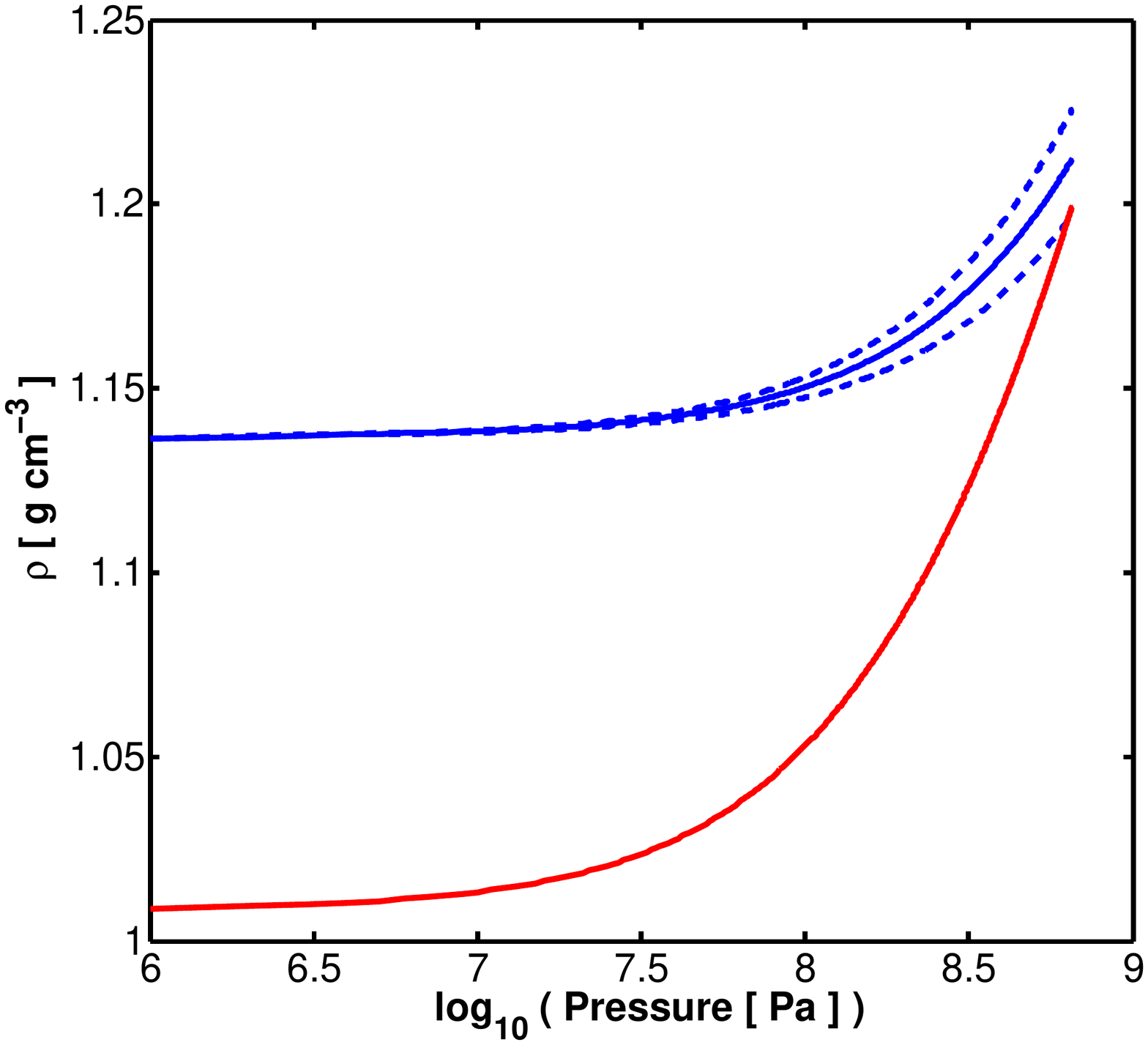}
\caption{\footnotesize{Bulk mass densities spanning the depth of the ocean for an isotherm of $275$\,K. Solid red curve is a water rich liquid assumed saturated with CO$_2$. The chosen isotherm is within the SI CO$_2$ clathrate hydrate stability field. Therefore the solubility of CO$_2$ is governed by the equilibrium with the clathrate phase. Solid blue curve is the bulk mass density of SI CO$_2$ clathrate hydrate for a bulk modulus of $8.5$\,GPa. Upper and lower dashed blue curves are the bulk mass densities of SI CO$_2$ clathrate hydrate for bulk moduli of $7$\,GPa and $11$\,GPa respectively.}}
\label{fig:DensityComparison1}
\end{figure}

For the case that more vigorous geological forces are at work resulting in a flux of internal CO$_2$ into the ocean that is kept higher than what the "gentle" mechanism prescribes, then the ocean may try to over-saturate with CO$_2$. The outcome of this over-saturation depends on the bulk mass densities of SI CO$_2$ clathrate grains and the ocean's water rich liquid. In fig.\ref{fig:DensityComparison1} the ocean's water rich liquid is considered saturated with CO$_2$. In the $\alpha$ domain saturation is the solubility of CO$_2$ when in equilibrium with the clathrate hydrate phase. The mass density correction to the pure liquid water mass density due to the dissolved CO$_2$ is derived using the work of \cite{Teng1997}.
We estimate the bulk mass density of a CO$_2$ SI clathrate hydrate grain as:
\begin{equation}
\rho_{clath} = \frac{46m_{w}+2y^{co_2}_{small}m_{co_2}+6y^{co_2}_{large}m_{co_2}}{V_{cell}}
\end{equation}
Here $m_{w}$ and $m_{co_2}$ are the masses of a water molecule and of a CO$_2$ molecule respectively. The definitions of the other parameters are the same as in subsection $2.2$. See also subsection $2.2$ for a discussion over the uncertainty in the clathrate hydrate bulk modulus.      

From fig.\ref{fig:DensityComparison1} we see that for $\alpha$ domain temperatures the SI CO$_2$ clathrate grain is more dense than the water rich liquid across the entire ocean's depth. Consequently, if the ocean tries to over-saturate (reaching CO$_2$ concentrations above the solubility in equilibrium with clathrates, see for example point C in fig.\ref{fig:IsothermSolubilityProfile}) \textit{the excess CO$_2$ outgassed from the ice mantle and into the ocean would form CO$_2$ SI clathrate grains. These grains will sink due to their high density and pile up on the ocean's bottom, rather then reach the atmosphere}.

\begin{figure}[ht]
\centering
\includegraphics[scale=0.6]{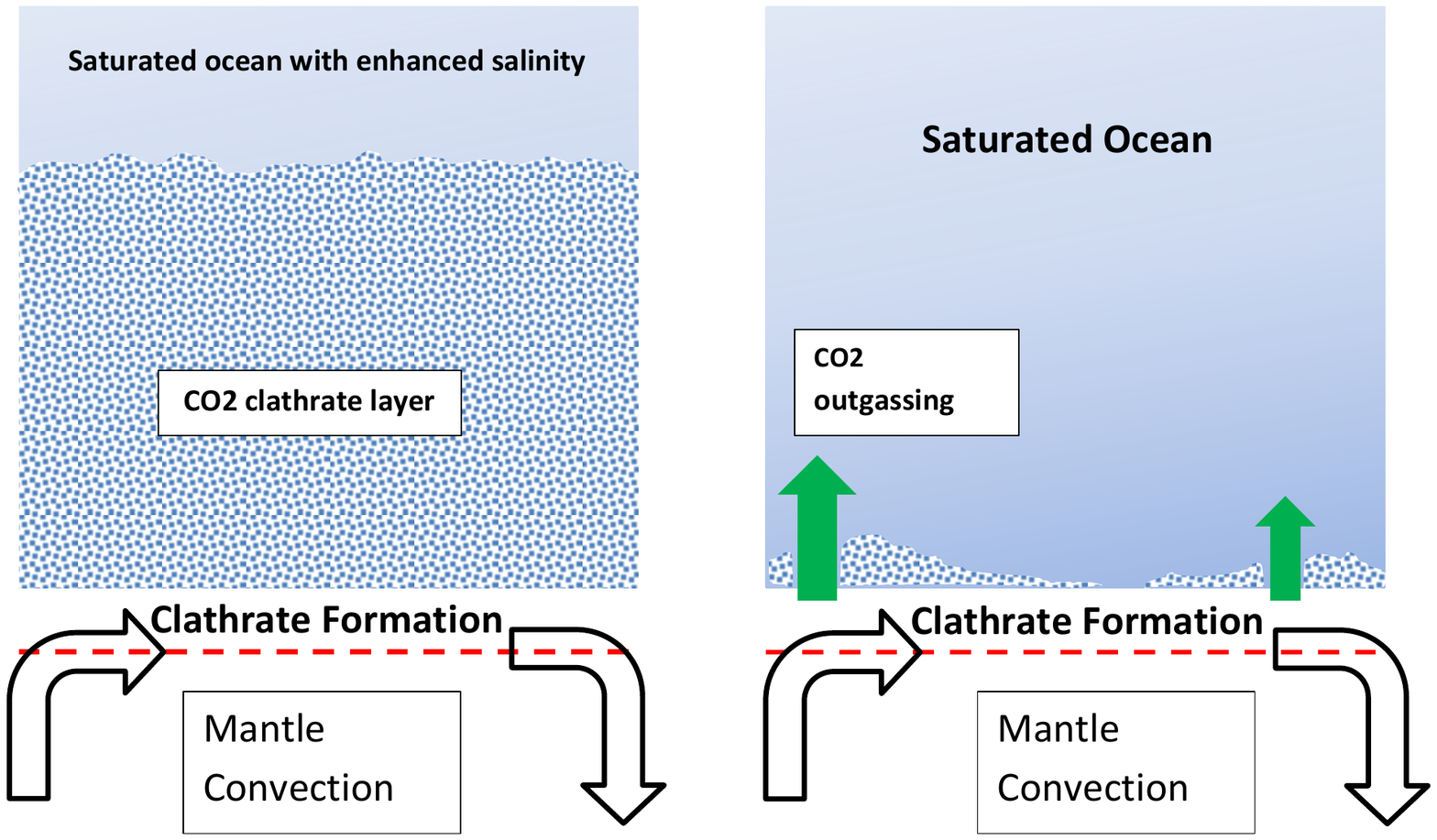}
\caption{\footnotesize{Here we present two illustrations in the $\alpha$ domain, for the case where the flux of CO$_2$ from the mantle and into the ocean was enough to saturate the ocean (The concentration of CO$_2$ reached the solubility value in equilibrium with clathrates). Any further CO$_2$ entering the ocean will thus sink as clathrate grains. In the right panel the flux of CO$_2$ is low enough so that even after billions of years little clathrate hydrate was accumulated around active CO$_2$ sources. In the left panel the flux was high enough so that the entire CO$_2$ SI clathrate hydrate stability field indeed solidified. In the latter case what remains of the ocean has a higher salinity than the original deep ocean.}}
\label{fig:HydrateBuildUp}
\end{figure}

As a simple example we consider a constant CO$_2$ flux from the mantle and into the ocean. Such a flux should eventually saturate the ocean initiating an inner oceanic "rain" of sinking clathrate grains. With time these will thicken the clathrate hydrate layer already composing the ocean's bottom surface.   
This constant flux may be low and consequently the SI CO$_2$ clathrate hydrate layer that will pile up on the ocean's bottom, in geological time, will be quite thin (see right panel in fig.\ref{fig:HydrateBuildUp}). On the other hand the constant flux may be high enough and the pile up of clathrate grains on the bottom surface fast enough so that in a geological time scale most of the \textit{ocean solidifies} as a single global clathrate layer (see left panel in fig.\ref{fig:HydrateBuildUp}). In the latter scenario any further outgassing of CO$_2$ will have to end up in the atmosphere.

The constant flux model can be quantified: Let us assume the ocean became saturated with CO$_2$ (with respect to equilibrium with clathrates) at $t=0$. The rate with which water molecules from the ocean solidify due to formation of clathrate grains is:
\begin{equation}
5.75F_{co_2}4\pi R^2_p
\end{equation}
where $R_p$ is the planetary radius and the constant flux of CO$_2$ from the mantle and into the ocean is $F_{co_2}$.  We also considered that in a full clathrate crystal every CO$_2$ molecule requires $5.75$ water molecules. Due to the growing hydrate layer the ocean's mass, $M_{ocean}$, will reduce with time according to:
\begin{equation}
\frac{dM_{ocean}}{dt}=-5.75F_{co_2}4\pi R^2_pm_w
\end{equation}
Here $m_w$ is the mass of a water molecule. Solving for the last equation one may obtain:
\begin{equation}
\frac{M_{ocean}(t)}{M_{ocean}(t=0)}=1-5.75\frac{F_{co_2}m_w}{\rho_{ocean}L_{ocean}(t=0)}t
\end{equation}
where $\rho_{ocean}$ is the ocean's bulk mass density and $L_{ocean}(t=0)$ is the ocean's initial depth for which we assume a value of $100$\,km. For example, a constant global CO$_2$ flux of $1.86\times 10^{11}$\,molec\,cm$^{-2}$\,s$^{-1}$ will transform ten percent of the oceans' initial mass into clathrate hydrate in $1$\,Gyr. 
The flux given here is global, it can be much higher locally in case the geological activity driving the CO$_2$ flux into the ocean is geographically confined to certain areas. Also, when we say total ocean solidification we still do mean inside the CO$_2$ SI clathrate thermodynamic stability field. This means that depending on the atmospheric pressure there could still remain a narrow liquid shell at the top of the former ocean composed of liquid water saturated in CO$_2$. This surviving aqueous layer will have an enhanced salinity due to salts not going into clathrates. Therefore, even if the ocean initially had very low concentrations of strong electrolytes the little salt that was present will be more concentrated in the remaining thin liquid layer. This may have consequences for the ability to form and sustain life. 

Finally, let us consider the $275$\,K isotherm in the $\alpha$ domain (see fig.\ref{fig:alphabetagamma}), and solve for the particular end scenario where the CO$_2$ flux from the mantle and into the ocean was high enough so that the entire clathrate stability field indeed solidified. We further assume that any additional CO$_2$ that outgassed from within ended up in the atmosphere, due to the exhaustion of the ocean's ability to sink CO$_2$ as clathrate. For this isotherm the CO$_2$ SI clathrate hydrate thermodynamic stability field spans the pressure range of $1.6$\,MPa to $0.75$\,GPa, though the ocean's bottom is at $0.68$\,GPa. Assuming a gravitational acceleration of $10^3$\,cm\,s$^{-2}$ the pressure range from the bottom of the ocean outward till clathrates cease to be stable corresponds to $57$\,km. Let us further assume the planetary radius is $8000$\,km \citep{Levi2014} then the mass of the ocean within the clathrate stability field is approximately $4.6\times 10^{25}$\,g. From this we know how many moles of water were in the original subsaturated ocean inside the clathrate stability field. Now in clathrate formation every mole of CO$_2$ requires $5.75$ moles of water, so the total mass of CO$_2$ stored in this \textit{maximum} clathrate hydrate layer is $2.0\times 10^{25}$\,g. This comes at the expense of the water in the ocean. This is the total capacity of this proposed CO$_2$ reservoir for our particular example. It is interesting to note that the carbon budgets in rocks and in the ocean for the Earth are: $7.4\times 10^{22}$\,g and $3.8\times 10^{19}$\,g, respectively \citep{CarbonCycle}.
If the CO$_2$ atmosphere that forms around the planet has a partial pressure of $1$\,bar ($10$\,bar) then the remaining liquid layer (what is left of the ocean after the entire clathrate stability field solidified) has a depth of $150$\,m ($60$\,m). Since clathrates do not occlude salt, the entire salt content of the original ocean now concentrates at the remaining liquid layer. Therefore, since the original ocean which had a depth of $\approx 100$\,km shrunk to a liquid reservoir whose depth is $\approx 100$\,m the latter layer experiences a three order of magnitude rise in salt concentration with respect to the initial ocean. 

\subsection{The $\beta$ Domain}

In this domain the ice layer composing the bottom surface of the ocean is outside of the thermodynamic stability field for CO$_2$ SI clathrate hydrate. Thus, even in the presence of CO$_2$ the top of the icy mantle convection cell (the ice layer composing the bottom of the ocean) is largely made of water ice VI. Therefore, the "gentle" outgassing mechanism proposed for the $\alpha$ domain can not operate here. It is uncertain whether the filled ice of CO$_2$ is stable at the $\beta$ domain range of temperatures \citep{Bollengier2013,Tulk2014}. Its existence at the bottom of the ocean in this domain is therefore speculative. We elaborate further on this issue in the discussion. 

In the $\beta$ domain there is a region of space right above the bottom of the ocean which is outside of the CO$_2$ SI clathrate thermodynamic stability field. In this region of space the lowest chemical potential for CO$_2$ is for a phase I solid of CO$_2$. Further out of this region CO$_2$ SI clathrates become thermodynamically stable. The latter may extend to the point where the lowest chemical potential for CO$_2$ turns to be the liquid form of CO$_2$. Understanding the deposition of CO$_2$ in the deep ocean for the $\beta$ domain one has to consider the mass densities of the different phases involved.  

\begin{figure}[ht]
\centering
\includegraphics[trim=0.2cm 4cm 0.15cm 5cm , scale=0.60, clip]{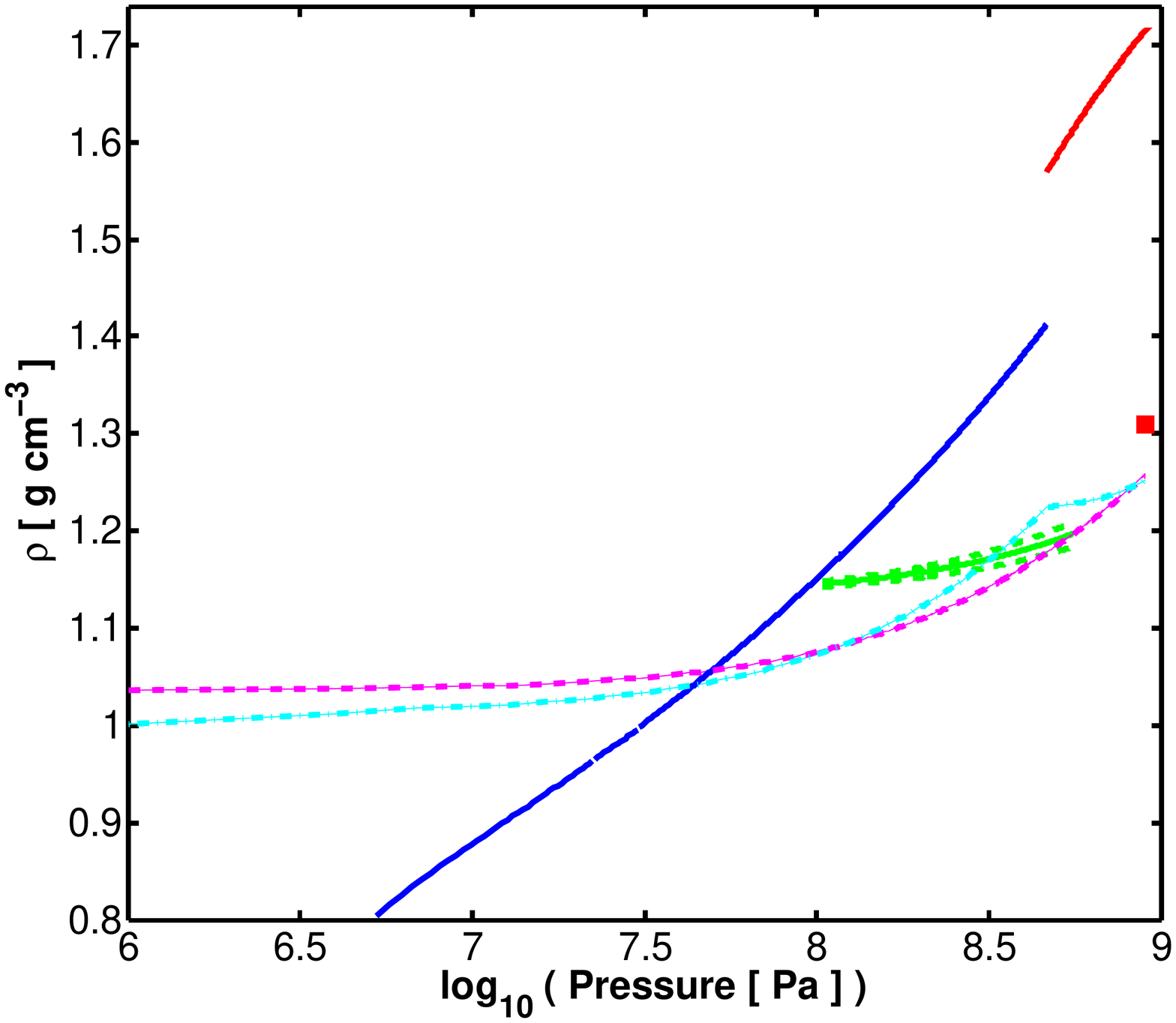}
\caption{\footnotesize{Bulk mass densities spanning the depth of the ocean for various phases and for an isotherm of $290$\,K. Solid red curve is the mass density for the phase I solid of CO$_2$. Solid blue curve is the mass density of liquid CO$_2$. Solid green curve is the mass density of SI CO$_2$ clathrate hydrate for a bulk modulus of $8.5$\,GPa. Upper and lower dashed green curves are the mass density of SI CO$_2$ clathrate hydrate for a bulk modulus of $7$\,GPa and $11$\,GPa, respectively. Dashed magenta is the mass density of water rich liquid saturated with CO$_2$ according to the equilibrium with the clathrate phase. Dashed-dotted cyan curve is the mass density of water rich liquid saturated with CO$_2$ according to the equilibrium with fluid CO$_2$, i.e. Henry's law. Red square is the mass density of pure water ice VI \citep{Kamb1965}. Each curve spans the thermodynamic stability field of the appropriate phase for the chosen isotherm.}}
\label{fig:DensityComparison2}
\end{figure}

In fig.\ref{fig:DensityComparison2} we plot the mass density for various phases of interest for the $\beta$ domain assuming an isotherm of $290$\,K. Each curve spans the thermodynamic stability of the given phase for the isotherm chosen. This figure sheds light on what is likely a complex deposition mechanism. Let's imagine an ocean initially subsaturated in CO$_2$. As solid convection in the ice mantle ensues CO$_2$ trapped within ice VI (or perhaps as filled-ice) comes into contact with the ocean. This CO$_2$ enters the ocean trying to saturate it. The solubility when in equilibrium with clathrates is a lower value than when in equilibrium with fluid CO$_2$ (see fig.\ref{fig:IsothermSolubilityProfile}). Therefore, after the ocean saturates with CO$_2$, to the value in equilibrium with clathrates, any further dissolution of CO$_2$ into the ocean from the interior would result in SI CO$_2$ clathrate grain formation. Clathrate grain formation would be restricted to the clathrate thermodynamic stability field. For our example isotherm the clathrate stability field extends $38$\,km, for an ocean which is $90$\,km deep. The clathrate stability field is elevated $28$\,km above the ice VI bottom and submerged $24$\,km below the ocean's surface.  

\begin{figure}[ht]
\centering
\includegraphics[trim=0.2cm 4cm 0.15cm 5cm , scale=0.60, clip]{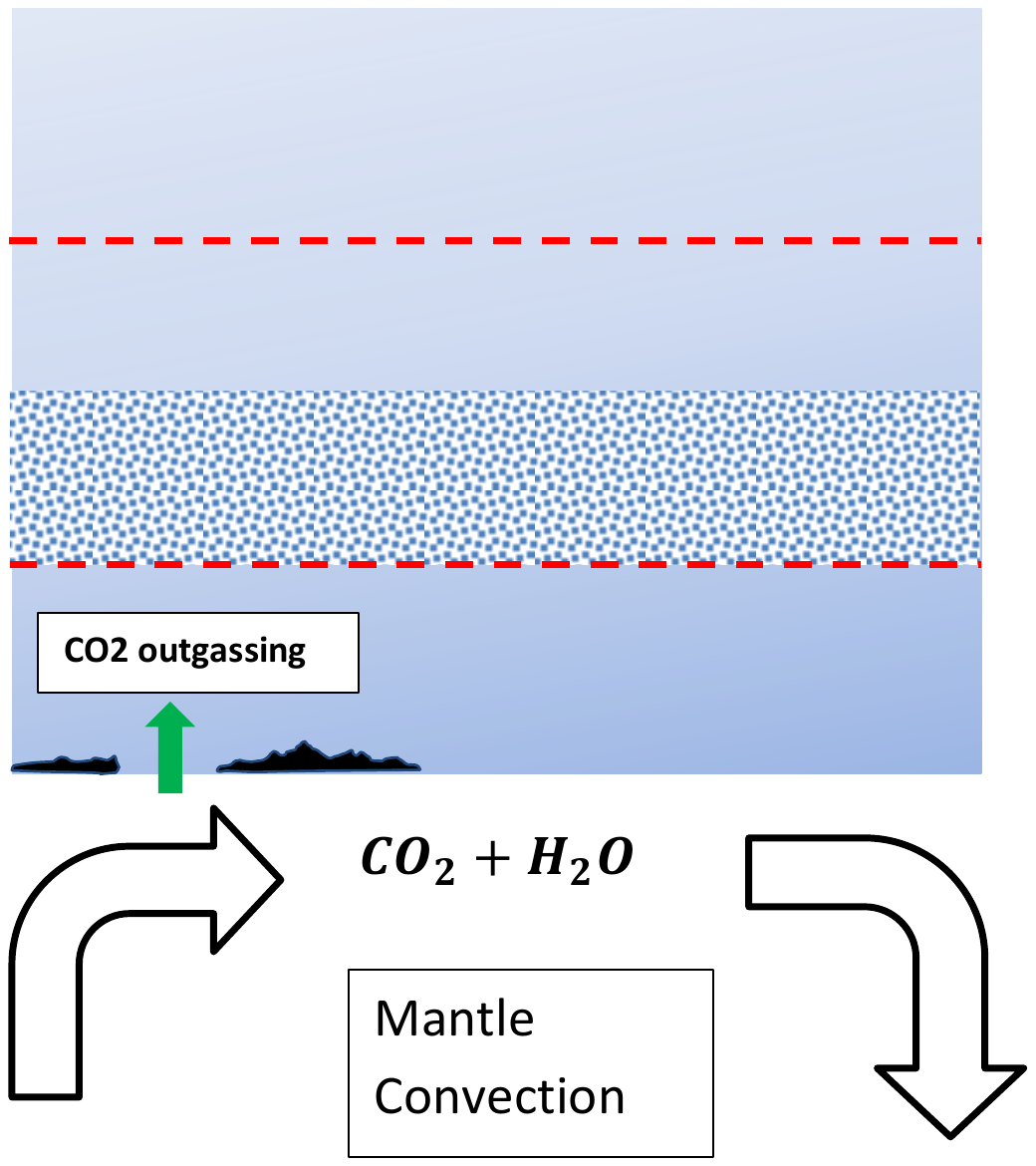}
\caption{\footnotesize{A mantle ice mixture of CO$_2$ and H$_2$O is exposed to the bottom of the ocean. Therefore the ocean becomes enriched with CO$_2$. Supersaturating the ocean with CO$_2$, with respect to the solubility when in equilibrium with the clathrate hydrate phase, will initiate clathrate grain formation within the clathrate hydrate thermodynamic stability field (between the two horizontal dashed red lines). The deep ocean becomes more dense than the grains because of the dissolved CO$_2$, thus making the clathrate grains float, forming a mid-ocean layer (patterned bar). A thick mid-ocean layer would separate the upper and lower parts of the ocean. Further enrichment of the lower ocean with CO$_2$ may result in the formation of dense solid CO$_2$ grains (blackened regions).}}
\label{fig:BetaStrata}
\end{figure}

Within the thermodynamic stability field of the SI CO$_2$ clathrate hydrate the clathrate grains are more dense than the surrounding water rich liquid. As a result, if supersaturation with respect to clathrates is forced, CO$_2$ clathrate grains would form and sink. However, very close to the high pressure boundary of the clathrate stability field the water rich liquid turns more dense than the clathrate grains (see fig.\ref{fig:DensityComparison2}). As a result SI CO$_2$ clathrate grains will begin to accumulate there, $28$\,km above the ocean's bottom. As more and more CO$_2$ is injected into the ocean from the interior, and as long as the clathrate layer is thin enough to allow CO$_2$ to diffuse across it, the thicker this elevated clathrate layer becomes. Eventually, if it becomes thick enough, it may isolate the deep ocean from the upper ocean. See illustration in fig.\ref{fig:BetaStrata}. Because a liquid layer separates this proposed mid-ocean solid SI CO$_2$ clathrate hydrate layer from the ice mantle it should experience only a mild shear stress, enhancing its stability. 

If the mid-ocean SI CO$_2$ clathrate hydrate layer indeed becomes thick enough to isolate the deep part of the ocean, then that part of the ocean may experience a further increase in the abundance of dissolved CO$_2$, in the case its transport from the mantle and into the ocean continues. This is because these $28$\,km of water rich liquid above the ocean's bottom are outside of the clathrate hydrate stability field. Now saturation with respect to pure CO$_2$ can be reached resulting in the formation of solid CO$_2$ grains. These grains are even more dense than water ice VI. Therefore, they probably become embedded in every crack and void forming in the ocean's bottom surface. These are likely since the ocean's bottom is also the top layer of the ice mantle convection cell, thus experiencing high stresses. In a previous paper we have discussed full ice mantle convection \citep{Levi2014}. If that is the case here as well then it is likely that at least some part of the ocean's ice VI bottom is reprocessed into the interior. In that case the embedded solid CO$_2$ will likely follow.

In the upper part of the ocean, above the mid-ocean clathrate hydrate layer, the solubility of CO$_2$ is governed by the equilibrium with the clathrate hydrate phase. Thus, the solubility of CO$_2$ is kept low enough to prohibit the formation of liquid CO$_2$ droplets (see fig.\ref{fig:IsothermSolubilityProfile}). Because of restrictions on the solubility of CO$_2$ when in equilibrium with clathrates it is unlikely that liquid CO$_2$ should form anywhere. 
In the event that liquid CO$_2$ does form, for example, between clathrate grain boundaries the mass density difference should drive it to flow out of the clathrate layer. Consequently, sinking into the deep ocean and transforming into the phase I solid of CO$_2$.   

In conclusion, if a thick mid-ocean SI CO$_2$ clathrate hydrate layer forms it would control the solubility of CO$_2$ in the upper ocean. Therefore, it would also control the atmospheric abundance of CO$_2$. \textit{Atmospheric observations would therefore mostly probe this layer rather than constrain deeper planetary fluxes}. Also, we find the $\beta$ domain resembles the $\alpha$ domain, where in both cases clathrates of CO$_2$ may dictate the abundance of CO$_2$ available to the atmosphere.

Finally, we wish to quantify the mass of CO$_2$ that can be stored in a full mid-ocean clathrate hydrate layer. We use the $290$\,K isotherm as an example (see fig.\ref{fig:alphabetagamma}). In addition, we assume a gravitational acceleration of $10^3$\,cm\,s$^{-2}$ and a planetary radius of $8000$\,km for our water planet \citep{Levi2014}. For $290$\,K the CO$_2$ SI clathrate hydrate thermodynamic stability field lies between the pressures of $0.55$\,GPa and $104$\,MPa corresponding to a layer width of $38$\,km. The total mass of this clathrate layer is approximately $3.58\times 10^{25}$\,g of which $1.07\times 10^{25}$\,g is CO$_2$.

\subsection{The $\gamma$ Domain}

In this domain the thermal profile in the deep-mid ocean is outside of the CO$_2$ SI clathrate hydrate thermodynamic stability field (see fig.\ref{fig:alphabetagamma}). Ocean saturation levels with CO$_2$ are now determined via equilibrium with non-clathrate phases. Following saturation any additional CO$_2$ convected outward from the mantle and into the ocean may pile up on the bottom surface of the ocean as solid and even liquid CO$_2$. 

Estimating the CO$_2$ storage capacity in the deep ocean is much more complicated in this domain. In the $\alpha$ and $\beta$ domains CO$_2$ is stored in the form of clathrate hydrate. This phase is more dense than the liquid ocean but less dense than ice VI and is therefore gravitationally stable. Liquid and solid CO$_2$ are more dense than ice VI (see fig.\ref{fig:DensityComparison2}). Therefore, in the $\gamma$ domain, gravity will put an upper bound on the storage capacity in the deep ocean. Hypothetically, disregarding this gravitational limit the maximum amount of CO$_2$ that can be stored in the ocean in the solid and liquid phases is approximately $8\times 10^{25}$\,g,
assuming $10^3$\,cm\,s$^{-2}$ for the acceleration of gravity and $8000$\,km for the planetary radius. 
    
In this section we have described ways in which CO$_2$ from the ice mantle may become locked deep in the ocean. Are these deep ocean reservoirs stable, or can the ocean mix thus making them available to the atmosphere?

\section{DEEP OVERTURNING CIRCULATION IN THE OCEAN}\label{sec:Circulation}

In this section we would like to briefly address the issue of global mixing in water planet oceans. 
When a fluid in a gravitational field is heated from below the gravitational energy acts against viscous dissipation to establish convection. When the heating source is at the same level or above the cooling source convection cannot develop. This is known as Sandstr\"{o}m's theorem \citep{Huang1999}. The ocean is such a system, it is both heated and cooled from its outer surface \citep{Kuhlbrodt2007}. Therefore, the Atlantic Overturning Circulation (AOC) rather then being a heat engine requires an external input of power to maintain a steady circulation. In the AOC warm and light surface water flows to the north pole. At the north pole and in sporadic locations the surface water cools enough and partially solidifies. Since ice Ih does not incorporate salt the remaining liquid becomes saltier and thus heavier. At this point the heavy water sinks to the bottom surface and spreads. At the tropics the cold and dense abyssal ocean upwells through a warmer and less dense liquid environment. This last arm of the AOC costs energy which is supplied by Lunar and Solar tidal forcing and winds. The power supplied by either tides or winds, that actually converts to upwelling, is estimated to have the same order of magnitude of about $1$\,TW. This is similar to the power required to maintain a steady state AOC \citep{Wunsch2004,Kuhlbrodt2007}. 

The pathways for the transfer of energy in Earth's oceanic circulation are still very much debated. However, it is clear that continental slopes and bottom surface topography play a major role in internal wave dissipation into turbulent mixing. Indeed vertical motions are enhanced above rough topography \citep{Kuhlbrodt2007}. Water planets do not have continental slopes. In Earth's oceans topographic features can be as high as the depth of the ocean (e.g. Mauna Kea). In water planets this is less likely due to the large depth scale of the ocean. 
Thus, the efficiency with which waves dissipate into turbulent mixing is probably lower in water planets' oceans than in the Atlantic. As a consequence of the subdued topography tidal forcing becomes less effective as a source of energy for maintaining a circulation.
 
The rate of direct input of mechanical energy from winds is proportional to wind speed and surface area. The surface area is not vastly different between our studied water planets and the Earth. Wind speeds are more difficult to assess, though the lack of continents in water planets may indicate lower wind speeds (we will return to this below when we discuss the wind-driven circulation). As a consequence of the above the power available to establish circulation in water planets' oceans is probably about $1$\,TW, as in the case of Earth. But how much power is actually needed to circulate a deep water planet ocean?

In the AOC surface water sinks to replace abyssal water and abyssal water is upwelled to replace surface water. One can think of circulation as a process where the outer oceanic water shell is continuously being exchanged with the deepest oceanic water shell.    
To derive the power required to maintain such a steady circulation we follow the arguments of \cite{Wunsch2004}. Let us consider a parcel of fluid of volume d$V$ and a potential density of $\sigma_p$. When the parcel is displaced vertically in a gravitational field through a medium of potential density $\sigma(z)$ one invests/gains a power of:
\begin{equation}
d\dot{U}_g=(\sigma_p-\sigma(z))\vec{g}\cdot\vec{w}dV
\end{equation} 
where $\vec{g}$ is the acceleration of gravity and $\vec{w}$ is the parcel's vertical velocity. For a constant vertical velocity the global energy rate is:
\begin{equation}\label{GlobalEnergyRate}
\dot{U}_g=4\pi R^2_p\vec{g}\cdot\vec{w}\int^0_{z_b}(\sigma_p-\sigma(z))dz
\end{equation}
where $R_p$ is the planetary radius, $z=0$ is the ocean's upper surface and $z_b$ is the ocean's bottom surface. Let us first estimate the vertical velocity of the circulation.

We consider a tangential outer surface current velocity of $v_e$. The latter describes the current regime in the upper boundary layer of the ocean of depth $d_e$. Due to the role Ekman transport plays in the general circulation we adopt for $v_e$ and $d_e$ the corresponding Ekman layer values. From conservation of mass we then have:
\begin{equation}
\frac{w}{v_e}\approx\frac{d_e}{R_p}
\end{equation}
The appropriate Ekman layer depth is \citep{Kundu2012fluid}:
\begin{equation}
d_e=\sqrt{\frac{2\nu_v}{f}}
\end{equation}
where $\nu_v$ is the vertical eddy viscosity and $f\equiv 2\Omega\sin\lambda$ is the Coriolis parameter. The Ekman velocity is \citep{Kundu2012fluid}:
\begin{equation}
v_e\approx\sqrt{\frac{2}{f\nu_v}}\frac{\tau}{\rho_w}
\end{equation}
Here $\rho_w$ is the bulk density of ocean water and $\tau$ is the shear stress the wind exerts on the surface water \citep{PhysicalOceanography}:
\begin{equation}
\tau = \rho_{air}C_DU^2_{10}
\end{equation} 
The drag coefficient $C_D$, appropriate for moderate to strong winds, is estimated to equal $2.6\times 10^{-3}$ \citep[see discussion on][page $490$]{Sverdrup1957oceans}. $U_{10}$ is the wind speed at $10$\,m above see level, and $\rho_{air}$ is the bulk density of the atmosphere. We therefore approximate the vertical velocity of ocean circulation as:
\begin{equation}
w\approx\frac{C_D}{\sin\lambda}\frac{\rho_{air}}{\rho_w}\frac{U^2_{10}}{\Omega R_p}
\end{equation} 
Let's test the last approximation with respect to circulation in Earth's oceans. For the bulk density of air we adopt $\rho_{air}=1.25\times 10^{-3}$\,g\,cm$^3$ \citep{PhysicalOceanography}. $R_p=6371$\,km and $\Omega=7.27\times 10^{-5}$\,s$^{-1}$ are Earth's radius and angular velocity respectively. We also adopt near surface wind speeds of $5-10$\,m\,s$^{-1}$ \citep{PhysicalOceanography}. Giving for mid-latitudes a vertical velocity between $10^{-5}$\,cm\,s$^{-1}$-$10^{-4}$\,cm\,s$^{-1}$. Indeed, vertical velocities for Earth's oceans far from sharp topographic features are inferred to fall within this range of values \citep{Wunsch2004}. 

Sinking of surface water to the abyss takes place where it costs the least amount of energy. For the reasons described above for the AOC this is at the poles. In water planets a substantial ice mantle separates between the ocean and the rocky interior. If this lack of interaction results in much lower salt concentrations in the ocean then sea-ice formation will not help promote subduction of surface water. In addition, in our studied water planets surface water are lighter than abyssal water because they contain much less dissolved carbon-dioxide. At the tropics due to the high surface temperatures surface water expands and enhances stability even further. Therefore it is energetically more efficient for water to sink at the poles in water planets as well. 
Let us estimate the external energy needed to maintain the down-welling arm of the circulation at the poles. 

For a vertical eddy viscosity of $300$\,cm$^2$\,s$^{-1}$ \citep{Yu1991} the Ekman layer is $24$\,m deep. This boundary layer at the ocean's surface is very well mixed. Therefore, the solubility of CO$_2$ may be assumed uniform throughout this layer. Let's assume for our water planet a $3$\,bar atmosphere of CO$_2$. According to Henry's law (see subsection $2.1$) this fits a dissolved CO$_2$ abundance of $0.0017$ for $25^\circ$C. This represents our surface tropical water that can now migrate to the poles. Closer to the poles the surface temperature decreases. Let's also assume that at the poles the exposed surface water is at the verge of freezing. Therefore, the solubility of CO$_2$ at the surface water further increases to about $0.0036$. In the cold abyss (assumed here as $275$\,K, see $\alpha$ domain in the previous section), and within the stability field of clathrates, the abundance of dissolved CO$_2$ would be about $0.016$. Converting the surface and abyssal water compositions to the corresponding solutions' bulk potential densities \citep{Teng1997}, for a $1$\,bar reference pressure, yields $\sigma_p-\sigma(z)=-0.0062$\,g\,cm$^{-3}$. Therefore, with the aid of Eq.($\ref{GlobalEnergyRate}$), the external power required to maintain a steady global circulation is at least:
\begin{equation}
\dot{U}_g > 349-3490\quad [TW]
\end{equation} 
where the range comes from the range of vertical velocities derived above. This is two to three orders of magnitude larger than what is actually available.

As in ocean's on Earth, lifting cold and dense deep water through the warmer shallow water generates gravitational potential energy. The rate of potential energy generation, using Eq.($\ref{GlobalEnergyRate}$), can crudely be described as \citep{Wunsch2004}:
\begin{equation}
\dot{U}_g \sim 4\pi R^2_pgw\frac{\partial\sigma}{\partial T}\Delta{T}L_{ocean} 
\end{equation} 
Let's consider a temperature variation of $\Delta{T}=10$\,K over the depth of the ocean, $L_{ocean}$. Raising the temperature from $275$\,K to $285$\,K, for a $1$\,bar reference, decreases the density of pure water by $4.2\times 10^{-4}$\,g\,cm$^{-3}$ \citep{wagner94}. Adopting Earth's parameters and a shallow $4$\,km ocean gives about $0.9$\,TW for the lower value derived for the vertical velocity. This agrees with the available external input of power as discussed above. For oceans that are at least an order of magnitude deeper, circulating water against the temperature gradient as in the case of Earth could prove too expensive energetically.      

In conclusion, \textit{driving a global oceanic circulation and ocean homogenization in water planets, with the same efficiency as it operates on Earth, requires much more external power}. Simply scaling up Earth's oceanic circulation and global vertical mixing for describing water planets' oceans is inappropriate. Due to the exploratory nature of this paper we do not dismiss vertical mixing entirely but rather explore the effects of various deep ocean vertical eddy diffusion coefficients. However, in light of the energy constrains just described it is reasonable to estimate for the vertical diffusion coefficient values that are lower than inferred values for Earth's deep oceans.

\section{CO$_2$ FLUXES BETWEEN THE ATMOSPHERE AND UPPER OCEAN}\label{sec:Fluxes}

Coupled ocean-atmosphere models for our studied water planets have not yet been developed. The principal difficulty is that every physical phenomenon we wish to consider based on analogy to the Earth needs to be re-evaluated. For example, basic issues like global ocean circulation may differ substantially between water planets and Earth's analogues. Recently coupled ocean-atmosphere models were tested for a planet with a rather shallow ocean (Earth-like) and no land mass \citep{Smith2006,Marshall2007}. This planetary case is intermediate between Earth and our studied water planets with their very deep oceans. In both works global circulation is assumed, which is energetically reasonable for a shallow ocean, however they still ended up with different results. \cite{Smith2006} argued that in the absence of land the meridional temperature gradient is much subdued, i.e. polar temperatures are high. On the contrary, \cite{Marshall2007} argued that the poleward heat transport in \cite{Smith2006} was much exaggerated, they further reported that sea-ice formed and was very stable throughout the entire time duration of their numerical run which was thousands of years. They found the ice caps to extend as far south as $60^\circ$ latitude and have temperatures as low as $250$\,K.
Both groups find marked wind patters of easterlies and westerlies as on Earth's surface. A weaker meridional temperature gradient would weaken the winds. However the lack of land mass is argued to reduce viscosity which would tend to cancel the former effect \citep{Marshall2007}. It is interesting to note that \cite{Marshall2007} found no polar easterlies in their model runs. 

\begin{figure}[ht]
\centering
\includegraphics[trim=0.15cm 4.3cm 0.2cm 3cm , scale=0.55, clip]{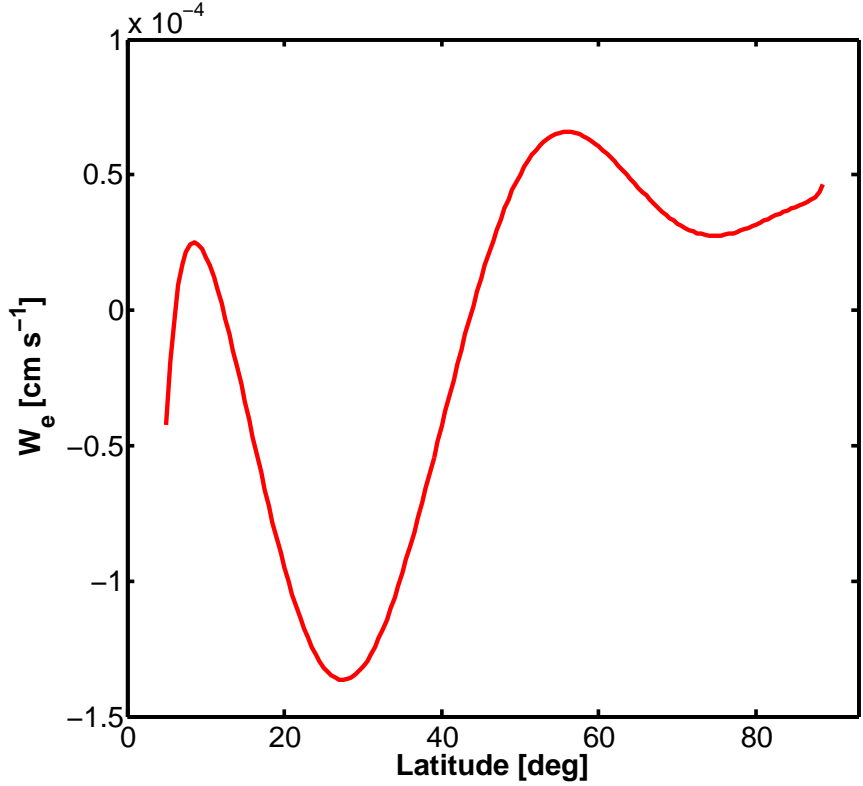}
\caption{\footnotesize{Vertical Ekman velocities versus latitude for an Earth sized planet without continents.}}
\label{fig:EkmanPumpVel}
\end{figure}

In light of these models it seems reasonable to assume that in water planets as well excess solar heating at the tropics would tend to form circulation cells in the atmosphere. Resulting surface wind patterns of easterlies and westerlies seem to be robust as well. These surface wind patterns are of paramount importance since they force surface water to either converge or diverge. The convergences and divergences drive vertical motions known as Ekman pumping in the upper ocean \citep{Gill1982}. The relation between the surface wind stress ($\tau$) and vertical flow in the ocean is:
\begin{equation}
\vec{w}_e=\frac{1}{2\Omega\rho_w}\vec{\nabla}\times\left(\frac{\vec{\tau}(\lambda)}{\sin(\lambda)}\right)=-\frac{\hat{r}}{2\Omega\rho_wR_p\cos(\lambda)}\frac{d}{d\lambda}\left[\tau(\lambda)\cot(\lambda)\right]
\end{equation}
where $\Omega=7.27\times 10^{-5}$\,s$^{-1}$ is the planetary rotation period, $\rho_w=1$\,g\,cm$^{-3}$ is the ocean surface bulk density, $R_p$ is the planet radius and $\lambda$ is the latitude.
We adopt the surface wind stress variation with latitude from \cite{Marshall2007} for a planet with no continents and convert it to vertical Ekman velocities in the upper ocean. The results are shown in fig.\ref{fig:EkmanPumpVel}. It is clear from fig.\ref{fig:EkmanPumpVel} that downwelling dominates the subtropics whereas upwelling of deeper ocean water is confined to the tropics and subpolar latitudes.

We do not wish to suggest that this profile of Ekman pumping is an exact representation of such flows in water planets. An exact Ekamn pumping profile for water planets would require solving a coupled ocean-atmosphere model for such a planetary case. The wind stress patterns described in \cite{Marshall2007} are obviously most appropriate for the planetary case they have solved for. However, since Ekman pumping is the result of basic atmospheric and ocean dynamics it should work in water planets as well. In addition, as long as the winds in water planets do not have substantially and fundamentally different magnitudes and gradients than those blowing across an Earth lacking continents, then vertical motions in the upper most ocean layers of water planets probably have similar magnitudes to those given in fig.\ref{fig:EkmanPumpVel}. Therefore, assuming similar dynamics, our goal in this section is to quantify atmosphere-ocean CO$_2$ fluxes considering polar sea-ice formation and wind-driven circulation.

\subsection{ Wind-Driven Circulation }\label{subsec:FluxesWindDriven}

\begin{figure}[ht]
\centering
\includegraphics[trim=0.15cm 12cm 0.2cm 3cm , scale=0.60, clip]{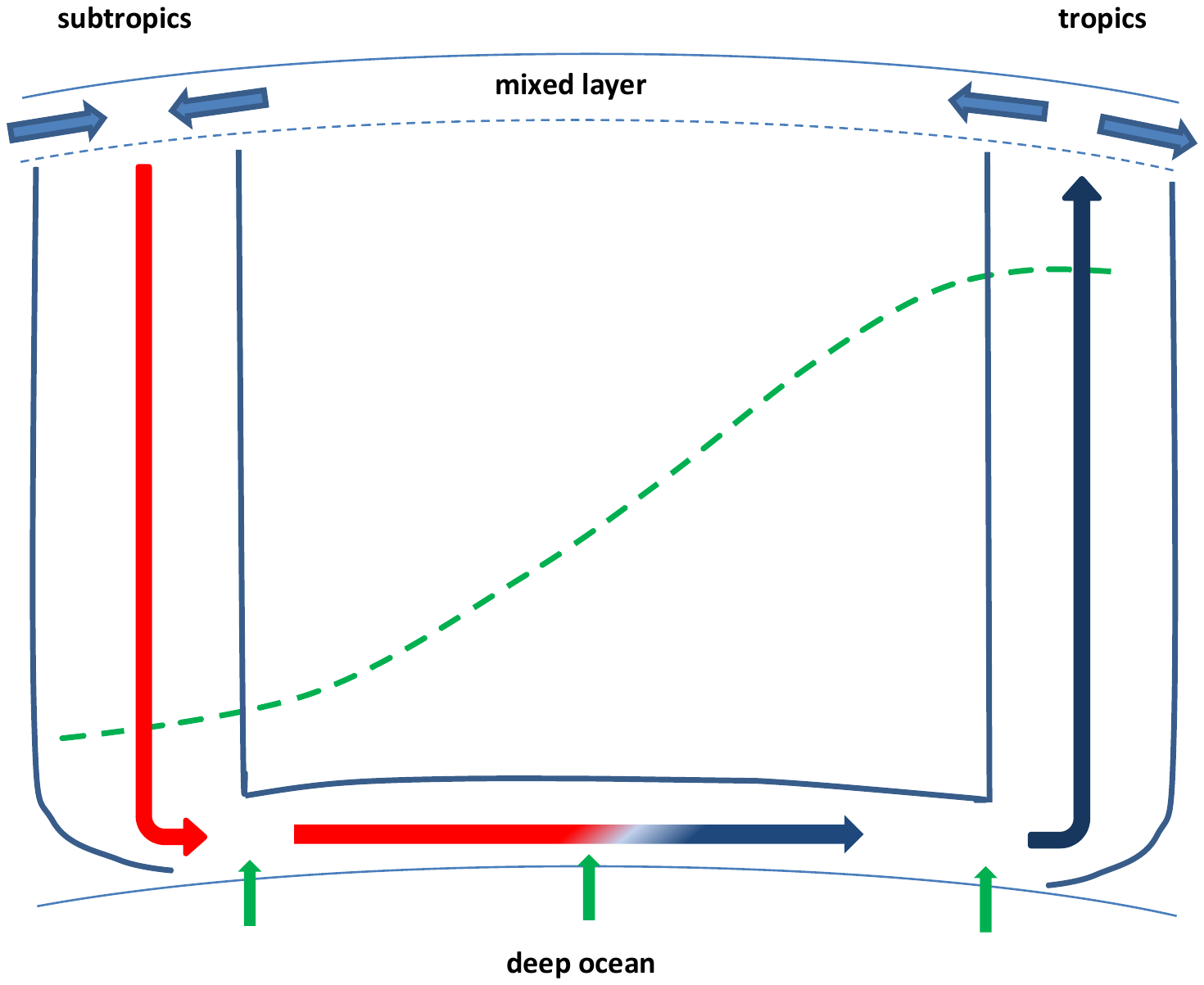}
\caption{\footnotesize{A diagram of the wind-driven circulation in the upper ocean between the tropics and the subtropics. In the subtropics convergence of the Ekman transport (approximately within the mixed layer) results in downwelling (Ekman pumping) of warm surface water. Therefore, the thermocline is depressed downward and so is the clathrate hydrate stability field (dashed green curve). Below the Ekman layer the flow turns geostrophic and Sverdrup balance drives a flow from the subtropics to the tropics. In the diagram we describe a flow tube whose horizontal segment confines the deepest geostrophic streamlines, which is the part of the flow which has diffusional contact (vertical green arrows) with the deep unmixed ocean. In the tropics divergence of the Ekman transport results in upwelling (Ekman suction) of cold water. The upwelled water is now enriched in CO$_2$ due to the contact with the deep ocean. Also the thermocline is lifted and so is the clathrate hydrate stability field. The upwelled water will experience degassing of CO$_2$ primarily when it ascends into the mixed layer.}}
\label{fig:WindDrivenCirculation}
\end{figure}

Circulation in the ocean is often divided between global and wind-driven. In this subsection we will only consider the latter. The setting in fig.\ref{fig:WindDrivenCirculation} is of particular importance to secondary outgassing in water planets following the evaporation of a primordial hydrogen and helium dominated atmosphere.  

In the subtropics, centred around latitude $\lambda_{sbt}\approx 28^\circ$, lies the peak in Ekman transport convergence (see fig.\ref{fig:EkmanPumpVel}). The resulting downwelling from the bottom of the local mixed layer has a velocity of $w_e=1.4\times 10^{-4}$\,cm\,s$^{-1}$. The cross section of our flow tube of interest in its vertical flow segment is $\delta_v$. The subtropic surface temperature of the ocean is $T_{sbt}$, which we assume to be $20^\circ$C. Due to Ekman pumping this will be the temperature in the downwelling water column. This causes a deepening of the local thermocline. Due to the sensitivity of the dissociation pressure of clathrates to the temperature the clathrate thermodynamic stability field will also locally deepen.

Let us first estimate the influx of CO$_2$ molecules in the downwelling arm of the circulation at the expense of the atmosphere. Consider an atmosphere with some partial pressure of carbon-dioxide, $P^{co_2}_{atm}$. The number density of carbon-dioxide dissolved at saturation at the ocean surface is proportional to the atmospheric partial pressure of CO$_2$. The coefficient of proportionality is directly related to Henry's solubility constant and is dependent on the oceanic surface temperature and independent of the atmospheric partial pressure of carbon-dioxide. One may therefore write for the subtropics:
\begin{equation}\label{SolubilitySubtropics}
n^{sat,sbt}_{co_2}(P=P^{co_2}_{atm})=\tilde{\beta}(T_{sbt}) P^{co_2}_{atm}
\end{equation}
where $\tilde{\beta}$ as a function of temperature is here derived from eq.($\ref{Henry}$) since the ocean surface is here assumed outside of the clathrate thermodynamic stability field 

A few tens of meters below the ocean surface is the well mixed layer where composition gradients are small, and the above number density may be assumed constant. Below the mixed layer, which approximately coincides with Ekman's layer, the vertical flow will carry this CO$_2$ inward with a velocity of $w_e$, yielding a carbon-dioxide influx of:
\begin{equation}
\tilde{\beta}(T_{sbt}) P^{co_2}_{atm}w_e
\end{equation} 
However, not all of the above flux comes at the expense of the atmosphere. Some of the CO$_2$ in the downwelling arm originates from CO$_2$ that dissolved at the tropics from upwelled water parcels and then transported to the subtropics. The number density of this can also be derived by use of Eq.($\ref{SolubilitySubtropics}$) except that the surface temperature at the subtropics must be replaced by the surface temperature at the tropics, $T_{trop}$. The influx of CO$_2$ at the subtropics at the expense of the atmospheric budget is therefore:
\begin{equation}
j^{in}_{co_2} = \left(\tilde{\beta}(T_{sbt})-\tilde{\beta}(T_{trop})\right) P^{co_2}_{atm}w_e
\end{equation}        

Outside of the clathrate hydrate stability field the solubility increases with the increasing pressure. Thus, along the downwelling arm the CO$_2$ remains dissolved in the descending water parcels.

Large scale flow below the Ekman layer is dominated by the Coriolis force and horizontal pressure gradients. This is known as geostrophic flow. Sverdrup examined the role of geostrophic flow in the wind-driven circulation and showed there is a return geostrophic mass transport from the subtropics to the tropics. In an ideal Sverdrupian flow there exists a depth, $L_{Sv}$, where horizontal pressure gradients vanish as does the geostrophic and any vertical flow. This depth of no motion is approximately $1$\,km below the ocean surface \citep[see chapter $11$ in][]{PhysicalOceanography}. The depth of no motion is of particular interest to us since it distinguishes between the wind-driven circulated upper ocean and the unmixed deep and cold abyss. Across this boundary CO$_2$ can be diffusionally exchanged between the two parts of the ocean. Since this exchange is our primary interest here we focus on the geostrophic flow at the vicinity of this boundary, and give the geostrophic part of the flow tube a cross section $\delta_h$ that encompasses the bulk of any such diffusional exchange:
\begin{equation}\label{TubeWidthH1}
\delta_h\sim\sqrt{D_{eddy}t}
\end{equation}
Here $D_{eddy}$ is the eddy diffusion coefficient for the deep ocean. We will discuss its value below. We estimate the time $t$ in the last equation as the maximum amount of time a fluid parcel confined to our flow tube stays in diffusional contact with the deep unmixed ocean. In other words, 
\begin{equation}\label{TimeHorizontalFlow}
t\sim\frac{L_g}{u_g}=\frac{R_p\Delta\lambda}{u_g}
\end{equation}   
where $u_g=0.1$\,m\,s$^{-1}$ is the geostrophic velocity (see chapter $10$ in \cite{PhysicalOceanography} and chapter $5$ in \cite{OceanDynamics}), $R_p=8000$\,km is the planetary radius \citep{Levi2014} and $\Delta\lambda= 30^\circ$ is the latitude difference between the subtropics and the tropics. $L_g=4200$\,km is the length of the horizontal geostrophic arm of the flow tube. Therefore, the time scale of fluid parcel cycling through our flow tube is:
\begin{equation}
T_{cyc}=2\frac{L_{Sv}}{w_e}+\frac{L_g}{u_g}=47~[yr]
\end{equation}

Considering that changes to the bulk density due to diffusion of CO$_2$ are much smaller than the difference between the vertical and horizontal flow velocities, mass conservation relates the cross sections of the flow tube in the vertical and horizontal arms:
\begin{equation}\label{WidthRelation}
w_e\delta_v\approx u_g\delta_h
\end{equation}
From Eqs.($\ref{TubeWidthH1}$) and ($\ref{TimeHorizontalFlow}$) we then have:
\begin{equation}
\delta_h\sim\sqrt{D_{eddy}\frac{R_p\Delta\lambda}{u_g}}
\end{equation}
which when inserted into Eq.($\ref{WidthRelation}$) yields:
\begin{equation}
\delta_v\sim\sqrt{D_{eddy}\frac{u_gR_p\Delta\lambda}{w^2_e}}
\end{equation} 

Fluid parcels in the downwelling arm of the flow tube reaching close to the boundary of no motion at $L_{Sv}$ have a CO$_2$ number density given by Eq.($\ref{SolubilitySubtropics}$). There the parcels begin to flow horizontally with geostrophic speed and exchange CO$_2$ with the unmixed deep ocean. When the parcels begin to upwell their dissolved CO$_2$ number density is $n^{out}_{co_2}$, which obeys:
\begin{equation}\label{UpwelledDensity}
n^{out}_{co_2} = n^{sat,sbt}_{co_2}(P=P^{co_2}_{atm})+\frac{1}{\delta_h}\int^{t+\delta t}_tF_{Sv}dt'
\end{equation} 
where $F_{Sv}$ is the flux of CO$_2$ between the parcel and the deep ocean. $t$ and $t+\delta t$ are the times a specific fluid parcel enters and leaves the geostrophic arm of the flow tube, respectively. Solving for $F_{Sv}$ we first need to solve for the transport of CO$_2$ across the deep unmixed ocean.

In a saturated ocean, the nature of the reservoir of CO$_2$ at the ocean's bottom controls the deep ocean's dissolved CO$_2$ number density, $n^{deep}_{co_2}$. We have quantified the various reservoirs in section \ref{sec:Reservoirs} of this paper. Here we estimate the transport of CO$_2$ across the deep ocean as a problem of eddy diffusion between two CO$_2$ number densities: 
$$
\frac{\partial n_{co_2}}{\partial t}=D_{eddy}\frac{\partial^2 n_{co_2}}{\partial z^2}
$$
$$
n_{co_2}(z,t=0)=n_{initial}
$$
$$
n_{co_2}(z=L_{ocean},t)= n^{deep}_{co_2}
$$
\begin{equation}\label{DeepOceanTransport}
n_{co_2}(z=0,t)=n^{sat,sbt}_{co_2}(P=P^{co_2}_{atm})
\end{equation}
where $z=0$ is here the boundary between the wind-driven circulation and the unmixed deep ocean. $L_{ocean}$ is the depth of the ocean. The solution for this system using separation of variables is \citep{Crank1956}:
$$
n_{co_2}(z,t)=n^{sat,sbt}_{co_2}(P=P^{co_2}_{atm})+\frac{n^{deep}_{co_2}-n^{sat,sbt}_{co_2}(P=P^{co_2}_{atm})}{L_{ocean}}z
$$
\begin{equation}\label{SineSeries}
+\sum^\infty_{n=1}C_n\exp\left\lbrace -\frac{n^2\pi ^2}{L^2_{ocean}}D_{eddy}t\right\rbrace \sin\left(\frac{n\pi z}{L_{ocean}}\right)
\end{equation}
where
\begin{equation}
C_n =
\begin{cases}
\frac{2\left(n^{deep}_{co_2}-n^{sat,sbt}_{co_2}(P=P^{co_2}_{atm})\right)}{n\pi} & \text{$n$ even}\\
\frac{2\left(2n_{initial}-n^{sat,sbt}_{co_2}(P=P^{co_2}_{atm})-n^{deep}_{co_2}\right)}{n\pi}  & \text{$n$ odd}
\end{cases}
\end{equation}
Sinusoidal series do not converge fast enough for small $t$, unlike expansions in terms of error functions \citep{Carslaw1959}, and can introduce fictitious fluxes. The solution in terms of complementary error functions is (see appendix \ref{subsec:AppendixC} for derivation):
\begin{align}
n_{co_2}(z,t) & = (n^{deep}_{co_2}-n_{initial})\sum^\infty_{r=0}erfc\left(\frac{(2r+1)L_{ocean}-z}{2\sqrt{D_{eddy} t}}\right) \nonumber \\ & +(n_{initial}-n^{sat,sbt}_{co_2}(P=P^{co_2}_{atm}))\sum^\infty_{r=0}erfc\left(\frac{2(r+1)L_{ocean}-z}{2\sqrt{D_{eddy} t}}\right) \nonumber \\
& + (n^{sat,sbt}_{co_2}(P=P^{co_2}_{atm})-n_{initial})\sum^\infty_{r=0}erfc\left(\frac{2rL_{ocean}+z}{2\sqrt{D_{eddy} t}}\right) \nonumber \\ & + (n_{initial}-n^{deep}_{co_2})\sum^\infty_{r=0}erfc\left(\frac{(2r+1)L_{ocean}+z}{2\sqrt{D_{eddy} t}}\right) + n_{initial}
\end{align} 

The vertical flux at the bottom of the wind-driven circulation is:
\begin{equation}
F_{Sv}=D_{eddy}\frac{\partial n_{co_2}}{\partial z}\bigg|_{z=0}
\end{equation}
which for the sinusoidal expansion solution gives:
\begin{equation}
F_{Sv}=D_{eddy}\left[ \frac{\pi}{L_{ocean}}\sum^\infty_{n=1}nC_n\exp\left\lbrace -\frac{n^2\pi ^2}{L^2_{ocean}}D_{eddy}t\right\rbrace +\frac{n^{deep}_{co_2}-n^{sat,sbt}_{co_2}(P=P^{co_2}_{atm})}{L_{ocean}} \right]
\end{equation}
and for the solution in terms of error functions:
\begin{align}
F_{Sv}= & \left(n_{initial}-n^{sat,sbt}_{co_2}(P=P^{co_2}_{atm})\right)\sqrt{\frac{D_{eddy}}{\pi t}} + \left(n^{deep}_{co_2}-n_{initial}\right)\sum^\infty_{r=0}2\sqrt{\frac{D_{eddy}}{\pi t}}e^{-\frac{(2r+1)^2L^2_{ocean}}{4D_{eddy}t}}  \nonumber \\ & +\left(n_{initial}-n^{sat,sbt}_{co_2}(P=P^{co_2}_{atm})\right)\sum^\infty_{r=1}2\sqrt{\frac{D_{eddy}}{\pi t}}e^{-\frac{r^2L^2_{ocean}}{D_{eddy}t}}
\end{align}

The integration in Eq.($\ref{UpwelledDensity}$), using the sinusoidal expansion, equals:
\begin{equation}
\frac{1}{\delta_h}\int^{t+\delta t}_tF_{Sv}dt'=\frac{D_{eddy}}{\delta_h}\left[ \frac{n^{deep}_{co_2}-n^{sat,sbt}_{co_2}(P=P^{co_2}_{atm})}{L_{ocean}}\delta{t} -\frac{\pi}{L_{ocean}}\sum^\infty_{n=1}\frac{nC_n}{\lambda_n}e^{-\lambda_nt}\left(e^{-\lambda_n\delta{t}}-1\right)    \right]
\end{equation} 
where we have defined:
\begin{equation}
\lambda_n\equiv\frac{n^2\pi^2}{L^2_{ocean}}D_{eddy}
\end{equation}
and for the error function expansion it is:
\begin{align}
& \frac{1}{\delta_h}\int^{t+\delta t}_tF_{Sv}dt' = \nonumber \\ 
& \frac{n^{deep}_{co_2}-n_{initial}}{\delta_h}\sum^\infty_{r=0}\frac{(2r+1)L_{ocean}}{\sqrt{\pi}}\left[\Gamma_{-\frac{1}{2}}-\gamma_{-\frac{1}{2}}\left(\frac{(2r+1)^2L^2_{ocean}}{4D_{eddy}(t+\delta{t})}\right)-\tilde{\Gamma}_{-\frac{1}{2}}\left(\frac{(2r+1)^2L^2_{ocean}}{4D_{eddy}t}\right)\right]  \nonumber \\
& - 2\frac{n^{sat,sbt}_{co_2}(P=P^{co_2}_{atm})-n_{initial}}{\delta_h}\sum^\infty_{r=1}\frac{rL_{ocean}}{\sqrt{\pi}}\left[\Gamma_{-\frac{1}{2}}-\gamma_{-\frac{1}{2}}\left(\frac{r^2L^2_{ocean}}{D_{eddy}(t+\delta{t})}\right)-\tilde{\Gamma}_{-\frac{1}{2}}\left(\frac{r^2L^2_{ocean}}{D_{eddy}t}\right)\right] \nonumber \\
& - 2\frac{n^{sat,sbt}_{co_2}(P=P^{co_2}_{atm})-n_{initial}}{\delta_h}\sqrt{\frac{D_{eddy}}{\pi}}\left[\sqrt{t+\delta{t}}-\sqrt{t}\right]
\end{align}
Here $\Gamma_{-\frac{1}{2}}$ is the gamma function of $-\frac{1}{2}$ and $\gamma_{-\frac{1}{2}}$ and $\tilde{\Gamma}_{-\frac{1}{2}}$ are the equivalent incomplete lower and upper gamma functions.

The $z=0$ boundary in the set of Eqs.($\ref{DeepOceanTransport}$) is not exact. It is more appropriate for describing the initial segment of the geostrophic arm of the flow tube. At more advanced segments the fluid parcels equilibrate with the upper layer of the deep unmixed ocean and the flux between the two ought to diminish substantially. Thus our $F_{Sv}$ represents a maximal flux, and should not be integrated over the entire length of time a parcel lingers in the geostrophic arm.    
A fluid parcel is in contact with the deep ocean for a time period of $L_g/u_g$, after which it begins to upwell to the mixed layer. In order to compensate for our flux being a maximal value we integrate it over half of the latter time, $\delta t= L_g/2u_g$.

We define for convenience, for the sinusoidal expansion:
$$
\eta^s_0(t)\equiv\frac{D_{eddy}L_g}{2u_g\delta_hL_{ocean}}n^{deep}_{co_2}-
$$ 
\begin{equation}
\frac{2L_{ocean}}{\pi^2\delta_h}\left[\sum^\infty_{n=1,3,5...}\frac{e^{-\lambda_nt}}{n^2}\left[2n_{initial}-n^{deep}_{co_2}\right]\left(e^{-\frac{\lambda_nL_g}{2u_g}}-1\right)  
+ \sum^\infty_{n=2,4,6...}\frac{e^{-\lambda_nt}}{n^2}n^{deep}_{co_2}\left(e^{-\frac{\lambda_nL_g}{2u_g}}-1\right)\right] 
\end{equation}

\begin{equation}
\eta^s_1(t) \equiv -\frac{D_{eddy}L_g}{2u_g\delta_hL_{ocean}} + \frac{L_{ocean}}{\delta_h}\sum^\infty_{n=1}\frac{2}{n^2\pi^2}\left(e^{-\lambda_nL_g/2u_g}-1\right)e^{-\lambda_nt} 
\end{equation}
We further define for convenience, for the error function expansion:
\begin{align}
& \eta^e_0(t)\equiv \nonumber \\
& \frac{n^{deep}_{co_2}-n_{initial}}{\delta_h}\sum^\infty_{r=0}\frac{(2r+1)L_{ocean}}{\sqrt{\pi}}\left[\Gamma_{-\frac{1}{2}}-\gamma_{-\frac{1}{2}}\left(\frac{(2r+1)^2L^2_{ocean}}{4D_{eddy}(t+\frac{L_g}{2u_g})}\right)-\tilde{\Gamma}_{-\frac{1}{2}}\left(\frac{(2r+1)^2L^2_{ocean}}{4D_{eddy}t}\right)\right]  \nonumber \\
& +2\frac{n_{initial}}{\delta_h}\sum^\infty_{r=1}\frac{rL_{ocean}}{\sqrt{\pi}}\left[\Gamma_{-\frac{1}{2}}-\gamma_{-\frac{1}{2}}\left(\frac{r^2L^2_{ocean}}{D_{eddy}(t+\frac{L_g}{2u_g})}\right)-\tilde{\Gamma}_{-\frac{1}{2}}\left(\frac{r^2L^2_{ocean}}{D_{eddy}t}\right)\right] \nonumber \\ & +2\frac{n_{initial}}{\delta_h}\sqrt{\frac{D_{eddy}}{\pi}}\left[\sqrt{t+\frac{L_g}{2u_g}}-\sqrt{t}\right] 
\end{align}
and,
\begin{align}
\eta^e_1(t) \equiv & -\frac{2}{\delta_h}  \sum^\infty_{r=1}\frac{rL_{ocean}}{\sqrt{\pi}}\left[\Gamma_{-\frac{1}{2}}-\gamma_{-\frac{1}{2}}\left(\frac{r^2L^2_{ocean}}{D_{eddy}(t+\frac{L_g}{2u_g})}\right)-\tilde{\Gamma}_{-\frac{1}{2}}\left(\frac{r^2L^2_{ocean}}{D_{eddy}t}\right)\right] \nonumber \\
& -\frac{2}{\delta_h}\sqrt{\frac{D_{eddy}}{\pi}}\left[\sqrt{t+\frac{L_g}{2u_g}}-\sqrt{t}\right]
\end{align} 
With these definitions we can rewrite Eq.($\ref{UpwelledDensity}$) as:
\begin{equation}
n^{out}_{co_2} = n^{sat,sbt}_{co_2}(P=P^{co_2}_{atm})\left\lbrace 1+\eta^{s,e}_1(t) \right\rbrace +\eta^{s,e}_0(t)
\end{equation}

While the water parcels were in contact with the abyssal cold ocean (i.e. a heat bath) they also exchanged heat and the ascending water will have the deep ocean temperature, $T_{deep}$.

As water parcels upwell (via Ekman suction) they experience a continuous decrease in hydrostatic pressure. 
Therefore, there will be a depth above which the fluid parcels will have to start degassing in order to maintain a number density of CO$_2$ equal to the local solubility. This solubility is the value with respect to equilibrium with a fluid of CO$_2$ (see dashed curves in fig.\ref{fig:IsothermSolubilityProfile}). This value for the solubility is higher than the value for the solubility when it is controlled by the presence of CO$_2$ clathrate hydrates (see solid curves in fig.\ref{fig:IsothermSolubilityProfile}). The low heat fluxes in our water planets of interest \citep[see $2$ Earth mass planets in][]{Levi2014} suggest the deep ocean reservoir of CO$_2$ is indeed in the form of clathrates, which would control the value of $n^{deep}_{co_2}$. 
Thus, although the ascending water parcels are relatively enriched with CO$_2$, their volatile content is stable against degassing. This will change dramatically when the upwelled water reaches the mixed layer at the tropics. There it will be rapidly mixed to the ocean surface where the pressure is low, and equal to $P^{co_2}_{atm}$, and the temperature is high. The difference between the number density of CO$_2$ in the upwelled water and what can stay dissolved at the hot tropical surface water must end up in the atmosphere.          
The outflux of CO$_2$ associated with this degassing is:
\begin{equation}\label{outfluxtropics}
j^{out}_{co_2}=w_e\left(n^{out}_{co_2}-\tilde{\beta}(T_{trop}) P^{co_2}_{atm}  \right)
\end{equation}

We now wish to derive the temporal evolution of the atmospheric carbon-dioxide partial pressure. 
The relation between the surface atmospheric pressure of CO$_2$ and the number of CO$_2$ molecules in the atmosphere is:
\begin{equation}\label{PressureMolecNumber}
4\pi R^2_pP^{co_2}_{atm}=m_{co_2}N^{co_2}_{atm}g
\end{equation}   
Here $m_{co_2}$ is the molecular mass of carbon-dioxide. The influx and outflux of carbon dioxide is related to the variation in the number of atmospheric CO$_2$ molecules in the following way:
\begin{equation}\label{GeneralEqWind}
dN^{co_2}_{atm}\approx 2\pi R_p\delta_v\left(j^{out}_{co_2}-j^{in}_{co_2}\right)N_{wdc}dt
\end{equation}
where $N_{wdc}$ is the number of wind-driven circulations operating. Each hemisphere will have two such circulations, one operating between the subtropics and the tropics and another between the subtropics and the subpolar region. In case sea-ice formation extends to low latitudes this could hinder somewhat the subtropic to subpolar circulations. In addition, the outflux of CO$_2$ from the higher latitude circulation will also be governed by Eq.($\ref{outfluxtropics}$) where the surface temperature at the tropics has to be replaced with the surface temperature at subpolar latitudes. Since the latter is probably lower than at the tropics the outflux from the higher latitude circulations will cease while it continues from lower latitudes. Therefore, the low latitude wind-driven circulation is dominant and we adopt $N_{wdc}=2$.

After some algebraic steps we obtain the relation:
\begin{equation}\label{MasterEquation1}
\frac{dP^{co_2}_{atm}}{dt}=\frac{\delta_vN_{wdc}w_em_{co_2}g}{2R_p}\left[\eta^{s,e}_0(t)+\eta^{s,e}_1(t)\tilde{\beta}(T_{sbt}) P^{co_2}_{atm}\right]-Q_c
\end{equation}
where we have manually added a term, $Q_c$, to account for any possible atmospheric erosion of CO$_2$. 

The steady state condition is:
\begin{equation}
\lim_{t \to \infty}\frac{dP^{co_2}_{atm}}{dt}=0
\end{equation} 
which gives a steady state CO$_2$ partial pressure of: 
\begin{equation}\label{MasterSteady1}
P^{co_2}_{atm,steady}=\frac{n^{deep}_{co_2}}{\tilde{\beta}(T_{sbt})}-\frac{4R_pL_{ocean}Q_c}{N_{wdc}m_{co_2}gD_{eddy}L_g\tilde{\beta}(T_{sbt})}
\end{equation}
Clearly, when $Q_c=0$, a steady state is reached when the number density of dissolved CO$_2$ downwelling at the subtropics equals the number density of dissolved CO$_2$ as forced by the abyss.

Before solving Eq.($\ref{MasterSteady1}$) we need to estimate the deep unmixed ocean vertical eddy diffusivity, $D_{eddy}$. 
The diffusion coefficient of dissolved CO$_2$ in water is $10^{-5}$\,cm$^2$\,s$^{-1}$ \citep{Zeebe2011}. This diffusion coefficient though is molecular in nature and would not be very efficient in mixing the ocean. However it is a good approximation for a lower bound value.

In Earth's ocean the dominant vertical mixing is caused by eddies. Much like conduction via eddies in Earth's ocean is much more efficient than molecular conduction \citep{Defant1961}. Experiments show that vertical mixing, i.e. vertical eddy diffusivity, is very large where oceanic flow interacts with ocean boundaries \citep[see discussion in chapter $8$ in][]{PhysicalOceanography}. The lowest open-ocean value for the vertical eddy diffusivity was found experimentally to be $0.1$\,cm$^2$\,s$^{-1}$ \citep{Ledwell1998}. This value was found for a depth which is below the local mixed layer, though not as deep as the case we are trying to solve for. In addition, a water planet lacks continents and has a much subdued topography. This lack of boundaries, and the energy constraints of the previous section, mean that the lowest value for $D_{eddy}$ found for Earth's ocean is very probably a good estimation for an upper bound value for our desired case study.

\begin{figure}[ht]
\centering
\mbox{\subfigure{\includegraphics[width=7cm]{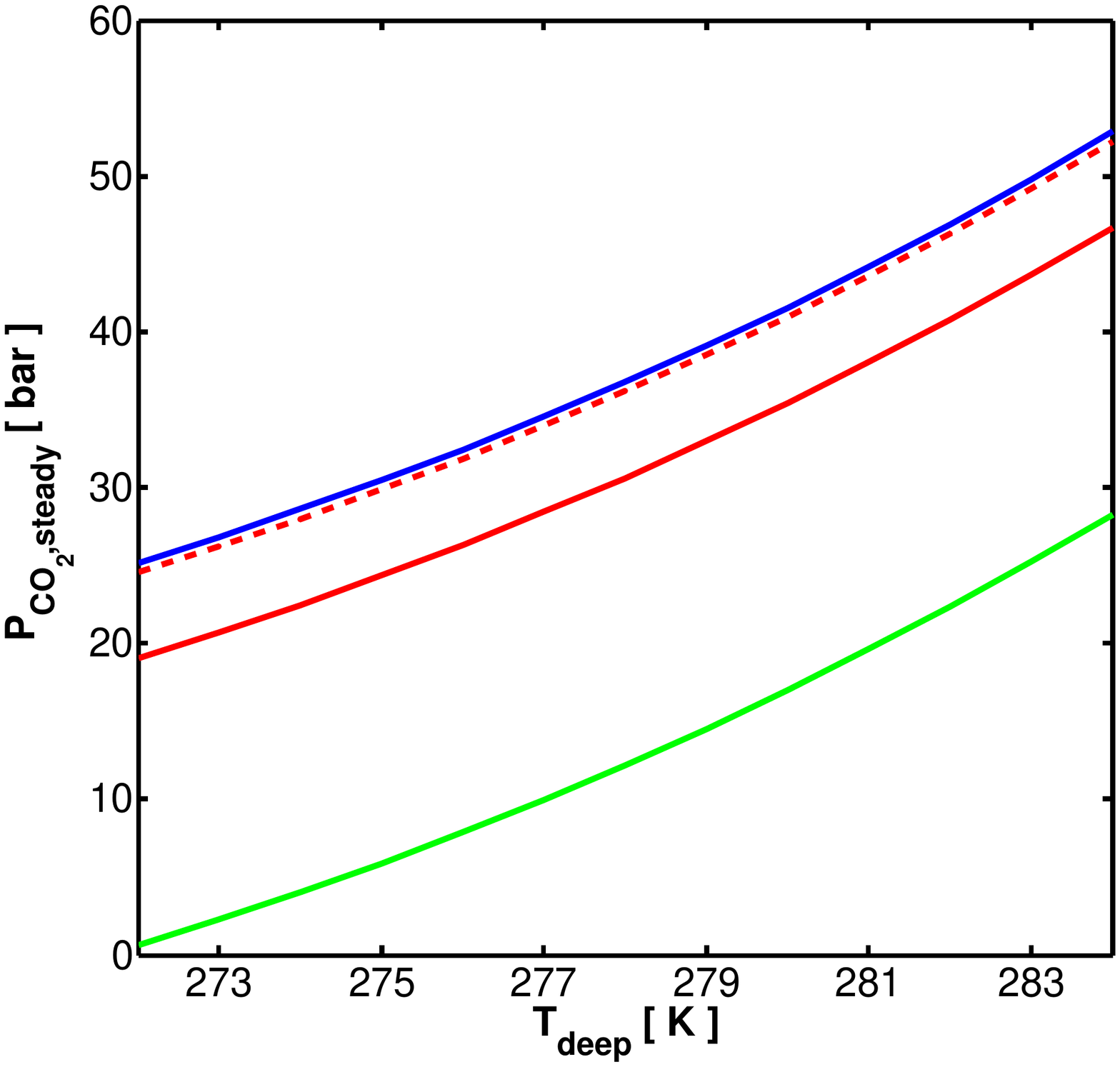}}\quad \subfigure{\includegraphics[width=7cm]{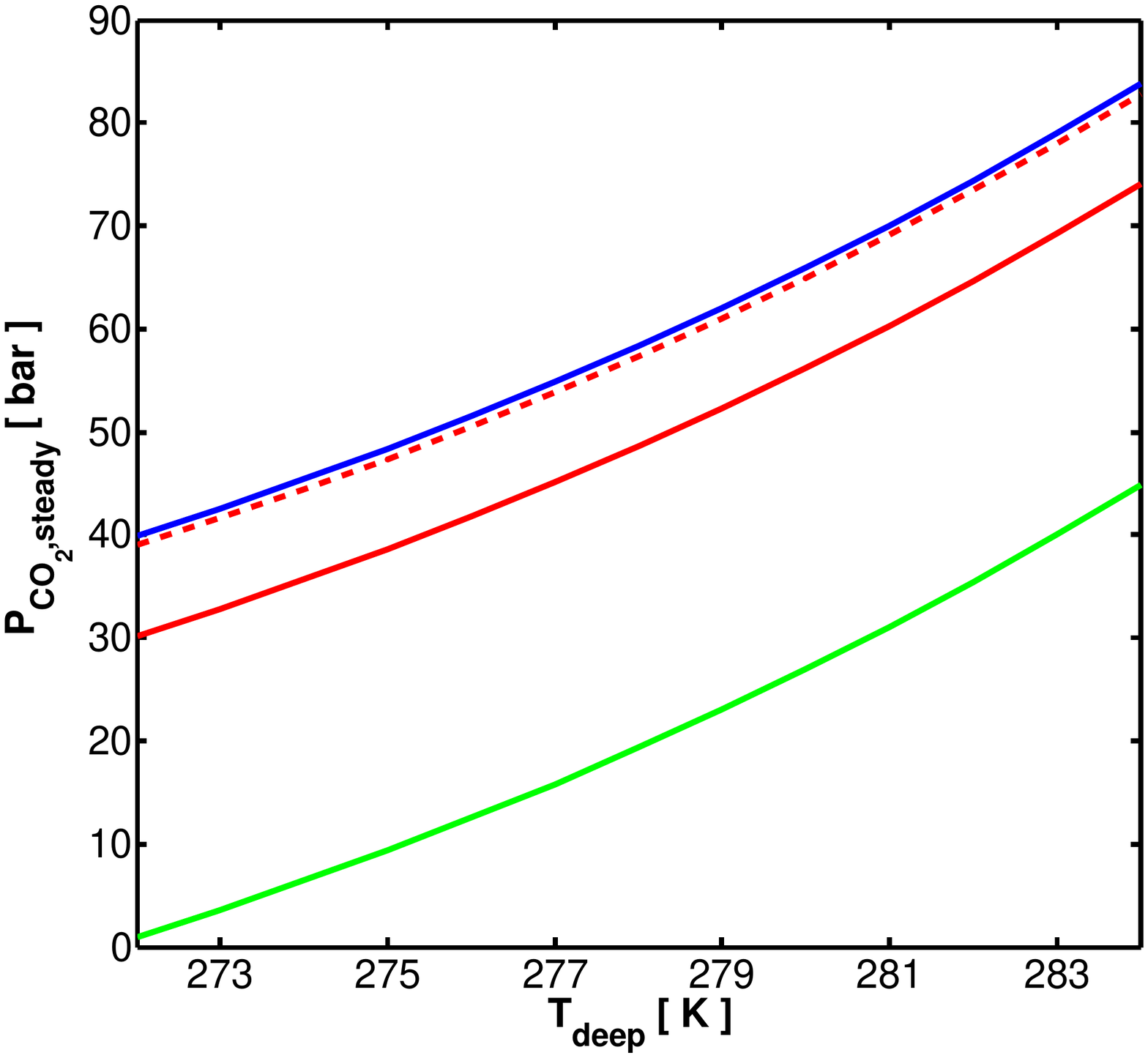}}}
\caption{\footnotesize{The steady state partial atmospheric pressure of carbon-dioxide as a function of the deep unmixed ocean temperature. Here it is assumed that no sea-ice forms. In the right (left) panel a subtropic surface water temperature of $40^\circ$C ($20^\circ$C) is adopted. Solid blue curve is for $Q_c=0$, i.e. no loss of carbon-dioxide due to erosion of the atmosphere. Dashed red curve is for $Q_c=0.01$\,bar\,Myr$^{-1}$ and $D_{eddy}=10^{-2}$\,cm$^2$\,s$^{-1}$. The solid red curve is for $Q_c=0.1$\,bar\,Myr$^{-1}$ and $D_{eddy}=10^{-2}$\,cm$^2$\,s$^{-1}$. The solid green curve is for $Q_c=0.04$\,bar\,Myr$^{-1}$ and $D_{eddy}=10^{-3}$\,cm$^2$\,s$^{-1}$. }}
\label{fig:SteadyStatePressure}
\end{figure}

In fig.\ref{fig:SteadyStatePressure} we solve for the steady state partial atmospheric pressure of carbon-dioxide as a function of the deep unmixed ocean temperature, and for two subtropic surface water temperatures. 
A higher subtropic surface water temperature results in more carbon-dioxide in the atmosphere in steady state. This is because the surface water is hot and is therefore outside of the clathrate hydrate stability field. This means the solubility there decreases with the increasing temperature. However, a steady state requires that the downwelling water at the subtropics have a dissolved concentration of carbon-dioxide equal (for $Q_c=0$) to that at the abyss. Thus, a higher subtropic surface temperature will require a higher atmospheric pressure to obtain the same solubility as when the surface temperature is lower.

If abyssal ocean temperatures fall within the clathrate hydrate thermodynamic stability field (see domains $\alpha$ and $\beta$ in section \ref{sec:Reservoirs}) it is the SI CO$_2$ clathrate hydrates that control the concentration of CO$_2$ in the ocean. In this case, as we have shown in subsection \ref{subsec:SolubilityInside}, the higher the temperature the more carbon-dioxide is dissolved in the water in equilibrium with its clathrates. Therefore, the higher the deep ocean temperature is the higher $n^{deep}_{co_2}$ is, which forces a higher partial atmospheric pressure of carbon-dioxide in steady state. 

Consider the ratio $Q_c/D_{eddy}$, where $Q_c$ is in bar\,Myr$^{-1}$ and $D_{eddy}$ is in cm$^2$\,s$^{-1}$. The phenomena contained in the term $Q_c$ start affecting the steady state atmospheric pressure when $Q_c/D_{eddy}>10$. When $Q_c/D_{eddy}>40$ the effect of $Q_c$ on the steady state partial atmospheric pressure of CO$_2$ becomes dominant. We find that for $Q_c/D_{eddy}>100$ the CO$_2$ atmosphere is completely eroded.

The values we find for the steady state atmospheric partial pressure of carbon-dioxide are substantially higher than for present day Earth's atmosphere. However one should consider that water planets are very rich in volatiles. \cite{Luger2015} showed that a total evaporation of a H/He rich atmosphere yields at most a $2$ Earth mass water planet. Consider such a planet and assume half of its mass is ice. Considering cometary composition to be a good approximation for the icy planetesimals that form the icy envelope of water planets, the CO$_2$ abundance by number in the icy envelope is therefore in the range of $1$-$10$\% \citep{Despois2005}. A $20$\,bar atmosphere of CO$_2$ is thus only about $0.01-0.1$\% of the total ice mantle budget of CO$_2$. Also, a $Q_c=0.1$\,bar\,Myr$^{-1}$ means that a system whose age is $10$\,Gyr lost at this rate between $1-10$\% of its primordial CO$_2$ budget.

\begin{figure}[ht]
\centering
\includegraphics[trim=0.5cm 4cm 0.15cm 5cm , scale=0.60, clip]{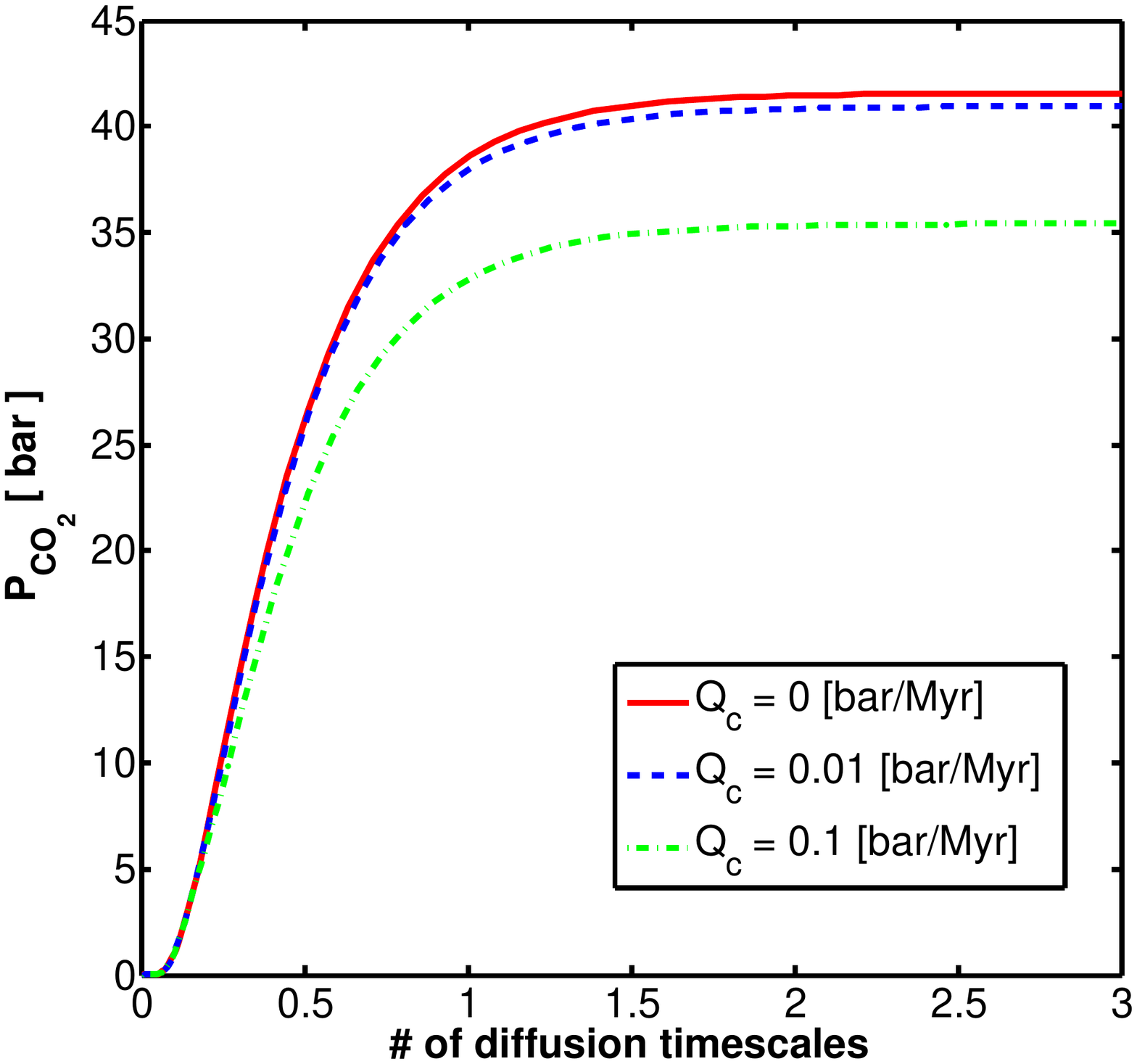}
\caption{\footnotesize{Time evolution of the partial pressure of atmospheric CO$_2$ for the case of a secondary atmosphere outgassing. The subtropic temperature is here assumed to be $20^\circ$C and the deep ocean temperature is assumed to be $280$\,K. The vertical eddy diffusion coefficient is taken to be $10^{-2}$\,cm$^2$\,s$^{-1}$ which for an ocean depth of $80$\,km gives a diffusional time scale of $204$\,Myr. The solid red curve is for $Q_c=0$\,bar\,Myr$^{-1}$. The dashed blue curve is for $Q_c=0.01$\,bar\,Myr$^{-1}$ and the dashed-dotted green curve is for $Q_c=0.1$\,bar\,Myr$^{-1}$. It is assumed here that no sea-ice forms.}}
\label{fig:PCO2TimeEvolution1}
\end{figure}

In fig.$\ref{fig:PCO2TimeEvolution1}$ we plot the temporal evolution of the secondary outgassing of a carbon-dioxide atmosphere. Clearly, the rate controlling step is the eddy diffusion from the bottom of the deep unmixed ocean up to the wind-driven circulation layer. Therefore, the time scale for reaching a steady state atmosphere is:
\begin{equation}
t_{steady}\sim\frac{L^2_{ocean}}{D_{eddy}}
\end{equation}
For $L_{ocean}=80$\,km this gives $20$\,Myr for $D_{eddy}=10^{-1}$\,cm$^s$\,s$^{-1}$ and $20$\,Gyr for $D_{eddy}=10^{-4}$\,cm$^s$\,s$^{-1}$. Thus, depending on the vertical eddy diffusivity across the deep ocean even an old planetary system may have a non-steady state secondary atmosphere.       

\begin{figure}[ht]
\centering
\includegraphics[trim=0.15cm 11cm 0.2cm 3cm , scale=0.60, clip]{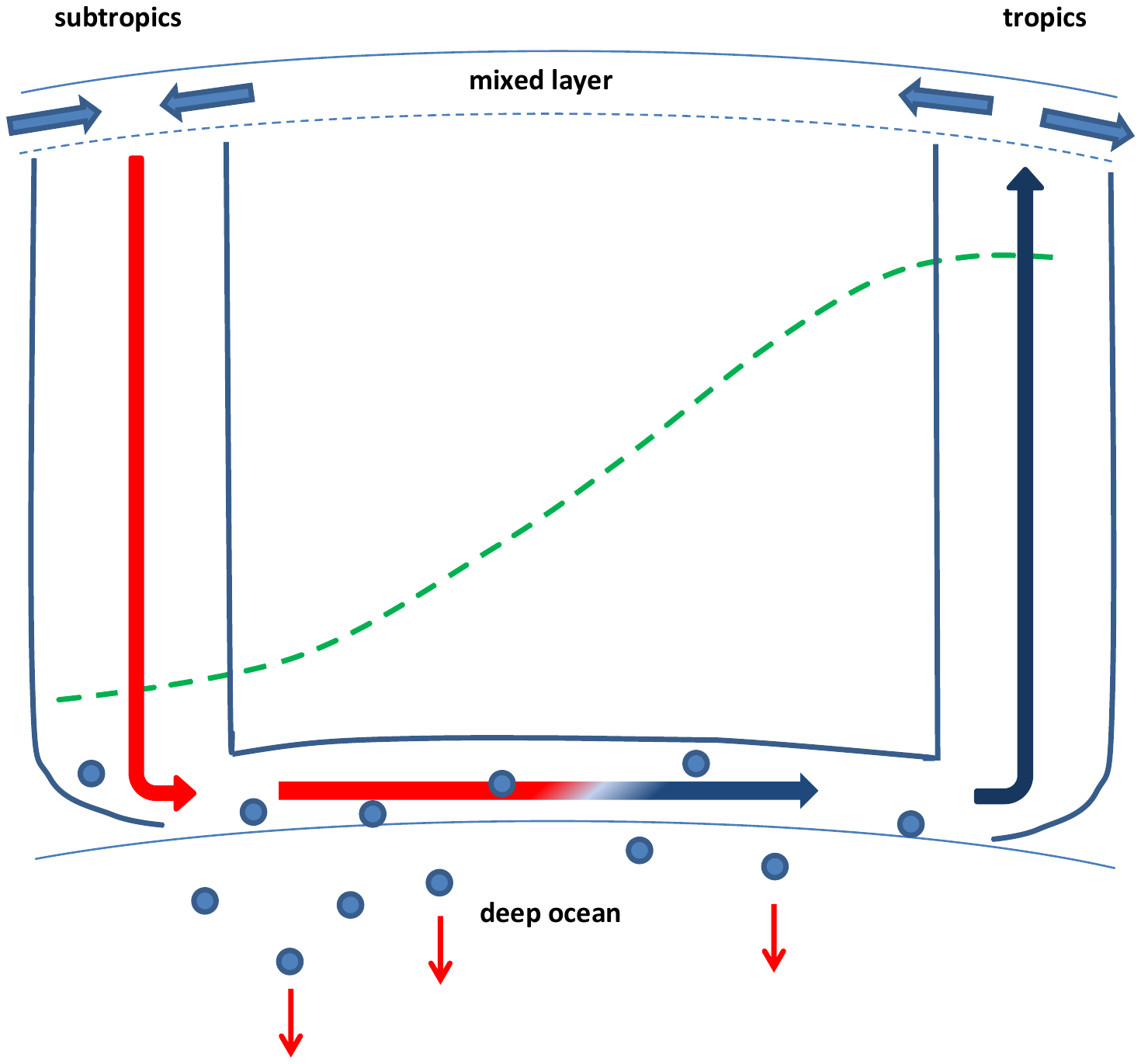}
\caption{\footnotesize{A diagram of the wind-driven circulation in the upper ocean between the tropics and the subtropics. See fig.\ref{fig:WindDrivenCirculation} for details of the flow tube described. Here we describe the case where the partial pressure of CO$_2$ in the atmosphere increases above its steady state value. Now as water parcels downwell at the subtropics they have enough dissolved CO$_2$ to support SI CO$_2$ clathrate grain formation upon entering the clathrate hydrate stability field (below the dashed green curve). The clathrate grains being denser than the surrounding fluid sink and clear the circulation. The water parcels that upwell at the tropics have a CO$_2$ concentration controlled by the equilibrium with the clathrate grains that formed. This concentration is lower than what was in the downwelling water parcels, thus the circulation strives to restore the steady state atmospheric partial pressure of carbon-dioxide.}}
\label{fig:WindDrivenCirculation2}
\end{figure}

A steady state pressure also implies that the atmosphere-ocean system ought try to restore it, not only during an outgassing period but also in circumstances where perturbations have increased the partial pressure of CO$_2$ above this value. In fig.\ref{fig:WindDrivenCirculation2} we describe the dynamics of this latter scenario for the case where $Q_c=0$.

In the case that the atmospheric partial pressure of CO$_2$ is higher than its steady state value, for $Q_c=0$, then the number density of dissolved carbon-dioxide in the surface water at the subtropics is higher than $n^{deep}_{co_2}$. This provides the necessary supersaturation required in order to form clathrate hydrate grains directly from the dissolved carbon-dioxide. Remember that $n^{deep}_{co_2}$ is here taken to be the value in saturation with respect to clathrate hydrate formation in the abyss. Nucleation and grain growth will commence when the downwelling fluid parcels enter the part of the flow tube which is within the SI CO$_2$ clathrate hydrate thermodynamic stability field (under the dashed green curve in fig.\ref{fig:WindDrivenCirculation2}).

\begin{figure}[ht]
\centering
\includegraphics[trim=0.15cm 4cm 0.15cm 5cm , scale=0.55, clip]{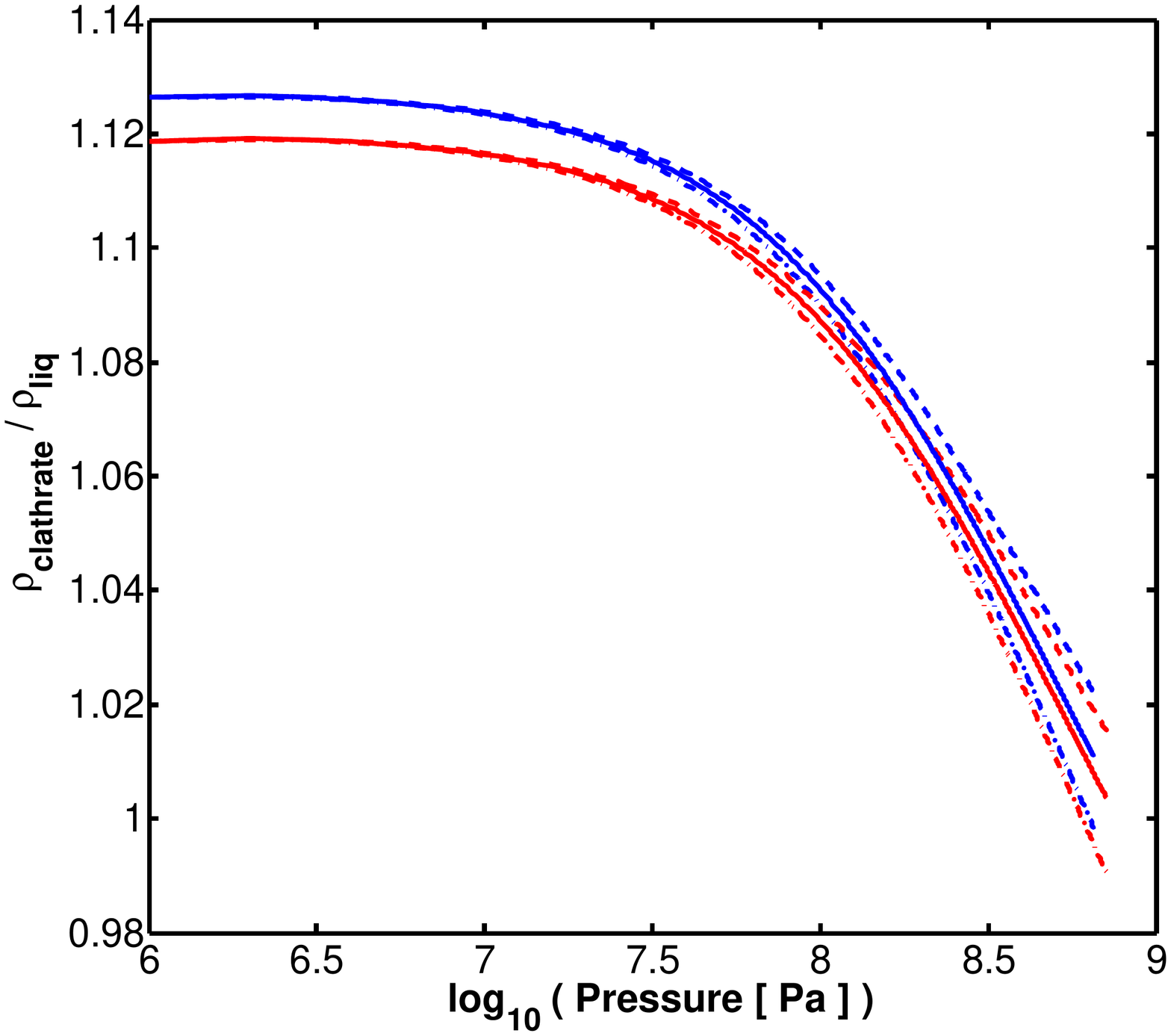}
\caption{\footnotesize{Ratio of the bulk mass density of SI CO$_2$ clathrate hydrate to the bulk mass density of a liquid water solution saturated with CO$_2$ as a function of pressure, spanning the pressure range in water planets' oceans. The liquid water solution is modelled using the equation of state for pure water from \cite{wagner02} and corrected for the solubility of CO$_2$ using the correlation from \cite{Teng1997}.
Curves plotted represent two different isotherms: $275$\,K (blue) and $280$\,K (red). We vary the bulk modulus of SI CO$_2$ clathrate hydrate between: $7$\,GPa (dashed curves), $8.5$\,GPa (solid curves) and $11$\,GPa (dashed-dotted curves). See subsection 2.2 for a discussion over the bulk modulus of SI CO$_2$ clathrate hydrate. The isothermal curves are truncated at the appropriate ocean bottom pressure.}}
\label{fig:ClathVH2ODensity}
\end{figure}

In fig.$\ref{fig:ClathVH2ODensity}$ we show that although the compressibility of liquid water is higher than that of CO$_2$ SI clathrate hydrate, even at the high pressure at the bottom of the water planet ocean the SI CO$_2$ clathrate hydrate grain is still expected to be more dense than the surrounding water. The exception is the case where we adopt for the bulk modulus of the CO$_2$ SI clathrate hydrate a value of $11$\,GPa. See discussion on the clathrate bulk modulus in subsection $2.2$. For this, somewhat extreme case, sinking clathrate grains turn buoyantly neutral $60$\,km below the ocean surface, whereas the ocean is $71$\,km deep. We assume here a gravitational acceleration of $g=10^3$\,cm\,s$^{-2}$.     
Therefore, it is very likely that clathrate grains forming in the liquid water from dissolved CO$_2$ will tend to sink to the bottom of the ocean. This mode of clathrate grain formation has been verified experimentally to yield grains more dense than freshwater and Earth's seawater \citep[see][and references therein]{Warzinski2000}.

The sinking velocity of clathrate grains, due to their negative buoyancy, was measured in a field study by \cite{Riestenberg2005}. In their experiment carbon-dioxide was injected into the ocean at a depth of about $1$\,km (approximately Sverdrup's depth of no motion), and a formation of CO$_2$ SI clathrate layer on the liquid carbon dioxide particles was reported. The radius of the particles formed following the injection was approximately $0.6$\,cm. The grains descent to the abyss was followed using a remotely operated vehicle, and a sinking velocity of $5$\,cm\,s$^{-1}$ was derived. This velocity includes turbulent mixing in the ocean and in the locality where the experiment was conducted. In their case study only $40\%$ of the carbon dioxide droplet was converted to clathrate hydrate, the rest remained as liquid CO$_2$ confined to the droplet core. At the depth where the experiment was conducted liquid CO$_2$ is less dense than water. Since in our case the grains form directly from dissolved carbon dioxide they are expected to be pure clathrate hydrate and therefore more dense than those formed in the experiment of \cite{Riestenberg2005}. The experimental velocity is therefore a lower bound for our case of study. 
With the aid of Stoke's drag force we can scale the experimental velocity to other particle sizes:
\begin{equation}
v(a)\approx 5\left(\frac{a}{0.6[cm]}\right)^2\quad [cm~s^{-1}]
\end{equation}
A grain with a radius of $a=100$\,$\mu$m will have a descent velocity of approximately $10^{-3}$\,cm\,s$^{-1}$ and will take about $100$\,yr to traverse the $100$\,km deep ocean in its way to the bottom. The descent time scale will also be affected by grain coagulation, nevertheless we still expect the sinking time scale to remain much shorter than any geological time scale. More importantly this sinking velocity is at least an order of magnitude larger than any expected vertical motion in the deep unmixed ocean (see section \ref{sec:Circulation}).

In conclusion, the SI CO$_2$ clathrate hydrate grains that form in the geostrophic arm of our flow tube of interest sink rather rapidly away from the wind-driven circulation. These grains take with them any excess in dissolved CO$_2$ above the value in equilibrium with the clathrate hydrate grains. Therefore, the CO$_2$ solubility in the water parcels upwelling in the tropics can not exceed $n^{deep}_{co_2}$. Replacing $n^{out}_{co_2}$ in Eq.($\ref{outfluxtropics}$) with $n^{deep}_{co_2}$ and resolving for the temporal evolution of the partial atmospheric pressure of CO$_2$ we obtain:
\begin{equation}\label{FillEvolutionTime}
P^{co_2}_{atm}(t) = \left(P^{co_2}_{atm}(t=0)-\frac{n^{deep}_{co_2}}{\tilde{\beta}(T_{sbt})}\right)e^{-t/\tau}+\frac{n^{deep}_{co_2}}{\tilde{\beta}(T_{sbt})}
\end{equation} 
where $P^{co_2}_{atm}(t=0)$ is an initial carbon-dioxide atmospheric pressure, and the time scale for restoring steady state is: 
\begin{equation}
\tau\equiv \frac{2R_p}{u_g\Re\tilde{\beta}(T_{sbt})m_{co_2}g N_{wdc}}\approx\left(\frac{16}{\Re[m]}\right)\quad[kyr]
\end{equation} 
Here $u_g\Re$ is the geostrophic flux of water that are supersaturated in CO$_2$ entering the clathrate thermodynamic stability field stretching a distance $\Re$ above Sverdrup's depth of no motion. Even a relatively small value for $\Re$ such as $1$\,m gives $10^4$\,yr. In other words, steady state pressure is restored rapidly following perturbations that try to increase it.

\subsection{ Sea-Ice Formation }\label{subsec:FluxesSeaIce}    

\cite{Marshall2007} predict subfreezing surface temperatures as low as $250$\,K at the poles. Under such subfreezing conditions solidification of surface water in the form of ice Ih is favoured even in the presence of gaseous CO$_2$ \citep[][see supplement as well for water consumption rates]{Nguyen2015}. Therefore, the formation of SI CO$_2$ clathrate hydrate directly from an aqueous CO$_2$ solution, the mode described in \cite{Englezos1987}, is less probable. It is more likely that ice Ih of subfreezing structure initially forms at the ocean's surface which would then transform to CO$_2$ clathrate hydrate when exposed to an atmosphere supersaturated in CO$_2$. This transformation between the phases is facilitated by a disordered layer of water molecules that is required in order to match both phases' crystal structures and provides the necessary low free energy surface for the heterogeneous nucleation to proceed \citep{Nguyen2015}. It is very important to note that \textit{an atmosphere supersaturated in CO$_2$ refers here to an atmosphere in which the partial pressure of CO$_2$ is higher than the appropriate dissociation pressure of the clathrate}. The higher the supersaturation the greater is the driving force that transforms the ice Ih grains to CO$_2$ clathrate grains.
 
\cite{Moudrakovski2001} experimented on clathrate hydrates of Xe and found that there is a time interval of tens to hundreds of seconds, following the exposure to Xe, until the first clathrate hydrate crystals form. He referred to this as the induction time for clathrate hydrates. Later experiments on clathrate hydrates of CH$_4$ and CO$_2$ disputed the existence of such a time interval \citep{Staykova2003,Genov2004}. A phenomenological theory describing the rate of transformation of ice Ih grains into clathrate hydrate was presented in \cite{Staykova2003}. The theory considers an initial rapid transformation on the ice grain boundaries followed by a slower transformation controlled by enclathration reaction rates and permeation rates of water and carbon-dioxide molecules through a forming clathrate hydrate layer surrounding the ice grain. This theory was further developed to include transformation of ice into clathrate hydrate inside grain cracks \citep{Genov2004}. \cite{Genov2004} performed experiments on the rate of transformation of ice Ih to CO$_2$ clathrate hydrate which were then used to estimate the various coefficients of their kinetic theory. The experiments were conducted in the temperature range of $193$\,K-$272$\,K, which fully covers our temperature range of interest. We adopt the theory of \cite{Staykova2003} and \cite{Genov2004} in order to derive the mole fraction of ice Ih converted to CO$_2$ clathrate hydrate, $\alpha(t)$. This parameter is a complicated function of the pressure of CO$_2$, temperature and of the ice morphology.

Understanding sea-ice formation on Earth is important in order to model climate trends. It has therefore rightfully received the attention of climatologists aspiring to incorporate it within atmospheric and oceanic models. When modelling sea-ice formation on Earth one usually has to consider for example brine pockets and a layer of accumulated snow \citep{Maykut1971,Semtner1976}. The most uptodate models can also account for algae accumulations \citep{Hunke2015}. Expeditions to the poles provide many of the required parameter values for modelling sea-ice \citep[e.g.][]{Lewis2011,Ackley2015}. If polar sea-ice indeed forms on water planets its modelling is complicated not only by the lack of field and observational data but also by the different chemical environment. First, we do not expect brine to exist in any significant quantity in sea-ice on water planets. Secondly, high concentrations of clathrate forming molecules in the atmosphere would render clathrate hydrates an important phase, one that must be considered when modelling the thermodynamics of sea-ice. In this subsection we consider sea-ice to be a composite of ice Ih and clathrate of CO$_2$.

A layer of ice Ih forming under subfreezing atmospheric conditions is initially quite thin. This means that it can support a high conductive heat flux which would cause the freezing of underlying layers. The evolution of sea-ice is therefore governed mostly by conduction of heat \citep[e.g.][]{Maykut1971}:
\begin{equation}
\rho_{comp}C_{comp}\frac{\partial T}{\partial t}=\frac{\partial}{\partial z}\left(\kappa_{comp}\frac{\partial T}{\partial z}\right)+Q
\end{equation}
Here $\rho_{comp}$, $C_{comp}$ and $\kappa_{comp}$ are the mass density, heat capacity and thermal conductivity of the composite. $Q$ represents a heat source due to the continuous phase change between ice Ih and clathrate.

The solution to the above heat transport equation is greatly complicated by the formation of clathrates. As shown in fig.\ref{fig:ClathVH2ODensity} (the lower pressure part is of relevance here) the mass density of SI CO$_2$ clathrate hydrates is higher than that of both ice Ih and liquid water. Since the clathrate will form mostly at the interface with the atmosphere, a sea-ice block would be heavier at its top. This is in contradiction to sea-ice forming on Earth and is an unstable configuration. This means the initial and surface conditions become complex functions of time which depend on the stability against flipping. 
This dynamical instability though leads us to suggest a simplification, that is that sea-ice on water planets tends to form rather uniformly. This is because the instability would tend to rotate the block of ice and expose the lower bottom, of the less dense ice Ih, to the atmosphere. Now this fresh ice Ih will also begin to transform into clathrate. Therefore, compositional gradients tend to diminish, and the mass density for the sea-ice composite (derived in appendix \ref{subsec:AppendixB}) can approximate the entire block of ice:
\begin{equation}
\rho_{comp}=\left(1-\phi^0_{pore}\right)\left[\rho_{Ih} + \alpha\left(\rho_{pore}-\rho_{Ih}\right)+\zeta\alpha\left(\rho_{clath}-\rho_{pore}\right)\right]+\phi^0_{pore}\rho_{pore}
\end{equation}
Here $\phi^0_{pore}$ is the \textit{initial} porosity of the formed sea-ice, $\zeta\approx 1.133$ is the expansion factor when a mole of ice Ih converts to SI CO$_2$ clathrate hydrate, $\rho_{Ih}$ and $\rho_{clath}$ are the homogeneous ice Ih and clathrate hydrate bulk mass densities and $\rho_{pore}$ is the mass density of the pore filling material.
Let us consider a few of the interesting features of such a sea-ice composite.

\begin{figure}[ht]
\centering
\includegraphics[trim=0.15cm 3cm 0.2cm 4.5cm , scale=0.55, clip]{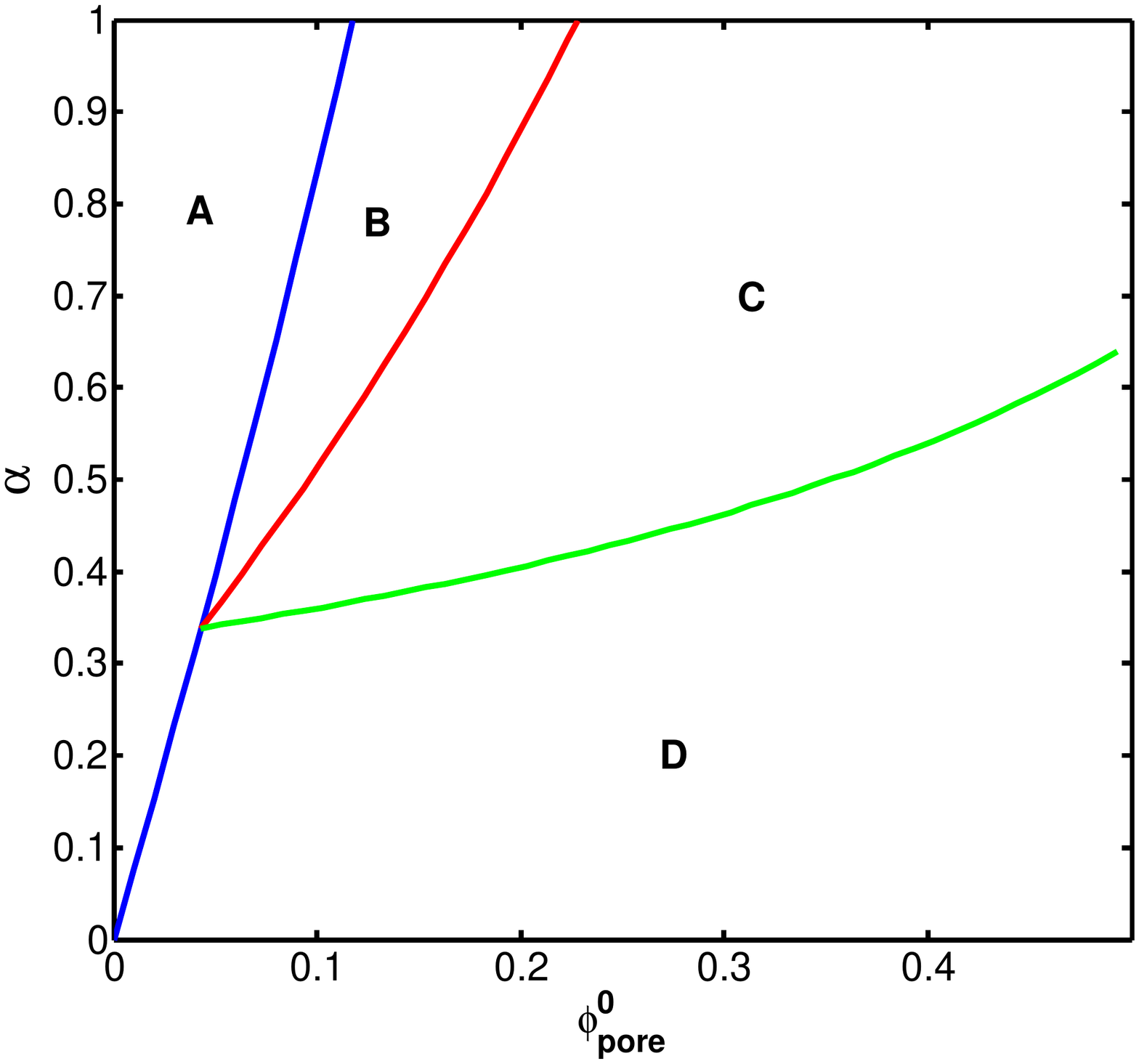}
\caption{\footnotesize{The mole fraction of water ice Ih converted to SI CO$_2$ clathrate hydrate, $\alpha$, versus the \textit{initial} porosity of the forming sea-ice. In region A the pore space has closed off, due to the expansion of the water network when ice Ih grains convert to clathrate hydrate. $\alpha$ values within this region can be obtained by slow molecular diffusion of CO$_2$ through the ice matrix rather than permeability via the pores. In region B both pore volume restrictions and the condition for sinking of the sea-ice are met, assuming that the pore filling material is vacuum. In case the pore filling material is ice Ih region B is extended to include region C as well. Region D upholds the pore volume restriction, however, will not allow for sinking of the sea-ice composite.}}
\label{fig:SeaIceStability}
\end{figure}

The mole fraction converted to clathrate hydrate, $\alpha$, naturally falls between $0$ and $1$. However, the requirement that the porosity at any given moment must be positive combined with Eq.($\ref{porespacerelation}$) from appendix \ref{subsec:AppendixB}, yields:
\begin{equation}
\alpha < \frac{\phi^0_{pore}}{(\zeta -1)(1-\phi^0_{pore})}\equiv\alpha_{max}
\end{equation} 
This criterion shuts off any further transformation of the ice Ih grains into clathrate hydrate when the expansion due to the transformation fills up the pore space. Beyond this value for $\alpha$ any further transformation is controlled not by the permeability of gas into the ice block but rather by macroscopic scale diffusion times of atmospheric CO$_2$ through the solid matrix, which should slow down the transformation process considerably. Complying with the above criterion means being below the solid blue curve in fig.\ref{fig:SeaIceStability}. From the figure we see that when $\phi^0_{pore}>0.117$ this pore space restriction is effectively lifted.

For the sea-ice to be able to sink into the ocean its mass density should be larger than that of the surface ocean water (i.e. $\rho_w<\rho_{comp}$). This results in a lower bound value for $\alpha$ of:
\begin{equation}\label{alphalowerbound}
\alpha > \frac{\rho_w-\rho_{Ih}+\phi^0_{pore}(\rho_{Ih}-\rho_{pore})}{(1-\phi^0_{pore})\left\lbrace\rho_{pore}(1-\zeta)-\rho_{Ih}+\zeta\rho_{clath}\right\rbrace}\equiv\alpha_{min}
\end{equation} 
For the scenario that only atmospheric gas (of negligible mass density) fills the pores, sinking of the sea-ice requires being above the solid red curve in fig.\ref{fig:SeaIceStability}.
However, the differential buoyancy between the ocean's liquid water and the SI CO$_2$ clathrate hydrate can drive a Darcy-type flow of the aqueous "magma" into the sea-ice composite. When entering the pores the liquid would solidify into ice Ih. The velocity associated with this flow may be estimated as \citep[see subsection $4.5$ in][]{schubert2001}:
\begin{equation}
V_{Darcy}=\frac{1}{32}\hat{\delta}^2\frac{g}{\mu_w}\left(\rho_{clath}-\rho_w\right)
\end{equation}
where $\hat{\delta}$ is the diameter of a pore tube which we assume to be in the range of $10-100\mu$m, from the size of ice grains that do not have the time to ripen. The dynamic viscosity of ocean water near freezing is $\mu_w=1.792$centipoise \citep{Cho1999}. Thus, crossing a $1$\,m length scale (depth of sea-ice on the Earth) takes between $1-110$\,hr. Some filling of residual pore space with ice Ih is therefore reasonable. We note that while CO$_2$ trapped in pores partially converts to SI CO$_2$ clathrate hydrate the higher dissociation pressure of N$_2$ clathrate hydrate would tend to keep N$_2$ in gaseous form. The later filling of the pores with liquid water and its solidification would probably expel this N$_2$ back into the atmosphere. Finally, assuming the pore filling material is ice Ih, the sinking of the sea-ice composite will be possible in the parameter space above the solid green curve in fig.\ref{fig:SeaIceStability}.   
Complying with both the lower and upper bounds for $\alpha$ requires an initial porosity of at least $0.043$.

Let's assume that the sea-ice formation is restricted to between the pole and latitude $\lambda_{sp}$, and that the surface atmospheric temperature in this region is $T_{sp}$. Therefore, the volume of the sea-ice sheet is:
\begin{equation}
2\pi R^2_p\left(1-\sin\lambda_{sp}\right)h_{ice}
\end{equation}  
where $h_{ice}$ is the sea-ice sheet thickness. The sea-ice interacts with the atmosphere and its SI CO$_2$ clathrate hydrate mole fraction increases. When this fraction reaches the value $\alpha_{min}$ (see Eq.$\ref{alphalowerbound}$) the floating sea-ice becomes gravitationally unstable. Therefore, the volume of the SI CO$_2$ clathrate hydrate in the sea-ice composite just prior to it sinking is:
\begin{equation}
2\pi R^2_p\left(1-\sin\lambda_{sp}\right)h_{ice}\zeta\alpha_{min}\left(1-\phi^0_{pore}\right)
\end{equation}   
where we have used Eq.($\ref{HydrateVolumeFraction}$) from appendix \ref{subsec:AppendixB}. Hence, the total number of CO$_2$ molecules trapped in clathrate cages within the ice sheet is:
\begin{equation}
2\pi R^2_p\left(1-\sin\lambda_{sp}\right)h_{ice}\zeta\alpha_{min}\left(1-\phi^0_{pore}\right)\frac{46}{V_{cell}}\frac{1}{5.75}
\end{equation}
where we have assumed full occupancy of the clathrate cages.
In case the time duration needed to reach $\alpha_{min}$ is $\Delta\tau$, the number of CO$_2$ molecules removed from the atmosphere per unit time is:
\begin{equation}\label{GeneralEqSeaIce}
\frac{dN_{ice-sink}}{dt} = \frac{2\pi R^2_p\left(1-\sin\lambda_{sp}\right)h_{ice}\zeta\alpha_{min}\left(1-\phi^0_{pore}\right)}{\Delta\tau}\frac{46}{V_{cell}}\frac{1}{5.75}
\end{equation}  
Following the sinking of the sea-ice a new sea-ice layer forms and the process repeats.

Consider, for example, the important process of a secondary atmospheric outgassing. If the initial partial atmospheric pressure of carbon-dioxide is below the dissociation pressure value for the SI CO$_2$ clathrate hydrate, for the temperature $T_{sp}$, clathrate enriched sea-ice does not form. Therefore secondary outgassing is first controlled by the wind driven circulation described in the previous subsection. As the partial atmospheric pressure of carbon-dioxide exceeds the dissociation pressure for the subpolar temperature a clathrate hydrate formation driving force begins to operate. As the atmospheric pressure of carbon-dioxide continues to increase so does the driving force. This results in a decreasing value for the time interval $\Delta\tau$. As steady state is approached the latter will reach some asymptotic value, $\Delta\tau_{asy}$. This asymptotic value is the time scale that the sea-ice remains afloat in steady state atmospheric conditions. 

We still need to estimate the sea-ice thickness, $h_{ice}$, at the time of sinking. Sea-ice grows by solidification of water at its base, releasing the latent heat of fusion, $L_f$. This heat must be conducted outward through the ice before the ice can continue to thicken. The liquid water at the base of the sea-ice layer must be on the verge of freezing, $T=T_{melt}$. Confined from below by the melting temperature of ice Ih and from above by the atmospheric temperature the conductive heat flux in the sea-ice is:
\begin{equation}
\kappa_{comp}\frac{T_{melt}-T_{sp}}{h_{ice}}
\end{equation}
Hence its rate of growth is:
\begin{equation}
\frac{dh_{ice}}{dt}\sim\kappa_{comp}\frac{T_{melt}-T_{sp}}{L_f\rho_{Ih}}\frac{1}{h_{ice}}
\end{equation}
which yields the result:
\begin{equation}
h_{ice}\sim\sqrt{2\kappa_{comp}\frac{T_{melt}-T_{sp}}{L_f\rho_{Ih}}\Delta\tau}
\end{equation}
Let's test the last scaling on Earth's sea-ice. The thermal conductivity of ice Ih at $T=250$\,K is $0.024\times 10^7$\,erg\,s$^{-1}$\,cm$^{-1}$\,K$^{-1}$ \citep{slack1980}, its bulk mass density is $0.917$\,g\,cm$^{-3}$ and its latent heat of fusion is $334\times 10^7$\,erg\,g$^{-1}$ \citep{feistel06}. Considering a temperature difference across the ice of $40$\,K and two winter months for the time interval yields an ice thickness of $1.8$\,m. This agrees well with satellite measurements for the thickness of Arctic sea-ice \citep{Ricker2014}. 

However, it is important to note that as time goes on the sea-ice will \textit{not} thicken indefinitely. This is important mostly when the partial atmospheric pressure of carbon-dioxide is very close to its clathrate dissociation pressure, a condition for which $\Delta\tau\rightarrow\infty$. Even if the summer months are disregarded, some heat from the abyssal ocean should cap the sea-ice thickness. This heat flux is hard to estimate without a global hydrodynamic solution, and for Earth is usually estimated from sea-ice thickness measurements assuming the system is in equilibrium \citep[see][and references therein]{Ackley2015}. There is however an interesting compensation in the case of sea-ice on water planets. On the one hand, an inefficient global circulation and the thick planetary ice mantle ought result in low heat fluxes at the sea-ice bottom. This would suggest a thicker ice sheet. On the other hand, the much lower thermal conductivity of clathrates would tend to encourage a thinner ice sheet. We will return to speculate about the sea-ice thickness below.  

Lastly, we wish to remark on the thermal conductivity of the sea-ice composite, $\kappa_{comp}$. During the conversion of the continuously shrinking ice Ih grains to SI CO$_2$ clathrate hydrate the latter expanded at the expense of the pore space \citep{Staykova2003,Genov2004}. Therefore, as the composite evolves it may be approximated as a continuous clathrate hydrate solid within which small ice Ih spheres are embedded. The thermal conductivity of such a composite was solved for by Maxwell \citep{Bird2007transport}, and for our case has the following form:
\begin{equation}
\frac{\kappa_{comp}}{\kappa_{clath}}=1+\frac{3\phi_{Ih}}{\left(\frac{\kappa_{Ih}+2\kappa_{clath}}{\kappa_{Ih}-\kappa_{clath}}\right)-\phi_{Ih}}
\end{equation}
where the volume fraction of the ice Ih spheres was derived in appendix \ref{subsec:AppendixB} (see Eq.$\ref{IhVolumeFraction}$).
In the last relation $\kappa_{Ih}$ is the thermal conductivity of ice Ih taken from \cite{slack1980}. The thermal conductivity of SI CO$_2$ clathrate hydrate, $\kappa_{clath}$, is not known experimentally. Molecular dynamics simulations suggest it is about $15$\% smaller than the thermal conductivity of SI CH$_4$ clathrate hydrate \citep{Jiang2010}. The thermal conductivity of CH$_4$ clathrate hydrate is known experimentally and here we adopt the values reported in \cite{krivchikov2006}.  

In a water planet the formation of polar sea-ice and the wind driven circulation operate simultaneously. Therefore, the evolution of the atmosphere is governed by the sum of all the fluxes of the mechanisms we have described:
\begin{equation}\label{GeneralPCO2Evolution}
\frac{dN^{co_2}_{atm}}{dt}= 2\pi R_p\delta_v\left(j^{out}_{co_2}-j^{in}_{co_2}\right)N_{wdc}-S_w\frac{dN_{ice-sink}}{dt}-Q_c\frac{4\pi R^2_p}{m_{co_2}g}
\end{equation}
where the different parameters were defined above in Eqs.($\ref{GeneralEqWind}$), ($\ref{MasterEquation1}$) and ($\ref{GeneralEqSeaIce}$). We have also inserted a probability function $S_w$ that acts as a switch, turning the sea-ice atmospheric CO$_2$ sink mechanism on and off. We will explain its origin and form in more detail below.

We expect the partial atmospheric pressure of CO$_2$ to reach a steady state after a very long time:
\begin{equation}
\lim_{t \to \infty}\frac{dN^{co_2}_{atm}}{dt}=0
\end{equation}
Inserting the relations:
\begin{equation}
\lim_{t \to \infty}\eta_0=\frac{D_{eddy}L_g}{2u_g\delta_hL_{ocean}}n^{deep}_{co_2}\quad , \quad \lim_{t \to \infty}\eta_1=-\frac{D_{eddy}L_g}{2u_g\delta_hL_{ocean}}
\end{equation}
into Eq.($\ref{GeneralPCO2Evolution}$) the steady state atmospheric pressure of CO$_2$ is found to obey:
\begin{equation}\label{GeneralPCO2Steady}
0=\pi R_pN_{wdc}\tilde{\beta}(T_{sbt})D_{eddy}\frac{L_g}{L_{ocean}}\left[\frac{n^{deep}_{co_2}}{\tilde{\beta}(T_{sbt})}-P^{co_2}_{atm}\right]-S_w\frac{dN_{ice-sink}}{dt}-Q_c\frac{4\pi R^2_p}{m_{co_2}g}
\end{equation} 
We now turn to solve the last equation numerically for various system parameters. 

\begin{figure}[ht]
\centering
\mbox{\subfigure{\includegraphics[width=7cm]{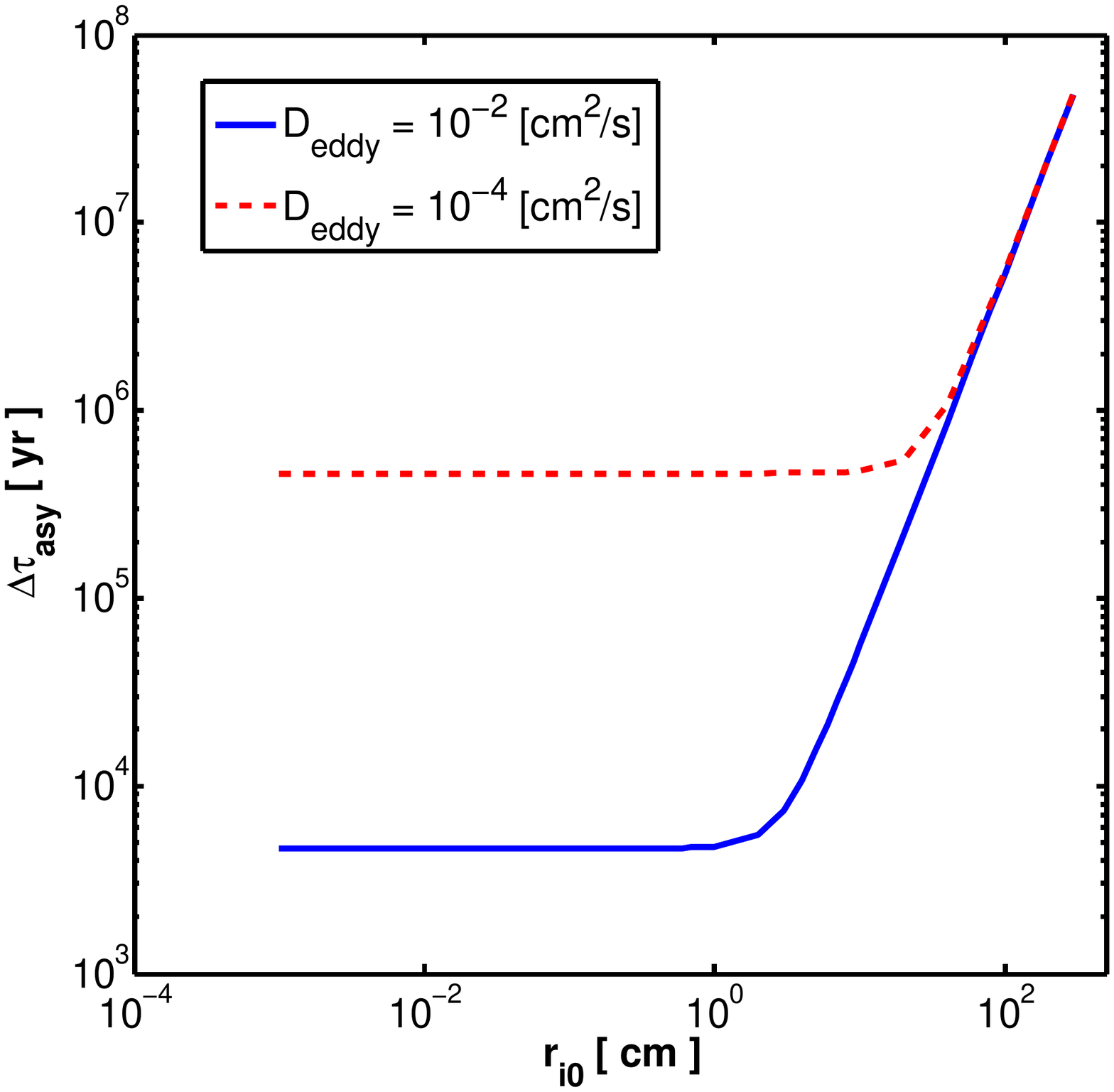}}\quad \subfigure{\includegraphics[width=7cm]{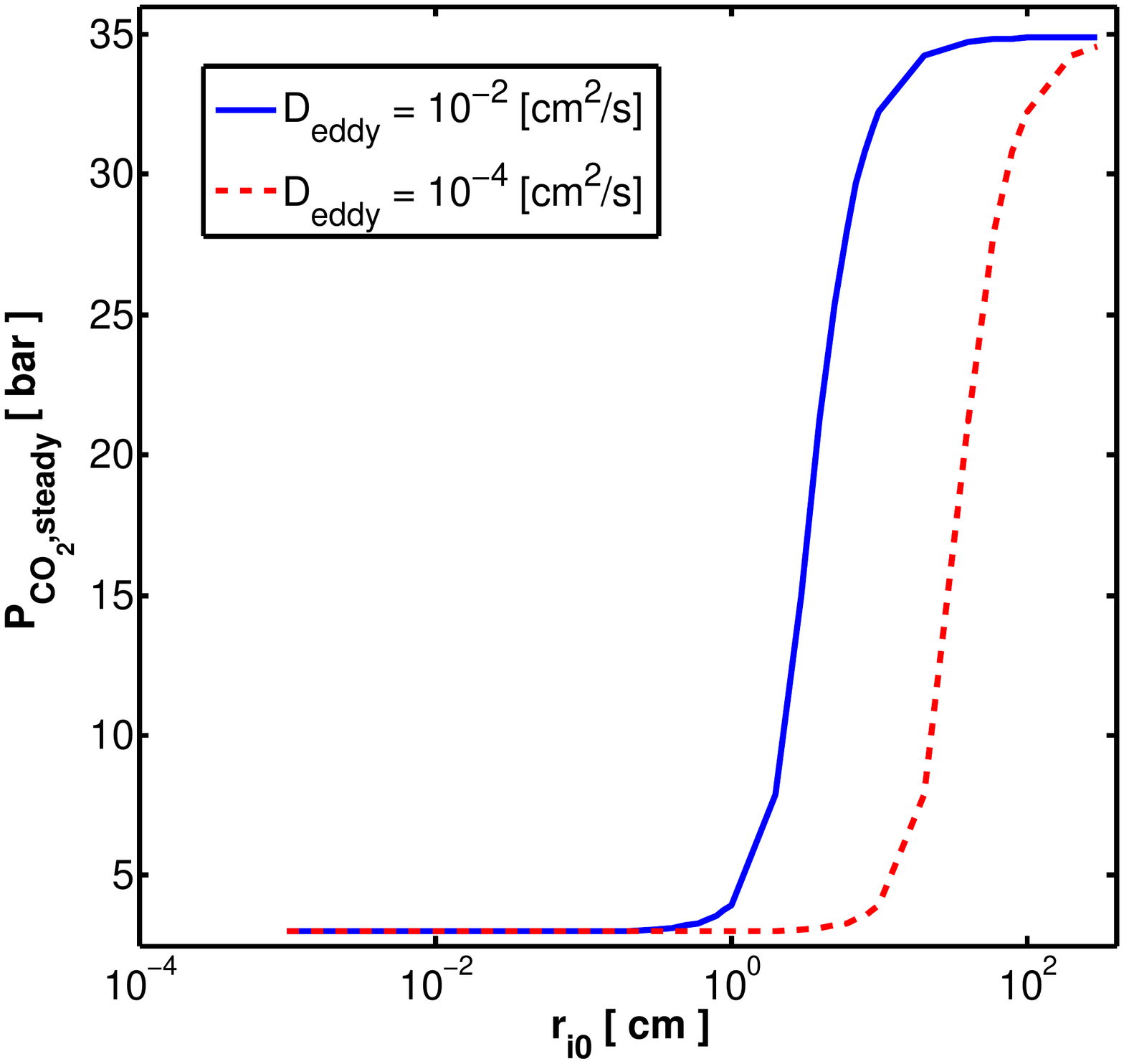}}}
\caption{\footnotesize{\textbf{The case where $S_w=1$}: (right panel) The steady state atmospheric CO$_2$ partial pressure dependence on the initial ice Ih grain radius. (left panel) The corresponding time the sea-ice remains afloat under steady-state atmospheric conditions. In both panels the solid blue and dashed red curves correspond to $D_{eddy}=10^{-2}$\,cm$^2$\,s$^{-1}$ and $10^{-4}$\,cm$^2$\,s$^{-1}$ respectively. Other system parameters adopted are: $T_{sp}=240$\,K, $T_{sbt}=20^\circ$C, $T_{deep}=4^\circ$C, $\phi^0_{pore}=0.1$, $\rho_{pore}=0$\,g\,cm$^{-3}$ and $\lambda_{sp}=60^\circ$. Also we assume $Q_c=0$\,bar\,Myr$^{-1}$}.}
\label{fig:Swunity}
\end{figure}

We start by explaining the role of the probability function $S_w$.
In fig.\ref{fig:Swunity} we solve for the steady state atmospheric partial pressure of CO$_2$ (see eq.$\ref{GeneralPCO2Steady}$) as a function of the  initial ice Ih grain radius. That is, the initial ice Ih grain size that forms when the liquid water first comes in contact with the sub-freezing polar atmospheric temperature, $T_{sp}$. For the corresponding steady state pressures we also plot the time durations the sea-ice remains afloat, $\Delta\tau_{asy}$. We further assume that $S_w=1$. The initial grain size is important in determining the rate of conversion from ice Ih into SI CO$_2$ clathrate hydrate \citep{Staykova2003,Genov2004}. The smaller the initial ice Ih grain size is the faster is the conversion. Thus, the period of time the sea-ice slab remains afloat is shorter. This results in a more efficient mechanism for removing CO$_2$ from the atmosphere.  
Assuming $T_{sp}=240$\,K the dissociation pressure of SI CO$_2$ clathrate hydrate is approximately $3$\,bar. In the case that no sea-ice forms a steady CO$_2$ atmospheric pressure of $35$\,bar is established. From the figure it is clear that the smaller grain sizes we have solved for remove atmospheric CO$_2$ so efficiently that the resulting steady state atmospheric pressure of CO$_2$ is only slightly above the dissociation pressure of SI CO$_2$ clathrate hydrate, for the subpolar temperature. This fact means that the driving force to form SI CO$_2$ clathrate hydrate can be rather small and still sink atmospheric CO$_2$ efficiently. However, for this to be true the time duration during which the sea-ice slab needs to remain afloat and in contact with the atmosphere is very long ($\Delta\tau_{asy}>1$\,kyr).

Clearly, in water planets sea-ice is not anchored in place, like some sea-ice on the Earth. If sea-ice on water planets floats for a long period of time it may migrate to warmer climate, where its clathrates could dissociate and release the enclathrated CO$_2$ back to the atmosphere. Therefore, removal of atmospheric CO$_2$ by sinking sea-ice becomes inactive, i.e. $S_w=0$. If this happens the pressure of CO$_2$ in the atmosphere would build up and increase the driving force to form clathrates. As a result $\Delta\tau_{asy}$ would decrease until it becomes small enough so that the sea-ice becomes buoyantly unstable before drifting out of the subpolar region. This leads to the criterion:
\begin{equation}\label{MigrationCriterion}
\Delta\tau_{asy}\lessapprox\frac{R_p}{v_{div}}\left(\frac{\pi}{2}-\lambda_{sp}\right)\equiv\Delta\tau^{limit}_{asy}
\end{equation}
Here $v_{div}$ is the divergence velocity that is responsible for the drift of ice slabs out of the polar region. Ice slabs roughly follow atmospheric isobars that constantly change, in addition to colliding with each other. Therefore, sea-ice motion has an important stochastic component \citep{SeaIceDrift}.
The free drift velocity of sea-ice is easier to estimate. It is mostly due to winds, and is at most $2\%$ of the geostrophic wind speed \citep{Thorndike1982}. For the latter we adopt $10$\,m\,s$^{-1}$ \citep[see][for derived subpolar wind velocities]{Marshall2007}.
However, the free drift velocity is probably somewhat higher than the divergence velocity, and is perhaps more appropriate for describing a mean free path velocity.   
Hence, for our adopted velocity the above criterion is somewhat too stringent, forcing the atmospheric pressure of CO$_2$ to higher values.

Whether the last criterion is satisfied or not switches $S_w$ between $1$ and $0$, respectively.
We model it using a smooth step function:
\begin{equation}\label{Swmodel}
S_w=\exp\left\lbrace-\left(\frac{\Delta\tau_{asy}}{\Delta\tau^{limit}_{asy}}\right)^d\right\rbrace
\end{equation}
Rather then a sharp latitudinal limit, the surface temperature ought gradually increase toward lower latitudes. As a consequence the local dissociation pressure of clathrates increases and a smooth step function is more appropriate. Below, we test for the sensitivity of the steady-state atmospheric pressure to our adopted model for $S_w$ by varying the power $d$.
 
For $\lambda_{sp}=60^\circ$ we have $\Delta\tau^{limit}_{asy}\approx 1$\,yr. This gives a sea-ice thickness, $h_{ice}$, of $2$\,m at most. We use this criterion to cap the maximal sea-ice thickness for purposes of modelling its role in sinking atmospheric CO$_2$. Now with the sea-ice migration taken into account let's resolve fig.\ref{fig:Swunity}.

\begin{figure}[ht]
\centering
\mbox{\subfigure{\includegraphics[width=7cm]{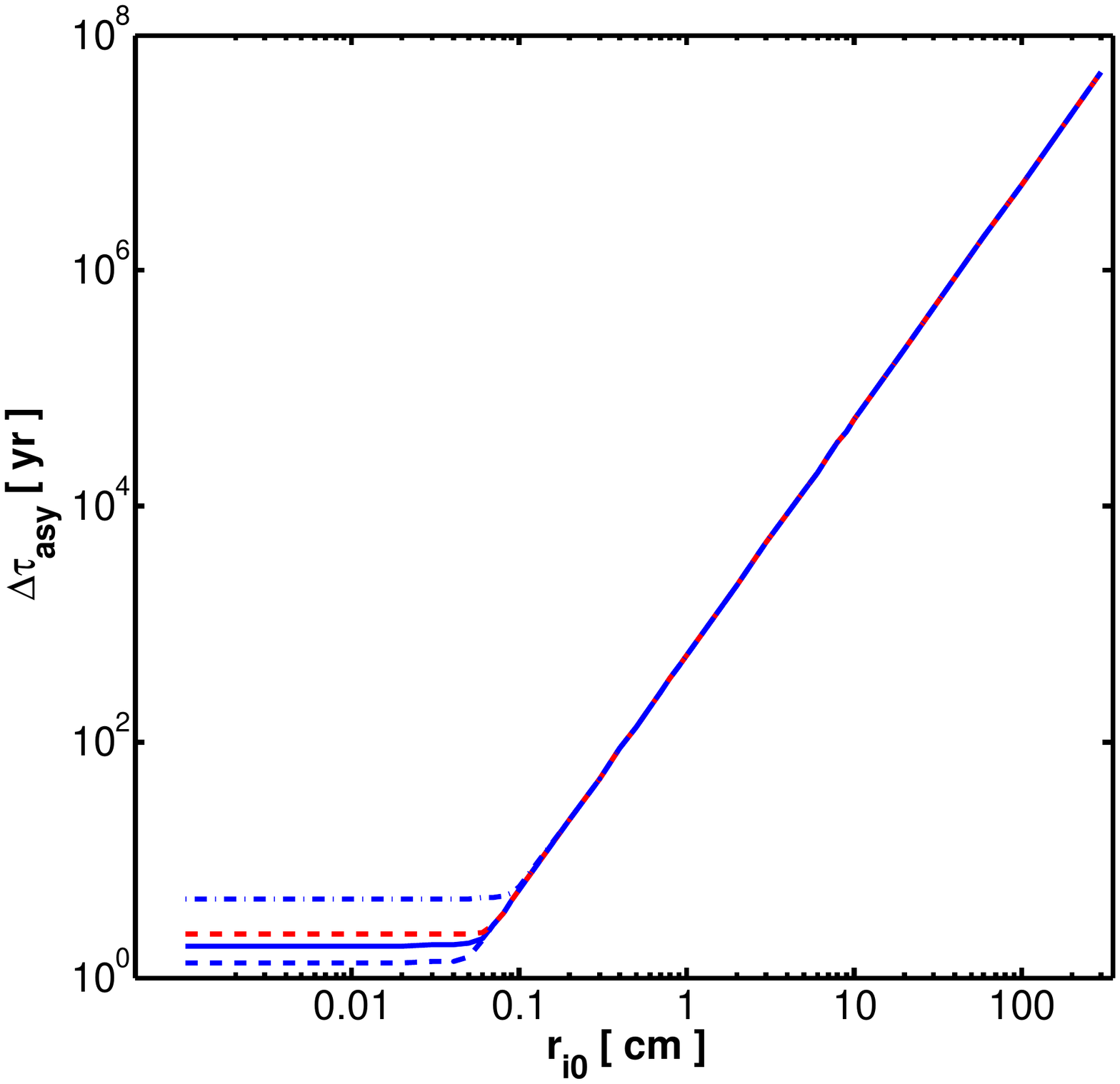}}\quad \subfigure{\includegraphics[width=7cm]{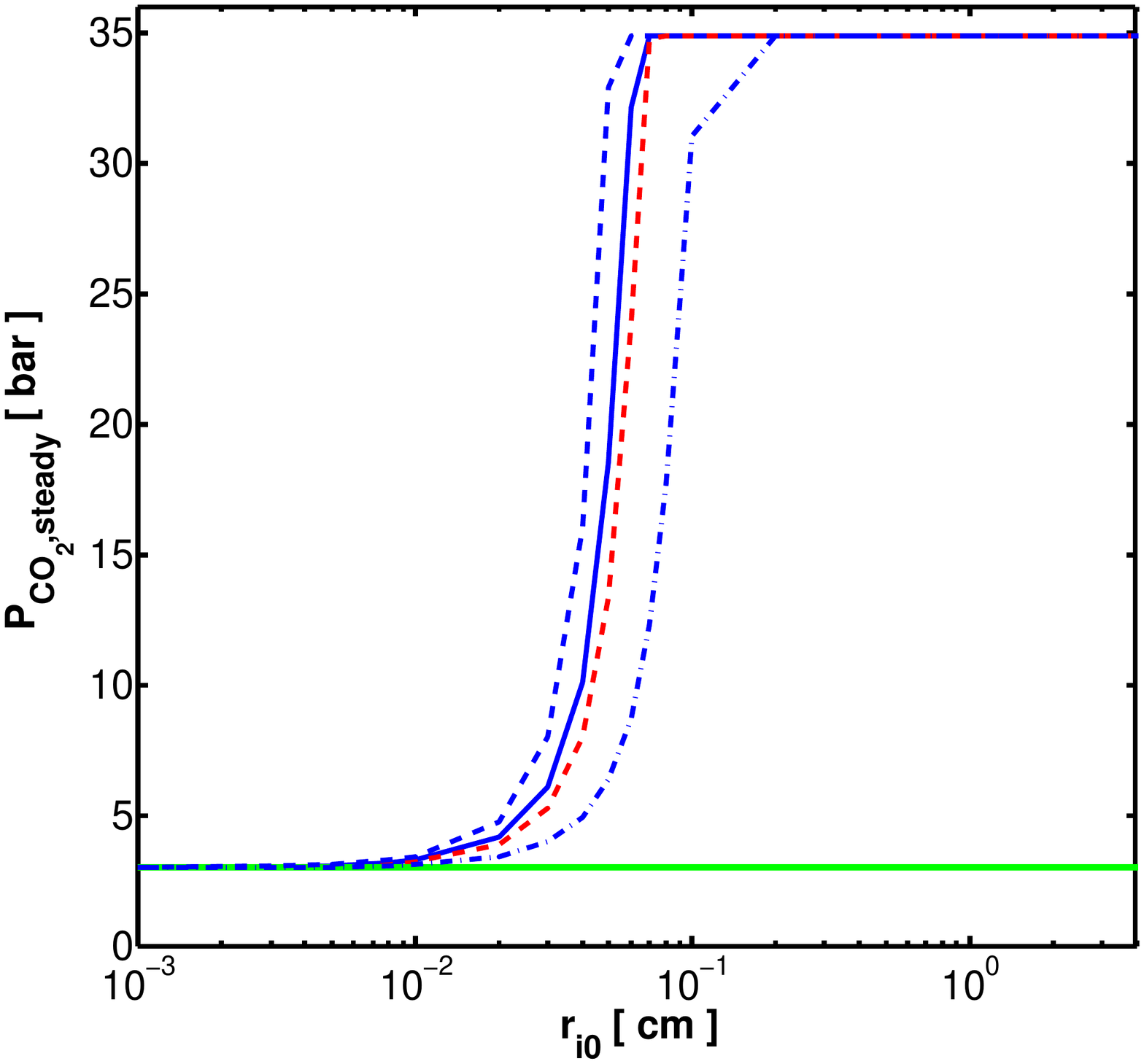}}}
\caption{\footnotesize{(right panel) The steady state atmospheric CO$_2$ partial pressure dependence on the initial ice Ih grain radius, the horizontal green curve is the dissociation pressure for SI CO$_2$ clathrate hydrate for a temperature of $T_{sp}=240$\,K. (left panel) The corresponding time the sea-ice remains afloat under steady-state atmospheric conditions. In both panels the dashed red curve correspond to $D_{eddy}=10^{-4}$\,cm$^2$\,s$^{-1}$ and $d=2$ in the model for $S_w$ in Eq.($\ref{Swmodel}$). All the blue curves assume $D_{eddy}=10^{-2}$\,cm$^2$\,s$^{-1}$. Dashed-dotted, solid and dashed blue curves assume $d=1$, $d=2$ and $d=3$ in the model for $S_w$ in Eq.($\ref{Swmodel}$), respectively. Other system parameters adopted are: $T_{sp}=240$\,K, $T_{sbt}=20^\circ$C, $T_{deep}=4^\circ$C, $\phi^0_{pore}=0.1$, $\rho_{pore}=0$\,g\,cm$^{-3}$ and $\lambda_{sp}=60^\circ$. Also we assume $Q_c=0$\,bar\,Myr$^{-1}$.}}
\label{fig:SwReal}
\end{figure}

In fig.\ref{fig:SwReal} we resolve for the steady state CO$_2$ partial atmospheric pressure as a function of the initial ice Ih grain size. For an initial ice Ih grain size smaller than $100\mu$m the global CO$_2$ atmospheric pressure is only a few tenths of a percent higher than the clathrate dissociation pressure for the subpolar temperature $T_{sp}$. For these initial grain sizes even such a small clathrate formation driving force is sufficient to make the time it takes the sea-ice to become buoyantly unstable and sink, $\Delta\tau_{asy}$, obey the criterion of Eq.($\ref{MigrationCriterion}$). When the initial ice Ih grain size is a few hundred micrometer, diffusional and reaction limitations begin to hinder the ability of the ice Ih grain to convert to CO$_2$ clathrate fast enough. Therefore, migration of clathrate rich sea-ice to warmer climate begins to choke the sink mechanism. Hence, the atmospheric CO$_2$ pressure continues to increase above the clathrate dissociation pressure for $T_{sp}$. If the initial grain size is close to $1$\,mm even a further build up in the atmospheric pressure of CO$_2$ cannot provide a sufficient driving force to form clathrates fast enough to counteract the increased inner grain diffusion time-scale. Remember, the conversion into clathrate hydrate requires that the CO$_2$ molecules diffuse across a clathrate hydrate layer in order to reach the inner ice Ih grain. Thus sea-ice migration to warmer climates becomes dominant and the CO$_2$ atmospheric pressure quickly bounces to the values dictated by the CO$_2$ saturated deep ocean alone.

From fig.\ref{fig:SwReal} we see that varying the vertical deep ocean eddy diffusion by two orders of magnitude effects the steady-state atmospheric pressure of CO$_2$ less than varying $d$, from the model for $S_w$, between $1$ and $3$. This implies our results our somewhat sensitive to the adopted model for $S_w$. However, the transition between an ice-cap controlled atmospheric pressure to one controlled by the saturated deep ocean falls in the grain size regime of a few hundred micrometers. This result is insensitive to how sharp the step function $S_w$ is.     

Clearly, the initial ice Ih grain size distribution is of great importance. Unfortunately, the range of a few hundred micrometers does not allow for an easy determination of whether the sea-ice CO$_2$ removal mechanism is dominant or not. Grains falling in this range are indeed observed when ice forms from supercooled water \citep{Arakawa1954}. Here we have used experimentally determined ice Ih to clathrate conversion rates. These rates were obtained by using artificially sieved spherical ice grain samples. However, when supercooled water freezes, disk crystals are first formed. A more flattened shape would turn the inner ice Ih grain more accessible to diffusing CO$_2$. Therefore, increasing the rates of conversion from ice Ih to clathrate. \cite{Shimada1997} observed that for pure supercooled water disk crystals, though a few mm in diameter, are only $100-200\mu$m thick. However, in their experiments CO$_2$ was not introduced. It is interesting to note that \cite{Shimada1997} found that an ice crystal thickness of $100\mu$m is attained after about $10^3$\,s (for their type II disc crystal). According to \cite{Genov2004} this time scale is of the same order of magnitude as that which is required to form a surface clathrate hydrate layer that may coat the ice Ih grain. Therefore, it is very probable that CO$_2$ clathrates become incorporated into the grain during its period of growth. The addition of clathrates would also decrease the grain's thermal conductivity keeping it thinner, and its interior more accessible. 
These complications though are beyond the scope of this work, and need to be addressed experimentally. 

Additional complications arise from our supercooled environment being highly influenced by waves, winds and resulting turbulences of various scales. 
It is however clear that sea-ice of the right morphology is able to decrease the atmospheric pressure of CO$_2$ by an order of magnitude from the wind driven value. Observations may therefore provide the answer. The considerable jump in atmospheric pressure over a relatively narrow grain size domain may actually increase the confidence in interpreting future atmospheric observations.                     

\begin{figure}[ht]
\centering
\includegraphics[trim=0.15cm 3cm 0.08cm 5.1cm , scale=0.55, clip]{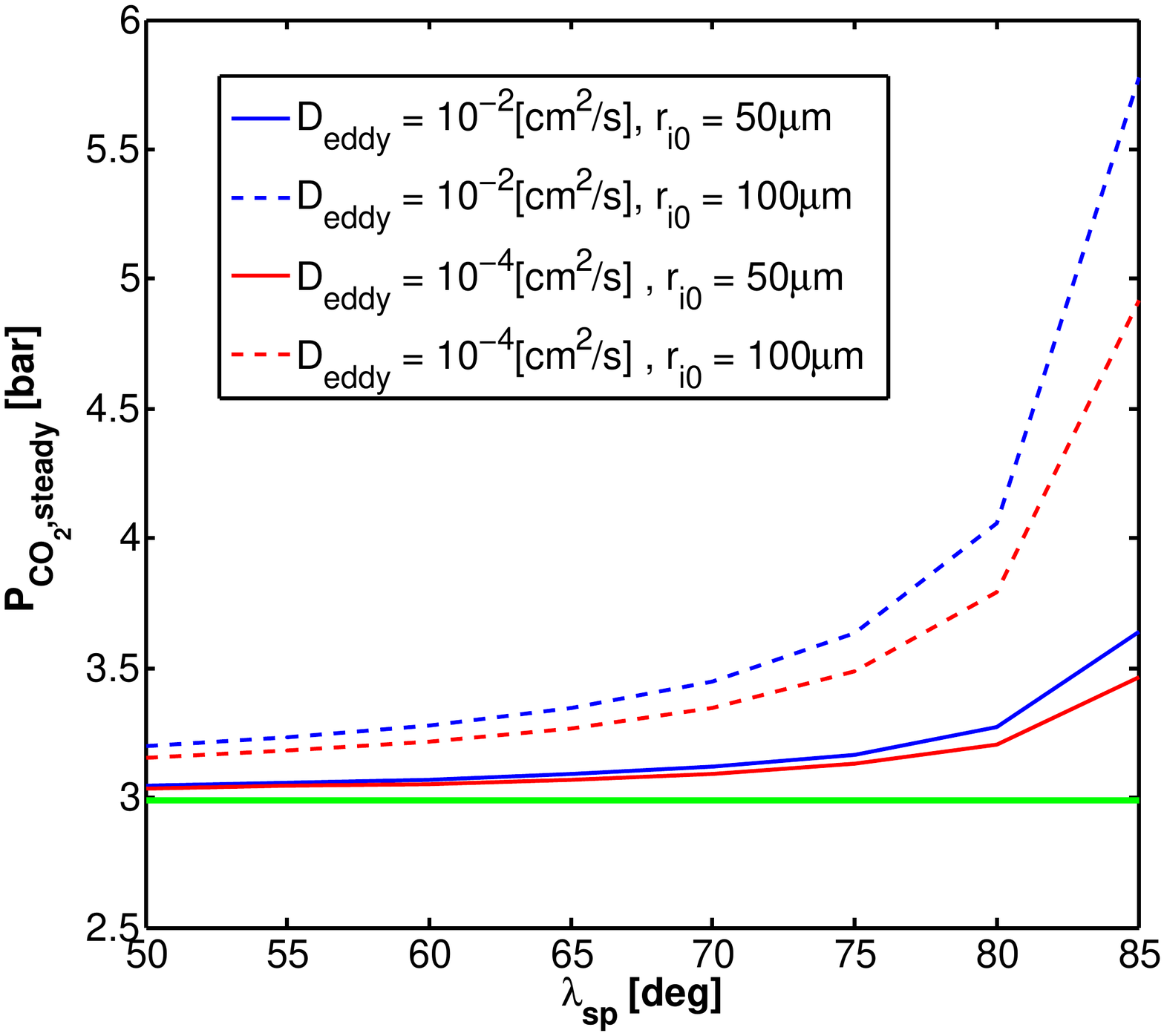}
\caption{\footnotesize{Steady-state atmospheric CO$_2$ partial pressure versus the latitudinal extent of the polar ice cap. Dashed (blue and red) curves are for an initial ice Ih grain size of $100\mu$m. Solid (blue and red) curves are for an initial ice Ih grain size of $50\mu$m. Blue and red curves correspond to $D_{eddy}=10^{-2}$\,cm$^2$\,s$^{-1}$ and $10^{-4}$\,cm$^2$\,s$^{-1}$ respectively.  
The horizontal green curve is the dissociation pressure of SI CO$_2$ clathrate hydrate, assuming $T_{sp}=240$\,K. Above this horizontal curve CO$_2$ clathrates form in the sea-ice. We also assume $d=2$. See other system parameters in the caption to fig.\ref{fig:SwReal}. }}
\label{fig:PCO2PoleLatitudeVar1}
\end{figure}

In fig.\ref{fig:PCO2PoleLatitudeVar1} we plot the steady-state atmospheric partial pressure of CO$_2$ as a function of the latitude to which the sea-ice cap extends. Clearly diminishing the extent of the sea-ice cap reduces the ability of sea-ice formation to remove atmospheric CO$_2$. 
It is however interesting to note that the atmospheric pressure of CO$_2$ is relatively insensitive to the ice cap area as long as that is more than about $10^\circ$ around the pole. The smaller the initial ice Ih grain, the less sensitive is the steady state pressure to the extent of the ice cap. In addition, it is reasonable to expect that a higher deep ocean vertical diffusivity would tend to force a higher volatile steady-state atmospheric pressure. This is seen in the figure where we have varied the vertical diffusivity coefficient of the deep ocean by two orders of magnitude. However, changing the grain size radius by a mere factor of two results in even a larger effect on the atmospheric pressure. This is reasonable because the diffusivity coefficient will mostly determine the time for the secondary outgassed volatiles to cross the deep unmixed ocean whereas the steady-state condition greatly depends on the formation of clathrate-rich sea ice, which due to the sea-ice migration criterion is highly dependent on the sea-ice grain morphology.         

Varying the distance between a water planet and its star probably changes the extent of the ice cap.   
The results of fig.\ref{fig:PCO2PoleLatitudeVar1} suggest that the planet's atmospheric CO$_2$ content may resist any considerable change as long as the ice cap stretches further out than $10^\circ$ from the pole. If that is not the case the system may collapse into the pure wind driven atmosphere we have already discussed above.

\begin{figure}[ht]
\centering
\mbox{\subfigure{\includegraphics[width=7cm]{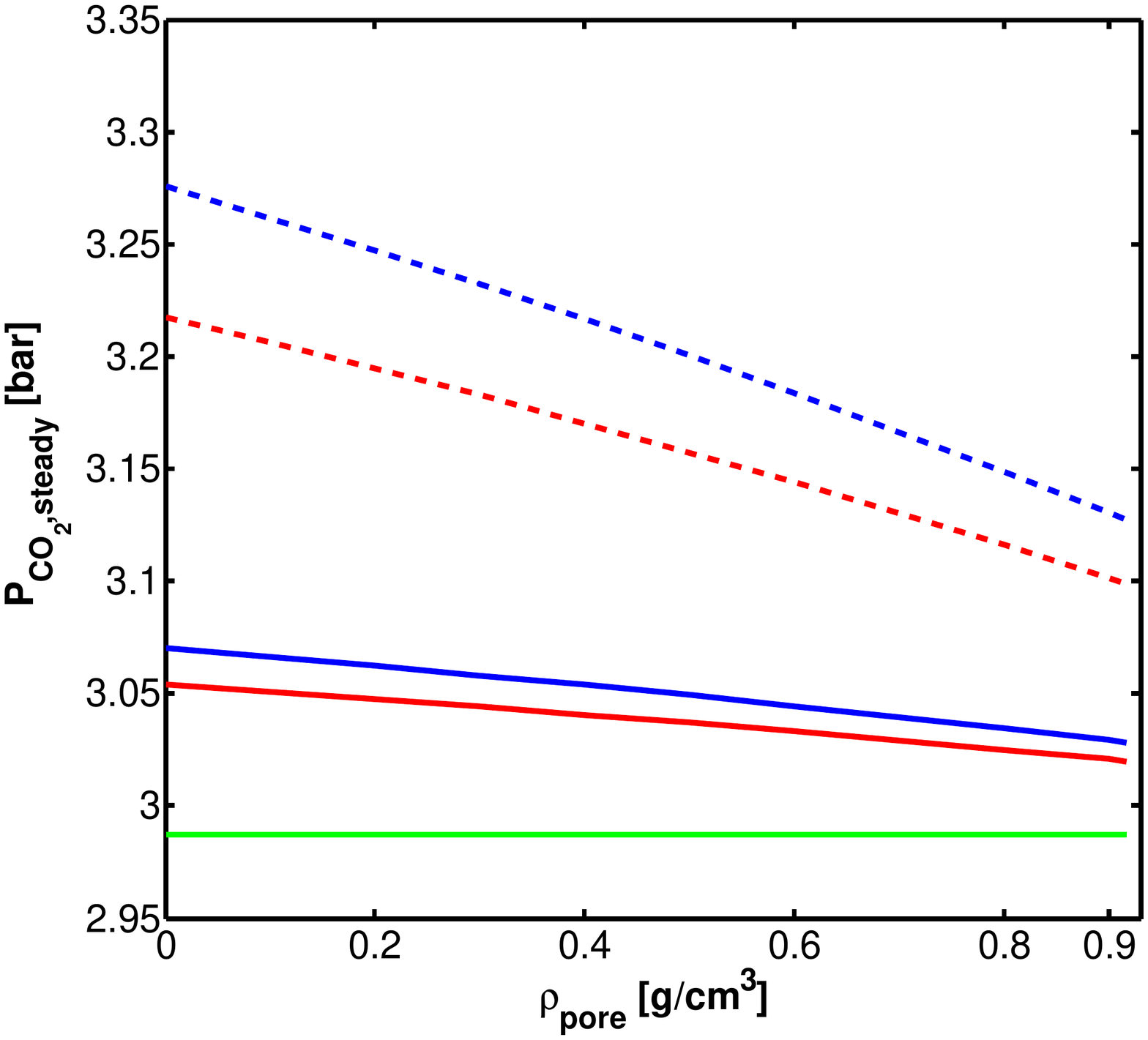}}\quad \subfigure{\includegraphics[width=7cm]{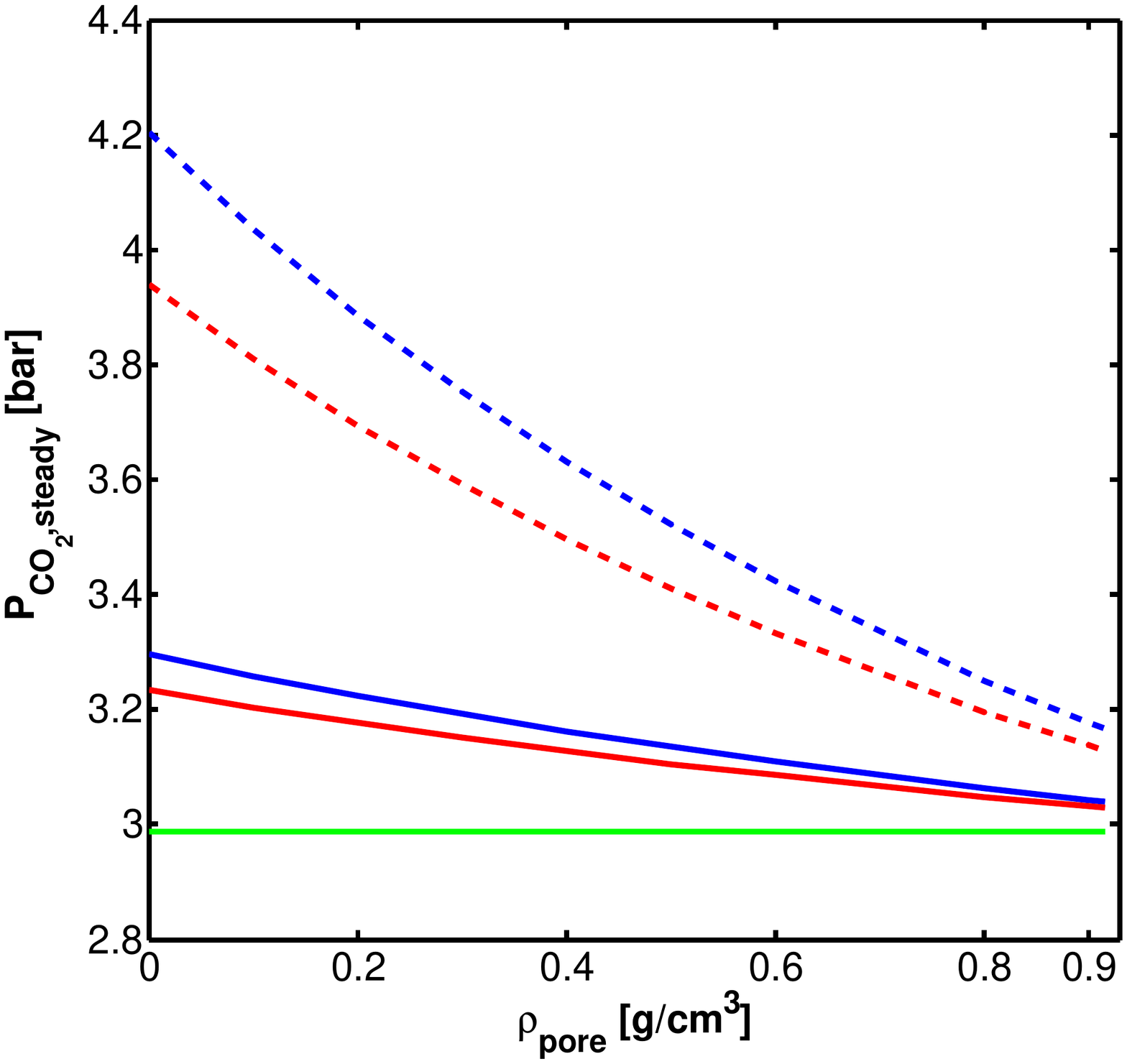}}}
\caption{\footnotesize{Steady-state atmospheric CO$_2$ partial pressure versus the mass density of the pore-filling material. Dashed (blue and red) curves are for an initial ice Ih grain size of $100\mu$m. Solid (blue and red) curves are for an initial ice Ih grain size of $50\mu$m. Blue and red curves correspond to $D_{eddy}=10^{-2}$\,cm$^2$\,s$^{-1}$ and $10^{-4}$\,cm$^2$\,s$^{-1}$ respectively.  
The horizontal green curve is the dissociation pressure of SI CO$_2$ clathrate hydrate, assuming $T_{sp}=240$\,K. Above this horizontal curve CO$_2$ clathrates form in the sea-ice. \textbf{Right panel} is for an initial sea-ice porosity of $\phi^0_{pore}=0.2$. \textbf{Left panel} is for an initial sea-ice porosity of $\phi^0_{pore}=0.1$. We also assume $d=2$. See other system parameters in the caption to fig.\ref{fig:SwReal}.}}
\label{fig:PCO2PoreFillerVar1}
\end{figure}

In fig.\ref{fig:PCO2PoreFillerVar1} we plot the steady-state atmospheric partial pressure of CO$_2$. We examine the influence of varying the mass density of the pore filling material between that of ice Ih and vacuum. One expects that the denser pore filler would make the sinking of the sea-ice more effective, and therefore the steady state atmospheric CO$_2$ pressure lower. On the right (left) panel we assume an initial porosity, $\phi^0_{pore}$, of $0.2$ ($0.1$). Over the range of mass density we test for the pore filling material, and for the lower initial porosity, the steady-state atmospheric pressure changes by $1.3\%$ and $4.4\%$ for the smaller and larger initial grain sizes respectively. For the higher initial porosity, the steady-state atmospheric pressure changes by $7.3\%$ and $25.5\%$ for the smaller and larger initial grain sizes respectively, over the mass density range examined.
The lower the initial porosity, the less sensitive the CO$_2$ atmospheric pressure becomes to the pore-filler density.    
The sea-ice becomes negatively buoyant when reaching a ice Ih to clathrate hydrate conversion fraction of $\alpha_{min}$. As the value for the initial sea-ice porosity increases, so does the \textit{difference} in the value for $\alpha_{min}$ between a pore filling density of vacuum and of ice Ih (see fig.\ref{fig:SeaIceStability}).

\begin{figure}[ht]
\centering
\includegraphics[trim=0.15cm 3cm 0.2cm 5cm , scale=0.60, clip]{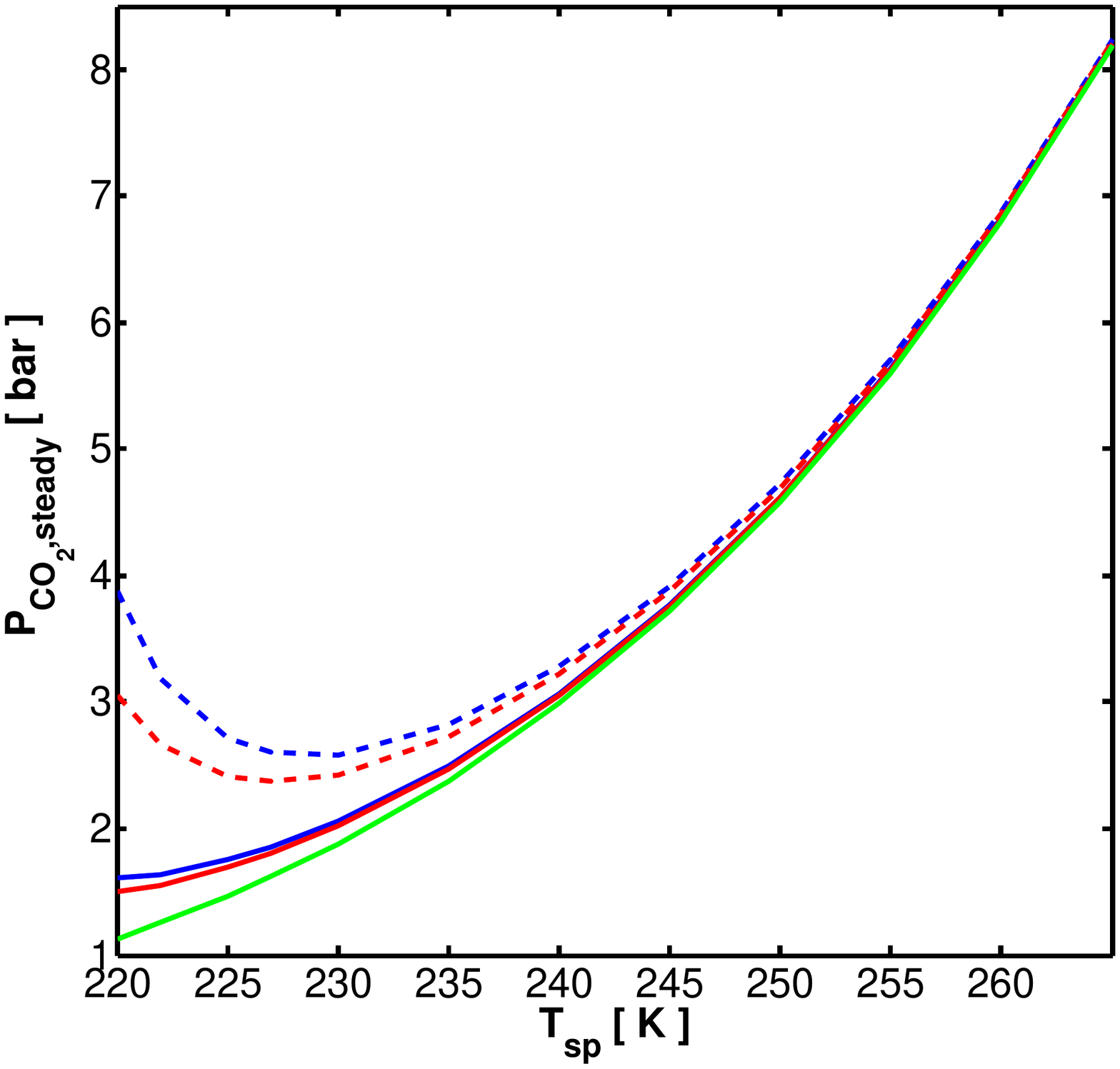}
\caption{\footnotesize{The steady-state partial atmospheric pressure of CO$_2$ as a function of the average polar surface temperature.
Dashed (blue and red) curves are for an initial ice Ih grain size of $100\mu$m. Solid (blue and red) curves are for an initial ice Ih grain size of $50\mu$m. Blue and red curves correspond to $D_{eddy}=10^{-2}$\,cm$^2$\,s$^{-1}$ and $10^{-4}$\,cm$^2$\,s$^{-1}$ respectively.  
The green curve is the dissociation pressure of SI CO$_2$ clathrate hydrate for the corresponding average polar temperature, $T_{sp}$.
Other system parameters adopted are: $T_{sbt}=20^\circ$C, $T_{deep}=4^\circ$C, $\phi^0_{pore}=0.1$, $\rho_{pore}=0$\,g\,cm$^{-3}$ and $\lambda_{sp}=60^\circ$. Also we assume $Q_c=0$\,bar\,Myr$^{-1}$ and $d=2$.}}
\label{fig:SteadyStateWithIce}
\end{figure}

In fig.\ref{fig:SteadyStateWithIce} we plot the steady-state partial atmospheric pressure of CO$_2$ as a function of the subpolar surface temperature. 
The time scale to reach a steady-state is of the order of the eddy diffusion time scale across the deep unmixed ocean. Therefore, $T_{sp}$ is an average over many planetary years rather then a seasonally dependent variable.
For our chosen system parameters we see that when the subpolar temperature is higher than about $240$\,K the steady-state atmospheric pressure of CO$_2$ is only slightly above the SI CO$_2$ clathrate hydrate dissociation pressure. This small pressure difference means a low driving force to form clathrates. It is the relatively higher temperatures that keep the rate of conversion from ice Ih to SI CO$_2$ clathrate high, and the sea-ice sink mechanism efficient.
For the lower subpolar temperatures examined the low conversion rate between the two water ice phases begins to prevent the ice from reaching buoyant instability before drifting out of the subpolar region. The system responds by increasing the atmospheric pressure of CO$_2$, in other words increasing the driving force hence accelerating the phase conversion. This extra driving force is even more necessary when considering larger initial ice Ih grains. Consequently for these larger grains the steady state atmospheric pressure has a minimum. We will return to this point in the discussion section below.

\begin{figure}[ht]
  \begin{minipage}{\textwidth}
  \centering
    \includegraphics[width=.4\textwidth]{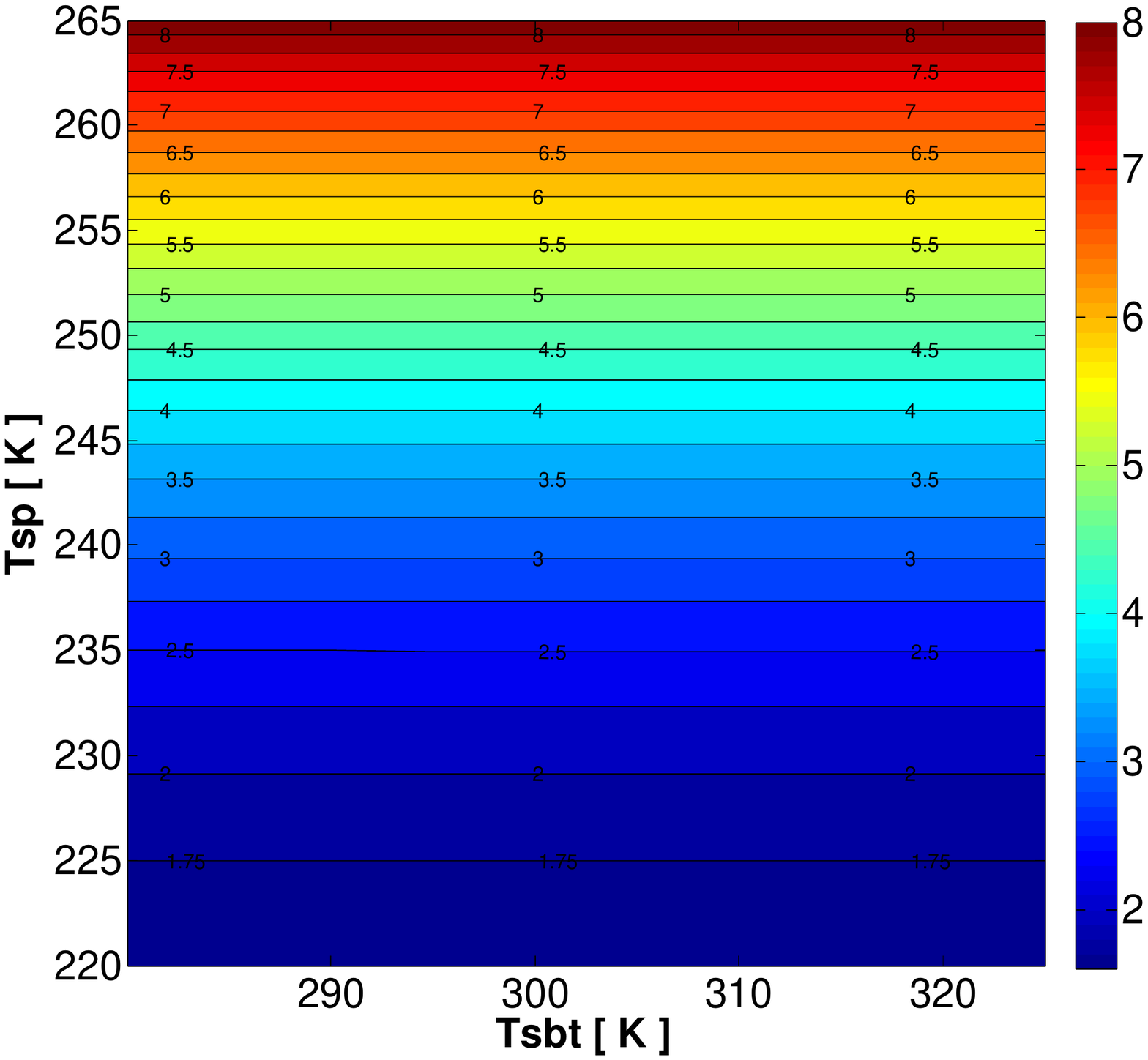}\quad
    \includegraphics[width=.4\textwidth]{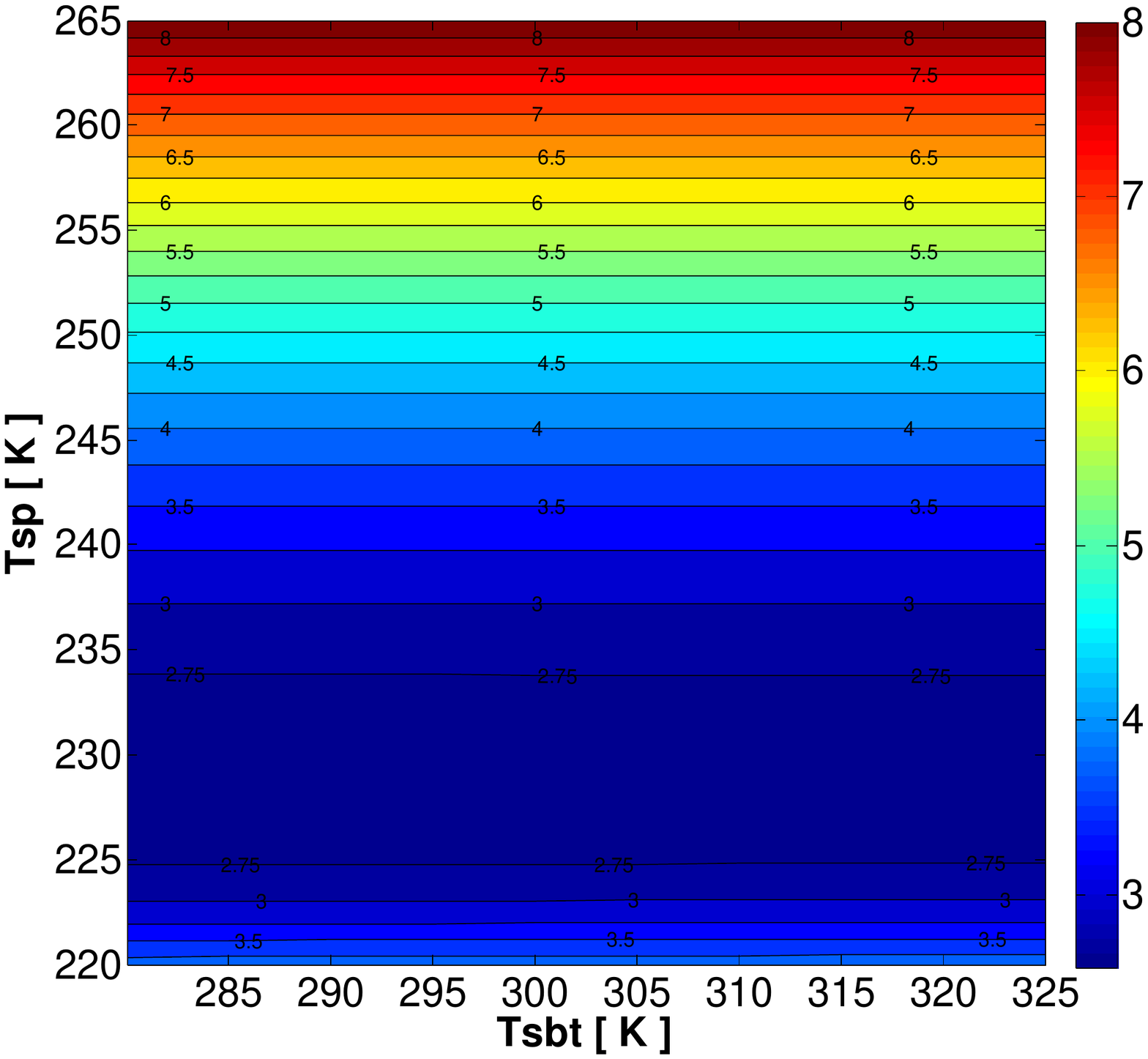}\\
    \includegraphics[width=.4\textwidth]{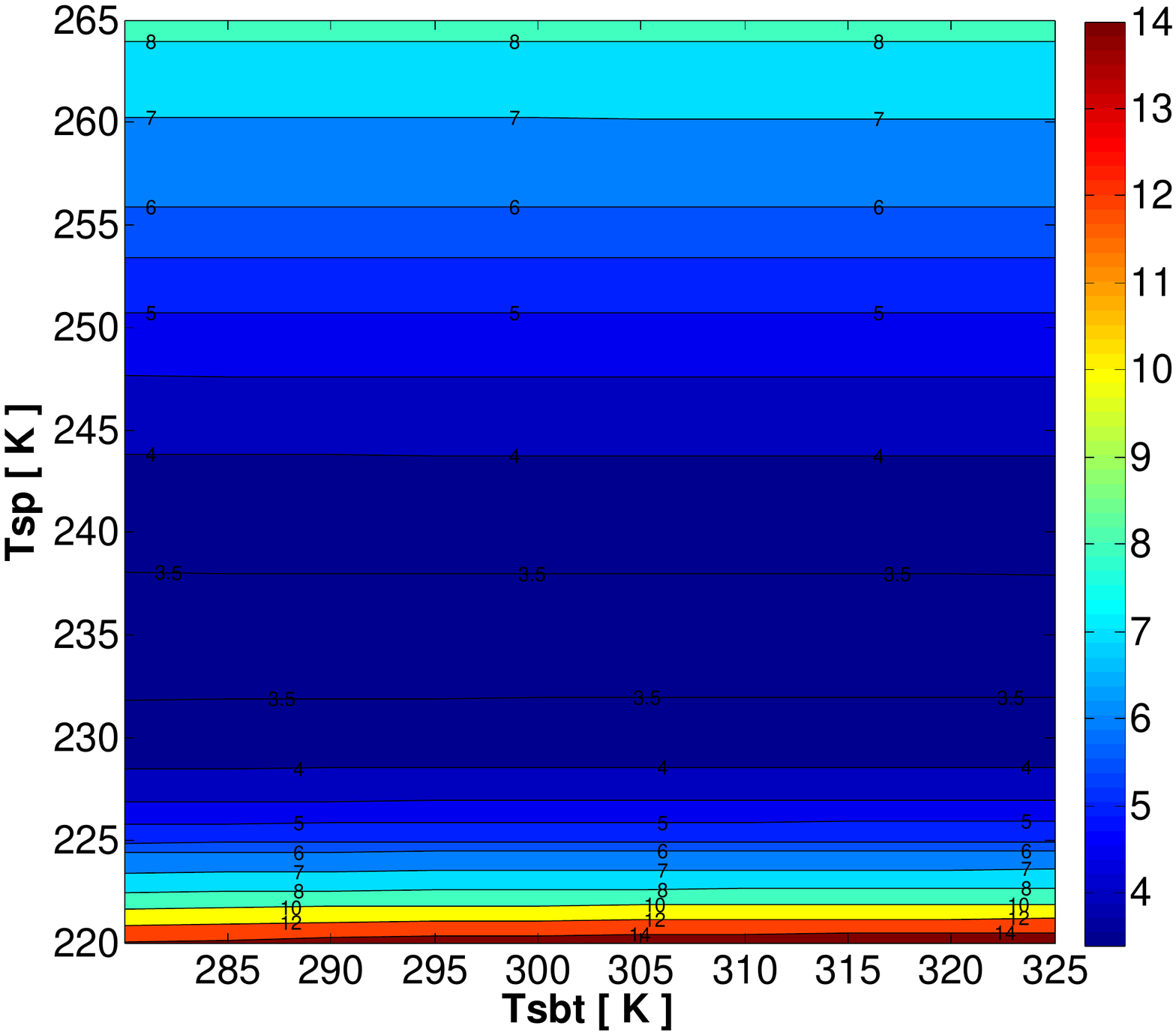}\quad
    \includegraphics[width=.4\textwidth]{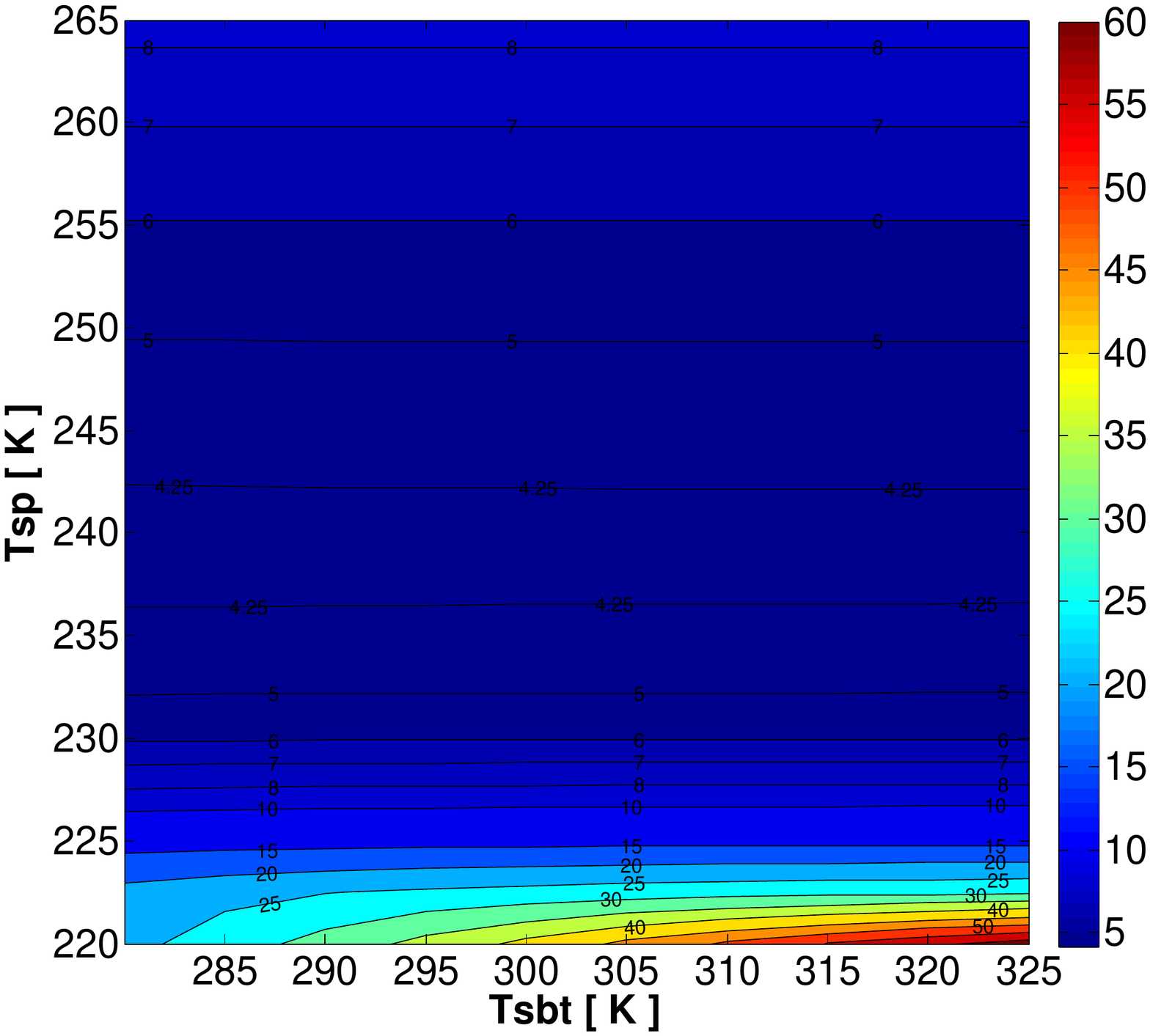}
    \caption{\footnotesize{Isobars of atmospheric CO$_2$ as a function of the subpolar, $T_{sp}$, and subtropic, $T_{sbt}$, oceanic surface temperatures. (Upper Left) Initial ice Ih grain radius of $50\mu$m. (Upper Right) Initial ice Ih grain radius of $100\mu$m. (Lower Left) Initial ice Ih grain radius of $150\mu$m. (Lower Right) Initial ice Ih grain radius of $200\mu$m. Other system parameters adopted are: $D_{eddy}=10^{-2}$\,cm$^2$\,s$^{-1}$, $T_{deep}=4^\circ$C, $\phi^0_{pore}=0.1$, $\rho_{pore}=0$\,g\,cm$^{-3}$ and $\lambda_{sp}=60^\circ$. Also we assume $Q_c=0$\,bar\,Myr$^{-1}$ and $d=2$.}}
    \label{fig:TspTsbtCold1}
  \end{minipage}\\[1em]
\end{figure}

\begin{figure}[ht]
  \begin{minipage}{\textwidth}
  \centering
    \includegraphics[width=.4\textwidth]{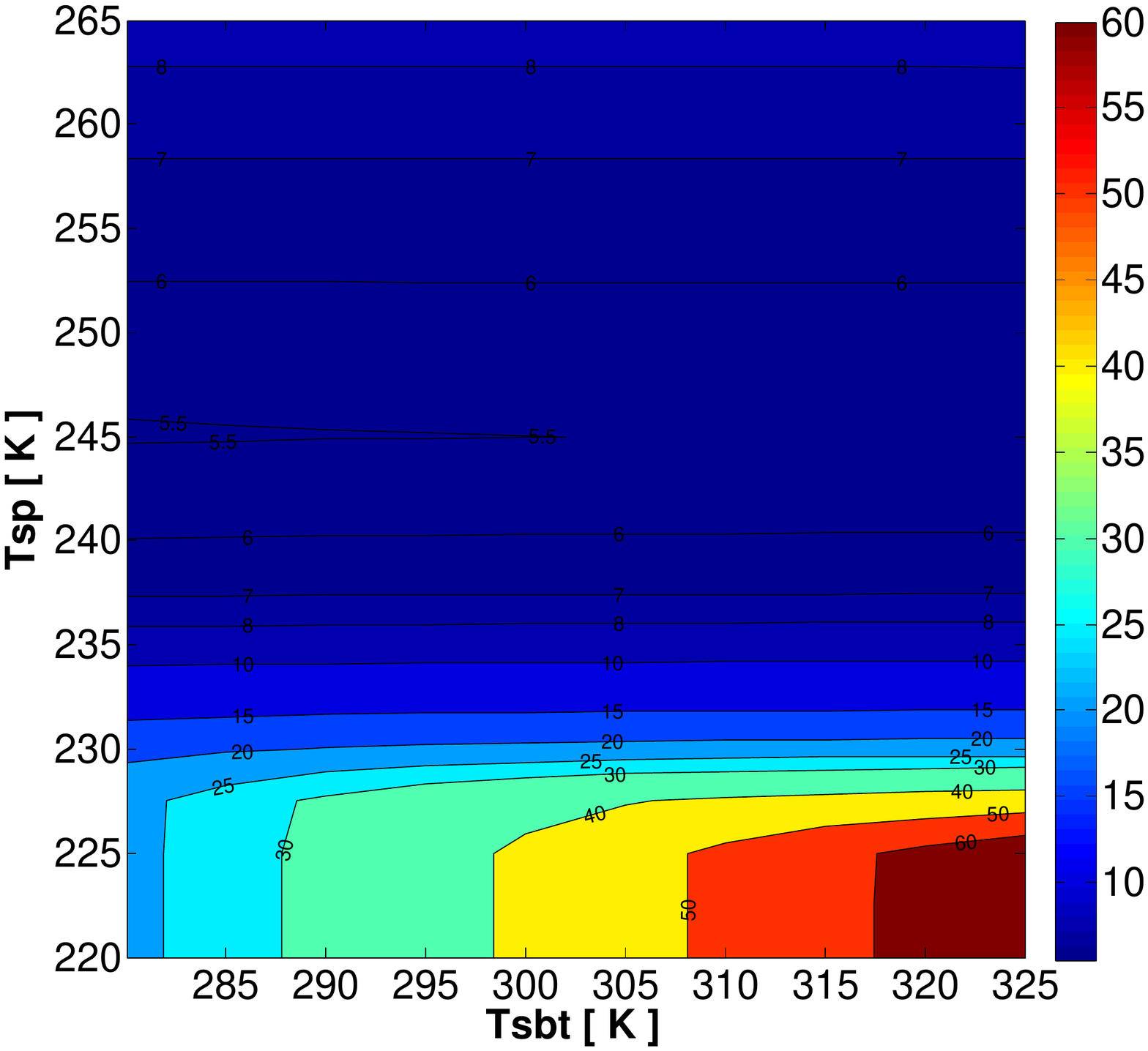}\quad
    \includegraphics[width=.4\textwidth]{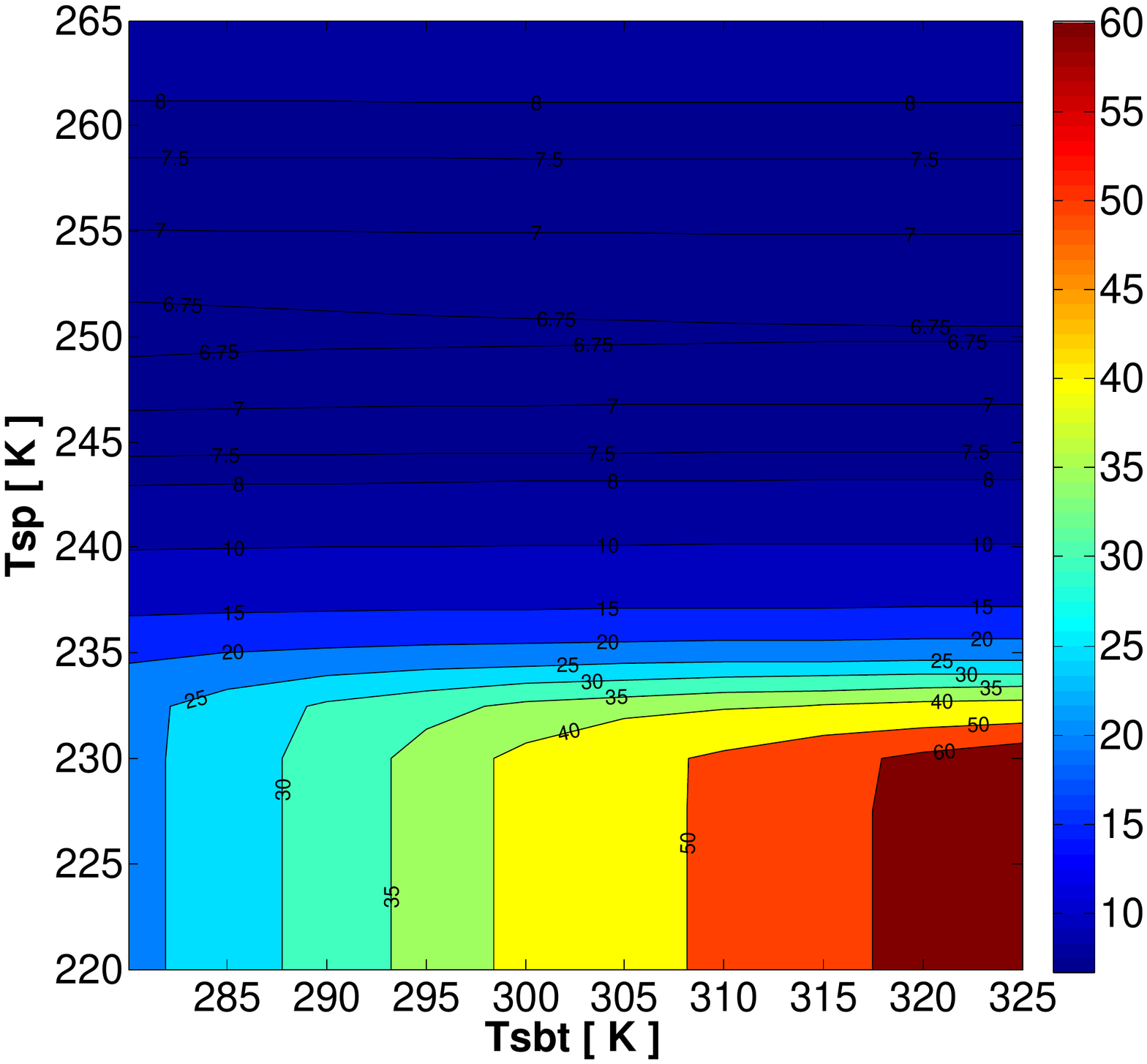}\\
    \includegraphics[width=.4\textwidth]{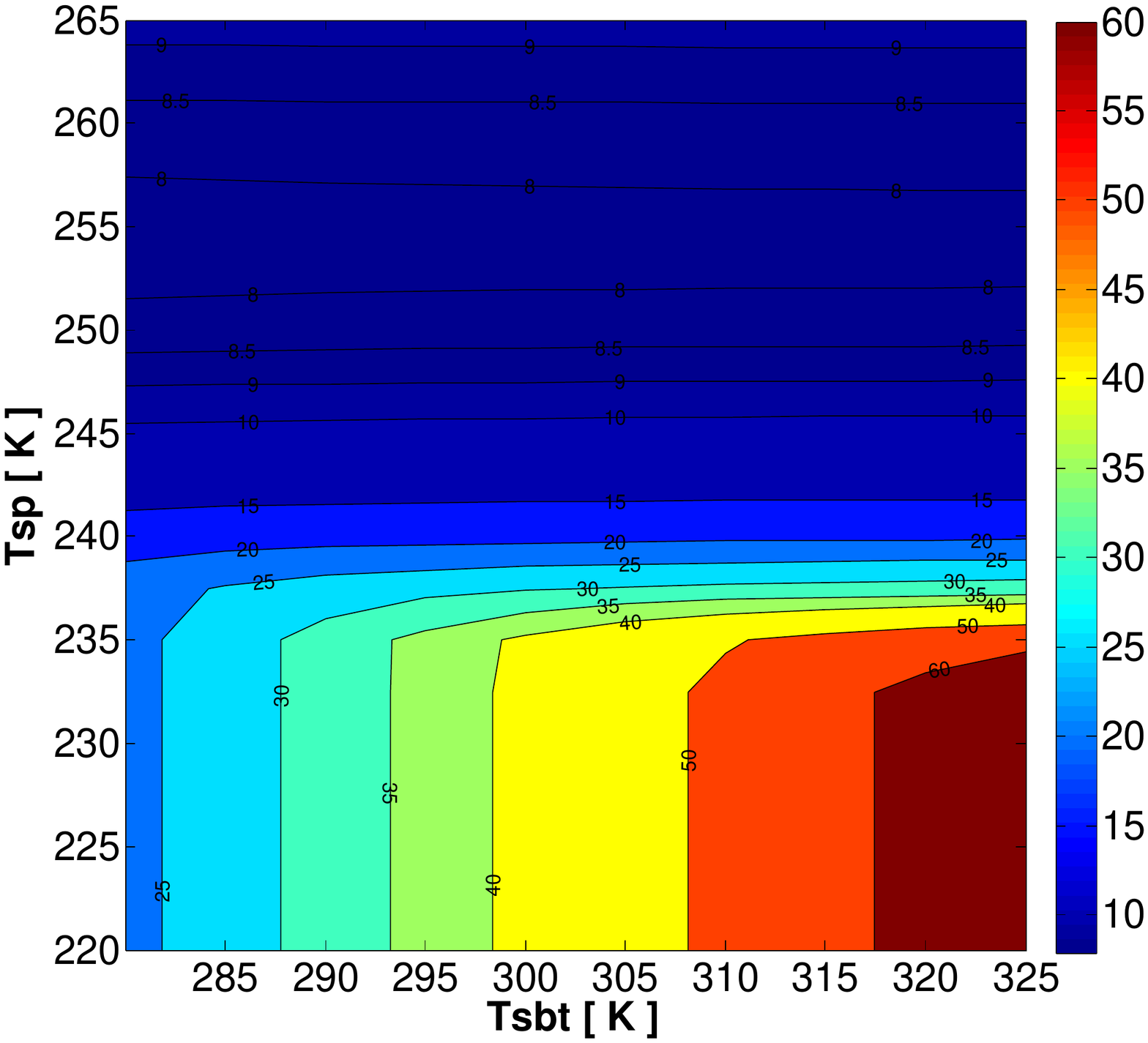}\quad
    \includegraphics[width=.4\textwidth]{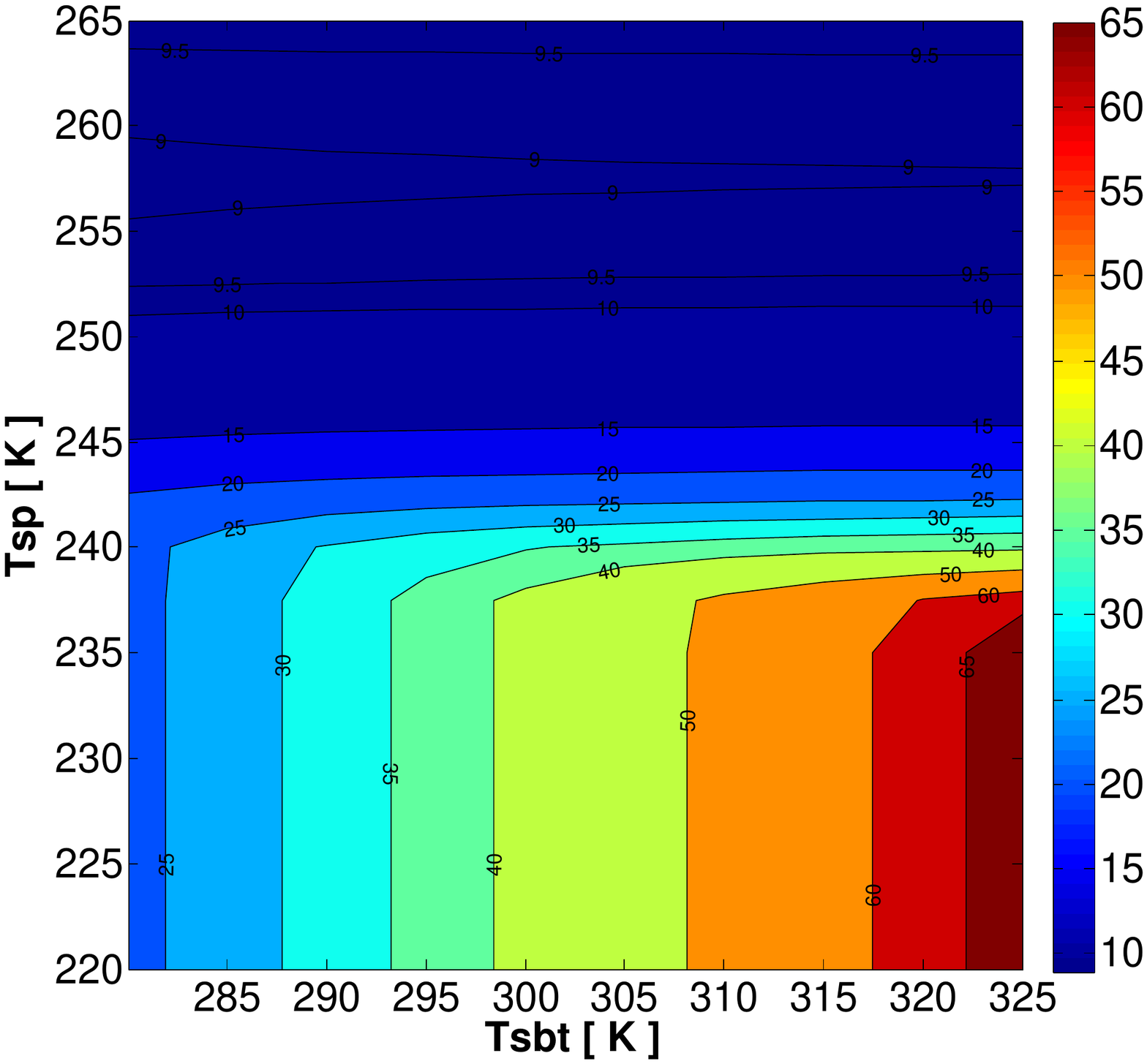}
    \caption{\footnotesize{Isobars of atmospheric CO$_2$ as a function of the subpolar, $T_{sp}$, and subtropic, $T_{sbt}$, oceanic surface temperatures. (Upper Left) Initial ice Ih grain radius of $300\mu$m. (Upper Right) Initial ice Ih grain radius of $400\mu$m. (Lower Left) Initial ice Ih grain radius of $500\mu$m. (Lower Right) Initial ice Ih grain radius of $600\mu$m. Other system parameters adopted are: $D_{eddy}=10^{-2}$\,cm$^2$\,s$^{-1}$, $T_{deep}=4^\circ$C, $\phi^0_{pore}=0.1$, $\rho_{pore}=0$\,g\,cm$^{-3}$ and $\lambda_{sp}=60^\circ$. Also we assume $Q_c=0$\,bar\,Myr$^{-1}$ and $d=2$.}}
    \label{fig:TspTsbtCold2}
  \end{minipage}\\[1em]
\end{figure}

\begin{figure}[ht]
  \begin{minipage}{\textwidth}
  \centering
    \includegraphics[width=.4\textwidth]{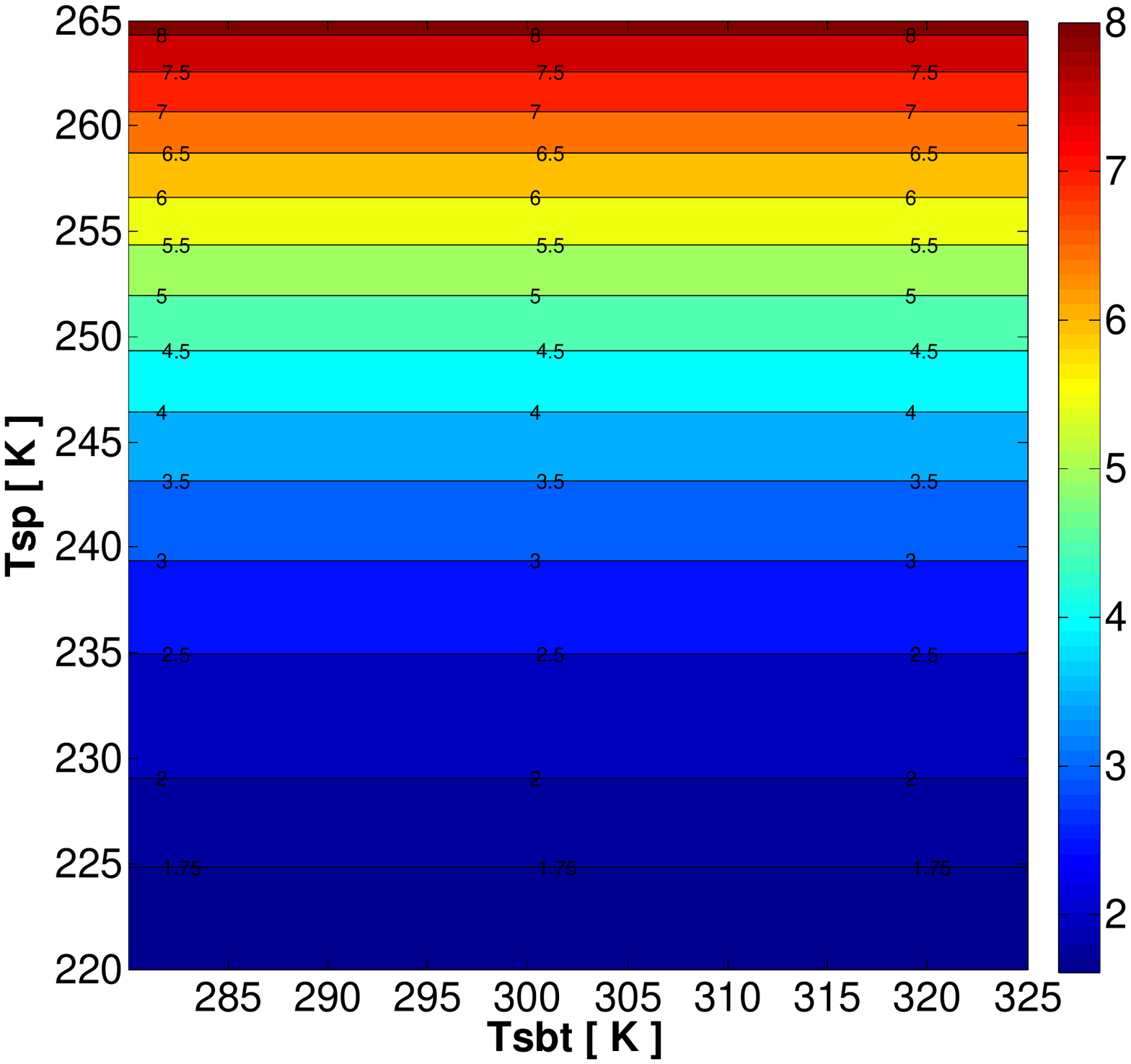}\quad
    \includegraphics[width=.4\textwidth]{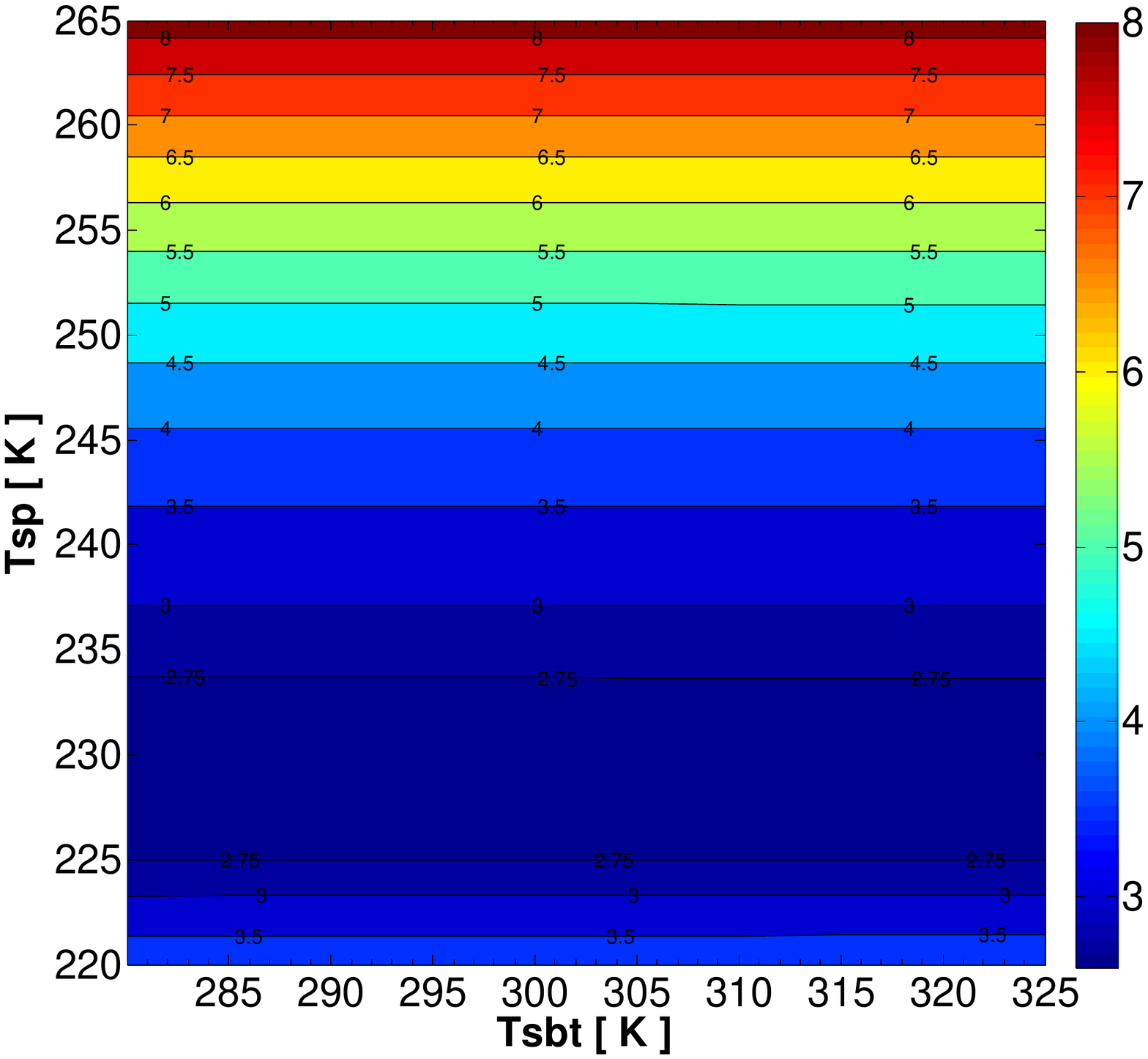}\\
    \includegraphics[width=.4\textwidth]{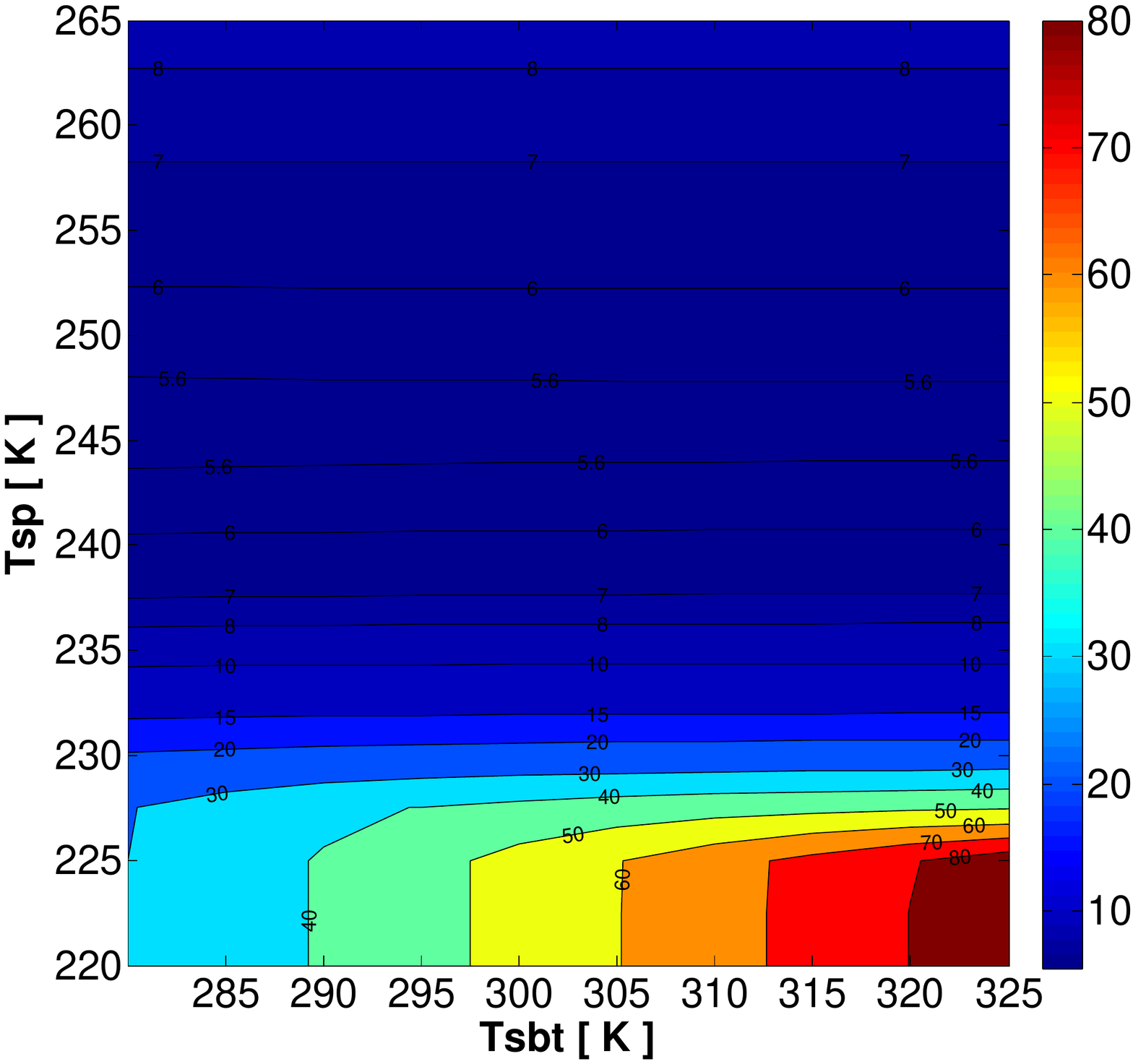}\quad
    \includegraphics[width=.4\textwidth]{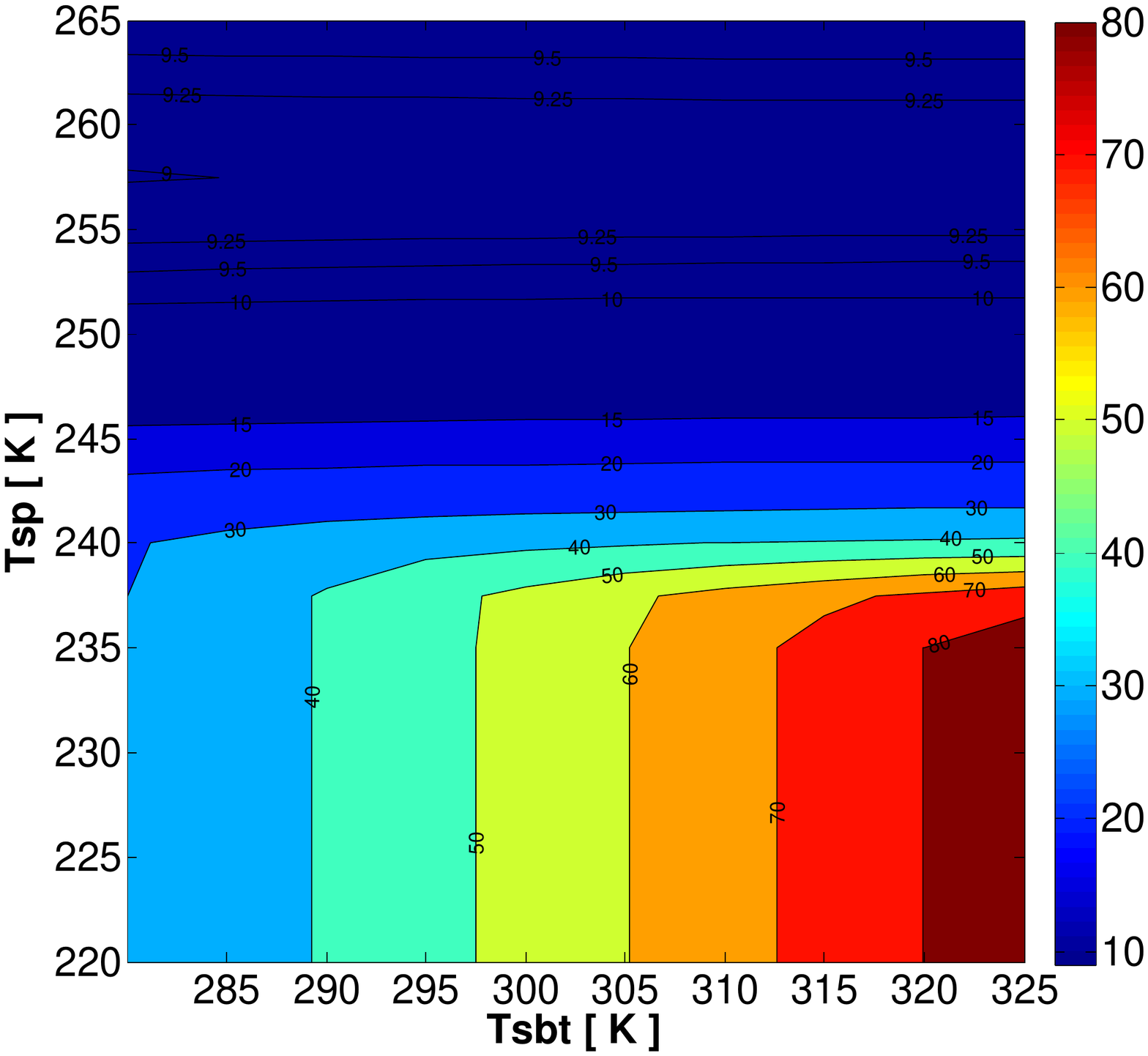}
    \caption{\footnotesize{Isobars of atmospheric CO$_2$ as a function of the subpolar, $T_{sp}$, and subtropic, $T_{sbt}$, oceanic surface temperatures. (Upper Left) Initial ice Ih grain radius of $50\mu$m. (Upper Right) Initial ice Ih grain radius of $100\mu$m. (Lower Left) Initial ice Ih grain radius of $300\mu$m. (Lower Right) Initial ice Ih grain radius of $600\mu$m. Other system parameters adopted are: $D_{eddy}=10^{-2}$\,cm$^2$\,s$^{-1}$, $T_{deep}=8^\circ$C, $\phi^0_{pore}=0.1$, $\rho_{pore}=0$\,g\,cm$^{-3}$ and $\lambda_{sp}=60^\circ$. Also we assume $Q_c=0$\,bar\,Myr$^{-1}$ and $d=2$.}}
    \label{fig:TspTsbtHot1}
  \end{minipage}\\[1em]
\end{figure}

To further our understanding of this intricate system we plot in the following figures steady-state isobars of atmospheric CO$_2$. These are plotted as a function of the subpolar and subtropic oceanic surface temperatures. In figs.\ref{fig:TspTsbtCold1} and \ref{fig:TspTsbtCold2} we vary the initial ice Ih grain radius for a deep ocean temperature of $4^\circ$C. In fig.\ref{fig:TspTsbtHot1} we resolve for some grain size cases for a hotter deep ocean, $8^\circ$C.

For the smaller \textit{initial} grain sizes the isobars are solely dependent on the subpolar temperature, throughout the entire temperature spectrum examined. Meaning, the sea-ice mechanism for sinking atmospheric CO$_2$ is the dominant effect. For the $100\mu$m and $150\mu$m cases the minimum in the atmospheric pressure is clearly seen. The fact that the isobars remain horizontal for these two cases at low $T_{sp}$ is proof that the increase in atmospheric pressure for low $T_{sp}$ is due to the system trying to increase the driving force to form clathrate hydrates rather then the wind-driven circulation taking over.

The wind-driven circulation tries to equilibrate the atmospheric pressure of CO$_2$ with the dissolved CO$_2$ throughout the deep ocean. This circulation is likely to become the dominant effect for the higher subtropic temperatures, which push the pressure upward, and for the lower subpolar temperatures, for which conversion to clathrate hydrate may become kinetically slow. Indeed this temperature criterion is where the dependency of the isobars on the subtropic temperature, $T_{sbt}$, begins to appear. For the case of the initial grain size of $200\mu$m, and for our highest examined $T_{sbt}/T_{sp}$ ratio, the wind driven circulation becomes the dominant effect.

For the larger initial grain sizes the isobars become more and more dependent on the subtropic surface temperature, as the wind driven circulation takes over. However, the isobars persist to be horizontal for the higher subpolar temperatures. In addition they keep close to the dissociation pressure of SI CO$_2$ clathrate hydrate, which is approximately $8.2$\,bar for $265$\,K.

For the case of the hotter deep ocean the response of the isobars to the different oceanic surface temperatures and grain sizes is similar to the case of the colder deep ocean. When the sea-ice sink mechanism is dominant the deep ocean temperature hardly affects the observed atmosphere. On the contrary, when the wind-driven circulation becomes dominant it equilibrates the atmospheric CO$_2$ to a value which is dependent on the deep ocean temperature.

\section{DISCUSSION}

Our model for the SI CO$_2$ clathrate hydrate yields cage occupancies as a function of pressure and temperature. For the pressure range of a few tens of bars some experiments suggest that the small cage occupancy is lower than predicted by our model (see figs.\ref{fig:OccupPorb} and \ref{fig:CO2AbundClath}). Cage occupancies are part of our model for the solubility of CO$_2$ in water while in equilibrium with the clathrate hydrate phase. This possible discrepancy at a few tens of bars suggests our derived solubility may be exaggerated by approximately $8$\%, for this pressure range. However, our geophysical model relies more on the solubility in equilibrium with the clathrate phase for pressures above hundreds of bars. Hence we do not expect this uncertainty to introduce a considerable error into the geophysical model.

Recent experiments \citep{Bollengier2013,Tulk2014} indicate that rather then separating to CO$_2$ ice and water ice under high pressure the SI clathrate hydrate of CO$_2$ transforms into a new phase with a similar crystallographic structure as of the filled ice of methane. This new phase is probably stable up to $1$\,GPa. Since at the moment little is known of this new phase quantifying its influence is impossible and more experimental data is needed. Because this new filled ice phase becomes unstable at pressures above $1$\,GPa it probably does not play a significant role in the transport of CO$_2$ from the deep mantle outward. It may play a role in setting the mechanical properties of the upper boundary layer of the ice mantle convection cell, and thus the dynamics of the ocean bottom. A filled ice sub-layer may act as extra storage for CO$_2$, in addition to the clathrate layer. CO$_2$ that may become available to the ocean if it tries to unsaturate.     

In the $\beta$ domain (see fig.\ref{fig:alphabetagamma}), if the ocean's bottom is composed of filled-ice of CO$_2$, then one should ask what solubility of CO$_2$ is enforced in the overlying ocean due to this new phase. If this solubility is higher than the value we estimate for the equilibrium with the clathrate hydrate phase, then an elevated clathrate layer should still form. This clathrate layer then controls the solubility in the ocean, clearing any supersaturation by forming clathrate grains. Only if the newly discovered filled-ice phase enforces a lower solubility, then supersaturation with respect to clathrates will not be achieved and a mid-ocean clathrate layer will not form. It is interesting to note here that for the case of CH$_4$ when going from a cage clathrate to filled-ice the guest to water-host abundance ratio increases \citep{lovedaynat01}. This is also true for the filled-ice of hydrogen, though with more complexity at intermediate pressures due to multiple cage occupancies \citep[see][and references therein]{Qian2014}. If this trend is also true for CO$_2$ it may suggest the necessary supersaturation required for clathrate formation within the ocean can be achieved. 

A full dynamical investigation of this hypothesized mid-ocean clathrate layer is beyond the scope of this work. However, if such a layer can form and become thick and stable it may isolate internal processes from the upper ocean. As a consequence, atmospheric observations may predominantly act as probes into the nature of this layer. Thus potentially severing connections between the deep mantle and atmospheric observations. 
               
In subsection \ref{subsec:FluxesWindDriven} we also introduce a possible loss of atmospheric CO$_2$ due to erosion. This is measured in bars lost per Myr, and denominated in the text as $Q_c$.
\cite{Pierrehumbert2010principles} has suggested that a high flux of energetic photons in a close orbit around an M-dwarf star could result in some loss of CO$_2$ from the planetary atmosphere. A small planet the size of Mars could be stripped off entirely of its primordial CO$_2$ atmosphere due to the action of solar winds. In the latter case $8$\,bar of CO$_2$ could be lost in a $1$\,Gyr \citep{Pierrehumbert2010principles}, giving $Q_c\approx 0.01$\,bar\,Myr$^{-1}$. The efficiency of erosion of CO$_2$ out of the atmosphere must be compared with the efficiency of the internal reservoirs of CO$_2$ in the deep ocean to replenish what is lost. For $Q_c$ in bars Myr$^{-1}$ and a deep ocean vertical eddy diffusion coefficient, $D_{eddy}$, in cm$^2$ s$^{-1}$ we find that only when $Q_c/D_{eddy}>40$ the effect of atmospheric erosion becomes prominent. For $Q_c/D_{eddy}>100$ the atmosphere can be completely eroded of its CO$_2$ content. A value larger than $100$ for $Q_c\approx 0.01$\,bar\,Myr$^{-1}$ is only possible if the deep ocean vertical eddy diffusion coefficient is close to its minimal possible value. One should also consider that very high values for the ratio $Q_c/D_{eddy}$ are probably more likely for a young M-dwarf star while it is still very active. Therefore, for more mature planetary systems the internal outgassing of CO$_2$ should control the pressure of CO$_2$ in the planet's atmosphere.

We note that if cometary composition is a good approximation for the primordial composition of a water planet's ice mantle, then a $20$\,bar CO$_2$ atmosphere, around a $2$M$_E$ planet with a $50$\% ice mass fraction, represents a very small fraction of about $0.01-0.1$\% of the total ice mantle budget of CO$_2$. A caveat to this approximation is the possibility that high-pressure chemistry in the deep ice mantle may change the primordial partitioning of carbon between the different carbon bearing molecules, according to the redox state of the mantle. This issue was not yet addressed and may change the given percentage margin we estimate here.

Partial filling of the sea-ice composite pore space by water would expel back into the atmosphere the pore space gaseous content. These are the  molecular species that while in the pore space did not experience any forcing to become enclathrated. For example, this includes O$_2$ and N$_2$, which have clathrate dissociation pressures much higher than that for CO$_2$ \citep{Kuhs2000}. Therefore, removal of atmospheric gas by the sinking of sea-ice works selectively on CO$_2$.

As we explain in subsection \ref{subsec:FluxesSeaIce} the partial atmospheric pressure of CO$_2$ likely has a minimum value as a function of the subpolar surface temperature (see fig.\ref{fig:SteadyStateWithIce} and its related text for more detailed information).
The minimum in the atmospheric pressure of CO$_2$ and the subpolar surface temperature corresponding to this minimum depend on the sea-ice grain morphology. This is clearly seen in figs.\ref{fig:TspTsbtCold1}-\ref{fig:TspTsbtHot1} where we have plotted atmospheric isobars of CO$_2$ as a function of the subtropic and subpolar surface temperatures.             
Whether this intricate behaviour produces a negative or a positive feedback mechanism requires coupling our model to a radiative-convective atmospheric model. We will address this issue quantitatively in future work. The greenhouse effect increases the surface temperature when more CO$_2$ enters the atmosphere. However, increased Rayleigh scattering takes over at some threshold causing cooling of the surface when more CO$_2$ is added. Therefore, the greenhouse effect has a maximum \citep{Kasting1993}. Various investigations of this phenomenon place the threshold at about $8$\,bar of CO$_2$ \citep{Kasting1993,Kopparapu2013,Kitzmann2015}. This does not include the effect of clouds which are difficult to account for in 1-D radiative-convective models for the atmosphere \citep{Kopparapu2013}. The threshold also depends on the type of star, and other molecular species in the atmosphere. Nevertheless, it is interesting to note that the threshold may fall somewhat above the minimal value we find for the partial atmospheric pressure of CO$_2$, versus the subpolar surface temperature, for initial ice Ih grains smaller than $400\mu$m (see figs.\ref{fig:TspTsbtCold1}-\ref{fig:TspTsbtHot1}). Thus a water rich planet experiencing a reduction in stellar irradiation, causing a drop in high latitude surface temperatures, may respond by increasing the abundance of CO$_2$ in its atmosphere while the greenhouse effect is still dominant. If the drop in high latitude surface temperatures is too big, so as to cause a shut down of the sea-ice sink mechanism, the abundance of CO$_2$ in the atmosphere may spiral to values where Rayleigh scattering becomes dominant causing further cooling of the surface.

In this work we find there are two end scenarios for the steady state atmospheric pressure of CO$_2$: one controlled by the polar sea-ice and the other by the wind-driven circulation and the deep ocean CO$_2$ saturation values. The transition between these two end scenarios happens when the initial sea-ice grain sizes are on the order of hundreds of micrometers (see fig.\ref{fig:SwReal}). This grain size is not unreasonable, and is found in natural environments \citep[e.g.][]{klapp2007}. Therefore, experiments for our studied system are required in order to resolve this issue. 
  
The likely SI CO$_2$ clathrate hydrate bottom of the ocean has a mass density allowing it to sit stably at the deep ocean, and moderate the abundance of CO$_2$ dissolved in the ocean. In other words, it makes sure the ocean stays saturated if it tries to unsaturate. We therefore conclude that a sub-bar CO$_2$ atmosphere around water-rich ocean exoplanets is less likely. Our results suggest the atmosphere has two discrete states: one of a few bars of CO$_2$, probably no less than $2$\,bar and a second discrete state where the planet is surrounded by tens of bars of CO$_2$. Which state is materialized depends on what ocean-atmosphere flux mechanism is dominant, polar sea-ice in the first case and wind-driven circulation in the second case.    

The pressure-temperature conditions within the ice mantle of a $2$M$_E$ planet are probably not high enough to induce the dissociation of CH$_4$. The incorporation of CH$_4$ in filled-ice aids in its transport across the ice mantle, if solid state convection is established \citep{Levi2014}. Therefore, it is likely that CH$_4$ locked in the deep ice mantle would reach the bottom of the ocean, in ocean exoplanets, and then the atmosphere. This outgassed CH$_4$ may potentially effect our model. For example, its existence in the atmosphere may effect the composition and dynamics of the sea-ice. If the partial atmospheric pressure of CH$_4$ exceeds the dissociation pressure of SI CH$_4$ clathrate hydrate for the subpolar surface temperature (e.g. $8.2$\,bar for $240$\,K), then CH$_4$ as well may become enclathrated within the sea-ice. Because the mass density of SI CH$_4$ clathrate hydrate \citep[$0.912$\,g\,cm$^{-3}$, see][]{cox} is less than that of liquid water it will contribute to the buoyancy of the sea-ice. Thus, sea-ice migration to warmer climate may become enhanced, resulting in an increased abundance of CO$_2$ in the atmosphere. Such a mechanism may limit the CH$_4$/CO$_2$ ratio in the atmosphere. Such an analysis may help astronomers pinpoint ocean exoplanets.

Another way in which CH$_4$ may influence our model is by becoming a part of the clathrate layer in the deep ocean. This will effect the abundance of CO$_2$ in this layer. Now, some clathrate hydrate cages will be filled with CO$_2$ and others with CH$_4$. As a consequence the solubility of CO$_2$ in equilibrium with the clathrate phase may change, influencing the solubility of CO$_2$ in the overlying ocean, and its atmospheric abundance. It would also be important to know how the addition of CH$_4$ varies the mass density of the ocean's bottom clathrate layer, especially, whether this decreases its mass density below that of the water-rich liquid. 
We hope to address this ternary system in the future.

\section{SUMMARY}

In this paper we focus on water planets, which we define to be planets whose water mass fraction is large enough to form an external mantle composed of high pressure water ice polymorphs and that lack a substantial H/He atmosphere. In particular we consider such planets in their habitable zone so their outermost condensed mantle is a vast and deep liquid ocean.

We study the solubility of CO$_2$ within the parameter space of such oceans, both outside (subsection \ref{subsec:SolubilityOutside}) and inside (subsection \ref{subsec:SolubilityInside}) the thermodynamic stability field of the SI CO$_2$ clathrate hydrate phase. We show that outside of the SI CO$_2$ clathrate hydrate thermodynamic stability field the solubility can be modelled using Henry's law, for the entire P-T space of interest for water planet oceans. The upper bound of this parameter space is the melting curve of water ice VI when in saturation with CO$_2$. We find that in order for Henry's law to match the inferred solubility from experimental data for the ice VI melt depression it is necessary to consider the fugacity of solid CO$_2$. Near the bottom of the ocean, CO$_2$ transforms from a fluid into its phase I solid. We find that in the region of stability of the phase I solid of CO$_2$ the solubility of CO$_2$ in the aqueous solution decreases with increasing pressure. Throughout this work we use Henry's law only for interpolating between solubility data points.

We model the solubility of CO$_2$ in water when in equilibrium with its clathrate hydrate phase. Our model uses results from molecular simulations in addition to macroscopic parameters. This approach allows us to overcome the fact that equations of state for the H$_2$O-CO$_2$ system do not cover our entire parameter space of interest. Our model for the solubility, when in equilibrium with the clathrate phase, predicts the solubility to be insensitive to the pressure. It also predicts that inside the clathrate hydrate stability field the solubility \textit{increases} with the \textit{increasing} temperature. These behaviours are verified experimentally. Our model can accurately describe experimental data for the solubility in the pressure range where such experiments were conducted (up to a few hundred bars). Our model also predicts solubilites at the bottom of the ocean (pressure of $\approx 10$\,kbar) that fall within the error in the inferred solubility from experiments on the melt depression of ice VI. Therefore, inside the clathrate hydrate stability field as well, our solubility model is interpolative rather then extrapolative.

We plot the phase diagram of the SI CO$_2$ clathrate hydrate, over the entire pressure-temperature space of interest for water planet oceans (see fig.\ref{fig:PhaseDiagram} in section \ref{sec:PhaseDiagram}). We show that phases which are reservoirs for CO$_2$ have direct contact with the bottom of the ocean. These phases aid in the flux of CO$_2$ from the ice mantle and into the ocean. This is because of the nature in which carbon is stored within the solid matrix. In the Earth carbon can become stably locked in Rock. However, in a water planet, CO$_2$ stored in a clathrate hydrate layer will leak CO$_2$ into the ocean if that becomes subsaturated.  

We investigate the possibility for storing CO$_2$ deep in the ocean (see section \ref{sec:Reservoirs}).
We argue that CO$_2$ outgassed from the interior may accumulate at the bottom of the ocean in three possible sink stratification cases (denoted as: $\alpha$, $\beta$ and $\gamma$ in fig.\ref{fig:alphabetagamma}). Which of the three cases occurs depends on the temperature profile mostly in the deep to mid ocean. If these sinks become exhausted any further substantial CO$_2$ outgassing from the interior accumulates in the atmosphere.

For temperatures less than $278$\,K (at the deep ocean), the SI CO$_2$ clathrate hydrate phase is stable at pressures higher than the melting pressure of water ice VI. We call this the $\alpha$ stratification domain (see figure \ref{fig:alphabetagamma}).
Therefore, in this domain, as CO$_2$ is transported outward it will enter the clathrate thermodynamic stability field and transform the water ice V/VI layer into a CO$_2$ SI clathrate hydrate layer. This clathrate layer then becomes the ocean's bottom surface, consequently making physical contact with the overlying ocean. If the ocean concentration of CO$_2$ is less than the value at equilibrium with clathrate hydrate then the CO$_2$ from the clathrates diffuses into the ocean. As the concentration of CO$_2$ in the ocean increases and approaches saturation the clathrate layer composing the bottom of the ocean stabilizes and further outgassing cannot continue via this diffusional mechanism. We regard this mechanism as a "gentle" outgassing mechanism, since it does not require any geologically active surface regions directly forcing CO$_2$ into the ocean. In fig.\ref{fig:OceanOutgassing} we give an illustration of this mechanism.
Field experiments show that the rate of clathrate dissolution in a subsaturated aqueous environment is diffusion limited \citep{Rehder2004}. Therefore, we find that for a deep unmixed ocean the "gentle" mechanism requires at least $200$\,Myr to bring the ocean close to saturation, if it was initially poor in CO$_2$.

Still in the $\alpha$ domain, it is possible a geologic forcing would yield a CO$_2$ outgassing flux into the ocean which exceeds that predicted by the "gentle" mechanism. In that case after CO$_2$ concentration in the ocean reaches the saturation value in equilibrium with the clathrate phase, any further CO$_2$ outgassing would sediment to the bottom of the ocean as CO$_2$ clathrate grains. If the outgassing flux of CO$_2$ forced into the ocean is low, little clathrate hydrate would pile up on the bottom of the ocean in a geological time scale (illustrated in the right panel in fig.\ref{fig:HydrateBuildUp}). In the case where the outflux is high, then in a geological time scale, most of the ocean would solidify into a CO$_2$ SI clathrate hydrate layer (illustrated in the left panel in fig.\ref{fig:HydrateBuildUp}). In this latter scenario a thin near-surface water rich liquid layer will survive solidification where the pressure is too low to stabilize clathrate hydrate. This narrow aqueous layer will be saturated in CO$_2$ and have a enhanced salinity compared with that of the pre-solidified ocean. 
We find a constant global mantle CO$_2$ outgassing flux into the ocean of the order of $10^{11}$\,molec CO$_2$\,cm$^{-2}$\,s$^{-1}$ will transform ten percent of the oceans' initial mass into clathrate hydrate in $1$\,Gyr. In the $\alpha$ domain we calculate that, for our choice of planets, a mass of the order of $10^{25}$\,g of CO$_2$ can be stored in the clathrate hydrate layer before this sink is exhausted. This is two orders of magnitude more than the carbon stored in rocks on the Earth \citep[see][]{CarbonCycle}. In addition, for the extreme scenario where the entire clathrate layer solidifies the surviving surface liquid layer can be as shallow as $100$\,m deep with a three orders of magnitude enhancement of its salinity compared to the original ocean. 

The region confined between a ocean bottom temperature higher than $278$\,K and a deep ocean temperature not exceeding $294$\,K defines the $\beta$ domain (see fig.\ref{fig:alphabetagamma}). For this domain the dissociation pressure of the SI CO$_2$ clathrate hydrate will be lower than the melting pressure of water ice VI. In this domain mantle CO$_2$ (either embedded in ice VI or as filled-ice) comes into contact with the ocean, and tries to saturate it. First reaching the value of saturation when in equilibrium with clathrates. If more CO$_2$ is driven into the ocean clathrate grains will form within this phase's stability field. For the deep ocean temperatures in the $\beta$ domain the clathrate grains become less dense than the surrounding water rich liquid near the high-pressure boundary of their thermodynamic stability field. A SI CO$_2$ clathrate hydrate layer may thus accumulate, elevated above the ocean's bottom of high pressure ice polymorphs. This elevated solid layer made of clathrates should control the solubility of CO$_2$ in the ocean lying above it, and therefore its accessibility to the atmosphere (see illustration in fig.\ref{fig:BetaStrata}).   
For a $290$\,K isotherm, and our choice of planetary parameters, we find the maximum mass of CO$_2$ that can be stored in this mid-ocean layer is approximately $10^{25}$\,g. For this isotherm the clathrate layer may extend as much as $38$\,km, starting at an elevation of $28$\,km above the ocean's ice VI bottom and ending $24$\,km below the ocean's surface.

In the case where the thermal profile in the ocean is everywhere higher than $294$\,K the ocean falls into the $\gamma$ domain (see fig.\ref{fig:alphabetagamma}).
In this regime the SI CO$_2$ clathrate hydrate is nowhere stable. In this case the ocean may saturate with CO$_2$ to a concentration appropriate in the absence of clathrate hydrates. When saturation is reached, and if outgassing into the ocean continues, the CO$_2$ will first accumulate on the bottom of the ocean as phase I solid followed by a liquid CO$_2$ layer when the former is exhausted. The liquid CO$_2$ layer will terminate at the pressure where it becomes less dense than the ocean's water-rich liquid.    
For the planets we are considering these two layers represent a total sink that can contain as much as $10^{26}$\,g of CO$_2$ at the deep ocean.
However, considering solid CO$_2$ is more dense than ice VI (see fig.\ref{fig:DensityComparison2}) gravity will probably limit the extent of such layers considerably. 
We argue though that such high deep ocean temperatures are less likely since a thick water ice mantle underlying the ocean translates to low heat fluxes at the ocean's bottom \citep[see section $5$ in][]{Levi2014}.

An overturning circulation in the ocean creates potential energy, and therefore requires an external energy source to operate. It was shown for the case of Earth that the power supplied by winds and tides is sufficient to support the Atlantic overturning circulation \citep{Wunsch2004}. In section \ref{sec:Circulation} we estimate the power needed to run an oceanic overturning circulation. In addition to having to lift cold and dense water through warm water, in water planets surface water relatively poor in dissolved CO$_2$ needs to be sunk through deep water more heavily loaded with CO$_2$ and thus more dense. We find that running a general circulation in deep water planet oceans with vertical velocities similar to those in Earth's circulation model requires energy two to three orders of magnitude higher than what is likely available.  
This means the deep ocean in water planets is likely unmixed or inefficiently mixed, and that the deep ocean deposits of CO$_2$ are stable. This also means that the lowest measured vertical eddy diffusion for Earth's oceans, $0.1$\,cm$^2$\,s$^{-1}$, is likely an upper bound value for the vertical eddy diffusion in water planet oceans. The lower bound value is the molecular diffusion value of $10^{-5}$\,cm$^2$\,s$^{-1}$. 

Given enough time the ocean's composition tries to equilibrate with the appropriate reservoir of CO$_2$ in the ocean's bottom. The low heat flux at the bottom of these oceans suggests this deep CO$_2$ reservoir is a layer of SI CO$_2$ clathrate hydrate. Given an ocean depth of $80$\,km, a vertical eddy diffusion of $10^{-2}$\,cm$^2$\,s$^{-1}$ yields a time scale of $200$\,Myr for the ocean's composition to reach equilibration with the ocean's bottom CO$_2$ reservoirs. 
This CO$_2$ then reaches the atmosphere.

We develop a model for the wind-driven circulation in water planets, and its resulting flux of CO$_2$ into the atmosphere. A diagram of this model is given in fig.\ref{fig:WindDrivenCirculation} in subsection \ref{subsec:FluxesWindDriven}. From this model we deduce the steady-state pressure of CO$_2$ in the atmosphere of a water planet. If no sea-ice forms, and if there are no active mechanisms eroding the atmosphere, steady state is attained when the downwelled subtropical surface water have a dissolved CO$_2$ concentration equal to that of the deep ocean. The resulting steady-state atmospheric pressure, for this case, is tens of bars of CO$_2$. The steady-state pressure of atmospheric CO$_2$ increases when: increasing the subtropical surface water temperature, and for higher deep ocean temperatures (see fig.\ref{fig:SteadyStatePressure}).  

We find that if the CO$_2$ atmospheric pressure is perturbed to values \textit{higher} than the steady state pressure, then the system tries to restore the steady state atmosphere. This is because, in this case, the water pushed inward by Ekman pumping in the subtropics becomes oversaturated with CO$_2$. This results in SI CO$_2$ clathrate grain formation when the circulating fluid parcels enter the thermodynamic stability field of this phase. The clathrate grains remove any excess CO$_2$ above the solubility value in equilibrium with the clathrate phase, and sink to the bottom of the ocean in a time scale of less than $100$\,yr. 
The resulting upwelled water due to Ekman suction at the tropics have a CO$_2$ concentration equal to the value of the deep ocean. This mechanism is illustrated in fig.\ref{fig:WindDrivenCirculation2}. We find that such perturbations in the CO$_2$ atmospheric pressure can be dampened by the wind-driven circulation in a time scale of the order of $10^4$\,yr.

The wind-driven circulation exposes deep water saturated in CO$_2$ to the atmosphere. The resulting atmospheric pressure of CO$_2$ may become higher than the dissociation pressure of SI CO$_2$ clathrate hydrate at the poles, depending on the temperature at high latitudes. If that is the case there is a driving force that transforms ice Ih sea-ice into a composite containing SI CO$_2$ clathrate hydrate. Since the latter phase is more dense than liquid water the sea-ice composite may sink in the ocean. This removes CO$_2$ from the atmosphere. We quantify this effect, and its influence on the steady-state atmospheric pressure of CO$_2$ (see subsection \ref{subsec:FluxesSeaIce}). 

The sea-ice forming at high latitudes in water planets is likely a composite, composed of: ice Ih, SI CO$_2$ clathrate hydrate and pore space. At the subfreezing conditions at the poles ice Ih grains first form on the ocean's surface, followed by a transformation of these grains into SI CO$_2$ clathrate hydrate. This phase transformation involves an expansion of the solid matrix, at the expense of the pore space. Therefore, enough initial pore space must be present to allow for this expansion. We find that for the sea-ice composite to become more dense than the surrounding water rich liquid a minimal initial porosity of $0.043$ is required. If the pore space is only filled with gas of negligible mass density the initial porosity must not be higher than $0.225$, otherwise the sea-ice composite will remain afloat on the ocean's surface. This upper bound on the initial porosity of the sea-ice is very much alleviated if the pores become partially filled with liquid water or ice Ih (see fig.\ref{fig:SeaIceStability} and related text for more information).  

Sea-ice rich in CO$_2$ clathrate is not anchored in place, and given time it may migrate to a warmer climate. At lower latitudes, where temperatures are higher, the clathrate dissociation pressure is higher. As a result the clathrates in the sea-ice slab would dissociate and release their caged CO$_2$ back into the atmosphere. Therefore, sea-ice may deposit atmospheric CO$_2$ in the deep ocean if it becomes buoyantly unstable in a time scale less than its drift time out of the subpolar region. This time criterion implies that the transformation of the sea-ice ice Ih grains into clathrate must be fast enough. The rate of this phase transformation depends on the ice Ih initial grain size distribution. We find that if the initial ice Ih grain size is tens of micrometers the sinking of atmospheric CO$_2$ in sea-ice slabs controls the global CO$_2$ atmosphere. In other words, the atmospheric pressure of CO$_2$ is to a very good approximation the dissociation pressure of SI CO$_2$ clathrate hydrate for the subpolar surface temperature. This atmospheric pressure may be as low as $2$\,bar of CO$_2$. This is much less than the tens of bars of CO$_2$ that accumulate in the atmosphere in case the wind-driven circulation is the only operating mechanism. If the initial ice Ih grain size, making up the sea-ice, is larger than about $1$\,mm, the phase transformation can not take place fast enough and the sea-ice slabs are buoyantly stable as they migrate out of the subpolar region. In this scenario the CO$_2$ abundance in the atmosphere quickly bounces to tens of bars as dictated by the wind-driven circulation and the deep saturated ocean. 

We further find that the atmospheric abundance of CO$_2$, when controlled by the polar sea-ice, is insensitive to the extent of the polar sea-ice cap, as long as the ice cap extends at least a few degrees around the pole. Therefore, even a relatively small ice cap may efficiently remove CO$_2$ from the atmosphere. It also means that changes in the surface area of the ice cap hardly establish any feedback mechanism between the ocean and atmosphere.

Because the atmospheric abundance of CO$_2$ is potentially controlled by the subpolar surface temperature, we have investigated how the system reacts to changes in this temperature. We wish to note that the subpolar surface temperature we refer to is an average value over a large time scale, and should not be confused with a seasonally changing value. We find that for subpolar surface temperatures higher than about $240$\,K, and an initial ice Ih grain size of $100\mu$m, the removal of atmospheric CO$_2$ by the sinking sea-ice is very efficient. Consequently, the partial atmospheric pressure of CO$_2$ will tend to follow closely the dissociation pressure of SI CO$_2$ clathrate hydrate for the subpolar temperature. This dissociation pressure increases with temperature. Therefore, as the subpolar region becomes hotter more CO$_2$ may end up in the atmosphere. Increasing the subpolar surface temperature from $240$\,K to $265$\,K would result in an increase in the atmospheric pressure of CO$_2$ from approximately $3$\,bar to approximately $8$\,bar, assuming an initial grain size of $100\mu$m. If the subpolar surface temperature drops below about $230$\,K the phase transition from ice Ih to CO$_2$ clathrate hydrate within the sea-ice slows down. The system's response would be to elevate the partial atmospheric pressure of CO$_2$ in order to increase the driving force responsible for the ice phase transition. As a result of this behaviour the partial atmospheric pressure of CO$_2$ likely has a minimum value as a function of the subpolar surface temperature (see fig.\ref{fig:SteadyStateWithIce} and its related text for more detailed information). \textit{The value of this minimum and its corresponding subpolar temperature depend on the sea-ice morphology} (see figs.\ref{fig:TspTsbtCold1}-\ref{fig:TspTsbtHot1}). \textit{This minimum in the atmospheric abundance of CO$_2$ introduces the possibility for a negative feedback mechanism, that may moderate climate change in ocean planets}.

\section{ACKNOWLEDGEMENTS}

We wish to thank Prof. Michael Follows for a helpful discussion and review. We also wish to thank our referee for a careful read of the manuscript and helpful comments. This work was supported by a grant from the Simons Foundation  (SCOL No.$290360$) to Dimitar Sasselov.

\section{APPENDIX}

\subsection{SOLID CO$_2$ FUGACITY}\label{subsec:AppendixA}

Modelling the phase diagram of SI CO$_2$ clathrate hydrate at pressures where carbon dioxide solidifies into its phase I requires the fugacity describing this phase. Deriving the fugacity requires an equation of state for the solid of interest. 
Experimental data for the phase I solid of carbon dioxide are not abundant. A few older compressibility experiments exist, describing the behaviour of solid carbon dioxide right above its melt curve in the region of interest for clathrates \citep[e.g.][]{Bridgman1938,Stevenson1957}, although, the reliability of the data provided by these older experiments was recently questioned \citep[see discussion in][]{Olinger1982}. For the sake of computational simplicity we have taken the experimental data of \cite{Olinger1982} and the compressibility data tabulated in \cite{Sterner1994} and modelled the compressibility, $Z$, of solid carbon dioxide of phase I in the temperature range of: $250$\,K$<T<450$\,K and pressure range of: $0.5$\,GPa$<P<5$GPa using a quadratic polynomial of the form:
\begin{equation}\label{compressibilitymodel}
Z(P,T)=a(T)+b(T)\left(\frac{P-\bar{P}}{\bar{P}}\right)+c(T)\left(\frac{P-\bar{P}}{\bar{P}}\right)^2
\end{equation}
where:
\begin{eqnarray}
a(T)=21.482-19.341\left(\frac{T-\bar{T}}{\bar{T}}\right)+20.079\left(\frac{T-\bar{T}}{\bar{T}}\right)^2-20.654\left(\frac{T-\bar{T}}{\bar{T}}\right)^3 \\
b(T)=18.742-18.182\left(\frac{T-\bar{T}}{\bar{T}}\right)+19.326\left(\frac{T-\bar{T}}{\bar{T}}\right)^2-18.974\left(\frac{T-\bar{T}}{\bar{T}}\right)^3 \\
c(T)=-1.8752+1.5015\left(\frac{T-\bar{T}}{\bar{T}}\right)-1.2495\left(\frac{T-\bar{T}}{\bar{T}}\right)^2+1.2038\left(\frac{T-\bar{T}}{\bar{T}}\right)^3 
\end{eqnarray} 
Here $\bar{P}=2.5$\,GPa and $\bar{T}=325$\,K are reference pressure and temperature respectively. 

The volume per CO$_2$ molecule in phase I solid ($V_{co_2}^{solid}$) is derived from the compressibility:
\begin{equation}
V_{co_2}^{solid}=\frac{Z(P,T)kT}{P}
\end{equation}
where $k$ is Boltzmann's constant.

For deriving the fugacity of CO$_2$ solid ($f_{co_2}^{solid}$) we first start with the fugacity coefficient defined as \citep{smithness}:
\begin{equation}\label{fugacitycoeffsolid}
\phi_{co_2}^{solid}\equiv\frac{f_{co_2}^{solid}}{P}
\end{equation}
The fugacity obeys the following relation \citep{smithness}:
\begin{equation}\label{fugacitydefinition}
V_{co_2}^{solid}dP=kTd\ln{f_{co_2}^{solid}}
\end{equation}
Then differentiating the logarithm of eq.($\ref{fugacitycoeffsolid}$) together with eq.($\ref{fugacitydefinition}$) gives after a few algebraic steps:
\begin{equation}
d\ln{\phi_{co_2}^{solid}}=\left(\frac{V_{co_2}^{solid}P}{kT}-1\right)\frac{dP}{P}=\left(Z(P,T)-1\right)\frac{dP}{P}
\end{equation} 
To obtain the fugacity coefficient for solid CO$_2$ at pressure $P$ and temperature $T$ we must integrate the last relation. Every P-T point may be reached by integrating from the desired temperature $T$ on the CO$_2$ melt curve, where the pressure is $P_{co_2}^{melt}(T)$, up to the desired pressure. The benefit in so doing is that on the melt curve the fugacity coefficient for the solid ($\phi_{co_2}^{solid}$) equals that for the liquid ($\phi_{co_2}^{liquid}$). The fugacity coefficient for the liquid, in turn, is reliably calculated using the Soave-Redlich-Kwong \citep{soave} equation of state. Therefore one may obtain:
\begin{equation}
\ln{\left(\frac{\phi_{co_2}^{solid}(P,T)}{\phi_{co_2}^{liquid}(P_{co_2}^{melt}(T),T)}\right)}=\int_{P_{co_2}^{melt}(T)}^P\left(Z(P,T)-1\right)\frac{dP}{P}
\end{equation}
Inserting the compressibility, as expressed in eq.($\ref{compressibilitymodel}$), reduces the last relation to:
$$
\frac{\phi_{co_2}^{solid}(P,T)}{\phi_{co_2}^{liquid}(P_{co_2}^{melt}(T),T)}=\left(\frac{P}{P_{co_2}^{melt}(T)}\right)^{a(T)-b(T)+c(T)-1}\times 
$$
\begin{equation}
\exp\left\lbrace\frac{b(T)-2c(T)}{\bar{P}}\left(P-P_{co_2}^{melt}(T)\right)+\frac{c(T)}{2\bar{P}^2}\left(P^2-P_{co_2}^{melt^2}(T)\right)\right\rbrace
\end{equation} 
The desired fugacity for solid carbon dioxide is then obtained from the above definition for the fugacity coefficient.

\subsection{SEA-ICE COMPOSITE DENSITY}\label{subsec:AppendixB}

Here we derive the mass density of the sea-ice forming at the poles of water planets. It is assumed to be made of ice Ih, SI CO$_2$ clathrate hydrate and pore space. Considering each occupies a volume of: $V_{Ih}$, $V_{clath}$ and $V_{pore}$, respectively, and a total system volume $V$ the mass density of the composite is:
\begin{equation}\label{CompositeDensityDef}
\rho_{comp}=\phi_{Ih}\rho_{Ih}+\phi_{clath}\rho_{clath}+\phi_{pore}\rho_{pore}
\end{equation}
In the last equation $\rho_{Ih}$ and $\rho_{clath}$ are the pure ice Ih and clathrate hydrate bulk mass densities. $\rho_{pore}$ is the mass density of the pore filling material, and:
\begin{eqnarray}
\phi_{Ih}\equiv \frac{V_{Ih}}{V} \nonumber     \\
\phi_{clath}\equiv \frac{V_{clath}}{V} \nonumber   \\
\phi_{pore}\equiv \frac{V_{pore}}{V} \nonumber \\
\end{eqnarray}
where $\phi_{pore}$ is the sea-ice porosity. We wish to relate the composite mass density to the mole fraction of water ice Ih converted to clathrate hydrate, $\alpha$. Considering that the water molecules first solidify as ice Ih (see discussion in subsection \ref{subsec:FluxesSeaIce}) we can write:
\begin{equation}
\alpha=\frac{\tilde{n}^{clath}_{h_2o}}{\tilde{n}^{clath}_{h_2o}+\tilde{n}^{Ih}_{h_2o}}
\end{equation}  
where $\tilde{n}^{Ih}_{h_2o}$ and $\tilde{n}^{clath}_{h_2o}$ are the number of moles of water in ice Ih occupying volume $V_{Ih}$ and in SI CO$_2$ clathrate hydrate occupying volume $V_{clath}$, respectively, which obey:
\begin{equation}
\tilde{n}^{Ih}_{h_2o}=\frac{\rho_{Ih}}{M_w}V_{Ih}
\end{equation}
\begin{equation}
\tilde{n}^{clath}_{h_2o}=\frac{46}{N_{A}}\frac{V_{clath}}{V_{cell}}
\end{equation} 
where we consider each SI clathrate hydrate unit cell volume (see Eq.$\ref{CellVolume}$) to consist of $46$ water molecules,  
$M_w$ is the molar weight of water and $N_{A}$ is Avogadro's number. After rearranging we obtain the relation:
\begin{equation}\label{relationhydratepore1}
\frac{1-\alpha}{\zeta\alpha}+1=\frac{1}{\phi_{clath}}\left(1-\phi_{pore}\right)
\end{equation}
where
\begin{equation}
\zeta\equiv\frac{\rho_{Ih}V_{cell}N_A}{46M_w}\approx 1.133
\end{equation}
The latter is the expansion factor when a mole of ice Ih converts to SI CO$_2$ clathrate hydrate. The numerical value for $\zeta$ is here derived from the tabulated data in \cite{Genov2004}. 

The field-emission scanning electron images in \cite{Staykova2003} show that the expansion (during conversion to clathrate hydrate) comes greatly at the expense of the pore volume. Therefore, if one approximates the total system volume $V$ to be constant one has:
\begin{equation}\label{relationhydratepore2}
\phi_{pore}-\phi^0_{pore}\approx\frac{1-\zeta}{\zeta}\phi_{clath}
\end{equation}
where $\phi^0_{pore}$ is the initial porosity of the formed sea-ice, before the conversion to clathrate hydrate initiates. Combining Eqs.($\ref{relationhydratepore1}$) and ($\ref{relationhydratepore2}$) yields:
\begin{equation}\label{HydrateVolumeFraction}
\phi_{clath}=\zeta\alpha\left(1-\phi^0_{pore}\right)
\end{equation} 
\begin{equation}\label{porespacerelation}
\phi_{pore}=\phi^0_{pore}+(1-\zeta)\alpha\left(1-\phi^0_{pore}\right)
\end{equation}
\begin{equation}\label{IhVolumeFraction}
\phi_{Ih}=(1-\alpha)\left(1-\phi^0_{pore}\right)
\end{equation}
From Eq.($\ref{porespacerelation}$) we see that for a complete conversion into clathrate hydrate ($\alpha=1$) to take place the initial porosity must be larger than about $0.117$. Indeed in \cite{Genov2004} the initial porosity was $0.33$ and so pore space restrictions were not an issue.   

Finally, inserting the last relations into Eq.($\ref{CompositeDensityDef}$) yields for the composite mass density:
\begin{equation}
\rho_{comp}=\left(1-\phi^0_{pore}\right)\left[\rho_{Ih} + \alpha\left(\rho_{pore}-\rho_{Ih}\right)+\zeta\alpha\left(\rho_{clath}-\rho_{pore}\right)\right]+\phi^0_{pore}\rho_{pore}
\end{equation}

\subsection{ERROR FUNCTION SERIES SOLUTION}\label{subsec:AppendixC}

Solving for the diffusion equation using separation of variables results in a series solution that converges slowly for time scales less than the system's squared length scale over the diffusion coefficient. The series solution developed here ought amend this problem.
Let's solve for the following system of equations:
\begin{align}
& \frac{\partial^2v}{\partial x^2}-\frac{1}{\kappa}\frac{\partial v}{\partial t}=0 \nonumber \\
& v = v_0 \quad t=0 \nonumber \\
& v = v_1 \quad x=0 \nonumber \\
& v = v_2 \quad x=l
\end{align} 
Following \cite{Carslaw1959} the Laplace transformation for the set of equations is:
\begin{align}\label{LaplaceSet}
& \frac{d^2\bar{v}}{dx^2}-q^2\bar{v}=-\frac{1}{\kappa}v_0 \nonumber \\
& \bar{v}(x=0)=\frac{v_1}{p} \nonumber \\
& \bar{v}(x=l)=\frac{v_2}{p}
\end{align} 
where bar denotes the Laplace transform and $q^2\equiv p/\kappa$. $p$ is the inverse time scale in the Laplace transformation. The general solution to Eqs.($\ref{LaplaceSet}$) is:
\begin{equation}
\bar{v} = Ae^{qx}+Be^{-qx}+\frac{v_0}{p}
\end{equation}
where,
\begin{align}
& A = \frac{v_2-v_0-(v_1-v_0)e^{-ql}}{p\left[e^{ql}-e^{-ql}\right]} \\
& B = \frac{(v_1-v_0)e^{ql}-v_2+v_0}{p\left[e^{ql}-e^{-ql}\right]}
\end{align} 
This solution is complex enough not to appear in tables for Laplace transforms. Therefore, we will use Taylor's expansion to develop the denominator in $A$ and $B$.
\begin{equation}
\frac{1}{p\left[e^{ql}-e^{-ql}\right]} = \frac{e^{-ql}}{p\left[1-e^{-2ql}\right]} = \frac{1}{p}\sum^\infty_{r=0}e^{-(2r+1)ql}
\end{equation}
Notice that for small time scales $p$ and hence $q$ are very large and the Taylor's expansion is adequate. Inserting the last expansion into the forms for $A$ and $B$ gives after some algebraic steps:
\begin{align}
\bar{v} &= \frac{v_2-v_0}{p}\sum^\infty_{r=0}e^{-q[(2r+1)l-x]} + \frac{v_0-v_1}{p}\sum^\infty_{r=0}e^{-q[2(r+1)l-x]} + \frac{v_1-v_0}{p}\sum^\infty_{r=0}e^{-q[2rl+x]}  \nonumber \\ 
&+\frac{v_0-v_2}{p}\sum^\infty_{r=0}e^{-q[(2r+1)l+x]} + \frac{v_0}{p}
\end{align}
Now each separate term is simple enough to be transformed back, yielding:
\begin{align}\label{ErrorSeries}
v(x,t) & = (v_2-v_0)\sum^\infty_{r=0}erfc\left(\frac{(2r+1)l-x}{2\sqrt{\kappa t}}\right) + (v_0-v_1)\sum^\infty_{r=0}erfc\left(\frac{2(r+1)l-x}{2\sqrt{\kappa t}}\right) \nonumber \\
& + (v_1-v_0)\sum^\infty_{r=0}erfc\left(\frac{2rl+x}{2\sqrt{\kappa t}}\right) + (v_0-v_2)\sum^\infty_{r=0}erfc\left(\frac{(2r+1)l+x}{2\sqrt{\kappa t}}\right) \nonumber \\
& + v_0
\end{align}
where,
\begin{equation}
erfc(y)=1-erf(y)=\frac{2}{\sqrt{\pi}}\int^\infty_ye^{-\xi^2}d\xi
\end{equation}

\begin{figure}[ht]
  \begin{minipage}{\textwidth}
  \centering
    \includegraphics[width=.4\textwidth]{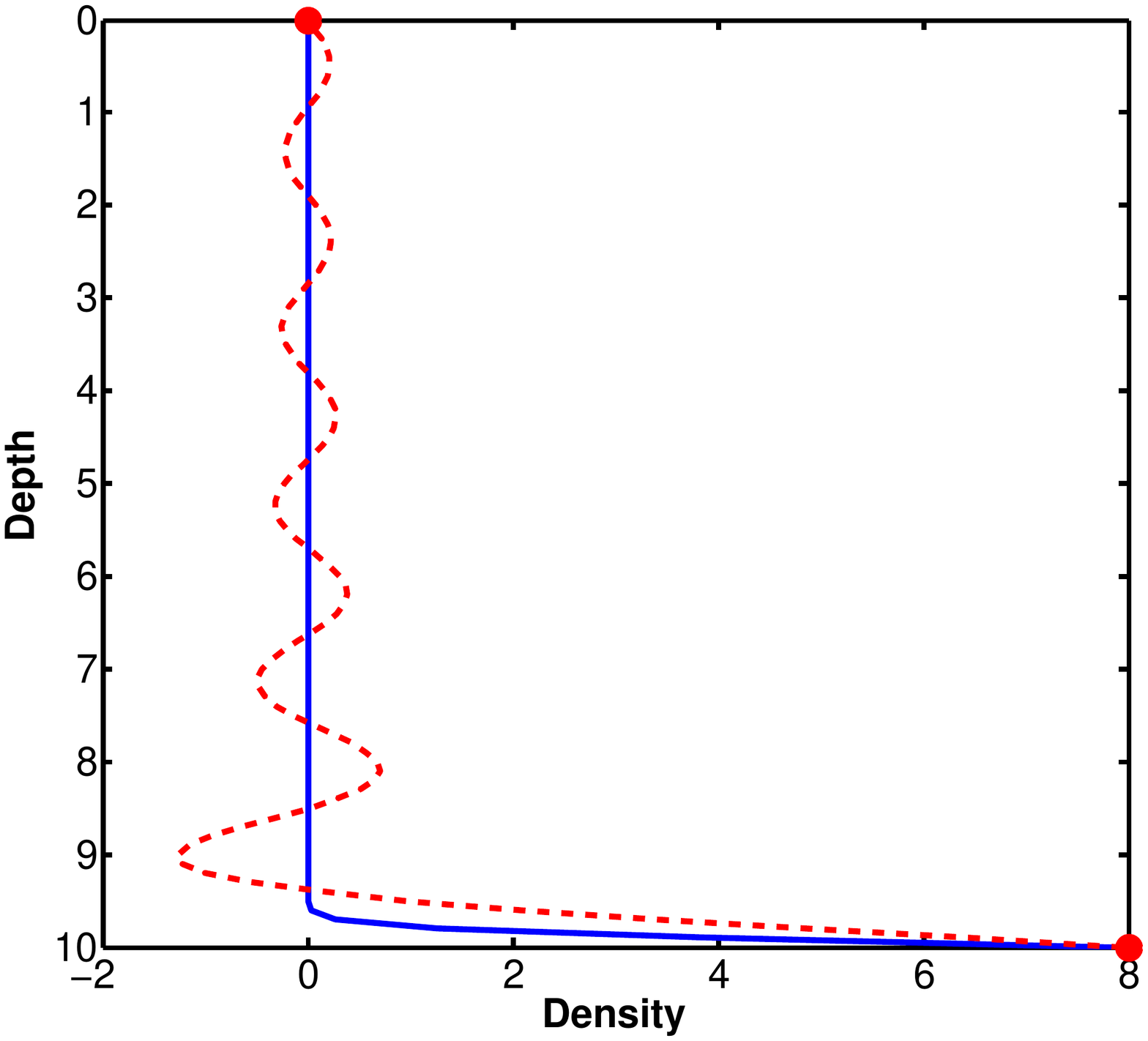}\quad
    \includegraphics[width=.4\textwidth]{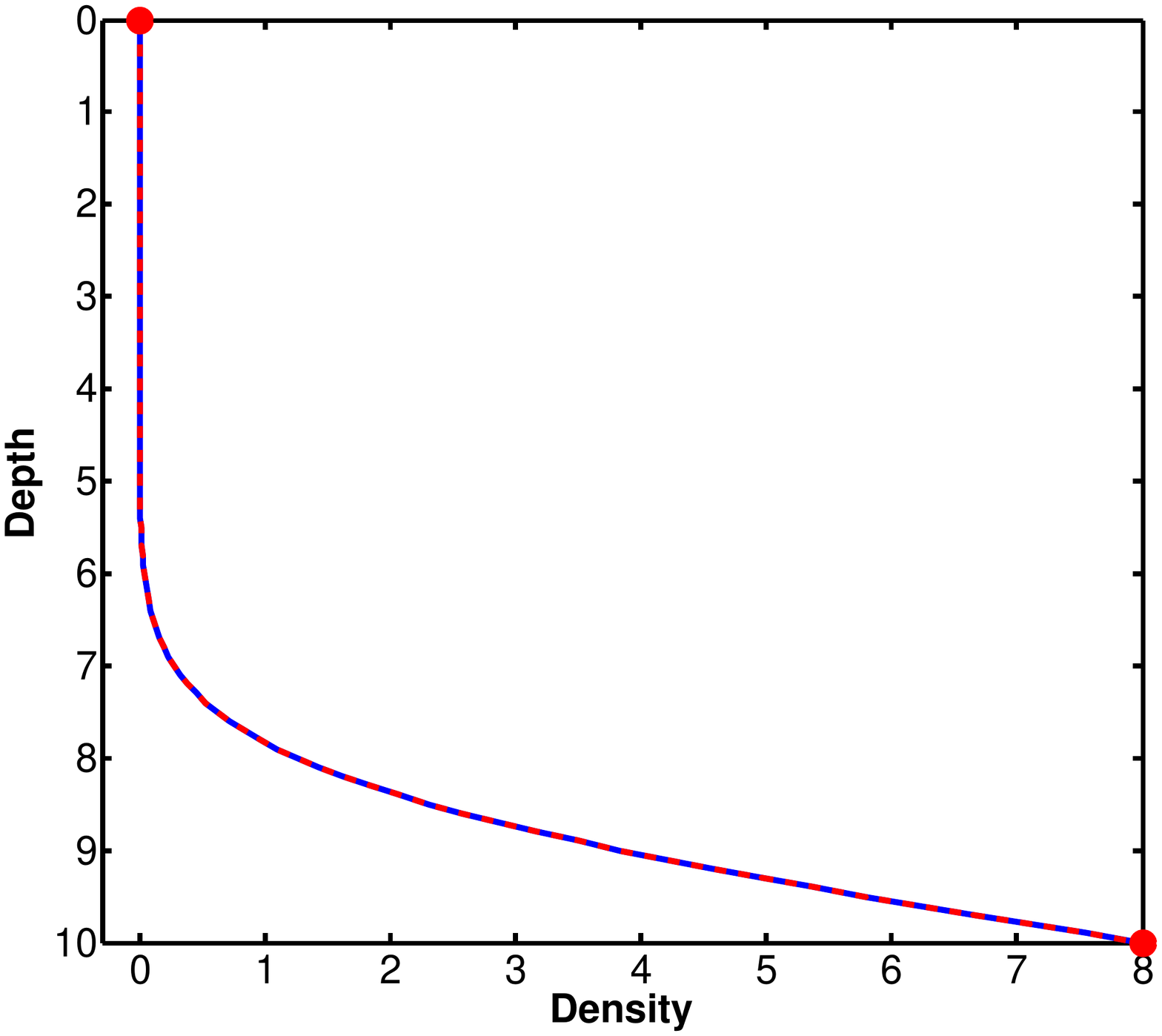}\\
    \includegraphics[width=.4\textwidth]{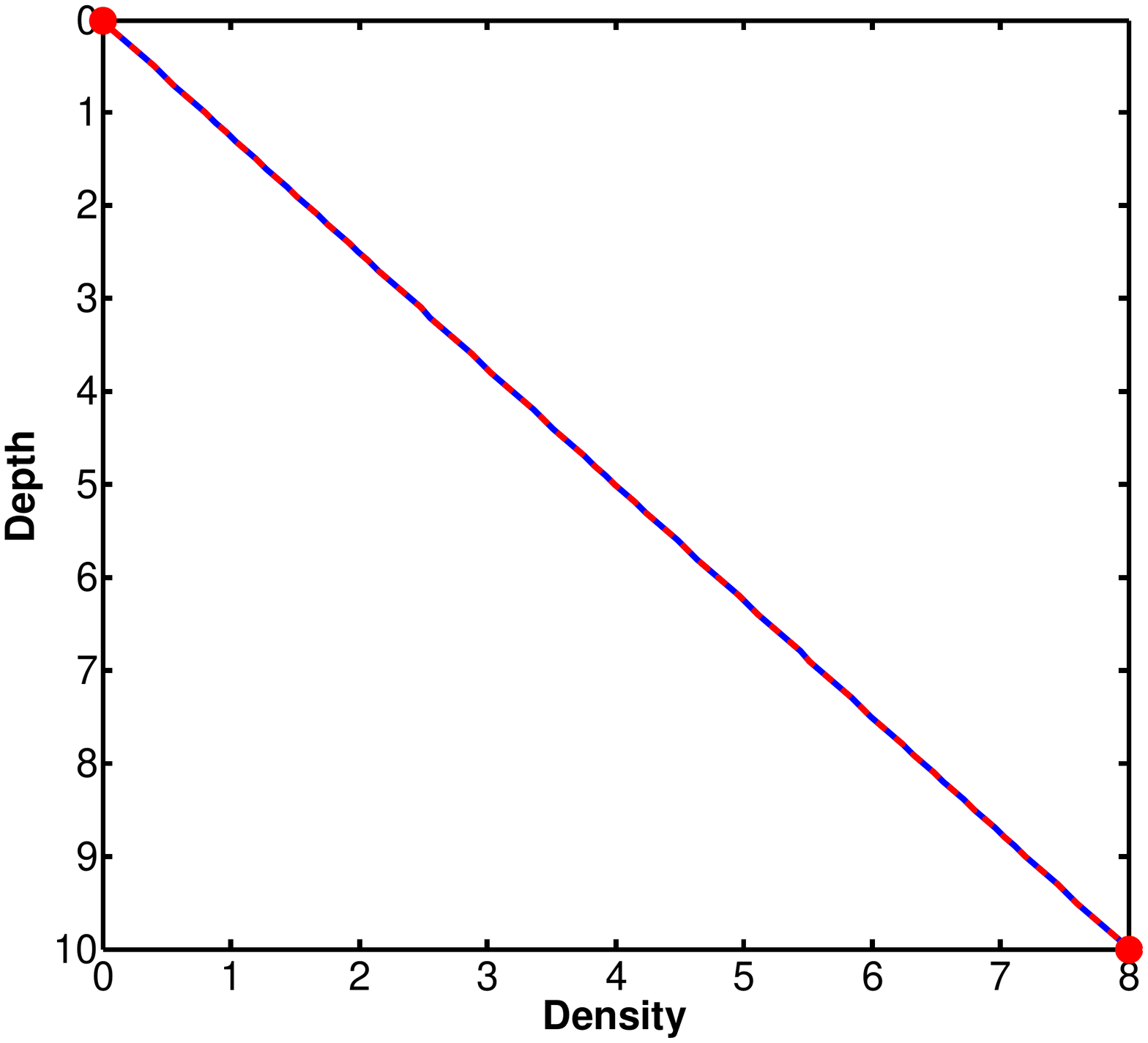}\quad
    \includegraphics[width=.4\textwidth]{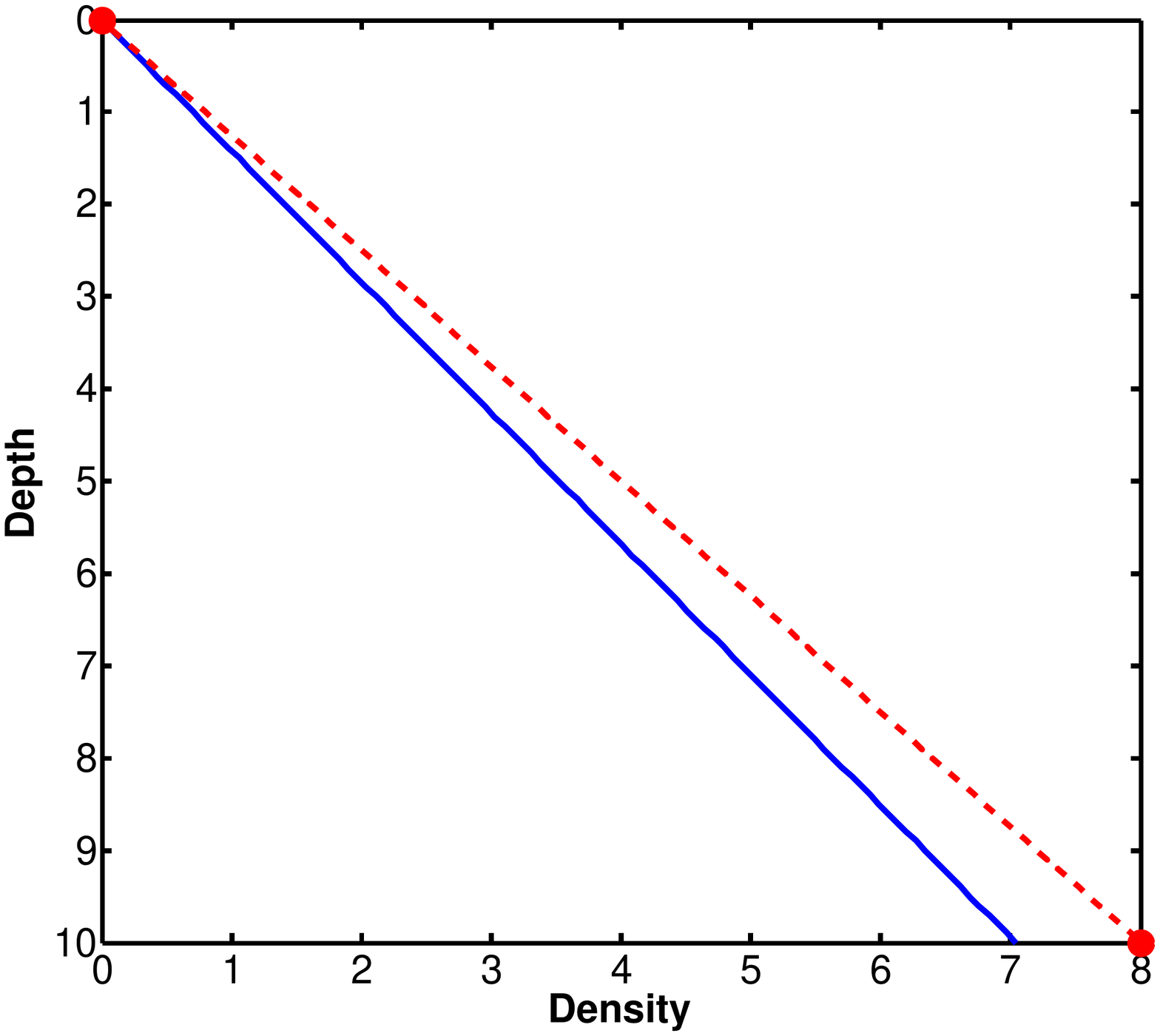}
    \caption{\footnotesize{Comparing between the convergences of our series solutions at different time steps. The dashed (red) curve is the sine solution from separation of variables (eq.$\ref{SineSeries}$) and the solid (blue) curve is using an expansion of complementary error functions (eq.$\ref{ErrorSeries}$). The two red dots mark our fixed boundary conditions. The initial condition is an ocean free of CO$_2$ (zero density). $t=10^{-4}t_{sc}$ in the upper left corner, $t=10^{-2}t_{sc}$ in the upper right corner, $t=t_{sc}$ in the lower left corner and $t=10^{2}t_{sc}$ in the lower right corner. We consider ten terms in each of the two series.}}
    \label{fig:SeriesCompare}
  \end{minipage}\\[1em]
\end{figure}

It is important to pinpoint possible fictitious fluxes due to a slow convergence of a series solution. Our system has a natural time scale of $t_{sc}=l^2/\kappa$. We compare in fig.\ref{fig:SeriesCompare} between the error function series (eq.$\ref{ErrorSeries}$) and the sine series (eq.$\ref{SineSeries}$) for various times. We see that for $t<<t_{sc}$ the sine series fluctuates around the boundaries which would cause erroneous fluxes if adopted. However, for $t>>t_{sc}$ the error function series requires more terms than the sine series to adequately represent the steady state solution. One should therefore switch between the series at $t\sim t_{sc}$.  

\bibliographystyle{apj}
\bibliography{amitmemo} 

\end{document}